\newif\iffigs\figstrue
\documentclass[a4paper,11pt]{article}
\usepackage[all]{xypic}
\usepackage{latexsym,amssymb,amsmath,lscape,graphics,setspace}
\usepackage{graphicx,tikz}        
\usepackage{longtable}
\usepackage{youngtab}
\usepackage{multirow}
\usepackage{color}
\usepackage{mathptmx}
\usepackage{slashed,epsfig}
\usepackage{amsfonts}
\usepackage{cite}

\newcommand{\mathsym}[1]{{}}
\newcommand{\unicode}[1]{{}}
\newtheorem{definizione}{Definition}[section]
\newtheorem{teorema}{Theorem}[section]

\newtheorem{statement}{Statement}[section]

\newcommand{\bd}{\begin{definizione}}
\newcommand{\ed}{\end{definizione}}

\def\IC{\relax\,\hbox{$\inbar\kern-.3em{\rm C}$}}
\def\IG{\relax\,\hbox{$\inbar\kern-.3em{\rm G}$}}
\def\IB{\relax{\rm I\kern-.18em B}}
\def\ID{\relax{\rm I\kern-.18em D}}
\def\IL{\relax{\rm I\kern-.18em L}}
\def\IF{\relax{\rm I\kern-.18em F}}
\def\IH{\relax{\rm I\kern-.18em H}}
\def\II{\relax{\rm I\kern-.17em I}}
\def\IN{\relax{\rm I\kern-.18em N}}
\def\IP{\relax{\rm I\kern-.18em P}}
\def\IQ{\relax\,\hbox{$\inbar\kern-.3em{\rm Q}$}}
\def\bfzero{\relax\,\hbox{$\inbar\kern-.3em{\rm 0}$}}
\def\IK{\relax{\rm I\kern-.18em K}}
\def\IG{\relax\,\hbox{$\inbar\kern-.3em{\rm G}$}}
 \font\cmss=cmss10 \font\cmsss=cmss10 at 7pt
\def\IR{\relax{\rm I\kern-.18em R}}
\def\ZZ{\relax\ifmmode\mathchoice
{\hbox{\cmss Z\kern-.4em Z}}{\hbox{\cmss Z\kern-.4em Z}} {\lower.9pt\hbox{\cmsss Z\kern-.4em Z}}
{\lower1.2pt\hbox{\cmsss Z\kern-.4em Z}}\else{\cmss Z\kern-.4em Z}\fi}
\def\bfone{\relax{\rm 1\kern-.35em 1}}

\def\diag{{\rm diag}}

\def\n010{N^{0,1,0}}
\def\inbar{\vrule height1.5ex width.4pt depth0pt}
\def\bfzero{\relax{\rm I\kern-.18em 0}}
\def\bfone{\relax{\rm 1\kern-.35em 1}}
\def\twomat#1#2#3#4{\left(\begin{array}{cc}
 {#1}&{#2}\nonumber \\ {#3}&{#4}\nonumber \\
\end{array}
\right)}

\def\o#1#2{{{#1}\over{#2}}}
\DeclareFontFamily{U}{rsf}{} \DeclareFontShape{U}{rsf}{m}{n}{
  <5> <6> rsfs5 <7> <8> <9> rsfs7 <10-> rsfs10}{}
\DeclareMathAlphabet\Scr{U}{rsf}{m}{n}

\newcommand{\ft}[2]{{\textstyle\frac{#1}{#2}}}
\def\tilde{\widetilde}

\def\1bar{1\hskip -.275cm -}
\def\2bar{2\hskip -.275cm -}
\def\3bar{3\hskip -.275cm -}

\newsavebox{\uuunit}
\sbox{\uuunit}
                 {\setlength{\unitlength}{0.825em}
                      \begin{picture}(0.6,0.7)
                                      \thinlines
                                      \put(0,0){\line(1,0){0.5}}
                                      \put(0.15,0){\line(0,1){0.7}}
                                      \put(0.35,0){\line(0,1){0.8}}
                                     \multiput(0.3,0.8)(-0.04,-0.02){10}{\rule{0.5pt}{0.5pt}}
                      \end {picture}}

\makeatletter \@addtoreset{equation}{section} \makeatother

\def\bfone{\relax{\rm 1\kern-.35em 1}}

\def\bfone{\relax{\rm 1\kern-.35em 1}}
\font\cmss=cmss10 \font\cmsss=cmss10 at 7pt

\newcommand{\su}{\mathfrak{su}}

\newcommand{\uu}{\mathfrak{u}}

\def\bfone{\relax{\rm 1\kern-.35em 1}}
\def\inbar{\vrule height1.5ex width.4pt depth0pt}
\def\IC{\relax\,\hbox{$\inbar\kern-.3em{\rm C}$}}
\def\ID{\relax{\rm I\kern-.18em D}}
\def\IF{\relax{\rm I\kern-.18em F}}
\def\IH{\relax{\rm I\kern-.18em H}}
\def\II{\relax{\rm I\kern-.17em I}}
\def\IN{\relax{\rm I\kern-.18em N}}
\def\IP{\relax{\rm I\kern-.18em P}}
\def\IQ{\relax\,\hbox{$\inbar\kern-.3em{\rm Q}$}}
\def\IR{\relax{\rm I\kern-.18em R}}
\font\cmss=cmss10 \font\cmsss=cmss10 at 7pt
\def\ZZ{\relax\ifmmode\mathchoice
{\hbox{\cmss Z\kern-.4em Z}}{\hbox{\cmss Z\kern-.4em Z}} {\lower.9pt\hbox{\cmsss Z\kern-.4em Z}}
{\lower1.2pt\hbox{\cmsss Z\kern-.4em Z}}\else{\cmss Z\kern-.4em Z}\fi}

\def\ni{\noindent}
\def\tilde{\widetilde}
\def\bar{\overline}

\def\hat{\widehat}

\def\Coe#1.#2.{{#1\over #2}}

\def\coe#1.#2.{\relax{\textstyle {#1 \over #2}}\displaystyle}

\def\to{\rightarrow}
\def\notin{\hbox{{$\in$}\kern-.51em\hbox{/}}}

\def\IE{\relax{{\rm I\kern-.18em E}}}

\def\IGam{\relax{{\rm I}\kern-.18em \Gamma}}

\def\IA{\relax{\hbox{{\rm A}\kern-.82em {\rm A}}}}

\def\diag{{\rm diag}}

\begin{document}
\begin{titlepage}
\begin{center}
\begin{flushright}
ARC-17-6
\end{flushright}
\vskip 0.2cm
\vskip 0.2cm {\Large The K\"ahler Quotient Resolution of $\mathbb{C}^3/\Gamma$ singularities, \\[5pt]
the McKay correspondence  and  D=3 $\mathcal{N}=2$ Chern-Simons gauge theories }\\[2cm]
{Ugo~Bruzzo${}^{\; a,e}$, Anna Fino${}^{\; b,e}$ and  Pietro~Fr\'e${}^{\; c,d,e}$, \\[10pt]
{${}^a$\sl\small SISSA, Scuola Internazionale Superiore di Studi Avanzati,    \\
via Bonomea 265, 34136 Trieste \ Italy \\ INFN -- Sezione di Trieste\\}
\emph{e-mail:} \quad {\small {\tt ugo.bruzzo@sissa.it}}\\
\vspace{5pt} {${}^b$\sl \small Dipartimento di Matematica G. Peano, Universit\'a di Torino\\ Via Carlo Alberto 10,
10123 Torino \ Italy\\}
\emph{e-mail:} \quad {\small {\tt annamaria.fino@unito.it}}\\
\vspace{5pt}
{${}^c$\sl\small Dipartimento di Fisica, Universit\'a di Torino\\INFN -- Sezione di Torino \\
via P. Giuria 1, \ 10125 Torino \ Italy\\}
\emph{e-mail:} \quad {\small {\tt pfre@unito.it}}\\
\vspace{5pt}
{{\em $^{d}$\sl\small  National Research Nuclear University MEPhI\\ (Moscow Engineering Physics Institute),}}\\
{\em Kashirskoye shosse 31, 115409 Moscow, Russia}~\quad\\
\quad \\[-4pt]
 \centerline{$^{(e)}$ \it Arnold-Regge Center,} \centerline{\it via P. Giuria 1,  10125 Torino,
Italy}}
\vspace{15pt}
\begin{abstract}
We advocate that the generalized Kronheimer
construction of the K\"ahler quotient crepant resolution
$\mathcal{M}_\zeta \longrightarrow \mathbb{C}^3/\Gamma$ of an
orbifold singularity where $\Gamma\subset \mathrm{SU(3)}$ is a
finite subgroup naturally defines the field content and the
interaction structure of a superconformal Chern-Simons Gauge Theory.
This latter is supposedly the dual of an M2-brane solution of $D=11$
supergravity with $\mathbb{C}\times\mathcal{M}_\zeta$ as transverse
space. We illustrate and discuss many aspects of this type of
constructions emphasizing that the equation $\pmb{p}\wedge\pmb{p}=0$
which provides the K\"ahler  analogue of the holomorphic sector in the
hyperK\"ahler moment map equations canonically  defines the
structure of a universal superpotential in the CS theory.
Furthermore the kernel $\mathcal{D}_\Gamma$ of the above equation
can be described as the orbit with respect to a quiver Lie group
$\mathcal{G}_\Gamma$ of a special locus $L_\Gamma \subset
\mathrm{Hom}_\Gamma(\mathcal{Q}\otimes R,R)$ that has also a
universal definition. We provide an extensive discussion of the
relation between the coset manifold
$\mathcal{G}_\Gamma/\mathcal{F}_\Gamma$, the gauge group
$\mathcal{F}_\Gamma$ being the maximal compact subgroup of the
quiver group, the moment map equations and the first Chern
classes of the so named tautological vector bundles that are in
one-to-one correspondence with the nontrivial irreps of $\Gamma$.
These first Chern classes are represented by (1,1)-forms on $\mathcal{M}_\zeta$ and
provide a basis for the cohomology group $H^2(\mathcal{M}_\zeta)$.
We also discuss the relation with conjugacy classes of $\Gamma$ and
we provide the explicit construction of several examples emphasizing
the role of a generalized McKay correspondence. The case of the ALE
manifold resolution of $\mathbb{C}^2/\Gamma$ singularities is
utilized as a comparison term and new formulae related with the
complex presentation of Gibbons-Hawking metrics are exhibited.
\end{abstract}
\end{center}
\end{titlepage}
\tableofcontents \noindent {}
\newpage
\section{Introduction}
\label{introibo}
The issue of  of $D=3$ $\mathcal{N}=2$
Chern-Simons gauge theories was reconsidered by one of us (P.F.) and P.A. Grassi from two complementary viewpoints \cite{pappo1}:
\begin{description}
  \item[a)] The constructive viewpoint of their formulation in terms of Integral Forms in superspace
  that  brings together, within a unified (super)cohomological scheme, all the existing formulation of supersymmetric field
  theories, namely the component approach, the
  rheonomic approach  and the superspace approach.
  \item[b)] The field content and the interaction structure of the theories that are supposed to
  represent gauge duals of  M2-brane solutions of $D=11$ supergravity having a $\mathbb{C}^3/\Gamma$ quotient singularity in
  their transverse space. Such a situation, as we recall in the present paper, arises by considering
  the classical case of an $\mathrm{AdS_4}\times \mathbb{S}^7$ solution where $\mathbb{S}^7$,
  viewed as a Hopf-fibration over $\mathbb{P}^3$, is
  modded by a discrete subgroup of $\Gamma \subset \mathrm{SU(3)}\subset \mathrm{SU(4)}$.
\end{description}
 The physical motivations and the perspectives of the study encompassed in point b) have been
extensively discussed in \cite{pappo1}. Here we are more concerned with the mathematical aspects of these
theories and we aim at proving the following:
\begin{statement}
\label{turiddu} There is a one-to-one map between the field-content and the interaction structure of a $D=3$,
$\mathcal{N}=2$ Chern-Simons gauge theory and the generalized Kronheimer algorithm of solving quotient
singularities $\mathbb{C}^3/\Gamma$ via a K\"ahler quotient based on the McKay correspondence. All items on
both sides of the one-to-one correspondence are completely determined by the structure of the finite group
$\Gamma$ and by its specific embedding into $\mathrm{SU(3)}$.
\end{statement}
An ultra short summary of the results that we are going to present
is the following. From the field--theoretic side the essential items
defining the theory are:
\begin{enumerate}
  \item The K\"ahler manifold $\mathcal{S}$ spanned by the
  Wess-Zumino multiplets. This is the $3|\Gamma|$ dimensional
  manifold $\mathcal{S}_\Gamma \, = \,
  \mathrm{Hom}_\Gamma\left(\mathcal{Q}\otimes R,R\right)$ where $\mathcal{Q}$
  is the representation of
  $\Gamma$ inside $\mathrm{SU(3)}$ and $R$ denotes the regular representation
  of the discrete group.
  \item The gauge group $\mathcal{F}_{\Gamma}$. This latter is identified
  as the maximal compact subgroup $\mathcal{F}_{\Gamma}$ of
  the complex quiver group $\mathcal{G}_\Gamma$ of complex
  dimension $|\Gamma|-1$, to be discussed later.
  Since the action of $\mathcal{F}_{\Gamma}$ on $\mathcal{S}_\Gamma$
  is defined by construction, the gauge
  interactions of the Wess-Zumino multiplets are also fixed and the associated moment maps are equally
  uniquely determined.
  \item The Fayet-Iliopoulos parameters. These are in a one-to-one association with the $\zeta$ levels of the
  moment maps corresponding  to the center of the gauge Lie algebra
  $\mathfrak{z}\left[\mathbb{F}_{\Gamma}\right]$\footnote{In this paper
  we follow the convention that the names of the Lie groups are denoted with
  calligraphic letters $\mathcal{F},\mathcal{G},\mathcal{U}$, the corresponding Lie
  algebras being denoted by mathbb letters $\mathbb{F},\mathbb{G},\mathbb{U}$.}. The
  dimension of this center is $r$ which is the number of nontrivial conjugacy classes of the discrete group
  $\Gamma$ and also of its nontrivial irreps.
  \item The superpotential $\mathcal{W}_\Gamma$. This latter is a cubic function uniquely associated with a
  quadratic constraint $[A,B]=[B,C]=[C,A]=0$ which characterizes the generalized Kronheimer construction,
  defines a K\"ahler subvariety $\mathbb{V}_{|\Gamma|+2}\subset \mathcal{S}_\Gamma$ and admits a universal
  group theoretical description in terms of  the quiver group $\mathcal{G}_\Gamma$.
  \item In presence of all the above items the manifold of vacua of the gauge theory, namely of extrema at
  zero of its scalar potential, is just the minimal crepant resolution of the singularity $\mathcal{M}_\Gamma
  \rightarrow \frac{\mathbb{C}^3}{\Gamma}$, obtained as K\"ahler quotient of $\mathbb{V}_{|\Gamma|+2}$ with respect to
  the gauge group $\mathcal{F}_{\Gamma}$.
  \item The Dolbeault cohomology of the space of vacua $\mathcal{M}_\Gamma$ is predicted by the finite group
  $\Gamma$ structure in terms of a grading of its conjugacy classes named \textit{age grading}.
  The construction of the homology cycles and exceptional divisors that are Poincar\'e dual to such
  cohomology classes is the most exciting issue in the present list of \textit{geometry}
  $\Longleftrightarrow$ \textit{field theory} correspondences, liable to give rise to many
  interesting physical applications. In this context a central item is the notion of tautological
  bundles associated with the
  nontrivial irreps of $\Gamma$ that we discuss at length in the sequel \cite{degeratu}.
\end{enumerate}
\section{The general form of $\mathcal{N}=2$ Chern-Simons Gauge Theories}
\label{genN2theo} To discuss the one-to-one map  in
statement \ref{turiddu}, we begin by recalling, from the results of
\cite{pappo1}, the general structure of the component lagrangian for
an $\mathcal{N}=2$, $D=3$ Chern-Simons gauge theory. The results of
\cite{pappo1} correspond to a generalization and more geometrical
transcription of the general form of   coupling of matter with D=3  gauge
theories  constructed in 1999 in a series of three papers
\cite{Fabbri:1999ay,ringoni,Fre:1999gok}, using auxiliary fields and
the rheonomic approach. In 1999, the motivation to consider this
type of theories was the reinterpretation
\cite{Fre:1999gok,ringoni,Fabbri:1999mk,Fabbri:1999ay,Fabbri:1999hw,Billo:2000zs},
within the AdS/CFT scheme of the Kaluza-Klein spectra of
supergravity localized on backgrounds $\mathrm{AdS_4
}\times\mathcal{M}_7$ that were calculated in the years 1982-1985
\cite{freurub,round7a,osp48,squash7a,biran,gunawar,kkwitten,noi321,spec321,multanna,
pagepopeM,dafrepvn,pagepopeQ,univer,bosmass,frenico,castromwar}. In
all those cases the manifold $\mathcal{M}_7$ was smooth, actually a
Sasakian coset manifold. As we better explain, in this paper we
are interested to apply the same ideas to the case where the metric
cone ove $\mathcal{M}_7$ is an orbifold $\mathbb{C}\times
\mathbb{C}^3/\Gamma$, denoting by $\Gamma$ a finite subgroup of
$\mathrm{SU(3)}$.
\par The  lagrangian of $\mathcal{N}=2$ Chern-Simons Gauge Theory, as systematized in \cite{pappo1}, takes the following form:
\begin{eqnarray}
\mathcal{L}_{CSoff}&=& - \, \alpha \, \mbox{Tr} \,\left(\mathfrak{F} \, \wedge \, \mathcal{A} \, + \,
\frac{2}{3}\, \mathcal{A}\wedge \, \mathcal{A}\wedge \, \mathcal{A}\right)\,+\,\left( \frac{1}{2} \,
g_{ij^\star} \, \Pi^{m\mid i} \, \nabla \bar{z}^{j^\star} \, + \,  \bar{\Pi}^{m\mid j^\star} \, \nabla
z^{i}\right)\, \wedge \, e^n \, \wedge \,  e^p \, \epsilon_{mnp}\nonumber\\ \null &&  - \, \frac{1}{6} \, \,
g_{ij^\star} \, \Pi^{m\mid i} \, \bar{\Pi}^{m\mid j^\star} \, e^r \,  \wedge \, e^s \, \wedge \,  e^t \,
\epsilon_{rst}\nonumber\\ && +\,{\rm i}\frac{1}{2} \, g_{ij^\star} \, \left(\bar{\chi}^{j^\star} \, \gamma^m
\,\nabla \chi^{i} \, + \,  \bar{\chi}^{i}_c \, \, \gamma^m \, \nabla \chi^{i^\star}_c\right) \, \wedge \, e^n
\, \wedge \,  e^p \, \epsilon_{mnp} \nonumber\\ && \left( \, - \, \frac{1}{3} \, M^\Lambda \,\left(
\partial_i k^j_\Lambda \, g_{j\ell^\star} \, \bar{\chi}^{\ell^\star} \,\chi^i \,+ \,  \partial_{i^\star}
k^{j^\star}_\Lambda \, g_{j\ell^\star} \bar{\chi}^\ell_c \, \chi^{i^\star}_c \right) \, + \, \frac{\alpha}{3}
\, \left( \bar{\lambda}^\Lambda \, \lambda^\Sigma \, + \, \bar{\lambda}^\Lambda_c \, \lambda^\Sigma_c\right)
\, \mathbf{\kappa}_{\Lambda\Sigma}\right.\nonumber\\ &&\left. \, +\, {\rm i}\, \frac{1}{3} \, \left(
\bar{\chi}^{j^\star}_c \, \lambda^\Lambda \, k^i_\Lambda \, - \, \bar{\chi}^{i}_c \, \lambda^\Lambda \,
k^{j^\star}_\Lambda\right) \, g_{ij^\star} \, \right.\nonumber\\ &&\left. + \, \frac{1}{6} \, \left(
\partial_i\partial_j \mathcal{W}\, \bar{\chi}^i_c \, \chi^j \, + \, \partial_{i^\star}\partial_{j^\star}
{\overline{\mathcal{W}}}\, \bar{\chi}^{i^\star} \, \chi^{j^\star}_c\right) \right)\, \wedge \, e^n \, \wedge
\,  e^p \, \epsilon_{mnp} \nonumber\\ &&-V\left(M,D,\mathcal{H},z,\bar{z}\right)  \, \epsilon_{mnp} \, e^m \,
\wedge \, e^n \, \wedge \, e^p \label{pastiglialeone}
\end{eqnarray}
where:
\begin{enumerate}
  \item The complex scalar fields $z^i$ span a K\"ahler manifold $\mathcal{S}$, $g_{ij^\star}$ denoting its
  K\"ahler metric.
  \item $\Pi^{m|i}$ are auxiliary fields that are identified with the world volume derivatives of the scalar
  $z^i$ by their own equation of motion.
  \item The one--forms $e^m$ denote the dreiben of the world volume.
  \item $\mathcal{A}^\Lambda$ is the gauge-one form of the gauge group
  $\mathcal{F}_{\Gamma}$.
  \item $\lambda^\Lambda$ are the gauginos, namely the spin $\ft 12$
  partners of the gauge bosons $\mathcal{A}^\Lambda$
  \item $\chi^i$ are the chiralinos, namely the spin $\ft 12$ partners
  of the Wess-Zumino scalars $z^i$.
  \item $M^\Lambda$ are the real scalar fields in the adjoint of
  the gauge group that complete the gauge
  multiplet together with the gauginos and the gauge bosons.
  \item $\mathcal{W}(z)$ is the superpotential.
  \item $k^i_\Lambda$ are the Killing vectors of the K\"ahler metric of
  $S$, associated with the generators
  of the gauge group.
  \item $\kappa_{\Lambda\Sigma}$, made of constants denotes the
  Cartan Killing metric on the Lie algebra $\mathbb{F}_\Gamma$ of the gauge
  group $\mathcal{F}_\Gamma$
\end{enumerate}
The scalar potential in terms of physical and auxiliary fields is the following one:
\begin{eqnarray} V\left(M,D,\mathcal{H},z,\bar{z}\right)&=&
\left(\frac{\alpha}{3} \,  M^\Lambda \,\mathbf{\kappa}_{\Lambda\Sigma}\,
 - \, \frac{1}{6}\,
\mathcal{P}_\Sigma(z,\bar{z})\, + \, \frac{1}{6} \, \zeta_{I} \,
\mathfrak{C}_\Sigma^I \right) \, D^\Sigma \,
+ \, \frac{1}{6} M^\Lambda \, M^\Sigma \, k^i_\Lambda \, k_\Sigma^{j^\star}
\, g_{ij^\star} \nonumber\\ &&
+\,\frac{1}{6} \left(\mathcal{H}^i \, \partial_i \mathcal{W} \, + \,
\mathcal{H}^{\ell^\star} \,
\partial_{\ell^\star} {\overline{\mathcal{W}}} \right) \, - \,
\frac{1}{6} \, g_{i\ell^\star} \,\mathcal{H}^i \,
\mathcal{H}^{\ell^\star} \label{karamella}
\end{eqnarray}
where $\mathcal{P}_\Sigma(z,\bar{z})$ are the moment maps associated
with each generator of the gauge-group, $\zeta_{I}$ are the
Fayet-Iliopoulos parameters associated with each generator of the
center of the gauge Lie algebra
$\mathfrak{z}\left(\mathbb{F}_\Gamma\right)$, $\mathcal{H}^i$ are
the complex auxiliary fields of the Wess-Zumino multiplets and
$D^\Lambda$ are the auxiliary scalars of the vector multiplets. By
$\mathfrak{C}^I_\Sigma$ we denote the projector onto a basis of
generators of the Lie Algebra center
$\mathfrak{z}\left[\mathbb{F}_{\Gamma}\right]$.
\par
 In these theories the gauge multiplet does not propagate and it is essentially made of lagrangian
multipliers for certain constraints. Indeed  the auxiliary fields,
the gauginos and the vector multiplet scalars have algebraic field
equations so that they can be eliminated by  solving such equations
of motion. The vector multiplet auxiliary scalars $D^\Lambda$ appear
only as lagrangian multipliers of the constraint\footnote{As it is
customary for all metrics $\mathbf{\kappa}^{\Lambda \Sigma}$ with
upper indices denotes the inverse of the Cartan Killing metric
$\mathbf{\kappa}_{\Delta \Pi}$ with lower indices.}:
\begin{equation}\label{Msolvo} M^\Lambda \, = \,
\frac{1}{2\alpha} \, {\mathbf{\kappa}}^{\Lambda \Sigma}\,\left(   \mathcal{P}_\Sigma \, -\, \zeta_I \,
\mathfrak{C}_\Sigma^I \right)
\end{equation}
while the variation of the auxiliary fields
$\mathcal{H}^{j^\star}$ of the Wess Zumino multiplets yields:
\begin{equation}\label{eliminoH}
\mathcal{H}^{i} \, = \, g^{ij^\star} \, \partial_{j^\star} \, \overline{W} \quad ; \quad
\overline{\mathcal{H}}^{j^\star} \, = \, g^{ij^\star} \, \partial_{i} \, {W}
\end{equation}
On the other
hand, the equation of motion of the field $M^\Lambda$ implies:
\begin{equation}\label{gospadi} D^\Lambda \, =
\, - \, \frac{1}{\alpha} \, \mathbf{\kappa}^{\Lambda \Gamma} g_{ij^\star} \, k_\Gamma^{i} \,
k^{j^\star}_\Sigma\, M^\Sigma \, = \, - \, \frac{1}{2\,\alpha^2} \,g_{ij^\star}\, \mathbf{\kappa}^{\Lambda
\Gamma}\, k_\Gamma^{i} \, k^{j^\star}_\Sigma\,\mathbf{\kappa}^{\Sigma \Delta}\,\left(   \mathcal{P}_\Delta \,
-\, \zeta_I \,  \mathfrak{C}_\Delta^I \right)
\end{equation}
which finally resolves all the auxiliary
fields in terms of functions of the physical scalars.
\par Upon use of both constraints (\ref{Msolvo}) and
(\ref{eliminoH}) the scalar potential takes the following positive definite form:
\begin{eqnarray}
V(z,\bar{z}) & = & \frac{1}{6} \, \left( \partial_i \mathcal{W} \, \partial_{j^\star} \overline{\mathcal{W}}
\,g^{ij^\star} \, + \, \mathbf{m}^{\Lambda\Sigma} \, \left(\mathcal{P}_\Lambda \, - \, \zeta_I \,
\mathfrak{C}_\Lambda^I\right) \, \left(\mathcal{P}_\Sigma \, - \, \zeta_J \, \mathfrak{C}_\Sigma^J \right)
\right)\nonumber\\ \mathbf{m}^{\Lambda\Sigma}(z,\bar{z}) & \equiv &
\frac{1}{4\alpha^2}\,\mathbf{\kappa}^{\Lambda\Gamma} \, \mathbf{\kappa}^{\Sigma\Delta} \, k_\Gamma^i \,
k_\Delta^{j^\star} \, g_{ij^\star} \label{quadraPot}
\end{eqnarray}
In a similar way the gauginos can be
resolved in terms of the chiralinos:
\begin{equation}\label{eliminogaugino} \lambda^\Lambda \, = \, -\,
\frac{1}{2\alpha} \, \mathbf{\kappa}^{\Lambda\Sigma} \, g_{ij^\star} \chi^i \, k^{j^\star}_\Sigma \quad ;
\quad \lambda^\Lambda_c \, = \, -\, \frac{1}{2\alpha} \, \mathbf{\kappa}^{\Lambda\Sigma} \, g_{ij^\star}
\chi^{j^\star} \, k^{i}_\Sigma
\end{equation} In this way if we were able to eliminate also the gauge one
form $\mathcal{A}$ the Chern-Simons gauge theory would reduce to a theory of Wess-Zumino multiplets with
additional interactions. The elimination of $\mathcal{A}$, however,  is not possible in the nonabelian case
and it is possible in the abelian case only through duality nonlocal transformations.  This is the corner
where interesting nonperturbative dynamics is hidden.
\subsection{A special class of ${\cal N}=2$ Chern-Simons gauge theory in three dimensions}\label{specN2gaugetheory} \label{specN2theo} In the realization of
the one-to-one map advocated in statement \ref{turiddu},
we are interested in theories where the Wess-Zumino
multiplets are identified with the non-vanishing entries of a
triple of matrices, named $A,B,C,$ and the
superpotential takes the following form:
\begin{eqnarray}\label{signorello}
    \mathcal{W}& = &\mathrm{const} \times\, \mbox{Tr}\left(A\left[B,C\right]\right)
    +\mbox{Tr}\left(B\left[C,A\right]\right)+\mbox{Tr}\left(C\left[A,B\right]\right)\nonumber\\
    &=&3 \, \mathrm{const} \times \, \left(\mbox{Tr}\left(A \, B \, C \right)-\mbox{Tr}\left(A \, C \, B
    \right)\right)
\end{eqnarray}
Because of the positive-definiteness of the K\"ahler metric $g^{ij^\star}$  and of the Killing metric
$\mathbf{m}^{\Lambda\Sigma}$ the zero of the potential, namely the vacua, are characterized by the two
conditions:
 \begin{eqnarray}
   \partial_i\mathcal{W} &=& 0 \quad \Rightarrow [A,B]=[B,C]=[C,A]=0 \label{holomorfocost} \\
   \mathcal{P}_\Lambda & = & \zeta_I \,
\mathfrak{C}_\Lambda^I
 \end{eqnarray}
which will have a distinctive interpretation in the K\"ahler quotient construction \`a la Kronheimer. Notice
that $\mathfrak{C}_\Lambda^I$ denotes the projector of the gauge Lie algebra onto its center, as we already
said.
\section{On superconformal Chern-Simons theories dual to M2-branes} In this short section we collect some issues and hints
relative to the construction of  superconformal gauge theories dual
to  orbifolds of the M2-brane transverse space with respect to
$\Gamma \subset \mathrm{SU(3)} \subset \mathrm{SU(4)}$. The most
important conclusion is that, as long as we require the existence of
a complex structure of $\mathbb{R}^8$ compatible with
$\mathrm{L_{168}}$\footnote{Following the notations of
\cite{miol168} by $\mathrm{L_{168}}$ we denote the simple group of
order 168 that is isomorphic with $\mathrm{PSL(2,7)}$ and which is
also the largest non abelian non solvable finite subgroup of
$\mathrm{SU(3)}$, according with the classification of
\cite{blicfeltus,blicfeltus2}.}, we reduce the singularity to:
\begin{equation}
\frac{\mathbb{C}^4}{\Gamma} \, \rightarrow \, \mathbb{C}\times \frac{\mathbb{C}^3}{\Gamma}\quad ; \quad
\Gamma \subset \mathrm{SU(3)}\label{galisco}
\end{equation}
From the point of view of supergravity and string theory the factorization of a $\mathbb{C}$ factor is
welcome. It provides the means to reduce $\mathrm{M2}$-branes to $\mathrm{D2}$-branes in $D=10$ type IIA
theory.
\par
As it is well known $\mathrm{SO(8)}$-gauged supergravity is obtained from d=11 supergravity compactified on:
\begin{equation}\label{sferasette}
 \mathrm{AdS_4} \times \mathbb{S}^7
\end{equation}
which is the near horizon geometry of an M2-brane with $\mathbb{R}^8$ transverse space. Indeed $\mathbb{R}^8$
is the metric cone on $\mathbb{S}^7$. The entire Kaluza-Klein spectrum which constitutes the spectrum of BPS
operators of the d=3 theory is organized in short representations of the supergroup:
\begin{equation}\label{ortosymplo}
 \mathrm{ Osp(8|4)}
\end{equation}
Our discussion leads to the conclusion that we can consider the compactification of supergravity on orbifolds
of the following type:
\begin{equation}\label{orbildus}
  \mathbb{C_{\mathrm{\Gamma}}S}^{7} \, = \, \frac{\mathbb{S}^7}{\Gamma}  \, \quad ; \quad \Gamma \subset
  \mathrm{SU(3)}
  \subset  \mathrm{SO(8)}
\end{equation}
The corresponding M2-brane solution has the orbifold:
\begin{equation}\label{rorbildo}
   \mathbb{C_{\mathrm{\Gamma}}R}^{8} \, = \, \frac{\mathbb{R}^8}{\Gamma}
\end{equation}
as transverse space.
\par
The massive and massless modes of the Kalauza Klein spectrum are easily worked out from the
$\mathrm{Osp(8|4)}$ spectrum of the 7-sphere. Indeed since the group $\Gamma$ is embedded by the above
construction into $\mathrm{SO(8)} \subset \mathrm{Osp(8|4)}$, it suffices to cut the spectrum to the $\Gamma$
singlets.
\subsection{Quotient singularities and CS gauge groups} What we said above can be summarized by saying that
we construct models where M2-branes are probing the singularity (\ref{rorbildo}) and we might be interested
in the smooth manifold obtained by blowing up the latter.
\par
From another viewpoint, according to \cite{Aharony:2008ug} the superconformal Chern-Simons theory with the
gauge  groups $\mathrm{SU(N)} \times \mathrm{SU(N)}$ at level $k$ is dual to supergravity on:
\begin{equation}\label{orbifoldponto}
  \mathrm{AdS_4} \times \frac{\mathbb{S}^7}{\mathbb{Z}_k}
\end{equation}
namely the association is at the orbifold point between the gauge-group structure and the $\Gamma$ discrete
group, so that:
\begin{equation}\label{assosiazia}
  \mathrm{SU(N)}\mid_{k\mbox{ level}} \times \mathrm{SU(N)}\mid_{k\mbox{ level}} \,
  \Leftrightarrow \, \mathbb{Z}_k \end{equation} Despite the difficulties in working out the blowup,
 it is therefore legitimate to ask the question: what is the CS gauge group corresponding
 to $\Gamma \, \subset \, \mathrm{SU(3)}$:
 \begin{equation}\label{rottamillo}
   \mbox{gauge CS ?} \, \, \Leftrightarrow \, \Gamma \,\subset
   \mathrm{SU(3)}
 \end{equation}
As it is implicit in the comparison between eq.\,(\ref{rottamillo}) and eq.\,(\ref{assosiazia}),   the statement
contained in \cite{Aharony:2008ug} is incomplete. It is not sufficient to say
$\frac{\mathbb{S}^7}{\mathbb{Z}_k}$. In order to derive the dual superconformal field theory it is essential
to specify the embedding of $\mathbb{Z}_k$ into the isometry group $\mathrm{SO(8)}$ of the seven sphere.
Different embeddings of the same discrete group can lead to different CS gauge theories.
\subsection{Some suggestions from ALE manifolds} Some useful suggestions on this conceptual link
can arise by comparing  with
the case of well known singularities like $\mathbb{C}^2/\Gamma$, the discrete group $\Gamma$ being one of the
finite subgroups of $\mathrm{SU(2)}$ falling into the ADE classification.  In that case the blowup of the
singularities can be done by means of a hyperK\"ahler quotient according to the Kronheimer construction
\cite{kro1,kro2}. The gauge group is essentially to be identified with the nonabelian extension
$\mathrm{U(1)} \to \mathrm{U(N)}$ of the group $\mathcal{F}$ one utilizes in the hyperK\"ahler quotient. The
group $\mathcal{F}_\Gamma$ is a product of $\mathrm{U(1)}$'s as long as the discrete group $\Gamma$ is  the cyclic group
$\mathbb{Z}_k$, yet it becomes nonabelian with factors $\mathrm{SU(k)}$ when $\Gamma$ is nonabelian. As far
as we know no one has constructed CS gauge theories corresponding to M2-branes that probe singularities of
the type $\mathbb{C}^2\times \mathbb{C}^2/\Gamma$ with  a  nonabelian $\Gamma$. This case might be a ground-zero
case to investigate.
\subsection{Moduli of the blowup and superconformal operators: inspirations from
geometry} It must be stressed that in the spectrum of the conformal field theory obtained at the orbifold
point (which corresponds to the Kaluza Klein spectrum in the case of smooth manifolds) there must be those
associated with the moduli of the blowup. By similarity with the case of superstrings at orbifold points, we
expect that these are \textit{twisted states}, namely states not visible in the Kaluza Klein spectrum on the
orbifold. Yet when the orbifold is substituted with its smooth counterpart, obtained  blowing of up  the
singularity, these states should appear as normal states in the supergravity Kaluza Klein spectrum.
\subsection{Temporary conclusion}
The above discussion shows that the embedding of $\Gamma$ is fundamental. We can treat
$\mathrm{L_{168}}\equiv \mathrm{PSL(2,\mathbb{Z}_7})$ and other discrete group $\Gamma$ singularities when
$\Gamma$ has a holomorphic action on $\mathbb{C}^4$. This happens when
\begin{equation}\label{cavicchius}
  \Gamma \subset \mathrm{SU(3) }\subset \mathrm{SO(8)}
\end{equation}
In this case the singularity is just $\mathbb{C}\times \mathbb{C}^3/\Gamma$ and
$\mathbb{C}^3/\mathrm{L_{168}}$ is the   blowup described by Markushevich in  \cite{marcovaldo}. We mention it again
in a later section. The classification of discrete subgroups of $\mathrm{SL(3,\mathbb{C})}$ was achieved at
the dawn of the XXth century in \cite{blicfeltus,blicfeltus2}. The largest nonabelian nontrivial group
appearing in this classification is the unique simple group with 168 elements named $L_{168}$ (see
\cite{miol168} and\cite{futuro} for a thorough discussion). The other possibilities are provided by a finite
list of cyclic and solvable groups reviewed for instance in \cite{62}.
\subsection{Quotient singularities and M2-branes} We come now to the mathematics which is of greatest interest
to us in order to address the physical problem at stake, \textit{i.e.}, the construction of CS theories dual
to M2-branes that have the metric cone on orbifolds $\mathbb{S}^7/\Gamma$ as transverse space. The first step
is to show that such metric cone is just $\mathbb{C}^4/\Gamma$. This is a rather simple fact but it is of the
utmost relevance since it constitutes the very bridge between the mathematics of quotient singularities,
together with their resolutions, and the physics of CS theories. The pivot of this bridge is the complex Hopf
fibration of the $7$-sphere. The argument leading to the above conclusion was provided in the paper
\cite{pappo1} and we do not deem it necessary to repeat it here. We just jump to the conclusion there
reached. The space $\mathbb{C}^4 -\{0\}$ can be regarded as the total space of the canonical  $
\mathbb{C}^\star$-fibration over $\mathbb{C}\mathbb{P}^3$:
\begin{eqnarray}\label{lignotto}
    \pi & : & \mathbb{C}^4 - \{0\} \, \rightarrow \, \mathbb{C}\mathbb{P}^3\nonumber\\
    \forall y \in\mathbb{C}\mathbb{P}^3 & : & \pi^{-1}(y) \sim \mathbb{C}^\star
\end{eqnarray}
By restricting to the unit sphere in $ \mathbb{C}^4 $ we obtain
  the  Hopf fibration of the seven sphere:
\begin{eqnarray}\label{offetto7}
    \pi & : & \mathbb{S}^7 \, \rightarrow \, \mathbb{C}\mathbb{P}^3\nonumber\\
    \forall y \in\mathbb{C}\mathbb{P}^3 & : & \pi^{-1}(y) \sim \mathbb{S}^1
\end{eqnarray}

\par
The consequence of such a  discussion is that if we have a finite subgroup $\Gamma \subset \mathrm{SU(4)}$,
which obviously is an isometry of $\mathbb{CP}^3$, we can consider its action both on $\mathbb{CP}^3$ and on
the seven sphere so that  we have:
\begin{equation}\label{raccimolato}
    \mathrm{AdS_4} \times \frac{\mathbb{ S}^7}{\Gamma} \, \to \partial\mathrm{AdS_4} \times
    \frac{\mathbb{C}^4}{\Gamma}
\end{equation}
We are therefore interested in describing the theory of M2-branes probing the singularity
$\frac{\mathbb{C}^4}{\Gamma}$. Hence an important guiding line in addressing mathematical questions comes
from their final use  in connection with M2-brane solutions of $D=11$ supergravity and  with the construction
of quantum gauge theories dual to such M2-solutions of supergravity.
\par
Recalling the results of \cite{pappo1} we start from the following diagram
\begin{equation}\label{caluffo2}
  K_3 \, \stackrel{\pi}{\longleftarrow} \, \mathcal{M}_7 \,  \stackrel{Cone}{ \hookrightarrow} \,  K_4 \,
  \stackrel{\mathcal{A}}{ \hookrightarrow} \, \mathbb{V}_q
\end{equation}
where $\mathcal{M}_7$ is the compact manifold on which D=11
supergravity is compactified and $\mathbb{V}_q$ denotes some
appropriate algebraic variety of complex dimension $q$. It is
required that $\mathcal{M}_7$ should be a Sasakian manifold.
\par
What Sasakian means is visually summarized in the following table.
\begin{center}
\begin{tabular}{|ccccc|}
  \hline
 base  of the fibration &   projection & $7$-manifold & inclusion & metric cone  \\
  $\mathcal{B}_6$ & $\stackrel{\pi}{\longleftarrow}$ &$\mathcal{M}_7$ &$\hookrightarrow $& $\mathcal{C}\left( \mathcal{M}_7\right)$ \\
  $\Updownarrow$ & $\forall p \in \mathcal{B}_6 \quad \pi^{-1}(p) \, \sim \, \mathbb{S}^1 \,$
   &$\Updownarrow$ &\null & $\Updownarrow$ \\
  K\"ahler $K_3$ &$\null$ & Sasakian&\null &K\"ahler Ricci flat $K_4$ \\
  \hline
\end{tabular}
\end{center}
First of all the $\mathcal{M}_7$ manifold must admit an $\mathbb{S}^1$-fibration over a K\"ahler three-fold
$K_3$:
\begin{equation}\label{fibratoS1}
    \pi \quad : \quad \mathcal{M}_7 \stackrel{\mathbb{S}^1}{\longrightarrow} \, K_3
\end{equation}
Calling $z^i$ the three complex coordinates of $K_3$ and $\phi$ the angle spanning $\mathbb{S}^1$, the
fibration  means that the metric of $\mathcal{M}_7$ admits the following representation:
\begin{equation}\label{fibrametricu}
    ds^2_{\mathcal{M}_7}\, = \, \left(d\phi - \mathcal{A}\right)^2 \, + \, g_{ij^\star} \, dz^i \otimes d{\bar z}^{j^\star}
\end{equation}
where the one--form $\mathcal{A}$ is some suitable connection one--form on the $\mathrm{U(1)}$-bundle
(\ref{fibratoS1}).
\par
Secondly the metric cone $\mathcal{C}\left( \mathcal{M}_7\right)$
over the manifold $\mathcal{M}_7$ defined by the direct product
$\mathbb{R}_+\times \mathcal{M}_7$ equipped with the following
metric:
\begin{equation}\label{gustoconetto}
    ds^2_{\mathcal{C}\left( \mathcal{M}_7\right)} \, =\,dr^2 + 4 \,e^2 \, r^2 \, ds^2_{\mathcal{M}_7}
\end{equation}
should also be a Ricci-flat complex K\"ahler $4$-fold. In the above equation $e$ just denotes a constant
scale parameter with the dimensions of an inverse length $\left[e\right] = \ell^{-1}$.
\par
Altogether the Ricci flat K\"ahler manifold $K_4$, which plays the role of transverse space to the M2-branes,
is a   line bundle over the base manifold $K_3$:
\begin{eqnarray}\label{convecchio}
    \pi &\quad : \quad& K_4 \, \longrightarrow \, K_3 \nonumber\\
     \forall p \in K_3 && \pi^{-1}(p) \, \sim \, \mathbb{C}
\end{eqnarray}
\par
In \cite{pappo1}, following \cite{Fabbri:1999hw},  it was emphasized that the fundamental geometrical clue to
the field content of the \textit{superconformal gauge theory} on the boundary is provided by the construction
of the K\"ahler manifold $K_4$ as a holomorphic algebraic variety in some higher dimensional affine or
projective space $\mathbb{V}_{q}$, plus a K\"ahler quotient. The equations identifying the algebraic locus in
$\mathbb{V}_{q}$ are related with the superpotential $\mathcal{W}$ appearing in the $d=3$ lagrangian, while
the K\"ahler quotient is related with the $D$-terms appearing in the same lagrangian. The coordinates
$u^\alpha$ of the space $\mathbb{V}_{q}$ are the scalar fields of the \textit{superconformal gauge theory},
whose vacua, namely the set of extrema of its scalar potential, should be in a one--to--one correspondence
with the points of $K_4$. Going from one to multiple M2--branes just means that the coordinates $z^i$ of
$\mathbb{V}_{q}$ acquire color indices under a proper set of color gauge groups and are turned into matrices.
In this way we obtain \textit{quivers}.
\par
This is the main link between the D=3 Chern-Simons gauge theories discussed in sections \ref{genN2theo},
\ref{specN2theo} and the geometry of the transverse space to the branes.
\par
Next in \cite{pappo1}  eq.\,(\ref{caluffo2}) was rewritten in slightly more general terms. The
$\mathrm{AdS_4}$ compactification of $D=11$ supergravity is obtained by utilizing as complementary
$7$-dimensional space a manifold $\mathcal{M}_7$ which occupies the above displayed position in the
inclusion--projection diagram (\ref{caluffo2}). The metric cone $\mathcal{C}(\mathcal{M}_7)$ enters the game
when, instead of looking at the vacuum:
\begin{equation}\label{vacuetto}
    \mathrm{AdS_4} \otimes \mathcal{M}_7
\end{equation}
we consider the more general M2-brane solutions of D=11 supergravity, where the D=11 metric is of the
following form:
\begin{equation}\label{m2branmet}
    ds^2_{11} \, = \, H(y)^{-\ft 23}\,\left(d\xi^\mu\otimes d\xi^\nu\eta_{\mu\nu}\right) - H(y)^{\ft 13} \,
    \left(ds^2_{\mathcal{M}_8} \right)
\end{equation}
 $\eta_{\mu\nu}$ being the constant Lorentz metric of $\mathrm{Mink}_{1,2}$ and:
\begin{equation}\label{metric8}
   ds^2_{\mathcal{M}_8} \, = \,dy^I\otimes dy^J \, g_{IJ}(y)
\end{equation}
being a Ricci-flat metric on an asymptotically locally Euclidean $8$-manifold $\mathcal{M}_8$. In eq.\,(\ref{m2branmet}) the symbol $H(y)$
denotes a harmonic function over the manifold $\mathcal{M}_8$, namely:
\newcommand*\DAlambert{\mathop{}\!\mathbin\Box}
\begin{equation}\label{cicio}
     \DAlambert_{g} H(y) \, = \, 0
\end{equation}
Eq.\,(\ref{cicio}) is the only differential constraint required in order to satisfy all the field equations of
$D=11$ supergravity in presence of  the standard M2-brain ansatz for the $3$-form field:
\begin{equation}\label{scarparotta}
    \mathbf{A}^{[3]} \, \propto \, H(y)^{-1} \left(d\xi^\mu\wedge d\xi^\nu\wedge d\xi^\rho \,
    \epsilon_{\mu\nu\rho}\right)
\end{equation}
In this more general setup the manifold $\mathcal{M}_8$ is  what substitutes the metric cone
$\mathcal{C}(\mathcal{M}_7)$. To see the connection between the two viewpoints it suffices to introduce the
radial coordinate $r(y)$ by means of the position:
\begin{equation}\label{radiatore}
    H(y) \, = \, 1 \, - \, \frac{1}{r(y)^6}
\end{equation}
The asymptotic region where $\mathcal{M}_8$ is required to be locally Euclidean is defined by the condition
$r(y) \to \infty$. In this limit the metric (\ref{metric8}) should approach the flat Euclidean metric of
$\mathbb{R}^8\simeq \mathbb{C}^4$. The opposite limit where $r(y)\to 0$ defines the near horizon region of
the M2-brane solution. In this region the metric (\ref{m2branmet}) approaches that of the space
(\ref{vacuetto}), the manifold $\mathcal{M}_7$ being a codimension one submanifold of $\mathcal{M}_8$ defined
by the limit $r\to 0$.
\par
To be mathematically more precise let us consider the harmonic function as a map:
\begin{equation}\label{haccusmap}
    \mathfrak{H}\quad : \quad \mathcal{M}_8 \, \rightarrow \, \mathbb{R}_+
\end{equation}
This view point introduces a foliation of $\mathcal{M}_8$ into a one-parameter family of $7$-manifolds:
\begin{equation}\label{romualdo}
  \forall h \in \mathbb{R}_+  \quad : \quad \mathcal{M}_7(h) \, \equiv \,
  \mathfrak{H}^{-1}(h) \subset \mathcal{M}_8
\end{equation}
In order to have the possibility of residual supersymmetries we are interested in cases where the Ricci flat
manifold $\mathcal{M}_8$ is actually a Ricci-flat K\"ahler $4$-fold.
\par
In this way the appropriate rewriting of eq.\,(\ref{caluffo2}) is as follows:
\begin{equation}\label{caluffo3}
K_3 \, \stackrel{\pi}{\longleftarrow}    \, \mathcal{M}_7 \quad
  \stackrel{\mathfrak{H}^{-1}}{ \longleftarrow} \quad  K_4 \quad
  \stackrel{\mathcal{A}}{ \hookrightarrow} \quad \mathbb{V}_q
\end{equation}
Next we recall the general pattern laid down in \cite{pappo1} that will be our starting point.
\paragraph{The $\mathcal{N}=8$ case with no singularities.} The prototype of the above inclusion--projection
diagram is provided by the case of the M2-brane solution with all preserved supersymmetries. In this case we
have:
\begin{equation}\label{caluffopiatto}
\mathbb{CP}^3 \quad \stackrel{\pi}{\longleftarrow}  \quad \mathbb{S}^7 \quad
  \stackrel{Cone}{ \hookrightarrow} \quad  \mathbb{C}^4 \quad
  \stackrel{\mathcal{A}=\mathrm{Id}}{ \hookrightarrow} \quad \mathbb{C}^4
\end{equation}
On the left we just have the projection map of the Hopf fibration
of the $7$-sphere. On the right we have the
inclusion map of the $7$ sphere in its metric cone
$\mathcal{C}(\mathbb{S}^7)\equiv \mathbb{R}^8\sim
\mathbb{C}^4$. The last algebraic inclusion map is just the identity map, since the algebraic variety
$\mathbb{C}^4$ is already smooth and flat and needs no extra treatment.
\paragraph{The singular orbifold cases.} The next orbifold cases are those of interest to us in this paper.
Let $\Gamma \subset \mathrm{SU(4)}$ be a finite discrete subgroup of $\mathrm{SU(4)}$. Then
eq.\,(\ref{caluffopiatto}) is replaced by the following one:
\begin{equation}\label{calufforbo}
\frac{\mathbb{CP}^3}{\Gamma} \quad \stackrel{\pi}{\longleftarrow}  \quad \frac{\mathbb{S}^7}{\Gamma} \quad
  \stackrel{Cone}{ \hookrightarrow} \quad  \frac{\mathbb{C}^4}{\Gamma} \quad
  \stackrel{\mathcal{A}=\mbox{?}}{ \hookrightarrow} \quad \mbox{?}
\end{equation}
The consistency of the above quotient is guaranteed by the inclusion $\mathrm{SU(4)}\subset \mathrm{SO(8)}$.
The question marks can be removed only by separating the two cases:
\begin{description}
  \item[A)] Case: $\Gamma \subset \mathrm{SU(2)} \subset \mathrm{SU(2)_I} \otimes \mathrm{SU(2)_{II}}
  \subset \mathrm{SU(4)}$. Here we obtain:
  \begin{equation}\label{cromatorosso}
    \frac{\mathbb{C}^4}{\Gamma} \, \simeq \, \mathbb{C}^2 \times \frac{\mathbb{C}^2}{\Gamma}
  \end{equation}
and everything is under full control for the Kleinian $\frac{\mathbb{C}^2}{\Gamma}$ singularities and their
resolution {\`a} la Kronheimer in terms of  hyperK\"ahler  quotients.
  \item[B)] Case: $\Gamma \subset \mathrm{SU(3)} \subset   \mathrm{SU(4)}$. Here we obtain:
  \begin{equation}\label{cromatonero}
    \frac{\mathbb{C}^4}{\Gamma} \, \simeq \, \mathbb{C}  \times \frac{\mathbb{C}^3}{\Gamma}
  \end{equation}
  and the study and resolution of the singularity $\frac{\mathbb{C}^3}{\Gamma}$ in a physicist friendly way
   is the main issue of the present paper. The comparison of case B) with the well known case A)
   will provide us with many important hints.
\end{description}
Let us begin by erasing the question marks in case A). Here we can write:
\begin{equation}\label{calufforboA}
\frac{\mathbb{CP}^3}{\Gamma} \quad \stackrel{\pi}{\longleftarrow}  \quad \frac{\mathbb{S}^7}{\Gamma} \quad
  \stackrel{Cone}{ \hookrightarrow} \quad  \mathbb{C}^2 \times \frac{\mathbb{C}^2}{\Gamma} \quad
  \stackrel{\mathrm{Id}\times\mathcal{A}_W}{ \hookrightarrow} \quad \mathbb{C}^2 \times \mathbb{C}^3
\end{equation}
In the first inclusion map on the right, $\mathrm{Id}$ denotes the identity map $\mathbb{C}^2\to \mathbb{C}^2$
while $\mathcal{A}_W$ denotes the inclusion of the orbifold $\frac{\mathbb{C}^2}{\Gamma}$ as a singular
variety in $\mathbb{C}^3$ cut out by a single polynomial constraint:
\begin{eqnarray}\label{inclusione}
  \mathcal{A}_W& : &  \frac{\mathbb{C}^2}{\Gamma}\rightarrow
  \mathbf{V}(\mathcal{I}^W_\Gamma)
  \subset \mathbb{C}^3\nonumber\\
\mathbb{C}\left[\mathbf{V}(\mathcal{I}_\Gamma )\right] &=& \frac{\mathbb{C}[u,w,z]}{W_\Gamma(u,w,z)}
\end{eqnarray}
where by $\mathbb{C}\left[{\mathbf{V}}(\mathcal{I}_\Gamma )\right]$ we denote the \textit{coordinate ring} of
the algebraic variety $\mathbf{V}$. As we recall in more detail in the next section, the variables $u,w,z$
are polynomial $\Gamma$-invariant functions of the coordinates $z_1,z_2$ on which $\Gamma$ acts linearly. The
unique generator $W_\Gamma(u,w,z)$ of the ideal $\mathcal{I}^W_\Gamma$ which cuts out the singular variety
isomorphic to $\frac{\mathbb{C}^2}{\Gamma}$ is the unique algebraic relation existing among such invariants.
In the next sections we discuss the relation between this algebraic equation and the embedding in higher
dimensional algebraic varieties associated with the McKay quiver and the  hyperK\"ahler  quotient.
\par
Let us now consider the  case B). Up to this level things go in a quite analogous way with respect to case
A). Indeed we can write
\begin{equation}\label{calufforboB}
\frac{\mathbb{CP}^3}{\Gamma} \quad \stackrel{\pi}{\longleftarrow}  \quad \frac{\mathbb{S}^7}{\Gamma} \quad
  \stackrel{Cone}{ \hookrightarrow} \quad  \mathbb{C} \times \frac{\mathbb{C}^3}{\Gamma} \quad
  \stackrel{\mathrm{Id}\times\mathcal{A}_\mathcal{W}}{ \hookrightarrow} \quad \mathbb{C} \times \mathbb{C}^4
\end{equation}
In the last inclusion map on the right, $\mathrm{Id}$ denotes the identity map $\mathbb{C}\to \mathbb{C}$
while $\mathcal{A}_\mathcal{W}$ denotes the inclusion of the orbifold $\frac{\mathbb{C}^3}{\Gamma}$ as a
singular variety in $\mathbb{C}^4$ cut out by a single polynomial constraint:
\begin{eqnarray}\label{inclusioneB}
  \mathcal{A}_\mathcal{W}& : &  \frac{\mathbb{C}^3}{\Gamma}\rightarrow
  \mathbf{V}(\mathcal{I}_\Gamma )
  \subset \mathbb{C}^4\nonumber\\
 \mathbb{C}\left [\mathbf{V}\left(\mathcal{I}_\Gamma \right)\right] & \sim &
 \frac{\mathbb{C}[u_1,u_2,u_3,u_4]}{\mathcal{W}_\Gamma(u_1,u_2,u_3,u_4)}
\end{eqnarray}
Indeed as we show in later sections for the case $\Gamma\,=\,
\mathrm{L_{168}}$, discussed by Markushevich, and  for all of its
subgroups\footnote{The group $\mathrm{L_{168}}$ has three maximal
subgroups, up to conjugation, namely two non conjugate copies of the
octahedral group $\mathrm{O_{24}} \sim \mathrm{S_4}$ and one non
abelian group of order $21$, denoted $\mathrm{G_{21}}$ that is
isomorphic to the semidirect product $\mathbb{Z}_3
\ltimes\mathbb{Z}_7$.}, including $\Gamma\,=\,\mathrm{G_{21}}
\subset \mathrm{L_{168}}$, the variables $u_1,u_2,u_3,u_4$ are
polynomial $\Gamma$-invariant functions of the coordinates
$z_1,z_2,z_3$ on which $\Gamma$ acts linearly. The unique generator
$\mathcal{W}_\Gamma(u_1,u_2,u_3,u_4)$ of the ideal
$\mathcal{I}_\Gamma$ which cuts out the singular variety isomorphic
to $\frac{\mathbb{C}^3}{\Gamma}$ is the unique algebraic relation
existing among such invariants. As for the relation of this
algebraic equation with the embedding in higher dimensional
algebraic varieties associated with the McKay quiver, things are now
more complicated.
\par
In the years 1990s up to 2010s there has been an intense activity in the mathematical community on the issue
of the crepant resolutions of $\mathbb{C}^3/\Gamma$ (see for
instance\cite{roanno,marcovaldo,giapumckay,crawthesis,giapumckay}) that has gone on almost unnoticed by
physicists since it was mostly formulated in the abstract language of algebraic geometry,  providing few clues
to the applicability of such results to gauge theories and branes. Yet, once translated into more explicit
terms, by making  use of coordinate patches,  and equipped with some additional ingredients of Lie group
character, these mathematical results turn out to be not only useful, but rather of outmost relevance for the
physics of M2-branes. In the present paper we aim at drawing the consistent, systematic scheme which emerges
in this context upon a proper interpretation of the mathematical constructions.
\par
So let us consider the case of smooth resolutions. In case A) the smooth resolution is provided by a manifold
$ALE_\Gamma$ and we obtain the following diagram:
\begin{equation}\label{caluffoALE}
 \mathcal{M}_7 \quad
  \stackrel{\mathfrak{H}^{-1}}{ \longleftarrow} \quad  \mathbb{C}^2 \times ALE_\Gamma \quad
  \stackrel{\mathrm{Id}\times qK}{ \longleftarrow} \quad \mathbb{C}^2
  \times\mathbb{V}_{|\Gamma|+1} \quad \stackrel{\mathcal{A}_{\mathcal{P}}}{\hookrightarrow}
  \quad \mathbb{C}^2 \times \mathbb{C}^{2|\Gamma|}
\end{equation}
In the above equation the map $\stackrel{\mathfrak{H}^{-1}}{ \longleftarrow}$ denotes the inverse of the
harmonic function map on $\mathbb{C}^2\times ALE_\Gamma$ that we have already discussed. The map
$\stackrel{\mathrm{Id}\times qK}{ \longleftarrow}$ is instead the product of the identity map $\mathrm{Id} \,
: \, \mathbb{C}^2 \to \mathbb{C}^2$ with the K\"ahler quotient map:
\begin{equation}\label{fraulein}
   qK \quad : \quad \mathbb{V}_{|\Gamma|+1} \, \longrightarrow \,\mathbb{V}_{|\Gamma|+1}\,  \,
  /\!\!/_{\null_K} \mathcal{F}_{|\Gamma|-1} \, \simeq \, ALE_\Gamma
\end{equation}
of an algebraic variety of complex dimension $|\Gamma|+1$ with respect to a suitable Lie group
$\mathcal{F}_{|\Gamma|-1}$ of real dimension $|\Gamma|-1$. Finally the map
$\stackrel{\mathcal{A}_{\mathcal{P}}}{\hookrightarrow}$ denotes the inclusion map of the variety
$\mathbb{V}_{|\Gamma|+1}$ in $\mathbb{C}^{2|\Gamma|}$. Let $y_1,\dots y_{2|\Gamma|}$ be the coordinates of
$\mathbb{C}^{2|\Gamma|}$. The variety $\mathbb{V}_{|\Gamma|+1}$ is defined by an ideal generated by
$|\Gamma|-1$ quadratic generators:
\begin{eqnarray}\label{congo}
    \mathbb{V}_{|\Gamma|+1} & = & \mathbf{V}\left( \mathcal{I}_\Gamma\right)\nonumber\\
\mathbb{C} \left[\mathbf{V}\left( \mathcal{I}_\Gamma\right) \right]&= & \frac{\mathbb{C}\left[y_1,\dots
y_{2|\Gamma|} \right]}{\left(\mathcal{P}_1(y),\mathcal{P}_2(y),\dots ,\mathcal{P}_{|\Gamma|-1}(y)\right)}
\end{eqnarray}
Actually the $|\Gamma|-1$ polynomials $\mathcal{P}_i(y)$ are the holomorphic part of the triholomorphic
moment maps associated with the triholomorphic action of the group $\mathcal{F}_{|\Gamma|-1}$ on
$\mathbb{C}^{2|\Gamma| }$ and the entire procedure from  $\mathbb{C}^{2|\Gamma|}$ to $ALE_\Gamma$ can be seen
as the  hyperK\"ahler  quotient:
\begin{equation}\label{cagliozzo}
    ALE_\Gamma \, = \,\mathbb{C}^{2|\Gamma|}/\!\!/_{\null_{HK}}\mathcal{F}_{|\Gamma|-1}
\end{equation}
yet we have preferred to split the procedure into two steps in order to compare case A) with case B) where
the two steps are necessarily distinct and separated.
\par
Indeed in case B) we can write the following diagram:
\begin{equation}\label{caluffoBalle}
 \mathcal{M}_7 \quad
  \stackrel{\mathfrak{H}^{-1}}{ \longleftarrow} \quad  \mathbb{C} \times Y_\Gamma \quad
  \stackrel{\mathrm{Id}\times qK}{ \longleftarrow} \quad \mathbb{C}
  \times\mathbb{V}_{|\Gamma|+2} \quad \stackrel{\mathrm{Id} \times \mathcal{A}_{\mathcal{P}}}{\hookrightarrow}
  \quad \mathbb{C} \times \mathbb{C}^{3|\Gamma|}
\end{equation}
In this case, just as in the previous one,  the intermediate step is provided by the K\"ahler quotient but
the map on the extreme right $\stackrel{\mathcal{A}_{\mathcal{P}}}{\hookrightarrow}$ denotes the inclusion
map of the variety $\mathbb{V}_{|\Gamma|+2}$ in $\mathbb{C}^{3|\Gamma|}$. Let $y_1,\dots y_{3|\Gamma|}$ be
the coordinates of $\mathbb{C}^{3|\Gamma|}$. The variety $\mathbb{V}_{|\Gamma|+2}$ is defined as the
principal branch of a set of quadratic algebraic equations that are group-theoretically defined. Altogether
the mentioned construction singles out  the holomorphic orbit of a certain group action to be discussed in
detail in the sequel. So we anticipate:
\begin{eqnarray}\label{belgone}
\mathbb{V}_{|\Gamma|+2} & = &  \mathcal{D}_\Gamma \equiv \mathrm{Orbit}_{\mathcal{G}_\Gamma}\left(
L_\Gamma\right)
\end{eqnarray}
where both the set $L_\Gamma$ and the complex  group $\mathcal{G}_\Gamma$ are  completely defined by the
discrete group $\Gamma$ defining the quotient singularity.
\section{Generalities on $\frac{\mathbb{C}^3}{\Gamma}$  singularities}
Recalling what we summarized above we conclude that the singularities relevant to our goals are of the form:
\begin{equation}\label{cialtrus}
   X \, = \, \frac{\mathbb{C}^3}{\Gamma}
\end{equation}
where the finite group $\Gamma \subset \mathrm{SU(3)}$  has a holomorphic action on $\mathbb{C}^3$. For this
case, as we mentioned above,  there is a series of general results and procedures developed in algebraic
geometry that we want to summarize in the  perspective of their use in physics.
\par
To begin with let us observe the schematic diagram sketched here below:
\begin{equation}
{\setlength{\unitlength}{1pt}
\begin{picture}(280,150)(-10,-70)
\thicklines \put(100,50){\circle*{20}} \put(100,35){\makebox(0,0){\Large $\Downarrow$}}
\put(100,25){\makebox(0,0){levels }}\put(100,15){\makebox(0,0){$\zeta$ }}\put(100,5){\makebox(0,0){of moment
map }}
 \put(100,50){\line(-1,-1){70}} \put(35,-15){\circle*{20}}
\put(100,50){\line(1,-1){70}}
\put(165,-15){\circle*{20}}\put(35,-15){\line(1,0){125}}
\put(100,65){\makebox(0,0){$r = \mbox{dim}\,
\mathfrak{z}\left[\mathbb{F}_\Gamma\right]$ center of the Lie
Algebra}} \put(225,-20){\makebox(0,0){$r = \# $ of nontrivial}}
\put(225,-30){\makebox(0,0){$\Gamma$ irred. represent.s}}
\put(-30,-20){\makebox(0,0){$r = \# $ of nontrivial}}
\put(-30,-30){\makebox(0,0) {$\Gamma$ conjugacy classes
}}\put(35,-33){\makebox(0,0){\Large
$\Downarrow$}}\put(165,-33){\makebox(0,0){\Large $\Downarrow$}}
\put(-30,-50){\makebox(0,0){age grading,}}
\put(-30,-60){\makebox(0,0) {exceptional divisors}}
\put(225,-50){\makebox(0,0){first Chern classes of}}
\put(225,-60){\makebox(0,0){tautological bundles}}
\put(35,-50){\circle{15}}\put(165,-50){\circle{15}}\put(35,-50){\circle{5}}\put(165,-50){\circle{5}}
\put(100,-51){\makebox(0,0){\LARGE
$\Leftrightarrow$}}\put(35,-50){\line(1,0){130}}
\end{picture}}
\label{salumaio}
\end{equation}
The fascination of the mathematical construction lying behind
the desingularization process, which has a
definite counterpart in the structure of the Chern-Simons gauge
theories describing M2-branes at the
$\mathbb{C}^3/\Gamma$ singularity, is the triple interpretation
of the same number $r$ which alternatively
yields:
\begin{itemize}
  \item The number of nontrivial conjugacy classes of the
  finite group $\Gamma$,
  \item The number of irreducible representations of the
  finite group $\Gamma$,
  \item The center of the  Lie algebra $\mathfrak{z}\left[\mathbb{F}_\Gamma\right]$ of the compact gauge group
  $\mathcal{F}_\Gamma$, whose structure, as we will see, is:
  \begin{equation}\label{sinuhe}
    \mathcal{F}_\Gamma \, = \, \bigotimes_{i=1}^r \mathrm{U(n_i)}
  \end{equation}
\end{itemize}
The levels $\zeta_I$ of the moment maps are the main ingredient of the singularity resolution. At
level$\zeta^I=0$ we have the singular orbifold $\mathcal{M}_0 \, = \, \frac{\mathbb{C}^3}{\Gamma}$, while at
$\zeta^i \neq 0$ we obtain a smooth manifold $\mathcal{M}_\zeta$ which develops a nontrivial homology and
cohomology. In physical parlance the levels $\zeta^I$ are the Fayet-Iliopoulos parameters appearing in the
lagrangian, while $\mathcal{M}_\zeta$ is the manifold of vacua of the theory, namely of extrema of the
potential, as we already emphasized.
\par
Quite generally, we find that each of the gauge factors $\mathrm{U(n_i)}$ is the structural group of a
holomorphic vector bundle of rank $n_i$:
\begin{equation}\label{broccoletti}
    \mathfrak{V}_i \, \stackrel{\pi}{\longrightarrow} \, \mathcal{M}_\zeta
\end{equation}
whose first Chern class is a nontrivial (1,1)-cohomology class of the resolved smooth
manifold:
\begin{equation}\label{cimedirapa}
    c_1\left(\mathfrak{V}_i\right) \, \in \, H^{1,1}\left(\mathcal{M}_\zeta\right)
\end{equation}
On the other hand a very deep theorem originally proved in the nineties by Reid and Ito \cite{giapumckay}
relates the dimensions of the cohomology groups $H^{q,q}\left(\mathcal{M}_\zeta\right)$ to the conjugacy
classes of $\Gamma$ organized according to the grading named \textit{age}. So named \textit{junior classes}
of \textit{age} = 1 are associated with $H^{1,1}\left(\mathcal{M}_\zeta\right)$ elements, while the so-named
\textit{senior classes} of \textit{age} = 2 are associated with $H^{2,2}\left(\mathcal{M}_\zeta\right)$
elements.
\par
The link that pairs irreps with conjugacy classes is provided by the relation, well-known in algebraic
geometry, between \textit{divisors} and \textit{line bundles}. The conjugacy classes of $\gamma$ can be put
into correspondence with the exceptional divisors created in the resolution $\mathcal{M}_\zeta
\stackrel{\zeta\to 0}{\longrightarrow} \frac{\mathbb{C}^3}{\Gamma}$ and each divisor defines a line bundle
whose first Chern class is an element of the $H^{1,1}\left(\mathcal{M}_\zeta\right)$ cohomology group.
\par
These line bundles labeled by conjugacy classes have to be compared with the line bundles created by the
K\"ahler quotient procedure that are instead associated with the irreps, as we have sketched above. In this
way we build the bridge between conjugacy classes and irreps.
\par
Finally there is the question whether the divisor is compact or not. In the first case, by Poincar\'e
duality, we obtain nontrivial $H^{2,2}\left(\mathcal{M}_\zeta\right)$ elements. In the second case we have
no new cohomology classes. The age grading precisely informs us about the compact or noncompact nature of
the divisors. Each senior class
corresponds to a cohomology class of degree 4, thus signaling the existence of a non-trivial closed
(2,2) form, and via Poincar\'e duality, it also corresponds to a compact component of the exceptional
divisor.
\par
The physics--friendly illustration of this general beautiful scheme, together with the explicit construction
of a few concrete examples is the main goal of the present paper. We begin with the concept of age grading.
\subsection{The concept of aging for conjugacy classes
of the discrete group $\Gamma$}
\label{vecchiardo}
According to the above quoted theorem that we shall explain below, the
\textit{age grading} of $\Gamma$ conjugacy classes allows to predict the Dolbeaults cohomology of the
resolved algebraic variety. It goes as follows.
\par
 Suppose that $\Gamma$ (a finite group) acts in a  linear way on
$\mathbb{C}^n$. Consider an element $\gamma \in \Gamma$ whose action is the following:
\begin{equation}\label{etoso}
  \gamma.\vec{z} \, = \, \underbrace{\left(
                                 \begin{array}{ccc}
                                   \ldots & \ldots &  \ldots \\
                                   \vdots &  \vdots& \vdots \\
                                  \ldots & \ldots &\ldots \\
                                 \end{array}
                               \right)
   }_{\mathcal{Q}(\gamma)}\, \cdot \, \left(\begin{array}{c}
                               z_1 \\
                               \vdots\\
                               z_n
                             \end{array}
    \right)
\end{equation}
Since in a finite group all elements have a finite order,  there exists  $r \in \mathbb{N}$,  such that $\gamma^r\,
= \, \mathbf{1}$. We define the age of an element in the following way. Let us diagonalize $D(\gamma)$,
namely compute its eigenvalues. They will be as follows:
\begin{equation}\label{gioffo}
  \left(\lambda_1 , \dots , \lambda_n\right ) \, = \, \exp\left[\frac{2\pi \, i}{r}
  \, a_i\right]  \quad; \quad r> a_i \in  \mathbb{N} \quad i\, = \, 1, \dots , n
\end{equation}
We define:
\begin{equation}\label{vecchione}
  \mathrm{age}\left(\gamma \right)\, = \, \frac{1}{r} \, \sum_{i=1}^n a_i
\end{equation}
Clearly the age is a property of the conjugacy class of the element, relative to the considered three-dimensional
complex representation.
\subsection{The fundamental theorem}
 In \cite{giapumckay} Y. Ito and M. Reid proved the following fundamental theorem:
\begin{teorema}
\label{reidmarktheo} Let $Y\rightarrow \mathbb{C}^3/\Gamma$ be a crepant \footnote{A resolution of singularities $X \rightarrow Y$  is crepant when the canonical bundle of $X$ is the pullback
of the canonical bundle of $Y$.}  resolution of a Gorenstein \footnote{A variety is Gorenstein when the canonical divisor is a Cartier divisor, i.e., a divisor corresponding
to a line bundle.}
singularity. Then we have the following relation between the de-Rham cohomology groups of the resolved smooth
variety $Y$ and the ages of $\Gamma$ conjugacy classes:
$$ \mathrm{dim} \, H^{2k} \left(Y\right) \, = \, \# \mbox{ of age $k$ conjugacy classes of  $\Gamma$}$$
\end{teorema}
On the other hand it happens that all odd cohomology groups are trivial:
\begin{equation}\label{oddi}
    \mathrm{dim} \, H^{2k+1} \left(Y\right) \, = \, 0
\end{equation}
This is true also in the case of $\mathbb{C}^2/\Gamma$ singularities, yet in $n=2,3$ the consequences of the
same fact are drastically different. In all complex dimensions $n$  the deformations of the K\"ahler class
are in one-to-one correspondence with the harmonic forms $\omega^{(1,1)}$, while those of the complex
structure are in correspondence with the harmonic forms $\omega^{(n-1,1)}$. In $n=2$ the harmonic
$\omega^{(1,1)}$ forms play the double role of K\"ahler class deformations and complex structure
deformations. This is the reason why we can do a  hyperK\"ahler  quotient and we have both moduli parameters
in the K\"ahler potential and in the polynomials cutting out the smooth variety. Instead in $n=3$
eq.\,(\ref{oddi}) implies that the polynomials constraints cutting the singular locus have no deformation
parameters. The parameters of the resolution occur only at the level of the K\"ahler quotient and are the
levels of the K\"ahlerian moment maps.
\par Given an algebraic representation of the variety $Y$ as the
vanishing locus of certain polynomials $W(x)\, = \, 0$, the algebraic $2k$-cycles  are the $2k$-cycles that
can be holomorphically embedded in $Y$.   The following statement
in $n=3$ is  elementary:
\begin{statement}
The Poincar\'e dual of any algebraic $2k$-cycle is   of type $(k,k)$.
\end{statement}
Its converse is known as the Hodge conjecture, stating that any cycle of type $(k,k)$
is a linear combination of algebraic cycles. This will hold true for the varieties we shall be considering.
\par
Thus we conclude that the so named \textit{junior conjugacy classes} (age=1) are in a
one-to-one correspondence with $\omega^{(1,1)}$-forms that span ${H^{1,1}}$, while \textit{conjugacy classes
of age 2} are in one-to-one correspondence with $\omega^{(2,2)}$-forms that span ${H^{2,2}}$.
\section{Comparison with ALE manifolds and comments}
 Let us compare the above
predictions  for the case B) of $\mathbb{C}^3/\Gamma$ singularities with the well known case A) of
$\mathbb{C}^2/\Gamma$ where $\Gamma$ is a Kleinian subgroup of $\mathrm{SU(2)}$ and the resolution of the
singularity leads to an ALE manifold \cite{mango,degeratu,kro1,kro2,Bertolini:2002pr,Bertolini:2001ma}.  As
we already stressed above,  this latter can be explicitly constructed by means of a  hyperK\"ahler  quotient,
according to Kronheimer's construction.
\par
In table \ref{kleinale} we summarize some well known facts about $Y\to X=\mathbb{C}^2/\Gamma$ which are the
following. Here $\chi$ denotes Hirzebruch's signature characteristic of the resolved manifold. 

\begin{table}[h!]\caption{Finite $\mathrm{SU(2)}$ subgroups versus ALE manifold
properties} \label{kleinale}{\small
\begin{center}
\begin{tabular}{||l|c|c|c|c|c||}\hline
$\null$ & $\null$ & $\null$ & $\null$ & $\null$ & $\null$ \\
$\Gamma$. & $W_\Gamma(u,w,z)$ & ${\cal R}=\frac{{\bf C}[u,w,z]}{\partial W}$
&$|{\cal R}|$ & ${\#} c.~c. $& $\tau\equiv\chi -1$  \\
$\null$ & $\null$ & $\null$ & $\null$ & $\null$ & $\null$ \\
\hline \hline
$\null$ & $\null$ & $\null$ & $\null$ & $\null$ & $\null$ \\
$A_k$&$u^2+w^2 - z^{k+1}$&$\{ 1, z,.. $&$k$&$k+1$&$k$ \\
$~$&$~$&  $.., z^{k-1} \}$&$~$&$~$&$~$ \\
$\null$ & $\null$ & $\null$ & $\null$ & $\null$ & $\null$ \\
\hline
$\null$ & $\null$ & $\null$ & $\null$ & $\null$ & $\null$ \\
$D_{k+2}$&$u^2 +w^2 z + z^{k+1}$&$\{ 1, w, z,w^2,$&$k+2$&$k+3$&$k+2$ \\
$~$&$~$&  $ z^2, ..., z^{k-1} \}$&$~$&$~$&$~$ \\
$\null$ & $\null$ & $\null$ & $\null$ & $\null$ & $\null$ \\
\hline
$\null$ & $\null$ & $\null$ & $\null$ & $\null$ & $\null$ \\
$E_6=$&$u^2+w^3 +z^{4}$&$\{ 1, w,  z,$&$6$&$7$&$6$ \\
${\cal T}$&$~$&  $wz, z^2,wz^{2} \}$&$~$&$~$&$~$ \\
$\null$ & $\null$ & $\null$ & $\null$ & $\null$ & $\null$ \\
\hline
$\null$ & $\null$ & $\null$ & $\null$ & $\null$ & $\null$ \\
$E_7=$&$u^2+w^3 +wz^{3}$&$\{ 1, w, z,w^2,$&$7$&$8$&$7$ \\
${\cal O}$&$~$&  $z^2,wz,w^2z \}$&$~$&$~$&$~$ \\
$\null$ & $\null$ & $\null$ & $\null$ & $\null$ & $\null$ \\
\hline
$\null$ & $\null$ & $\null$ & $\null$ & $\null$ & $\null$ \\
$E_8=$&$u^2+w^3 + z^{5}$&$\{ 1,w, z,z^2,wz,$&$8$&$9$&$8$ \\
${\cal I}$&$~$&  $z^3,wz^2,wz^3  \}$&$~$&$~$&$~$ \\
$\null$ & $\null$ & $\null$ & $\null$ & $\null$ & $\null$ \\
\hline
\end{tabular}\end{center}}\end{table}
\vskip 0.2cm
\begin{enumerate}
  \item As an affine variety the singular orbifold $X$ is described by a single polynomial  equation
  $W_\Gamma(u,w,z)\, =\,0$ in $\mathbb{C}^3$. This equation is simply given by a relation existing
  among the invariants of $\Gamma$ as we anticipated in the previous section.  Note that this is  the case also for $X \, = \,
  \frac{\mathbb{C}^3}{\mathrm{L_{168}}}$, as Markushevich has shown. He has found one polynomial constraint
  $W_{\mathrm{L_{168}}}(u_1,u_2,u_3,u_4)\, = \,0$ of degree 10 in $\mathbb{C}^4$ which describes $X$.
  We  were able to find   a similar result for the subgroup $\mathrm{G_{21}} \subset \mathrm{L_{168}}$
  and obviously also for the cases $\mathbb{C}^3/\mathbb{Z}_3$ and $\mathbb{C}^3/\mathbb{Z}_7$.
  In the $\mathrm{G_{21}}$  case the equation is   of order  16.  We will present  these results in a future publication.
  \item The resolved locus $Y$ in the case of ALE manifolds is described  by a deformed equation:
  \begin{eqnarray}
{ W}^{ALE}_{\Gamma} \left (  u,w,z ; \, \mathbf{t} \right )&= &W_{\Gamma}(u,w,z)\, + \,
\sum_{i=1}^{r} \ t_i \, {\cal P}^{(i)}(u,w,z) \nonumber\\
r &\equiv& \mbox{dim}\, \mathcal{R}_\Gamma \label{grouptheory21}
\end{eqnarray}
where
\begin{description}
\item[a)] { $W_{\Gamma}(u,w,z)$ is the simple singularity polynomial
corresponding to the finite subgroup $\Gamma\subset \mathrm{SU(2)}$} according to Arnold's classification of isolated critical points of functions \cite{arnolsimplicius}, named simple singularities in the literature.
\item[b)]{ ${\cal P}^{(i)}(u,w,z)$ is a basis spanning the chiral ring
\begin{equation}
\mathcal{R}_\Gamma=\frac{{\mathbb{C}}[u,w,z]}{\partial W_{\Gamma}} \label{grouptheory22}
\end{equation}
of polynomials in $u,w,z$ that do not vanish upon use of the vanishing relations $\partial_u \, W_{\Gamma}
\,= \, \partial_w \, W_{\Gamma} \, = \, \partial_z  \, W_{\Gamma} \, = \, 0$.}
\item[c)]{The complex parameters $t^i$ are the complex structure moduli and they
are in one--to--one correspondence with the set of complex level parameters ${ \ell}^{\bf X}_+$.}
\end{description}
   \item According to the general view put forward in the previous section, for ALE manifolds we have:
\begin{equation}\label{carezza}
    \mbox{dim} H^{1,1} \, = \, r \, \equiv \,\# \mbox{ nontrivial conjugacy classes of $\Gamma$}
\end{equation}
We also have:
\begin{equation}\label{casilinus}
    \mbox{dim} \mathcal{R}_\Gamma \, = \, r
\end{equation}
as one sees from table \ref{kleinale}. From the point of view of complex  geometry this is the
consequence of a special coincidence, already stressed in the previous section, which applies only to the
case of complex dimension $2$. As one knows, for Calabi-Yau $n$-folds complex structure deformations are
associated with $\omega^{n-1,1}\in H^{n-1,1}$ harmonic forms, while K\"ahler structure deformations, for all
$n$, are associated  with $\omega^{1,1}\in H^{1,1}$ harmonic forms. Hence when $n=2$, the $(1,1)$-forms play
a double role as complex structure deformations and as K\"ahler structure deformations. For instance, this is
well known in the case of $K3$. Hence in the $n=2$ case the number of \textit{nontrivial conjugacy classes}
of the group $\Gamma$ coincides both with the number of K\"ahler moduli and with number of complex structure
moduli of the resolved variety.
\item In the case of $Y \to X=\frac{\mathbb{C}^3}{\Gamma}$ the number of $(1,1)$-forms and hence of
K\"ahler moduli is still related with $r \, = \, \# \mbox{ \textit{junior conjugacy classes of }  $\Gamma$}$
but there are no complex structure deformations.
\end{enumerate}
\subsection{The McKay correspondence for $\mathbb{C}^2/\Gamma$} The   table of characters  $\chi^{(\mu)}_i$ of any finite group
$\gamma$ allows to reconstruct the decomposition coefficients of any representation along the irreducible
representations:
\begin{eqnarray}\label{multipillini}
    D&=&\bigoplus_{\mu =1}^{r} \, a_\mu \, D_\mu \nonumber\\
    a_\mu &=& \o{1}{g} \, \sum_{i =1}^{r} \, g_i \, \chi^{(D)}_{i} \, \chi^{(\mu) \, \star}_{i}
\end{eqnarray}
where $\chi^{(D)}$ is the character of $D$. For the finite subgroups $\Gamma\subset \mathrm{SU(2)}$ a particularly important case is the decomposition of
the tensor product of an irreducible representation $D_\mu$ with the defining 2-dimensional representation
${\cal Q}$. It is indeed at the level of this decomposition that the relation between these groups and the
simply laced Dynkin diagrams becomes explicit and  is named the McKay correspondence. This decomposition
plays a crucial role in the explicit construction of ALE manifolds according to Kronheimer. Setting
\begin{equation}
{\cal Q} \, \otimes \, D_\mu ~=~\bigoplus_{\nu =0}^{r} \, A_{\mu \nu} \, D_\nu \label{grouptheory16}
\end{equation}
where $D_0$ denotes the identity representation, one finds that the matrix ${\bar
c}_{\mu\nu}=2\delta_{\mu\nu}-A_{\mu\nu}$ is the {\it extended Cartan matrix} relative to the {\it  extended
Dynkin diagram} corresponding to the given group. We remind the reader that the extended Dynkin diagram of
any simply laced Lie algebra is obtained by adding to the {\it  dots} representing  the {\it  simple roots}
$\left \{ \, \alpha_1 \, ......\, \alpha_r \,  \right \}$ an {\it  additional dot} (marked black in figs.
\ref{dynfigure1A}, \ref{dynfigure2A}) representing the negative of the highest root $\alpha_0 \, = \,
\sum_{i=1}^{r} \, n_i \, \alpha_i$ ($n_i$ are the Coxeter numbers). Thus we see a correspondence between the
nontrivial conjugacy classes $\mathcal{C}_i$ (or equivalently the nontrivial irreps) of the group
$\Gamma(\mathbb{G})$ and the simple roots of $\mathbb{G}$. In this correspondence the extended Cartan matrix
provides  the Clebsch-Gordon coefficients (\ref{grouptheory16}), while the Coxeter numbers $n_i$ express the
dimensions of the irreducible representations. All these informations are summarized in Figs.
\ref{dynfigure1A}, \ref{dynfigure2A} where the numbers $n_i$ are attached to each of the dots: the number $1$
is attached to the extra dot since it stands for the identity representation.
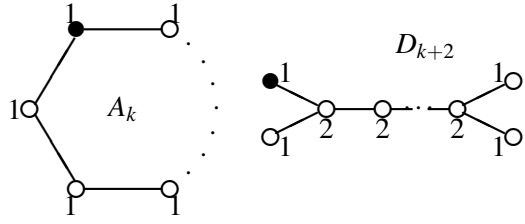
\begin{figure}[tb]
\begin{center}
{\setlength{\unitlength}{0.7pt}
\begin{picture}(280,110)(-60,-60)
\thicklines \put(25,43){\circle{8}} \put(-25,43){\circle*{8}} \put(-50,0){\circle{8}}
\put(-25,-43){\circle{8}} \put(25,-43){\circle{8}} \put(35,35){\makebox(0,0){$\cdot$}}
\put(43,25){\makebox(0,0){$\cdot$}} \put(48,13){\makebox(0,0){$\cdot$}} \put(50,0){\makebox(0,0){$\cdot$}}
\put(48,-13){\makebox(0,0){$\cdot$}} \put(43,-25){\makebox(0,0){$\cdot$}}
\put(35,-35){\makebox(0,0){$\cdot$}} \put(-21,43){\line(1,0){42}} \put(-47,3){\line(3,5){21}}
\put(-47,-3){\line(3,-5){21}} \put(-21,-43){\line(1,0){42}} \put(28,52){\makebox(0,0){1}}
\put(-28,52){\makebox(0,0){1}} \put(-58,0){\makebox(0,0){1}} \put(-28,-52){\makebox(0,0){1}}
\put(28,-52){\makebox(0,0){1}} \put(0,0){\makebox(0,0){$A_{k}$}} \put(75,-30){
\begin{picture}(150,50)(0,-30)
\thicklines \put(67,30){$D_{k+2}$} \multiput(30,0)(30,0){2}{\circle{8}} \put(30,-10){\makebox(0,0){2}}
\put(60,-10){\makebox(0,0){2}} \put(34,0){\line(1,0){22}} \put(64,0){\line(1,0){12}}
\put(80,0){\makebox(0,0){$\cdots$}} \put(84,0){\line(1,0){12}} \put(100,0){\circle{8}}
\put(100,-10){\makebox(0,0){2}} \put(0,15){\circle*{8}} \put(8,20){\makebox(0,0){1}} \put(0,-15){\circle{8}}
\put(8,-20){\makebox(0,0){1}} \multiput(130,15)(0,-30){2}{\circle{8}} \put(122,20){\makebox(0,0){1}}
\put(122,-20){\makebox(0,0){1}} \put(3,13){\line(2,-1){22}} \put(3,-13){\line(2,1){22}}
\put(103,2){\line(2,1){22}} \put(103,-2){\line(2,-1){22}}
\end{picture}}
\end{picture}
} \caption{\label{dynfigure1A} Extended Dynkin diagrams of the infinite series}
\end{center}
\end{figure}
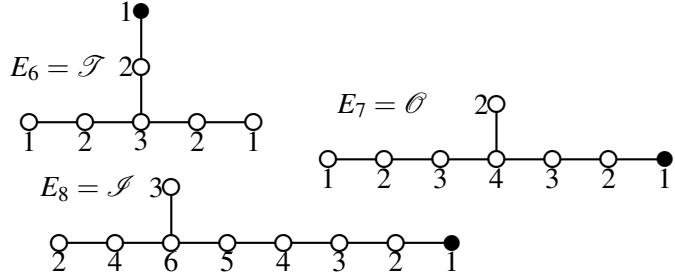
\begin{figure}[tb]
\begin{center}
{\setlength{\unitlength}{0.7pt}
\begin{picture}(300,135)(0,-10)
\thicklines \put(-10,25){$E_8 = {\cal I}$} \multiput(0,0)(30,0){7}{\circle{8}} \put(210,0){\circle*{8}}
\put(0,-10){\makebox(0,0){2}} \put(30,-10){\makebox(0,0){4}} \put(60,-10){\makebox(0,0){6}}
\put(90,-10){\makebox(0,0){5}} \put(120,-10){\makebox(0,0){4}} \put(150,-10){\makebox(0,0){3}}
\put(180,-10){\makebox(0,0){2}} \put(210,-10){\makebox(0,0){1}} \multiput(4,0)(30,0){7}{\line(1,0){22}}
\put(60,4){\line(0,1){22}} \put(60,30){\circle{8}} \put(52,30){\makebox(0,0){3}} \put(140,45){
\begin{picture}(180,50)(0,0)
\thicklines \put(5,25){$E_7= {\cal O}$} \multiput(0,0)(30,0){6}{\circle{8}} \put(180,0){\circle*{8}}
\put(0,-10){\makebox(0,0){1}} \put(30,-10){\makebox(0,0){2}} \put(60,-10){\makebox(0,0){3}}
\put(90,-10){\makebox(0,0){4}} \put(120,-10){\makebox(0,0){3}} \put(150,-10){\makebox(0,0){2}}
\put(180,-10){\makebox(0,0){1}} \multiput(4,0)(30,0){6}{\line(1,0){22}} \put(90,4){\line(0,1){22}}
\put(90,30){\circle{8}} \put(82,30){\makebox(0,0){2}}
\end{picture}}
\put(-20,80){
\begin{picture}(120,50)(0,15)
\thicklines \put(-10,25){$E_6= {\cal T}$} \multiput(0,0)(30,0){5}{\circle{8}} \put(0,-10){\makebox(0,0){1}}
\put(30,-10){\makebox(0,0){2}} \put(60,-10){\makebox(0,0){3}} \put(90,-10){\makebox(0,0){2}}
\put(120,-10){\makebox(0,0){1}} \multiput(4,0)(30,0){4}{\line(1,0){22}} \put(60,4){\line(0,1){22}}
\put(60,30){\circle{8}} \put(52,30){\makebox(0,0){2}} \put(52,60){\makebox(0,0){1}}
\put(60,34){\line(0,1){22}} \put(60,60){\circle*{8}}
\end{picture}}
\end{picture}
} \caption{\label{dynfigure2A} Exceptional extended Dynkin diagrams}
\end{center}
\end{figure}
\subsection{Kronheimer's construction} Given any finite subgroup  $\Gamma \subset \mathrm{SU(2)}$, we
consider a space $\mathcal{P}$ whose elements are two-vectors of $|\Gamma |\times |\Gamma |$ complex
matrices: $\left(A, B\right) \in \mathcal{P}$. The action of an element $\gamma\in \Gamma$ on the points of
$\mathcal{P}$ is the following:
\begin{equation}\label{gammaactiononp}
 \left(\begin{array}{c} A \cr B \end{array}\right) \, \stackrel{\gamma}{\longrightarrow}\,
 \left(\begin{array}{cc} u_{\gamma} & i\,{\bar v}_{\gamma} \\
 i\,v_{\gamma}& {\bar u}_{\gamma}\\
 \end{array}\right)\,
 \left( \begin{array}{c} R(\gamma)\,A
\,R(\gamma^{-1}) \\ R(\gamma)\, B \, R(\gamma^{-1}) \\
\end{array}\right)
\end{equation}
where the two-dimensional matrix on the right hand side is the realization of $\gamma$ inside the defining
two-dimensional representation $\mathcal{Q}\subset \mathrm{SU(2)}$, while $R(\gamma)$ is the regular,
$|\Gamma|$-dimensional representation. The basis vectors in $R$ named $e_\gamma$ are in one-to-one
correspondence with the group elements $\gamma \in \Gamma$ and transform as follows:
\begin{equation}
 R(\gamma) \, e_\delta ~=~e_{\gamma \cdot \delta} ~~~~~\forall \, \gamma \, , \, \delta \, \in \,
\Gamma
 \label{grouptheory15}
 \end{equation}
 Intrinsically, the space $\mathcal{P}$ is named as:
 \begin{equation}\label{ommaomma}
    \mathcal{P}\, \simeq \, \mathrm{Hom}\left(R,\mathcal{Q}\otimes R \right)
 \end{equation}
Next we introduce the space $\mathcal{S}$, which by definition is the subspace made of $\Gamma$-invariant
elements in $\mathcal{P}$:
\begin{equation}
\mathcal{S}\equiv\left\{p\in\mathcal{P} / \forall \gamma\in\Gamma, \gamma\cdot p = p\right\}\,\,
\label{carnevalediRio}
\end{equation}
Explicitly the invariance condition  reads as follows:
\begin{equation}\label{invariancecond}
\left(\begin{array}{cc} u_{\gamma} & i\,{\bar v}_{\gamma} \\
 i\,v_{\gamma}& {\bar u}_{\gamma}\\
 \end{array}\right)\, \left(\begin{array}{c} A \cr B \end{array}\right) \, =\,
\left( \begin{array}{c} R(\gamma^{-1})\,A
\,R(\gamma) \\ R(\gamma^{-1})\, B \, R(\gamma) \\
\end{array}\right)
\end{equation}
The decomposition (\ref{grouptheory16}) is very useful in order to determine the $\Gamma$-invariant vector
space (\ref{carnevalediRio}).
\par
A two-vector of matrices can be thought of also as a matrix of two-vectors: that is,
$\mathcal{P}=\mathcal{Q}\otimes{\rm Hom}(R,R)={\rm Hom}(R,\mathcal{Q}\otimes R)$. Decomposing the regular
representation, $R=\bigoplus_{\nu=0}^{r} n_{\mu} D_{\mu}$ into irreps, using eq.\,(\ref{grouptheory16}) and
Schur's lemma, we obtain:
\begin{equation}
\mathcal{S}=\bigoplus_{\mu,\nu} A_{\mu,\nu}{\rm Hom}(\mathbb{C}^{n_{\mu}},\mathbb{C}^{n_{\nu}})\,\, .
\label{defmuastratta}
\end{equation}
The dimensions of the irreps,  $n_{\mu}$ are dispayed in figs. \ref{dynfigure1A},\ref{dynfigure2A}. From
eq.\,(\ref{defmuastratta}) the real dimension of $\mathcal{S}$ follows immediately: $\dim\,
\mathcal{S}=\sum_{\mu,\nu}2 A_{\mu\nu} n_{\mu}n_{\nu}$ implies, recalling that $A=2\times\mathbf{1}-\bar c$
[see eq.\,(\ref{grouptheory16})] and that for the  extended Cartan matrix $\bar c n =0$:
\begin{equation}
\dim_{\mathbb{C}}\, \mathcal{S}=2\sum_{\mu}n_{\mu}^2= 2 |\Gamma |\,\, . \label{dimm}
\end{equation}
Intrinsically, one writes the space $\mathcal{S}$ as:
\begin{equation}\label{carillo}
    \mathcal{S}\, \simeq \, \mathrm{Hom}_{\,\Gamma}\left(R,\mathcal{Q}\otimes R \right)
\end{equation}
So we can summarize the discussion by saying that:
\begin{equation}\label{summillus}
    \mathrm{dim}_\mathbb{C} \, \left[\mathrm{Hom}_{\,\Gamma}\left(R,\mathcal{Q}\otimes R \right)\right] \, =
    \, 2 \, |\Gamma|
\end{equation}
The quaternionic structure of the flat manifolds $\mathcal{P}$ and $\mathcal{S}$ can be seen by simply
writing their elements as follows:
\begin{equation}
 p \,= \,\twomat{A}{iB^{\dagger}}{iB}{A^{\dagger}} \,\in \, \mathrm{Hom}\left(R,\mathcal{Q}\otimes R\right)
 \hskip 1cm A,B\in {\rm End}(R)\,\, .
\label{pquaternions}
\end{equation}
Then the  hyperK\"ahler  forms and the  hyperK\"ahler  metric are defined by the following formulae:
\begin{eqnarray}
  \Theta&=&\mathrm{Tr} (\mathrm{d}p^\dagger\wedge \mathrm{d}p) \,=\, \left(\begin{array}{cc}
                                                   {\rm i} \,\mathbf{K}  &   {\rm i} \overline{\pmb \Omega}  \\
                                                   {\rm i} \, \pmb{\Omega} & -{\rm i} \,\mathbf{K} \\
                                                 \end{array}\right)\nonumber \\
  ds^2 \times \mathbf{1}& = &\mathrm{Tr}(\mathrm{d}p^\dagger \otimes \mathrm{d}p)\label{palestrus}
\end{eqnarray}
In the above equations the trace is taken over the matrices belonging to ${\rm End}(R)$ in each entry of the
quaternion. From eq.\,(\ref{palestrus}) we extract the explicit expressions for the K\"ahler 2-form
$\mathbf{K}$ and the holomorphic 2-form $\pmb{\Omega}$ of the flat  hyperK\"ahler  manifold
$\mathrm{Hom}\left(R,\mathcal{Q}\otimes R\right)$. We have:
\begin{eqnarray}
\mathbf{K} &=& -{\rm i} \left[\mbox{Tr}\left(\mathrm{d}A^\dagger\wedge \mathrm{d}A\right) +
\mbox{Tr}\left(\mathrm{d}B^\dagger\wedge \mathrm{d}B\right) \right]
\, \equiv \, {\rm i} g_{\alpha{\bar \beta}} \, \mathrm{d}q^\alpha \wedge \mathrm{d}q^{\bar \beta} \nonumber\\
ds^2  &=& g_{\alpha{\bar \beta}} \,\mathrm{d}q^\alpha \otimes \mathrm{d}q^{\bar \beta}  \nonumber\\
\pmb{\Omega}& = & 2 \mbox{Tr}\left(\mathrm{d}A\wedge \mathrm{d}B\right) \, \equiv \,\Omega_{\alpha\beta} \,
\mathrm{d}q^\alpha \wedge \mathrm{d}q^{\beta}\label{ermengarda}
\end{eqnarray}
Starting from the above written formulae,  by means of an elementary calculation one verifies that both the
metric and the  hyperK\"ahler  forms are invariant with respect to the action of the discrete group $\Gamma$
defined in eq.\,(\ref{gammaactiononp}). Hence one can consistently reduce the space
$\mathrm{Hom}\left(R,\mathcal{Q}\otimes R\right)$ to the invariant space $\mathrm{Hom}_{\,
\Gamma}\left(R,\mathcal{Q}\otimes R\right)$ defined in eq.\,(\ref{carnevalediRio}). The  hyperK\"ahler
$2$-forms and the metric of the flat space $\mathcal{S}$, whose real dimension is $4|\Gamma|$,  are given by
eqs.(\ref{ermengarda}) where the matrices $A,B$ satisfy the invariance condition eq.\,(\ref{invariancecond}).
\subsubsection{Solution of the invariance constraint in the case of the cyclic
groups $A_k$} The space $\mathcal{S}$ can be easily described when $\Gamma$ is the cyclic group $A_{k}= \mathbb{Z}_{k+1},$ whose multiplication
 table can be read off. We can
immediately read it off from the matrices of the regular representation. Obviously, it is sufficient to
consider the representative of the first element $e_1$, as $R(e_j)=(R(e_1))^j$.
\par
One has:
\begin{equation}
R(e_1)=\left(\begin{array}{ccccc}0&0&\cdots &0&1\cr 1&0&\cdots &0&0\cr 0&1&\cdots &0&0\cr \vdots &\vdots
&\ddots &\vdots &\vdots\cr 0&0&\cdots &1&0\end{array}\right) \label{r1offdiag}
\end{equation}
Actually, the invariance condition eq.\,(\ref{invariancecond}) is best solved by changing basis so as to
diagonalize the regular representation, realizing explicitly its decomposition in terms of the $k$
unidimensional irreps. Let $\nu=e^{2\pi i\over {k+1}}$, be a $(k+1)$th root of unity  so that $\nu^{k+1}=1$.
The looked for change of basis is performed by means of the matrix:
\begin{eqnarray}\label{canovaccio}
    S_{ij} & = & \frac{1}{\sqrt{k+1}} \, \nu^{ij} \quad; \quad i,j \, = \, 0,1,2,\dots , k \nonumber\\
    \left(S^{-1}\right)_{ij}\, = \, \left(S^\dagger\right)_{ij} & = & \frac{1}{\sqrt{k+1}} \, \nu^{k+1-ij}
\end{eqnarray}
In the new basis we find:
\begin{eqnarray}\label{caramellina}
\widehat{R}(e_0)& \equiv & S^{-1}\, R(e_0)\, S \, = \, \left(\begin{array}{ccccc}
                                                                     1 & 0 & \ldots & 0 & 0 \\
                                                                     0 & 1 & 0 \ldots & 0 & 0 \\
                                                                     \vdots & \vdots & \ddots & \vdots &
                                                                     \vdots \\
                                                                     0 & 0 & \ldots & 1 & 0 \\
                                                                     0 & 0 & \ldots & 0 & 1
                                                                   \end{array}
     \right)\nonumber\\
    \widehat{R}(e_1)& \equiv & S^{-1}\, R(e_1)\, S \, = \, \left(\begin{array}{ccccc}
                                                                     1 & 0 & \ldots & 0 & 0 \\
                                                                     0 & \nu & 0 \ldots & 0 & 0 \\
                                                                     \vdots & \vdots & \ddots & \vdots &
                                                                     \vdots \\
                                                                     0 & 0 & \ldots & \nu^{k-1} & 0 \\
                                                                     0 & 0 & \ldots & 0 & \nu^{k}
                                                                   \end{array}
     \right)
\end{eqnarray}
Eq.\,(\ref{caramellina}) displays on the diagonal the representatives of $e_j$ in the one-dimensional irreps.
\par
In the above basis, the explicit solution of eq.\,(\ref{invariancecond}) is given  by
\begin{equation}
A=\left(\begin{array}{ccccc}0&u_0&0&\cdots &0\cr 0&0&u_1&\cdots &0\cr \vdots&\vdots& \vdots&\ddots &\vdots\cr
\vdots &\vdots &\vdots &\null & u_{k-1}\cr u_{k}&0&0&\cdots &0\end{array}\right) \hskip 0.3cm ;\hskip 0.3cm
B=\left(\begin{array}{ccccc}0&0&\cdots &\cdots&v_k\cr v_0&0&\cdots &\cdots &0\cr 0&v_1&\cdots &\cdots &0\cr
\vdots &\vdots &\ddots &\null &\vdots \cr 0&0&\cdots &v_{k-1}&0\end{array}\right) \label{invariantck}
\end{equation}
We see that these matrices are parameterized in terms of $2k+2$ complex, i.e. $4(k+1)=|A_{k}|$ real
parameters. In the $D_{k+2}$ case, where the regular representation is $4k$-dimensional, choosing
appropriately a basis, one can solve analogously eq.\,(\ref{invariancecond}); the explicit expressions are too
large, so we do not write them. The essential point is that the matrices $A$ and $B$ no longer correspond to
two distinct set of parameters, the group being nonabelian.
\subsection{The gauge group for the quotient and its moment maps}
The next step in the Kronheimer construction of the ALE manifolds is the determination of the group
$\mathcal{F}$ of triholomorphic isometries with respect to which we will perform the quotient. We borrow from
physics the nomenclature \textit{gauge group} since in a $\mathcal{N}=3,4$ rigid three-dimensional gauge
theory where the space $\mathrm{Hom}_{\, \Gamma}\left(R,\mathcal{Q}\otimes R\right)$ is the flat manifold of
hypermultiplet scalars, the triholomorphic moment maps of $\mathcal{F}$  emerge as scalar dependent nonderivative terms in the hyperino supersymmetry transformation rules generated by the \textit{gauging} of the
group $\mathcal{F}$.
\par
 Consider the action of $\mathrm{SU(|\Gamma |)}$ on $\mathrm{Hom}\left(R,\mathcal{Q}\otimes R\right)$ given,
 using the quaternionic notation
for the elements of $\mathrm{Hom}\left(R,\mathcal{Q}\otimes R\right)$, by
\begin{equation}
\forall g\in \mathrm{SU(|\Gamma |)}\quad , \quad g \quad :\quad
\left(\begin{array}{cc} A & i\,B^{\dagger} \\
i\,B & A^{\dagger}\\ \end{array}\right) \longmapsto
\left(\begin{array}{cc} gAg^{-1}& i \, g\, B^\dagger\,g^{-1}\\
i\, g\, B \, g^{-1} & g\, A^{\dagger}\, g^{-1}\end{array}\right)\label{sunaction}
\end{equation}
It is easy to see that this action is a triholomorphic isometry of $\mathrm{Hom}\left(R,\mathcal{Q}\otimes
R\right)$. Indeed both the  hyperK\"ahler  forms $\Theta$ and the metric $ds^2$ are invariant.
\par
Let $\mathcal{F}\subset \mathrm{SU(|\Gamma |)}$ be the subgroup of the above group which {\it commutes with
the action of $\Gamma$ on the space $\mathrm{Hom}\left(R,\mathcal{Q}\otimes R\right)$}, action which was
defined in eq.\,(\ref{gammaactiononp}). Then the action of $\mathcal{F}$ descends to $\mathrm{Hom}_{\,
\Gamma}\left(R,\mathcal{Q}\otimes R\right)\subset\mathrm{Hom}\left(R,\mathcal{Q}\otimes R\right)$ to give a
{\it triholomorphic isometry}: indeed the metric and the  hyperK\"ahler  forms on the space $\mathrm{Hom}_{\,
\Gamma}\left(R,\mathcal{Q}\otimes R\right)$ are just the restriction of those on
$\mathrm{Hom}\left(R,\mathcal{Q}\otimes R\right)$. Therefore one can take the  hyperK\"ahler  quotient of
$\mathrm{Hom}_{\, \Gamma}\left(R,\mathcal{Q}\otimes R\right)$ with respect to $\mathcal{F}$.
\par
Let $\{f_A\}$ be a basis of generators for $\mathbb{F}$, the Lie algebra of $\mathcal{F}$. Under the
infinitesimal action of $f=\mathbf{1}+\lambda^A f_A\in \mathbb{F}$, the variation of $p\in \mathrm{Hom}_{\,
\Gamma}\left(R,\mathcal{Q}\otimes R\right)$ is $\delta p= \lambda^A\delta_A p$, with
\begin{equation}
\delta_A p = \twomat{[f_A,A]}{i[f_A,B^{\dagger}]}{i[f_A,B]} {[f_A,A^{\dagger}]}\, \label{deltaam}
\end{equation}
The components of the momentum map are then given by
\begin{equation}
\mu_A=\mathrm{Tr}\,( q^\dagger\,\delta_A p)\,\,\,\equiv\,\,\,
\mathrm{Tr}\,\twomat{f_A\,\mu_3(p)}{f_A\,\mu_-(p)}{f_A\,\mu_+(p)}{f_A\,\mu_3(p)} \label{momentummatrix}
\end{equation}
so that the real and holomorphic maps $\mu_3:\mathrm{Hom}_{\, \Gamma}\left(R,\mathcal{Q}\otimes
R\right)\rightarrow\mathbb{F}^*$ and $\mu_+:\mathrm{Hom}_{\, \Gamma}\left(R,\mathcal{Q}\otimes
R\right)\,\rightarrow\,\mathbb{C} \times\mathbb{F}^*$ can be represented as matrix-valued maps:
\begin{eqnarray}
\mu_3(p)&=&-i\left([A,A^{\dagger}]+[B,B^{\dagger}]\right)\nonumber\\
\mu_+(p)&=&\left([A,B]\right)\,\, . \label{momentums}
\end{eqnarray}
In this way we get:
\begin{equation}\label{carontasco}
    \mu_A \, = \, \left(\begin{array}{cc}
                          \mathfrak{P}^3_A & \mathfrak{P}_A^- \\
                          \mathfrak{P}_A^+  & -\mathfrak{P}^3_A
                        \end{array}
     \right)
\end{equation}
where:
\begin{eqnarray}
  \mathfrak{P}^3_A &=& - {\rm i} \, \left[\mbox{Tr} \left(\left[ A \, , \, A^\dagger \right] \, f_A \right)
  + \mbox{Tr} \left(\left[ B^\dagger \, , \, B \right] \, f_A \right)\right]\nonumber \\
  \mathfrak{P}^+_A &=& \mbox{Tr} \left(\left[ A \, , \, B \right] \, f_A \right) \label{cubicolario}
\end{eqnarray}
Let $\mathfrak{Z}^\star$ be the dual of the center of $\mathbb{F}$.
\par
In correspondence with a level $\zeta=\{\zeta^3,\zeta^+\}\in{\mathbb R}^3\otimes\mathfrak{Z}^\star$ we can form
the  hyperK\"ahler  quotient:
\begin{equation}\label{grancassone}
    \mathcal{M}_{\zeta}\equiv\mu^{-1}(\zeta)\, /\!\!/_{\null_{HK}}\, \mathcal{F}
\end{equation}
{\it Varying $\zeta$ and $\Gamma$ all ALE manifolds can be obtained as $\mathcal{M}_{\zeta}$}.
\par
First of all, it is not difficult to check that $\mathcal{M}_{\zeta}$ is four-dimensional. Let us see how
this happens. There is a nice characterization of the group $\mathcal{F}$ in terms of the extended Dynkin
diagram associated with $\Gamma$. We have
\begin{equation}
\mathcal{F}=\bigotimes_{\mu=1}^{r+1} \mathrm{U(n_{\mu})}\,\bigcap \, \mathrm{SU(|\Gamma|)} \label{formofF}
\end{equation}
where the sum is extended to all the irreducible representations of the group $\Gamma$ and $n_\mu$ are their
dimensions. One should also take into account that the determinant of all the  elements must be one, since
$\mathcal{F}\subset \mathrm{SU(|\Gamma |)}$. Pictorially the group $\mathcal{F}$ has a $\mathrm{U(n_{\mu})}$
factor for each dot of the diagram, $n_{\mu}$ being associated with the dots as in figs.
\ref{dynfigure1A},\ref{dynfigure2A}. $\mathcal{F}$ acts on the various \textit{components} of
$\mathrm{Hom}_{\, \Gamma}\left(R,\mathcal{Q}\otimes R\right)$ that are in correspondence with the edges of
the diagram, see eq.\,(\ref{defmuastratta}), as dictated by the
 diagram structure. From eq.\,(\ref{formofF}) it is immediate to derive:
\begin{equation}\label{cargnolabato}
    \mbox{dim}\, \mathcal{F}=\sum_{\mu} n_{\mu}^2 -1 = |\Gamma |-1
\end{equation}
It follows that
\begin{equation}
\mbox{dim}_\mathbb{R}\,\mathcal{M}_{\zeta}=\dim_\mathbb{R}\, \mathrm{Hom}_{\,
\Gamma}\left(R,\mathcal{Q}\otimes R\right) -4\, \mbox{dim}_\mathbb{R}\, \mathcal{F} = 4|\Gamma | -(4|\Gamma
|-1)=4\, \label{dimxzeta}
\end{equation}
 Analyzing the construction we see that there are two steps.
In the first step, by setting the holomorphic part of the moment map to its level $\zeta$, we define an
algebraic locus in $\mbox{Hom}_\Gamma(\mathcal{Q}\otimes R,R)$. Next the K\"ahler quotient further reduces
such a locus to the necessary complex dimension 2. The two steps are united in one because of the
triholomorphic character of the isometries. As we already stressed in the previous section,
 in complex dimension 3 the isometries are not triholomorphic rather just holomorphic;  hence the holomorphic part
 of the moment map does not exist and the two steps have to be separated.
 There must be  another principle that leads to  impose those constraints that cut  out the algebraic locus
 $\mathbb{V}_{|\Gamma|+2}$ of which
 we perform  the K\"ahler quotient in the next step (see eq.\,(\ref{belgone})).
 The main question is to spell out such principles. As anticipated,  equation
 $\mathbf{p}\wedge \mathbf{p}\, = \, 0$
 is the one that does the job. We are not able to reduce the $3|\Gamma|^2$ quadrics on $3|\Gamma|$ variables
 to an ideal with $2|\Gamma|-2$ generators, yet we know that such reduction must
 exist. Indeed, by means of another argument
 that utilizes Lie group orbits we can show that  there is a variety of complex dimension 3, named
 $\mathcal{D}^0_\Gamma$ which is in the kernel of the equation $\mathbf{p}\wedge \mathbf{p} \, = \, 0$.
\subsubsection{The triholomorphic moment maps in the $A_k$ case of Kronheimer construction}
\label{CDdiscussia}
 The structure of $\mathcal{F}$ and the momentum map for its action are very simply worked
out in the $A_{k}$ case. An element $f\in \mathcal{F}$ must commute with the action of $A_{k}$ on
$\mathcal{P}$, eq.\,(\ref{gammaactiononp}), where the two-dimensional representation in the l.h.s. is given by:
\begin{equation}
 \Gamma (A_{k})\, \ni \, \gamma_\ell \, = \, {\cal Q}_\ell \, \equiv \, \twomat {e^{2\pi i
\ell/(k+1)}}{0}{0}{e^{-\, 2\pi i \ell /(k+1)}} \quad ;\quad\{ \ell=1,.....,k+1\}\, \label{grouptheory7}
\end{equation}
Then $f$ must have the form
\begin{equation}
f={\rm diag} (e^{i\varphi_0},e^{i\varphi_1},\ldots ,e^{i\varphi_{k}}) \hskip 0.2cm ; \hskip 0.2cm
\sum_{i=0}^k \varphi_{i}=0\,\, . \label{Fforck}
\end{equation}
Thus $\mathbb{F}$ is just the algebra of diagonal traceless $k+1$-dimensional matrices, which is
$k$-dimensional. Choose a basis of generators for $\mathbb{F}$, for instance:
\begin{eqnarray}
  f_1 &=&\diag(1,-1,0,\ldots,0) \nonumber \\
  f_2 &=& \diag (1,0,-1,0,\ldots,0)\nonumber \\
  \dots &=& \dots \nonumber \\
 f_{k}&=&\diag (1,0,0,\ldots,0,-1) \label{lavailmuso}
\end{eqnarray}
From eq.\,(\ref{cubicolario}) we immediately obtain the components of the momentum map:
\begin{eqnarray}
\mathfrak{P}^3_A&=&|u^0|^2-|u^k|^2-|v_0|^2+|v_k|^2 +\left( |u^{A-1}|^2-|u^A|^2-|v_{A-1}|^2 +|v_{A}|^2\right)
\nonumber \\
\mathfrak{P}^+_A&=&u^0 v_0 - u^k v_k +\left(u^{A-1}v_{A-1}\, - \, u^A \, v_A\right)\quad ,\quad (A=1,\dots
,k)\label{momentummapck}
\end{eqnarray}
\subsection{Level sets and Weyl chambers} If $\mathcal{F}$ acts freely on $\mu^{-1}(\zeta)$ then
$\mathcal{M}_{\zeta}$ is a smooth manifold. This happens or does not happen depending on the value of
$\zeta$. A simple characterization of $\mathfrak{Z}$ can be given in terms of the simple Lie algebra
$\mathbb{G}$ associated with $\Gamma$. There exists an isomorphism between $\mathfrak{Z}$ and the Cartan
subalgebra $\mathcal{H}_{CSA}\subset \mathbb{G}$. Thus we have
\begin{eqnarray}
\dim\, \mathfrak{Z}=\dim\,\mathcal{H}_{CSA} &=&{\rm rank}\,\mathbb{G}\nonumber\\
&=&\#{\rm \, of\,\,\,nontrivial\,\,\,conj.\,\,\,classes\,\,\,in}\,\,\,\Gamma \,\,\label{golosaidea}
\end{eqnarray}
The space $\mathcal{M}_{\zeta}$ turns out to be singular when, under
the above identification $\mathfrak{Z}\sim\mathcal{H}_{CSA}$, any of
the level components $\zeta^i\in {\mathbb R}^3\otimes \mathfrak{Z}$ lies
on a wall of a Weyl chamber. In particular, as the point
$\zeta^i=0$, ($i=1,\dots,r$) is identified with the origin of the
root space, which lies of course on all the walls of the Weyl
chambers, {\it the space $\mathcal{M}_0$ is singular}. Not too
surprisingly we will see in a moment that $\mathcal{M}_0$
corresponds to the {\it orbifold limit} $\mathbb{C}^2/\Gamma$ of a
family of ALE manifolds with boundary at infinity
$\mathbb{S}^3/\Gamma$.
\par
To verify this statement in general let us choose the natural basis for the regular representation $R$, in
which the basis vectors $e_{\delta}$ transform as in eq.\,(\ref{grouptheory15}). Define the space $L\subset
\mathcal{S}$ as follows:
\begin{equation}\label{thespacel}
L \,= \, \left\{\left(\begin{array}{c} C\\ D \end{array}\right)\in\mathcal{S}\,/\,C,D \,\,{\rm
are\,\,diagonal\,\,in\,\,the\,\,basis\,\,}\left\{e_{\delta}\right\} \right\}
\end{equation}
For every element $\gamma\in\Gamma$ there is a pair of numbers $(c_{\gamma},d_{\gamma})$ given by the
corresponding entries of $C,D$\,:\,\, $C\cdot e_{\gamma}=c_{\gamma}e_{\gamma}$, $D\cdot e_{\gamma}=d_{\gamma}
e_{\gamma}$. Applying the invariance condition eq.\,(\ref{invariancecond}), which is valid since
$L\subset\mathcal{S}$, we obtain:
\begin{equation}
\left(\begin{array}{c} c_{\gamma\cdot\delta}\\ d_{\gamma\cdot\delta} \end{array}\right) =
\left(\begin{array}{cc}u_{\gamma}&{i\bar v_{\gamma}}\\ {iv_{\gamma}}&{\bar u_{\gamma}} \end{array} \right)
\left(\begin{array}{c} c_{\delta}\\
d_{\delta} \end{array}\right)\label{orbitofthepair}
\end{equation}
We can identify $L$ with $\mathbb{C}^2$ associating for instance $(C,D)\in L \longmapsto (c_0,d_0)\in
\mathbb{C}^2$. Indeed all the other pairs $(c_{\gamma}, d_{\gamma})$ are determined in terms of
eq.\,(\ref{orbitofthepair}) once $(c_0,d_0)$ are given. By eq.\,(\ref{orbitofthepair}) the action of $\Gamma$ on
$L$ induces exactly the action of $\Gamma$ on $\mathbb{C}^2$ provided by  its two-dimensional defining
representation inside $\mathrm{SU(2)}$. It is quite easy to show the following fundamental fact: {\it each
orbit of $\mathcal{F}$ in $\mu^{-1}(0)$ meets $L$ in one orbit of $\Gamma$}. Because of the above
identification between $L$ and $\mathbb{C}^2$, this leads to conclude that {\it $\mu^{-1}(0)/\mathcal{F}$ is
isometric to $\mathbb{C}^2/\Gamma$}. Instead of reviewing the formal proofs of these statements as devised by
Kronheimer, we will verify them explicitly in the case of the cyclic groups, giving a description which sheds
some light on the {\it deformed} situation; that is we show in which way a nonzero level $\zeta^+$ for the
holomorphic momentum map puts $\mu^{-1}(\zeta)$ in correspondence with the affine hypersurface in
$\mathbb{C}^3$ cut out by the polynomial constraint (\ref{grouptheory21})  which is a deformation  of that
describing the $\mathbb{C}^2/\Gamma$ orbifold, obtained for $\zeta^+=0$.
\subsubsection{Retrieving the polynomial constraint from the  hyperK\"ahler  quotient in the  $\Gamma$=$A_k$ case.}
We can directly realize $\mathbb{C}^2/\Gamma$ as an affine algebraic surface in $\mathbb{C}^3$ by expressing
the coordinates $x$, $y$ and $z$ of $\mathbb{C}^3$ in terms of the matrices $(C,D)\in L$. The explicit
parametrization of the matrices in ${\cal S}$ in the $A_{k}$ case, which was given in eq.\,(\ref{invariantck})
in the basis in which the regular representation $R$ is diagonal, can be conveniently rewritten in the
\textit{natural basis} $\left\{e_{\gamma} \right\}$ via the matrix $S^{-1}$ defined in eq.\,(\ref{canovaccio}).
The subset $L$ of diagonal matrices $(C,D)$ is given by:
\begin{equation}
C=c_0\, {\rm diag}(1,\nu,\nu^2,\ldots,\nu^{k}),\hskip 12pt D=d_0\, {\rm
diag}(1,\nu^{k},\nu^{k-1},\ldots,\nu), \label{unouno}
\end{equation}
This is nothing but the fact that $ L \sim \mathbb{C}^2$. The set of pairs $\left(\begin{array}{c}\nu^m c_0\cr
\nu^{k-m}d_0\end{array} \right)$, $m=0,1,\ldots,k$ is an orbit of $\Gamma$ in $\mathbb{C}^2$ and determines
the corresponding orbit of $\Gamma$ in $L$. To describe $\mathbb{C}^2 / A_{k}$ one needs to identify a
suitable set of invariants $(u,w,z)\in \mathbb{C}^3$ such that
\begin{equation}\label{wgammus}
  0\,= \,  W_\Gamma (u,w,z) \, \equiv \, u^2+w^2\, - \, z^{k+1}
\end{equation}
To this effect we define:
\begin{equation}\label{transformillina}
    u= \ft 12 \left(x+y\right) \quad ; \quad w = -{\rm i}\ft 12 \, \left(x-y\right) \quad \Rightarrow
    \quad xy \, = \, u^2 + w^2
\end{equation}
and we make the following ansatz:
\begin{equation}
x=\det\, C \hskip 12pt ;\hskip 12pt y=\det\, D, \hskip 12pt ;\hskip 12pt z=\frac{1}{k+1} \mathrm{Tr} \,CD.
\label{identifyAk}
\end{equation}
This guess is immediately confirmed  by the study of the deformed surface. We know that there is a one-to-one
correspondence between the orbits of $\mathcal{F}$ in $\mu^{-1}(0)$ and those of $\Gamma$ in $L$. Let us
realize this correspondence  explicitly.
\par
Choose the basis where $R$ is diagonal. Then $(A,B) \in {\cal S}$ have the form of eq.\,(\ref{invariantck}).
The relation $xy=z^{k+1}$  holds also true when, in eq.\,(\ref{identifyAk}), the pair $(C,D)\in L$ is replaced
by an element $(A,B)\in \mu^{-1}(0)$.
\par
To see this, let us describe the elements $(A,B)\in\mu^{-1}(0)$. We have to equate the right hand sides of
eq.\,(\ref{momentums}) to zero. We note that:
\begin{equation}\label{candelinusA1}
    [A,B]=0 \quad \Rightarrow \quad v_i={u_0v_0\over u_i} \quad \forall i
\end{equation}
Secondly,
\begin{equation}\label{candelinusA2}
    [A,A^\dagger]+[B,B^\dagger]=0 \quad \Rightarrow \quad |u_i|=|u_j| \, \mbox{and} \,|v_i|=|v_j| \quad \forall i,j
\end{equation}
From the previous two equations we conclude that:
\begin{equation}\label{candelinusA3}
    u_j=|u_0|{\rm e}^{i\phi_j} \quad ; \quad v_j=|v_0|{\rm e}^{i\psi_j}
\end{equation}
Finally:
\begin{equation}\label{candelinusA4}
    [A,B]=0 \quad \Rightarrow \quad \psi_j=\Phi- \phi_j \quad \forall j
\end{equation}
where $\Phi$ is an arbitrary overall phase.
\par
In this way, we have characterized $\mu^{-1}(0)$ and we immediately check that the pair $(A,B)\in\mu^{-1}(0)$
satisfies $xy=z^{k+1}$ if $x=\det\,A$, $y=\det\,B$ and $z= 1/(k+1) \, \mathrm{Tr} \,AB$ as we have proposed
in eq.\,(\ref{identifyAk}).
\par
After this explicit solution of the momentum map constraint has been implemented we are left with $k+4$
parameters, namely the $k+1$ phases $\phi_j$, $j=0,1,\ldots k$, plus the absolute values $|u_0|$ and $|v_0|$
and the overall phase $\Phi$. So we have:
\begin{equation}\label{pralasso}
    {\rm dim}\,\mu^{-1}(0)={\rm dim}\,{\cal S}-3 \,{\rm dim}\,\mathcal{F}=
4|\Gamma|-3(|\Gamma|-1)=|\Gamma|+3
\end{equation}
where $|\Gamma|=k+1$.
\par
Now we perform the quotient of $\mu^{-1}(0)$ with respect to $ \mathcal{F} $. Given a set of phases $f_i$
such that $\sum_{i=0}^{k}f_i=0\, {\rm mod} \, 2\pi$ and given $f={\rm diag} ({\rm e}^{if_0},{\rm
e}^{if_1},\ldots,{\rm e}^{if_{k}})\in  \mathcal{F} $, the orbit of $ \mathcal{F} $ in $\mu^{-1}(0)$ passing
through  $\left(\begin{array}{c}A\cr B\end{array}\right)$ has the form $\left(\begin{array}{c}fAf^{-1}\cr
fBf^{-1}\end{array}\right)$.
\par
Choosing $f_j=f_0+j\psi+\sum_{n=0}^{j-1}\phi_n$, $j=1,\ldots,k$, with $\psi=-{1\over k}\sum_{n=0}^{k}\phi_n$,
and $f_0$ determined by the condition $\sum_{i=0}^{k}f_i=0\, {\rm mod} \, 2\pi$, one obtains
\begin{equation}
fAf^{-1}=a_0\left(\begin{array}{ccccc} 0 & 1 & 0 & \ldots & 0\cr 0 & 0 & 1 & \ldots & 0\cr
  & \ldots  & & \dots  &  \cr
0 & 0  &  \dots & 0 & 1 \cr 1 & 0 & 0 & \dots & 0 \cr \end{array}\right)\, ,  \hskip 10pt
fBf^{-1}=b_0\left(\begin{array}{ccccc} 0 & 0  &  \dots & 0 & 1 \cr 1 & 0 & 0 & \dots & 0 \cr 0 & 1 & 0 &
\ldots & 0\cr
  & \ldots  & & \dots  &  \cr
0 & \ldots & 0 & 1 & 0\cr \end{array}\right) \label{op}
\end{equation}
where $a_0=|u_0|{\rm e}^{i\psi}$ and $b_0=|v_0|{\rm e}^{i(\Phi-\psi)}$. Since the phases $\phi_j$ are
determined modulo $2\pi$, it follows that $\psi$ is determined modulo $2\pi\over {k+1}$. Thus we can say
$(a_0,b_0)\in {\mathbb{C}^2/\Gamma}$. This is the one-to-one correspondence between $\mu^{-1}(0)/\mathcal{F}$
and $\mathbb{C}^2/\Gamma$.
\par
Next we derive the deformed relation between the invariants $x,y,z$. It fixes the correspondence between the
resolution of the singularity performed in the momentum map approach and the resolution performed on the
hypersurface $xy=z^{k+1}$ in $\mathbb{C}^3$. To this purpose, we focus on the holomorphic part of the
momentum map, i.e.\ on the equation:
\begin{eqnarray}
    [A,B]=\Lambda_0 & = & {\rm
diag}(\lambda_0,\lambda_1,\lambda_2,\ldots,\lambda_{k})\in \mathfrak{Z}\otimes\mathbb{C}\label{cruspilliA}\\
\lambda_{0}&=&-\sum_{i=1}^{k}\lambda_i\label{cruspilliB}
\end{eqnarray}
Let us recall the expression (\ref{invariantck}) for the matrices $A$ and $B$. Naming $a_i=u_iv_i$,
eq.\,(\ref{cruspilliA}) implies:
\begin{equation}\label{minestrarancida}
    a_i=a_0+\lambda_i \quad ; \quad i=1,\ldots,k
\end{equation}
Let $\Lambda={\rm diag}(\lambda_1,\lambda_2,\ldots,\lambda_{k})$. We have
\begin{equation}
xy=\det A \, \det B= a_0\, \Pi_{i=1}^{k}(a_0+\lambda_i)=a_0^{k+1}\,\det \left(1+{1\over {a_0}}
\Lambda\right)=\sum_{i=0}^{k}a_0^{k+1-i}S_i(\Lambda) \label{def1}
\end{equation}
The $S_i(\Lambda)$ are the symmetric polynomials in the eigenvalues of $\Lambda$. They are defined by the
relation $\det (1+ t \Lambda)= \sum_{i=0}^{k} t^i S_i(\Lambda)$ and are given by:
\begin{equation}\label{symmipolli}
   S_i(\Lambda)=\sum_{j_1<j_2<\cdots<j_i}\lambda_{j_1}\lambda_{j_2} \cdots \lambda_{j_i}
\end{equation}
In particular, $S_0=1$ and $S_1=\sum_{i=1}^{k} \lambda_i$. Define $S_{k+1}(\Lambda)=0$, so that we can
rewrite:
\begin{equation}\label{svekolnyk}
    xy=
\sum_{i=0}^{k+1}a_0^{k+1-i}S_i(\Lambda)
\end{equation}
and note that
\begin{equation}
z={1\over {k+1}}\mathrm{Tr} AB=a_0+{1\over {k+1}}S_1(\Lambda).
\end{equation}
Then the desired deformed relation between $x$, $y$ and $z$ is obtained by substituting $a_0=z-{1\over k}S_1$
in (\ref{def1}), thus obtaining
\begin{eqnarray}
&&xy=\sum_{m=0}^{k+1}\sum_{n=0}^{k+1-m}\left(\begin{array}{c}k+1-m\cr n\end{array}\right) \,z^n
\left(-{1\over {k+1}}S_1\right)^{k+1-m-n}S_m \, z^n=\sum_{n=0}^{k+1}
t_{n+1} \, z^n \nonumber\\
&&\Longrightarrow\hskip 10pt t_{n+1}=\sum_{m=0}^{k+1-n}\left(\begin{array}{c}k+1-m\cr n\end{array}\right)
\left(-{1\over {k+1}}S_1\right)^{k+1-m-n} \label{def2}
\end{eqnarray}
Note in particular that $t_{k+2}=1$ and $t_{k+1}=0$, i.e.
\begin{equation}\label{cromagnolo}
    xy=z^{k+1}+\sum_{n=0}^{k}t_{n+1}z^n
\end{equation}
which means that the deformation proportional to $z^{k}$ is absent. This establishes a clear correspondence
between the momentum map construction and the polynomial ring ${\mathbb{C}[x,y,z]\over \partial W}$ where
$W(x,y,z)=xy-z^{k+1}$. Moreover, note that we have only used one of the
momentum map equations, namely $[A,B]=\Lambda_0$. The equation $[A,A^\dagger]+[B,B^\dagger]=\Sigma$ has been
completely ignored. This means that the deformation of the complex structure is described by the parameters
$\Lambda$, while the parameters $\Sigma$ describe the deformation of the K\"ahler structure. The relation
(\ref{def2}) can also be written in a simple factorized form, namely
\begin{equation}
xy=\Pi_{i=0}^{k}(z-\mu_i),
\end{equation}
where
\begin{eqnarray}
\mu_i&=&{1\over k}(\lambda_1+\lambda_2+\cdots +\lambda_{i-1} -2\lambda_i+\lambda_{i+1}+\cdots+\lambda_k),
\,\,\,\,
i=1,\ldots,k-1\nonumber\\
\mu_0&=&-\sum_{i=1}^k\mu_i={1\over k}S_1.
\end{eqnarray}
\section{Generalization of the correspondence: McKay quivers for $\mathbb{C}^3/\Gamma$ singularities}
One can generalize the extended Dynkin diagrams obtained in the above way by constructing McKay quivers,
according to the following definition:
\begin{definizione}
Let us consider the quotient $\mathbb{C}^n/\Gamma$, where $\Gamma$ is a finite group that acts on
$\mathbb{C}^n$ by means of the complex representation $\mathcal{Q}$ of dimension $n$ and let $\mathrm{D}_i$,
($i=1,\dots ,r+1$) be the set of irreducible representations of $\Gamma$ having denoted by $r+1$  the number
of conjugacy classes of $\Gamma$. Let the matrix $\mathcal{A}_{ij}$ be defined by:
\begin{equation}\label{quiverro2}
    \mathcal{Q}\otimes \mathrm{D}_i \, = \, \bigoplus_{j=1}^{r+1} \, \mathcal{A}_{ij}\,\mathrm{D}_j
\end{equation}
To such a matrix we associate a quiver diagram in the following way. Every irreducible representation is
denoted by a circle  labeled with a number equal to the dimension of the corresponding irrep. Next we write an
oriented line going from circle $i$ to  circle $j$ if $\mathrm{D}_j$ appears in the decomposition of
$\mathcal{Q}\otimes \mathrm{D}_i$, namely if the matrix element $\mathcal{A}_{ij}$ does not vanish.
\end{definizione}
The analogue of the extended Cartan matrix discussed in the case of $\mathbb{C}^2/\Gamma$ is defined below:
\begin{equation}\label{caspiterina}
    \bar{c}_{ij} \, = \, n \, \delta_{ij} \, - \, \mathcal{A}_{ij}
\end{equation}
and it has the same property, namely,  it admits the vector of irrep dimensions
\begin{equation}\label{dimvecco}
    \mathbf{n}\, \equiv \, \{1,n_1,\dots,n_r\}
\end{equation}
as a null vector:
\begin{equation}\label{nullitone}
    \bar{c}.\mathbf{n}\,= \, \mathbf{0}
\end{equation}
Typically the McKay quivers encode the information determining the
interaction structure of the dual gauge theory on the brane world
volume. Indeed the bridge between Mathematics and Physics is located
precisely at this point. In the case of a single $M2$-brane, the
$n|\Gamma|$ complex coordinates ($n$=2, or 3) of the flat K\"ahler
manifold $Hom_\Gamma(R,Q\otimes R)$ are the scalar fields of the
Wess-Zumino multiplets, the unitary group $\mathcal{F}$ commuting
with the action of $\Gamma$ is the \textit{gauge group}, the moment
maps of $\mathcal{F}$ enter the definition of the potential,
according to the standard supersymmetry formulae, recalled in
section \ref{genN2theo} and the holomorphic constraints defining the
$\mathbf{V}_{|\Gamma|+2}$ variety  have to be related with the
superpotential $\mathfrak{W}$ of the $\mathcal{N}=2$ theories in d=3
(\textit{i.e.} the $n$=3 case where the singular space is
$\mathbb{C} \times \mathbb{C}^3/\Gamma$). In the case of
$\mathcal{N}=4$ theories, also in d=3, (\textit{i.e.} the $n$=2 case
where the singular space is $\mathbb{C}^2 \times
\mathbb{C}^2/\Gamma$), the holomorphic constraints
$\mathcal{P}_i(y)$ are identified with the holomorphic part of the
tri-holomorphic moment map. When one goes to the case of multiple
$M2$-branes the gauge group is enlarged by color indices. This is
another story. The first step is to understand the case of one
$M2$-brane and here the map between Physics and Mathematics is
one-to-one.
\subsection{Representations of the quivers and K\"ahler quotients}
Let us now follow the same steps of the Kronheimer construction and derive the representations of the
$\mathbb{C}^3/\Gamma$ quivers. The key point is the construction of the analogues of the spaces
$\mathcal{P}_\Gamma$ in eq.\,(\ref{ommaomma}) and of its invariant subspace $\mathcal{S}_\Gamma$ in
eq.\,(\ref{carnevalediRio}). To this effect we introduce three matrices $|\Gamma|\times|\Gamma|$ named $A,B,C$
and set:
\begin{eqnarray}
    p \in \mathcal{P}_\Gamma \, \equiv \, \mbox{Hom}\left(R,\mathcal{Q}\otimes R\right) \, \Rightarrow\,
    p\,=\, \left(\begin{array}{c}
                   A \\
                   B \\
                   C
                 \end{array}
     \right) \label{homqg}
\end{eqnarray}
The action of the discrete group $\Gamma$ on the space $\mathcal{P}_\Gamma$ is defined in full analogy with
the Kronheimer case:
\begin{equation}\label{gammazione}
    \forall \gamma \in \Gamma: \quad \gamma\cdot p \,\equiv\, \mathcal{Q}(\gamma)\,\left(\begin{array}{c}
                  R(\gamma)\, A \, R(\gamma^{-1})\\
                   R(\gamma)\, B \, R(\gamma^{-1}) \\
                  R(\gamma)\, C \, R(\gamma^{-1})
                 \end{array}
     \right)
\end{equation}
where $\mathcal{Q}(\gamma)$ denotes the three-dimensional complex representation of the group element
$\gamma$, while $R(\gamma)$ denotes its $|\Gamma|\times|\Gamma|$-matrix image in the regular representation.
\par
In complete analogy with eq.\,(\ref{carnevalediRio}) the subspace $\mathcal{S}_\Gamma$ is obtained by setting:
\begin{equation}
\mathcal{S}_\Gamma \, \equiv \, \mbox{Hom}_\Gamma\left(R,Q\otimes R\right)\, = \,
\left\{p\in\mathcal{P}_\Gamma / \forall \gamma\in\Gamma, \gamma\cdot p = p\right\}\,\,
\label{carnevalediPaulo}
\end{equation}
\par
Just as in the previous case a three-vector of matrices can be thought  as a matrix of three-vectors: that
is, $\mathcal{P}_\gamma=\mathcal{Q}\otimes{\rm Hom}(R,R)={\rm Hom}(R,\mathcal{Q}\otimes R)$. Decomposing the
regular representation, $R=\bigoplus_{i=0}^{r} n_i D_i$ into irreps, using eq.\,(\ref{quiverro2}) and Schur's
lemma, we obtain:
\begin{equation}
\mathcal{S}_\Gamma=\bigoplus_{i,j} A_{i,j}{\rm Hom}(\mathbb{C}^{n_{i}},\mathbb{C}^{n_{j}})\,\,
\label{defmuastrattaG}
\end{equation}
The properties (\ref{caspiterina},\ref{dimvecco},\ref{nullitone}) of the matrix $A_{ij}$ associated with the
quiver diagram guarantee, in perfect analogy with eq.\,(\ref{dimm})
\begin{equation} \dim_{\mathbb{C}}\,
\mathcal{S}_\Gamma\simeq \, \mathrm{Hom}_{\,\Gamma}\left(R,\mathcal{Q}\otimes R \right)\,=3\sum_{i}n_{i}^2= 3
|\Gamma |\,\, . \label{dimmus}
\end{equation}
\subsection{The quiver Lie group, its maximal compact subgroup and the K\"ahler quotient} \label{VG2p1}
We address now the most important point, namely the reduction of the $3|\Gamma|$-dimensional complex manifold
$\mbox{Hom}_\Gamma\left(R,\mathcal{Q}\otimes R\right)$ to a $|\Gamma|+2$-dimensional subvariety  of which we
will perform the K\"ahler quotient in order to obtain the final 3-dimensional (de-singularized) smooth
manifold that provides the crepant resolution. The  inspiration about  how this can be done is provided by
comparison with the $\mathbb{C}^2/\Gamma$ case, \textit{mutatis mutandis}. The key formulae to recall are the
following ones: eq.\,(\ref{momentums}), (\ref{formofF}) and (\ref{thespacel}).
\par
From eq.\,(\ref{momentums}) we see that the analytic part of the triholomorphic moment map is provided by the
projection onto the gauge group generators of the commutator $\left[A\, , \, B\right]$. When the level
parameters are all zero (namely when the locus equation is not perturbed by the elements of the chiral ring)
the outcome of the moment map equation is simply the condition $\left[A\, , \, B\right]=0$. In the case of
$\mathbb{C}^3/\Gamma$ we already know that there are no deformations of the complex structure and that the
analogue of the holomorphic moment map constraint has to be a rigid parameterless condition.  Namely the
ideal that cuts out the $\mathbb{V}_{|\Gamma|+2}$ variety should be generated by a list of quadric
polynomials $\mathcal{P}_i(y)$ fixed once and for all in a parameterless way. It is reasonable to guess that
these equations should be a  generalization of the condition $\left[A\, , \, B\right]=0$. In the
$\mathbb{C}^3/\Gamma$ case  we have three matrices $A,B,C$ and the obvious generalization is given below:
\begin{equation}\label{ganimusco}
  \mathbf{p} \wedge \mathbf{p} \, = \,0
\end{equation}
where:
\begin{eqnarray}
  \mathbf{ p} & = & \left( \begin{array}{c}
                               A \\
                               B \\
                               C
                             \end{array}
  \right) \, \in \, \text{Hom}_{\Gamma}\left(R,\mathcal{Q}\otimes R\right)\nonumber\\
  p_1 &=& A \quad ; \quad  p_2 \, = \, B \quad ; \quad p_3 \, =\, C \label{luciopatta}
\end{eqnarray}
This is a short-hand for the following explicit equations
\begin{eqnarray}
 0 &=& \epsilon^{ijk}\mathbf{ p}_i \cdot \mathbf{p}_j \nonumber\\
 &\Updownarrow& \nonumber\\
  0 &=& \left[ A,B\right] \, = \, \left[ B,C\right] \, = \, \left[ C,A\right] \label{poffarbacchio}
\end{eqnarray}
Eq.\,(\ref{ganimusco}) is the very same equation numbered (1.18) in Craw's doctoral thesis \cite{crawthesis}. We will see in a
moment that it is indeed the correct equation reducing $\mbox{Hom}_\Gamma\left(R,\mathcal{Q}\otimes R\right)$
to a $|\Gamma|+2$-dimensional subvariety. The way to understand it goes once again through a detailed
comparison with the Kronheimer case.
\par
One has to discuss the construction of the gauge group and to recall the identification of the singular
orbifold $\mathbb{C}^2/\Gamma$ with the subspace named $L$  defined by eq.\,(\ref{thespacel}). Both
constructions have a completely parallel analogue in the $\mathbb{C}^3/\Gamma$ case and these provide the key
to understand why (\ref{ganimusco}) is the right choice.
\par
Before we do that let us provide the main link between the here considered mathematical constructions and
the Physics of three-dimensional Chern-Simons gauge theories. To this purpose let us go back to the results
of \cite{pappo1}. For those special $\mathcal{N}=2$ Chern-Simons gauge theories that are actually
$\mathcal{N}=3$, the superpotential $\mathcal{W}$ has the form displayed below:
\begin{equation}\label{pappoide}
    \mathcal{W}\, = \, - \frac{1}{8\alpha} \, \mathcal{P}_+^\Lambda \,
\mathcal{P}_+^\Sigma \, \kappa_{\Lambda\Sigma}
\end{equation}
where $\mathcal{P}_+^\Lambda$ denote the holomorphic parts of the
triholomorphic moment maps and $\kappa_{\Lambda\Sigma}$ is the
Killing metric of the gauge Lie algebra. When looking for extrema at
$V=0$ of the scalar potential, namely for classical vacua of the
gauge theory, taking into account the positive definiteness of the
scalar metric $g^{\alpha\beta^\star}$ of the Killing metric
$\kappa_{\Lambda\Sigma}$ and of the matrix
$\mathbf{m}^{\Lambda\Sigma}$ one obtains the following conditions:
\begin{eqnarray}
  \mathcal{P}^\Lambda_3 &=& \zeta_3^\Lambda \label{coniglio1}\\
  \mathcal{P}^\Lambda_+ &=& \zeta_+^\Lambda \label{coniglio1bis}\
\end{eqnarray}
where $\mathcal{P}^\Lambda_3$ denotes the real part of the
tri-holomorphic moment map. In mathematical language, the above
equations just define the level set $\mu^{-1}\left(\zeta\right)$
utilized in the hyperK\"ahler quotient.
\par
The same field theoretic mechanism is realized in a gauge theory whose scalar fields span the space
$\mathcal{S}_\Gamma$ for a $\mathbb{C}^3/\Gamma$ singularity, if we introduce the following superpotential:
\begin{equation}\label{cascurtosco}
  \mathcal{W }\, = \, \mbox{Tr}\left[p_x \,p_y \, p_z\right] \, \epsilon^{xyz}
\end{equation}
With this choice the conditions for the vanishing of the scalar potential are indeed the K\"ahler moment map
equations that we are going to discuss and eq.\,(\ref{ganimusco}).
\subsubsection{Quiver Lie groups}\label{quiverino} We are
interested in determining the subgroup
\begin{equation}\label{gstorto}
  \mathcal{G}_\Gamma \,\subset \, \mathrm{SL(|\Gamma|,\mathbb{C})}
\end{equation}
made by those elements that commute with the group $\Gamma$.
\begin{equation}\label{carciofillo}
   \mathcal{G}_\Gamma  \, = \, \left\{ g \in \mathrm{SL(|\Gamma|,\mathbb{C})} \quad | \quad \forall
   \gamma \in \Gamma \,\, : \,\,
   \left[ \mathrm{D}_\mathrm{R}\left(\gamma\right)\, , \, \mathrm{D}_{\mathrm{def}}\left(g\right)\,\right]
   \, = \, 0\right\}
\end{equation}
In the above equation $D_R(\null)$ denotes the regular representation while $\mathrm{D}_{\mathrm{def}}$
denotes the defining representation of the complex linear group. The two representations,
by construction, have the same dimension and this is the reason why equation (\ref{carciofillo}) makes sense.
\par
It is sufficient to impose the defining constraint for the generators of the group on a generic matrix
depending on $|\Gamma|^2$ parameters: this reduces it to a specific matrix depending on
$|\Gamma|$-parameters. The further condition that the matrix should have determinant one, reduces the number
of free parameters  to $|\Gamma|-1$.  In more abstract terms we can say that the group $\mathcal{G}_\Gamma$
has the following general structure:
\begin{equation}\label{caliente}
\mathcal{G}_\Gamma \, = \, \bigotimes_{\mu = 1}^{r+1} \mathrm{GL(n_\mu, \mathbb{C})}\bigcap
\mathrm{SL(|\Gamma|,\mathbb{C})}
\end{equation}
This is a perfectly analogous result to that displayed in eq.\,(\ref{formofF}) for the Kronheimer case. The
difference is that there we had unitary groups while here we are talking about general linear complex groups
with a holomorphic action on the quiver coordinates. The reason is that we have not yet introduced a K\"ahler
structure on the quiver space $\mathrm{Hom}_\Gamma\left(R\, ,\,  \mathcal{Q}\otimes R\right)$: we do it
presently and we shall realize that isometries of the constructed K\"ahler metric will be only those elements
of  $\mathcal{G}_\Gamma$ that are contained in the unitary subgroup mentioned below:
\begin{equation}\label{colasciutto}
\mathcal{F}_\Gamma \,\equiv \,\bigotimes_{\mu = 1}^{r+1} \mathrm{U(n_\mu)}\bigcap \mathrm{SU(|\Gamma|)} \,
\subset \, \mathcal{G}_\Gamma
\end{equation}
\subsubsection{The holomorphic quiver group and the reduction to $V_{|\Gamma|+2}$} Yet the group $\mathcal{G}_\Gamma$ plays an important role in understanding the rationale
of the holomorphic constraint (\ref{ganimusco}).  The key item is
the coset $\mathcal{G}_\Gamma/\mathcal{F}_\Gamma$.
\par
Let us introduce some notations. Relaying on eq.\,(\ref{homqg}) we define the diagonal  embedding:
\begin{equation}\label{caziloro}
  \mathbb{D} \quad : \quad \mathrm{GL(|\Gamma|,\mathbb{C})} \, \rightarrow \, \mathrm{GL(3|\Gamma|,\mathbb{C})}
\end{equation}
\begin{equation}\label{grossoD}
  \forall M \in \mathrm{GL(|\Gamma|,\mathbb{C})} \quad ; \quad  \mathbb{D}[M] \, \equiv \, \left(
  \begin{array}{c|c|c}
  M & 0 & 0 \\
  \hline
  0 & M & 0 \\
  \hline
  0 & 0 & M \\
  \end{array}
  \right)
\end{equation}
In this notation, the invariance condition that defines $\mathcal{S}_\Gamma \, = \,
\mathrm{Hom}_\Gamma(R,\mathcal{Q}\times R)$ can be rephrased as follows:
\begin{equation}\label{cascalippo}
  \forall \gamma \in \Gamma \quad : \quad \mathcal{Q}[\gamma] \, \mathbf{p} \, = \,
  \mathbb{D}[R^{-1}_\gamma] \,\mathbf{ p} \, \mathbb{D}[R_\gamma]
\end{equation}
It is clear that any  $|\Gamma| \times |\Gamma| $ - matrix $M$ that commutes with $R_\gamma$ realizes an
automorphism of the space $\mathcal{S}_\Gamma$, namely it maps it into itself. The group $\mathcal{G}_\Gamma$
is such an automorphism group. In particular equation (\ref{ganimusco}) or alternatively
(\ref{poffarbacchio}) is invariant under the action of $\mathcal{G}_\Gamma$. Hence the locus:
\begin{eqnarray}
  \mathcal{D}_\Gamma &\subset& \mathcal{S}_\Gamma \nonumber \\
  \mathcal{D}_\Gamma &\equiv& \left\{ \mathbf{p} \in \mathcal{S}_\Gamma \,\, |\,\, \left[A,B\right]\,=\,
  \left[B ,C\right]=\left[C,A\right] \, = \, 0 \right\} \label{granitacaffe}
\end{eqnarray}
is invariant under the action of $\mathcal{G}_\Gamma$. A priori the locus $\mathcal{D}_\Gamma$ might be
empty, but this is not so because there exists an important solution of the constraint (\ref{ganimusco})
which is the obvious analogue of the space $L_\Gamma$ defined for the $\mathbb{C}^2/\Gamma$-case in
eq.\,(\ref{thespacel}). In full analogy we set:
\begin{equation}\label{thespacellone}
\mathcal{S}_\Gamma\, \supset\, L_\Gamma \,\equiv \, \left\{\left(\begin{array}{c} A_0\\ B_0\\ C_0
\end{array}\right)\in\mathcal{S}_\Gamma \,\,\mid\,\,A_0,B_0,C_0 \,\,{\mbox{
are diagonal in the natural basis of R}\, : \,}\left\{e_{\delta}\right\} \right\}
\end{equation}
Obviously diagonal matrices commute among themselves and they do the same in any other basis where they are
not diagonal, in particular in the \textit{split basis}. By definition we name in this way the basis where
the regular representation R is split into irreducible representations. A general result in finite group
theory tells us that every $n_i$-dimensional irrep $\pmb{D}_i$ appears in R exactly $n_i$-times:
\begin{equation}\label{santachiara}
    R \, = \, \bigoplus_{i=0}^{r} n_i \, \pmb{D}_i \quad ; \quad \mbox{dim}\pmb{D}_i \equiv n_i
\end{equation}
In the split basis every element $\gamma\in \Gamma$ is given by a
block diagonal matrix of the following form:
\begin{equation}\label{mascherone}
 R(\gamma)\,=\,   \left(
       \begin{array}{c|c|c|c|c|c}
         \mathbf{1} & \mathbf{0} & \mathbf{\dots} & \mathbf{\dots} & \mathbf{0} & \mathbf{1} \\
         \hline
         \mathbf{0} &
         \begin{array}{ccc}
         a_{1,1} & \dots & a_{1,n_1} \\
         \vdots & \dots & \vdots \\
         a_{n_1,1} & \dots & a_{n_1,n_1} \\
         \end{array}
          & \mathbf{0} & \dots & \dots & \mathbf{0} \\
         \hline
         \vdots & \dots & \dots & \dots & \dots & \vdots \\
         \hline
         \vdots & \dots & \dots & \dots & \dots & \vdots \\
         \hline
         \mathbf{0} & \dots & \dots & \mathbf{0} & \begin{array}{ccc}
         b_{1,1} & \dots & b_{1,n_{r-1}} \\
         \vdots & \dots & \vdots \\
         b_{n_{r-1},1} & \dots & b_{n_{r-1},n_{r-1}} \\
         \end{array} & \mathbf{0}\\
         \hline
         \mathbf{0} & \dots & \dots & \dots & \mathbf{0} & \begin{array}{ccc}
         c_{1,1} & \dots & c_{1,n_r} \\
         \vdots & \dots & \vdots \\
         c_{n_r,1} & \dots & c_{n_r,n_r} \\
         \end{array} \\
       \end{array}
     \right)
\end{equation}
In appendix \ref{exemplaria} we provide the explicit form of the matrices $A_0$,$B_0$,$C_0$ in the split
basis and for the case of several groups $\Gamma$. In analogy to what was noticed for the Kronheimer case,
the space $L_\Gamma$ has complex dimension three (in Kronheimer case it was two):
\begin{equation}\label{pereoliato}
    \mbox{dim}_\mathbb{C} \,L_\Gamma\, = \,3
\end{equation}
Indeed if we fix the first diagonal entry of each of the three matrices, the invariance condition
(\ref{cascalippo}) determines all the other ones uniquely. In any other basis the number of parameters
remains three. Let us call them $(a_0,b_0,c_0)$. Because of the above argument and, once again, in full
analogy with the Kronheimer case, we can conclude that the space $L_\Gamma$ is isomorphic to the singular
orbifold $\mathbb{C}^3/\Gamma$, the $\Gamma$-orbit of a triple $(a_0,b_0,c_0)$ representing a point in
$\mathbb{C}^3/\Gamma$.
\par
The existence of the solution of the constraint
(\ref{ganimusco}) provided by the complex three-dimensional space $L_\Gamma$ shows that we can construct a
variety of dimension $|\Gamma|+2$ which is in the kernel of the constraint (\ref{ganimusco}). This is just
the orbit, under the action of $\mathcal{G}_\Gamma$ of $L_\Gamma$. We set:
\begin{equation}\label{gneccoD}
    \mathcal{D}_\Gamma \,  \equiv \,\mbox{Orbit}_{\mathcal{G}_\Gamma}\left(L_\Gamma\right)
\end{equation}
The counting is easily done.
\begin{enumerate}
  \item A generic point in $L_\Gamma$ has the identity as stability subgroup in $\mathcal{G}_\Gamma$.
  \item The group $\mathcal{G}_\Gamma$ has complex dimension $|\Gamma|-1$, hence we get:
  \begin{equation}\label{licciu}
    \mbox{dim}_{\mathbb{C}}\left(\mathcal{D}_\Gamma\right) \, = \,|\Gamma|-1+3 = |\Gamma|+2
  \end{equation}
\end{enumerate}
\par
In the sequel we define the variety $V_{|\Gamma|+2}$ to be equal to $\mathcal{D}_\Gamma^0$.
\subsubsection{The coset $\mathcal{G}_\Gamma/\mathcal{F}_\Gamma$ and the K\"ahler quotient}
It is now high time to introduce the K\"ahler potential of the original $3|\Gamma|$-dimensional complex flat
manifold $\mathcal{S}_\Gamma$. We set:
\begin{eqnarray}\label{kalerpotent}
    \mathcal{K}_{\mathcal{S}_\Gamma} & \equiv & \mbox{Tr} \left(\mathbf{p}^\dagger \, \mathbf{p}\right)
    \,=\, \mbox{Tr}\left(A^\dagger\, A\right)\, + \,\mbox{Tr}\left(B^\dagger\, B\right)
          \, + \,\mbox{Tr}\left(C^\dagger\, C\right)
\end{eqnarray}
Using the matrix elements of $A,B,C$ as complex coordinates of the  manifold and naming $\lambda_i$ the
independent parameters from which they depend  in a given explicit solution of the invariance constraint, the
K\"ahler metric is defined, as usual, by:
\begin{eqnarray}\label{giorgini}
    ds^2_{\mathcal{S}_\Gamma} & = & g_{\ell\bar{m}}\, d\lambda^\ell \otimes d\bar{\lambda}^{ \bar{m}}
\end{eqnarray}
where:
\begin{equation}\label{cricket}
    g_{\ell\bar{m}} \, = \, \partial_\ell \, \bar{\partial}_{\bar{m}} \,\mathcal{K}
\end{equation}
From eq.\,(\ref{kalerpotent}) we easily see that the K\"ahler potential is invariant under the unitary subgroup
of the quiver group defined by:
\begin{equation}\label{unitosubbo}
  \mathcal{F}_\Gamma \, = \, \left\{M \in \mathcal{G}_\Gamma \, \mid \, M\, M^\dagger \, = \, \mathbf{1}\, \right\}
\end{equation}
whose structure was already mentioned in eq.\,(\ref{colasciutto}).
The center $\mathfrak{z}\left(\mathbb{F}_\Gamma\right)$ of the Lie
algebra $\mathbb{F}_\Gamma$ has dimension $r$, namely the same as
the number of nontrivial conjugacy classes of $\Gamma$ and it has
the following structure:
\begin{equation}\label{centratore}
  \mathfrak{z}\left(\mathbb{F}_\Gamma\right) \, = \, \underbrace{\uu(1)\oplus\uu(1)\oplus\dots\oplus\uu(1)}_{r}
\end{equation}
In the appendices we provide the explicit form of $\mathbb{F}_\Gamma$ while working out examples.
\par
Since $\mathcal{F}_\Gamma$ acts as a group of isometries on the space $\mathcal{S}_\Gamma$ we might construct
the K\"ahler quotientof the latter with respect to the former, yet we may do better.
\par
In the case of an abelian $|\Gamma|$ the center $\mathfrak{z}[\mathbb{F}]=\mathbb{F}$ coincides with the
entire gauge algebra. We discuss in detail these cases in the sequel.
\par
Let us consider the inclusion map of the variety $\mathcal{D}_\Gamma$ into $\mathcal{S}_\Gamma$:
\begin{equation}\label{includendus}
  \iota \, : \, \mathcal{D}_\Gamma \, \rightarrow \, \mathcal{S}_\Gamma
\end{equation}
and let us define as K\"ahler potential and K\"ahler metric of the locus $\mathcal{D}_\Gamma$  the pull backs
of the K\"ahler potential (\ref{kalerpotent}) and of metric (\ref{giorgini}) of $\mathcal{S}_\Gamma$, namely
let us set:
\begin{eqnarray}
 \mathcal{ K}_{\mathcal{D}_\Gamma} &\equiv&\iota^\star\, \mathcal{ K}_{\mathcal{S}_\Gamma} \label{labio1} \\
  ds^2_{\mathcal{D}_\Gamma}&=& \iota^\star\,  ds^2_{\mathcal{S}_\Gamma} \label{labio2}
\end{eqnarray}
By construction, the isometry group $\mathcal{F}_{\Gamma}$ is inherited by the pullback metric on
$\mathcal{D}_\Gamma$ and we can consider the K\"ahler quotient:
\begin{equation}\label{gioiabella}
  \mathcal{M}_\zeta \, \equiv \, \mathcal{D}_\Gamma /\!\!/^\zeta_{\mathcal{F}_{\Gamma}}
\end{equation}
Let $f_I$ be a basis of generators of  $\mathcal{F}_{\Gamma}$ ($I=1,\dots,|\Gamma|-1$) and
let $Z_i$ ($i=1,\dots,|\Gamma|+2$) be a system of complex coordinates spanning the points of $\mathcal{D}_\Gamma$.
By means of the inclusion map we have:
\begin{equation}\label{craturato}
\forall Z \in  \mathcal{D}_\Gamma \quad : \quad  \iota (Z) \, = \, \mathbf{p}(Z) \, = \,\left(
\begin{array}{c}
A(Z) \\
B(Z) \\
C(Z) \\
\end{array}
\right)
\end{equation}
The action of the \textit{gauge group} $\mathcal{F}_\Gamma$ on $\mathcal{D}_\Gamma$ is implicitly defined by:
\begin{equation}\label{kratinus}
  \mathbf{p}(\delta_I Z) \, = \, \delta_I  \mathbf{p}(Z) \, = \,\left(
  \begin{array}{c}
  \left [f_I \, , \,  A(Z)\right] \\
  \left [f_I \, , \,  B(Z)\right] \\
  \left [f_I \, , \,   C(Z)\right] \\
  \end{array}
  \right)
\end{equation}
and the corresponding real moment maps are easily calculated:
\begin{equation}\label{prosperus}
 \mu_I (Z,\bar{Z})\, = \, \mbox{Tr}\left(f_I \,\left[A(Z),A^\dagger (\bar{Z}) \right]\right)\,
 +\,\mbox{Tr}\left(f_I \,\left[B(Z),B^\dagger (\bar{Z}) \right]\right)\,
 +\,\mbox{Tr}\left(f_I \,\left[C(Z),C^\dagger (\bar{Z}) \right]\right)
\end{equation}
One defines the level sets  by means of the equation:
\begin{equation}\label{carampa}
\mu^{-1}\left( \zeta\right)\, = \, \left\{ Z \, \in \mathcal{D}_\Gamma\, \,  \parallel \, \mu_I (Z,\bar{Z})
\, = \, 0 \quad \mbox{if $f_I \not\in  \mathfrak{Z} $} \quad ; \quad \mu_I (Z,\bar{Z}) \, = \, \zeta_I \quad
\mbox{if $f_I \in  \mathfrak{Z} $}\right\}
\end{equation}
which, by construction, are invariant under the gauge group $\mathcal{F}_\Gamma$ and we can finally set:
\begin{equation}\label{fammicuccia}
\mathcal{M}_\zeta \, \equiv  \, \mu^{-1}\left( \zeta\right)/\!\!/_{\mathcal{F}_\Gamma}\, \equiv \,
\mathcal{D}_\Gamma /\!\!/^\zeta_{\mathcal{F}_{\Gamma}}
\end{equation}
The real and complex dimensions of $\mathcal{M}_\zeta$ are easily calculated. We start from $|\Gamma|+2$
complex dimensions, namely from $2|\Gamma|+4$ real dimensions. The level set equation imposes  $|\Gamma|-1$
real constraints, while the quotiening by the group action takes other $|\Gamma|-1$ parameters  away.
Altogether we remain with $6$ real parameters that can be seen as $3$ complex ones. Hence the manifolds
$\mathcal{M}_\zeta$ are always   complex three-folds that, for generic values of $\zeta$, are smooth:
supposedly the crepant resolutions of the singular orbifold. For  $\zeta \,= \, 0$ the manifold
$\mathcal{M}_0$ degenerates into the singular orbifold $\mathbb{C}^3/\Gamma$, since the solution of the
moment map equation is given by the $\mathcal{F}_\Gamma$ orbit of the locus $L_\Gamma$, namely:
\begin{equation}\label{quadriglia}
  \mu^{-1}\left(0\right) \, = \, \mbox{Orbit}_{\mathcal{F}_\Gamma}\left(L_\Gamma\right)
\end{equation}
Comparing eq.\,(\ref{gneccoD}) with eq.\,(\ref{quadriglia}) we are
led to consider the following direct sum decomposition of the Lie
algebra:
\begin{eqnarray}
  \mathbb{G}_\Gamma &=& \mathbb{F}_\Gamma \oplus \mathbb{K}_\Gamma\\
  \left[\mathbb{F}_\Gamma \, , \, \mathbb{F}_\Gamma\right] &\subset & \mathbb{F}_\Gamma \quad ; \quad
\left[\mathbb{F}_\Gamma \, , \, \mathbb{K}_\Gamma\right] \,\subset \, \mathbb{K}_\Gamma \quad ; \quad
\left[\mathbb{K}_\Gamma \, , \, \mathbb{K}_\Gamma\right] \,\subset \, \mathbb{F}_\Gamma \label{salameiolecco}
\end{eqnarray}
where $\mathbb{F}_\Gamma$ is the maximal compact subalgebra and
$\mathbb{K}_\Gamma$ denotes its complementary orthogonal subspace
with respect to the Cartan Killing metric.
\par
A special feature  of all the quiver Groups and Lie Algebras is that $\mathbb{F}_\Gamma$ and
$\mathbb{K}_\Gamma$ have the same real dimension $|\Gamma|-1$ and one can choose a basis of hermitian
generators $T_I$ such that:
\begin{equation}\label{sacherdivuli}
    \begin{array}{ccccccc}
       \forall \pmb{\Phi} \in \mathbb{F}_\Gamma & : & \pmb{\Phi} & = & {\rm i} \times \sum_{I=1}^{|\Gamma|-1} c_I T^I & ; &
       c_I \in \mathbf{R} \\
       \forall \pmb{K} \in \mathbb{K}_\Gamma & : & \pmb{K} & = & \sum_{I=1}^{|\Gamma|-1} b_I T^I & ; &
       b_I \in \mathbf{R} \\
     \end{array}
\end{equation}
Correspondingly a generic element $g\in \mathcal{G}_\Gamma$ can be split as follows:
\begin{equation}\label{consolatio}
   \forall g \in \mathcal G_\Gamma \quad : \quad g=\mathcal{U} \, \mathcal{H} \quad  ; \quad \mathcal{U} \in
   \mathcal{F}_\Gamma \quad ; \quad  \mathcal{H} \in \exp\left[ \mathbb{K}_\Gamma\right]
\end{equation}
Using the above property we arrive at the following parametrization of the space $\mathcal{D}_\Gamma$
\begin{equation}\label{krumiro}
    \mathcal{D}_\Gamma \, = \, \mbox{Orbit}_{\mathcal{F}_\Gamma}\left(\exp\left[
    \mathbb{K}_\Gamma\right]\cdot L_\Gamma\right)
\end{equation}
where, by definition, we have set:
\begin{eqnarray}
  p\in \exp\left[
    \mathbb{K}_\Gamma\right]\cdot L_\Gamma  &\Rightarrow & p=\left\{\exp\left[-\pmb{K}\right]\, A_0
    \exp\left[\pmb{K}\right], \, \exp\left[-\pmb{K}\right]\, B_0\,\exp\left[\pmb{K}\right],\, \exp\left[-\pmb{K}\right]\, C_0
    \exp\left[\pmb{K}\right]\right\} \\
 \left\{ A_0, \, B_0,\,  C_0\right\} &\in & L_\Gamma \\
 \pmb{K} &=& \mathbb{K}_\Gamma
\end{eqnarray}
Relying on this, in the K\"ahler quotient we can invert the order of the operations. First we quotient the
action of the compact gauge group $\mathcal{F}_\Gamma$ and then we implement the moment map constraints. We
have:
\begin{equation}\label{cascapistola}
 \mathcal{D}_{\Gamma}/\!\!/_{\mathcal{F}_\Gamma}\, = \,\exp\left [\mathbb{K}_\Gamma\right]\cdot L_\Gamma
\end{equation}
Calculating the moment maps on $\exp\left [\mathbb{K}_\Gamma\right]\cdot L_\Gamma$ and imposing the moment
map constraint we find:
\begin{equation}\label{carampana}
\mu^{-1}\left( \zeta\right)/\!\!/_{\mathcal{F}_\Gamma}\, = \, \left\{ Z \, \in \exp\left
[\mathbb{K}_\Gamma\right]\cdot L_\Gamma\,  \parallel \, \mu_I (Z,\bar{Z}) \, = \, 0 \quad \mbox{if $f_I
\not\in  \mathfrak{Z} $} \quad ; \quad \mu_I (Z,\bar{Z}) \, = \, \zeta_I
 \quad \mbox{if $f_I \in  \mathfrak{Z} $}\right\}
\end{equation}
Eq.\,(\ref{carampana}) provides an explicit algorithm to calculate the K\"ahler potential of the final
resolved manifold if we are able to solve the constraints in terms of an appropriate triple of complex
coordinates. Furthermore for each level parameter $\zeta_a$ we have to find the appropriate one-parameter
subgroup of $\mathcal{G}_\Gamma$ that lifts the corresponding moment map from the $0$-value to the generic
value $\zeta$. Indeed we recall that the K\"ahler potential of the resolved variety is given by the
celebrated formula:
\begin{equation}\label{celeberro}
  \mathcal{K}_{\mathcal{M}}\, = \, \pi^\star \, \mathcal{K}_{\mathcal{N}} \, + \,\zeta_I \mathfrak{C}^{IJ} \,
  \pmb{\Phi}_J
\end{equation}
where, by definition:
\begin{equation}\label{cirimella}
  \pi \, : \, \mathcal{N} \, \rightarrow \, \mathcal{M}
\end{equation}
is the quotient map and $\exp[\zeta_I  \, \mathfrak{C}^{IJ} \, \pmb{\Phi}_I]\in \exp\left[
    \mathbb{K}_\Gamma\right] \subset\mathcal{G}_{\Gamma}$ is the element of the quiver
group which lifts the moment maps from zero to the values $\zeta_I$, while $\mathfrak{C}^{IJ}$ is a constant
matrix whose definition we discuss later on. Indeed the rationale behind formula (\ref{celeberro}) requires a
careful discussion, originally due to Hitchin, Karlhede, Lindstr\"om and Ro\v cek \cite{HKLR} which we  shall review in
the next section.
\section{Lessons from the Eguchi-Hanson case}
\label{toroandante} In order to give a concrete illustrative example of the Kronheimer construction we focus
on the simplest and oldest known ALE manifold, namely on the Eguchi-Hanson space\cite{eguccio}. To this
effect we begin by introducing a set of  Maurer Cartan forms on the three sphere $\mathbb{S}^3\sim
\mathrm{SU(2)}$:
\begin{eqnarray}
  \sigma_1 &=& - \,\frac{1}{2} (d\theta  \cos (\psi )+d\phi  \sin (\theta ) \sin (\psi ))\nonumber \\
  \sigma_2 &=& \frac{1}{2} (d\theta  \sin (\psi )-d\phi  \sin (\theta ) \cos (\psi ))\nonumber \\
  \sigma_3 &=& - \, \frac{1}{2} (d\phi  \cos (\theta )+d\psi )
\end{eqnarray}
which depend on three Euler angles $\theta,\phi, \psi$ and satisfy the Maurer Cartan equations in the form:
\begin{equation}\label{castrofinallo}
    d\sigma_i \, = \, \epsilon_{ijk} \, \sigma_j \wedge \sigma_k
\end{equation}
 Furthermore, let us introduce a radial coordinate $ m \leq r \leq +\infty $ and the following function:
\begin{equation}\label{caorso}
    G(r)\, = \, \sqrt{1\, - \, \left(\frac{m}{r}\right)^2}
\end{equation}
The Eguchi-Hanson metric is given by the following expression:
\begin{eqnarray}\label{carnepuzzosa}
    ds^2_{EH} & = &  G(r)^{-2} \, dr^2 \, + \, r^2 \left( \sigma_1^2 \, + \sigma_2^2 \right) \, + \,
    r^2 \, G(r)^2 \, \sigma_3^2\nonumber\\
    & = & \frac{1}{4} \left(\frac{\left(r ^4-a^4\right) (d\phi \cos (\theta )
    +d\psi)^2}{r ^2}+\frac{4 dr^2}{1-\frac{a^4}{r ^4}}+r ^2 \left(d\phi^2 \sin
   ^2(\theta )+d\theta^2\right)\right)\nonumber\\
\end{eqnarray}
Calculating  the curvature two-form of the above metric, we find that it is self-dual, while its Ricci tensor
vanishes. Hence the Eguchi-Hanson metric is an Euclidean vacuum solution of Einstein equations and it
describes a gravitational instanton. As $r \to \infty$ the Eguchi-Hanson metric approaches  a Euclidean
metric:
\begin{equation}\label{carnevaledilobbi}
ds^2_{EH} \,\stackrel{r\to\infty}{\Longrightarrow} \,    \frac{1}{2} r ^2 d\psi d\phi \cos (\theta
)+\frac{1}{4} r ^2 d\theta^2+\frac{1}{4}
   r ^2 d\psi ^2+\frac{1}{4} r ^2 d\phi^2+dr^2
\end{equation}
\par
Next, one describes  the Eguchi-Hanson space as a complex manifold $\mathcal{M}_{EH}$ and the Eguchi-Hanson metric
$\widehat{ds_{EH}^2}$ as a K\"ahler metric on $\mathcal{M}_{EH}$. To this effect let us introduce the
following two complex coordinates:
\begin{eqnarray}
  Z^1 &=& \left( r^4 -m^4 \right)^{\ft 14} \,\frac{\left(e^{i (\theta +\phi )}+i e^{i \theta }+e^{i \phi }-i\right)
  e^{-\frac{1}{2} i (\theta -\psi +\phi )}}{2 \sqrt{2}} \nonumber\\
  Z^2 &=& \left( r^4 -m^4 \right)^{\ft 1 4}\, \frac{\left(e^{i (\theta +\phi )}-i e^{i \theta }
  +e^{i \phi }+i\right) e^{-\frac{1}{2} i (\theta -\psi +\phi )}}{2 \sqrt{2}}
\label{pariginodisera}
\end{eqnarray}
By direct calculation we can verify that:
\begin{equation}\label{confettino}
    {ds_{EH}^2}\, = \, \frac{\partial}{\partial Z^i} \, \frac{{\partial}}{\partial \bar{Z}^{j^\star}}
    \, \mathcal{K}_{EH} \, dZ^i \otimes d\bar{Z}^{j^{\star}}
\end{equation}
where:
\begin{eqnarray}\label{cravattino}
    \mathcal{K}_{EH} & = & \sqrt{\tau^2+m^4}-m^2 \,\log \left(\sqrt{\tau^2+1}+m^4\right)+m^2\,\log (\tau)\nonumber\\
    \tau & \equiv & |Z^1|^2 \, + \, |Z^2|^2 \label{cravattino2}
\end{eqnarray}
Having derived the form of the K\"ahler potential for the Eguchi-Hanson metric we can now connect it to the
Kronheimer construction of the ALE manifolds by recalling eqs. (\ref{momentummapck}) and rewriting them in
the case $k=1$ for which the group $\mathcal{F} \, = \, \mathrm{U(1)},$ so that there is only one component of
the triholomorphic moment map:
\begin{eqnarray}
  \mathfrak{P}^3 &=& |u_0|^2 - |v_0|^2 +|v_1|^2 - |u_1|^2  \\
  \mathfrak{P}^+ &=& u^0 \, v_0 - u^1 \, v_1 \label{fedoci}
\end{eqnarray}
In this case it is convenient to redefine:
\begin{eqnarray}
  U &=& \left\{u_0,v_1\right\} \\
  V &=& \left\{v_0,u_1\right\}\label{giamblico}
\end{eqnarray}
so that eqs. (\ref{fedoci}) can be rewritten as follows:
\begin{eqnarray}
  \mathfrak{P}^3 &=& \mathcal{P}^3(U,V) \, \equiv \, \sum_{i=1}^2 |U_i|^2 \, - \, \sum_{i=1}^2 |V_i|^2 \\
  \mathfrak{P}^+&=& \mathcal{P}^+(U,V) \, \equiv \, \sum_{i=1}^2 U_i \, V_i \label{tritacarta}
\end{eqnarray}
Furthermore the action of the group $\mathcal{F}_{\mathbb{Z}_2} \, = \, \mathrm{U(1)}$ on the complex
coordinates $U,V$ is the following one:
\begin{equation}\label{cianfrusca}
    \mathrm{U(1)} \quad : \quad \left( U,V\right) \quad \Longrightarrow \quad \left(e^{i \varphi} U \, , \, e^{-i\varphi}
    V\right)
\end{equation}
Considering the quiver group $\mathcal{G}_{\mathbb{Z}_2}$ which is just the complexification of
$\mathcal{F}_{\mathbb{Z}_2}$   we obtain the transformation:
\begin{equation}\label{cianfrusca2}
    \mathcal{G}_{\mathbb{Z}_2} \quad : \quad \left( U,V\right) \quad \Longrightarrow
    \quad \left(e^{-\pmb{\Phi}} U \, , \, e^{\pmb{\Phi}}
    V\right)
\end{equation}
Relying on these preliminaries we are ready to perform the algebro-geometric  hyperK\"ahler  quotient
according to  formula (\ref{celeberro}). Introducing the level parameters we have to solve the equations:
\begin{eqnarray}
    \ell & = &\mathcal{P}^3\left(e^{-\pmb{\Phi}}U,e^{\pmb{\Phi}}V\right)\nonumber\\
    \mathfrak{s}& = & \mathcal{P}^+\left(e^{-\pmb{\Phi}}U,e^{\pmb{\Phi}}V\right)\, = \,
    \mathcal{P}^+\left(U,V\right) \label{geronimi}
\end{eqnarray}
As stated several times and recalled in the second line of the above equation,  the holomorphic part of the
moment map is invariant under the action of the quiver group. This is very useful for the solution of the
constraints. Indeed we can just choose a gauge condition like the following one:
\begin{equation}\label{gaugus}
    U_1 \, = \, V_2
\end{equation}
Furthermore, in the case $k=1$ the holomorphic level parameter $\mathfrak{s}$ can be just set equal to zero
without loss of generality,  since it simply amounts to a change of  coordinates. In this way we arrive at:
\begin{equation}\label{dnaEH}
    U_1 \, = \, V_2 \, \equiv \, \ft 12 Z^1 \quad ; \quad U_2 \, = \, V_1 \, \equiv \, \ft 12 Z^2
\end{equation}
and the first of equations (\ref{geronimi}) is solved by:
\begin{equation}\label{minerale}
    \pmb{\Phi} \, = \,  - \, \log \left[\frac{\ell \pm \sqrt{\ell^2 + 4 |U|^2 \, |V^2|}}{2 |V^2|}\right] \, =\,\,
     - \, \log \left[
    \frac{\ell \pm \sqrt{\ell^2 + |\mathbf{Z}|^4 }}{2 |\mathbf{Z}|^2}\right] \quad ; \quad
    |\mathbf{Z}|^2\equiv |Z^1|^2 + |Z^2|^2
\end{equation}
The restriction to the level surface of the ambient K\"ahler potential is calculated in an equally easy
fashion:
\begin{equation}\label{cagnesca}
    \mathcal{K}|_{\mathcal{N}} \, = \, e^{-2\,\pmb{\Phi}} \, |U|^2 \, + \, e^{2\,\pmb{\Phi}} \, |V|^2 \, = \,
    \sqrt{\ell^2 +  |\mathbf{Z}|^4 }
\end{equation}
Choosing one branch of the solution (\ref{minerale}) and applying the general formula (\ref{celeberro}) we obtain the K\"ahler potential of the manifold $\mathcal{M}$:
\begin{equation}\label{raviolo}
  \mathcal{K}_{\mathcal{M}} \, = \,   \sqrt{\ell^2 +  |\mathbf{Z}|^4 } \, - \, \ell \, \log \left[
    \frac{\ell \pm \sqrt{\ell^2 + |\mathbf{Z}|^4 }}{2 |\mathbf{Z}|^2}\right]
\end{equation}
For $\ell \, = \, m^2$, we see that the K\"ahler potential (\ref{raviolo}) obtained by means of the
 hyperK\"ahler  quotient advocated in the Kronheimer construction coincides with that of the Eguchi-Hanson
manifold displayed in eq.\,(\ref{cravattino}). This concludes the proof that the Eguchi-Hanson manifold is a
smooth resolution of the singularity $\mathbb{C}^2 /\mathbb{Z}_2$.
\subsection{The algebraic equation of the locus and the exceptional divisor}
First we consider the algebraic equation of the locus in $\mathbb{C}^3$ that corresponds to the Eguchi-Hanson
manifold. According to the discussion following eq.\,(\ref{identifyAk}) such an equation is provided by the
relation between the $\Gamma$ invariants:
\begin{equation}\label{ququzo}
     x \equiv \mbox{Det}\,A \quad ; \quad \mbox{Det}\, B \quad ; \quad z \equiv \ft 1 2 \, \mbox{Tr}
     \left(A\, B\right)
\end{equation}
Upon use of the gauge condition (\ref{gaugus}) and of the solution of the holomorphic moment map constraint
(\ref{dnaEH}) we have:
\begin{equation}\label{cicciamiccia}
    A \, = \, \left(
                \begin{array}{cc}
                  0 & \ft 12 Z_1 \\
                  \ft 12 Z_2 & 0 \\
                \end{array}
              \right)\quad ; \quad  B \, = \, \left(
                \begin{array}{cc}
                  0 & \ft 12 Z_2 \\
                  \ft 12 Z_1 & 0 \\
                \end{array}
              \right)
\end{equation}
so that:
\begin{equation}\label{cinematokko}
    x \, = \, - \, \ft 14 \, Z_1 \, Z_2 \quad ; \quad y \, = \, - \, \ft 14 Z_1 \, Z_2 \quad ; \quad z \, =
    \, \ft 1 4 \, Z_1 \, Z_2
\end{equation}
and the equation of the orbifold locus $\mathbb{C}^2/\Gamma$:
\begin{equation}\label{nocambio}
    xy \, = \, z^2
\end{equation}
remains unmodified. This happens because the holomorphic moment map has not been lifted away from zero and
similarly will happen in all the resolutions of the $\mathbb{C}^3/\Gamma$ singularities since, as we
stressed, there we have no complex structure deformations and the analogue of the holomorphic moment map
equation $\left[A\, , \, B\right]\, = \, \left[B\, , \, C\right]\, = \, \left[C\, , \, A\right]$ obtains no
deformation. Yet we know that by lifting the level of the real moment map we obtain the smooth Eguchi-Hanson
manifold which has a nontrivial homology 2-cycle, as foreseen by the general theorem \ref{reidmarktheo}. In
quasi polar coordinates these homology cycle is the two--sphere spanned by angles $\theta$ and $\phi$ when we
set $r\,=\,m$ and we disregard the angle $\psi$. Such a homology cycle disappears when $m \rightarrow 0$
hence it is the exceptional divisor generated by the minimal resolution of the singularity. Hence it is
interesting to see where such an exceptional divisor is located in the complex description of the Eguchi-Hanson manifold obtained from the Kronheimer construction. To this effect it is convenient to recall the
relation between divisors and line bundles.
\subsubsection{Divisors and line bundles}
A {\em prime divisor} in a complex manifold or algebraic variety $X$ is an irreducible closed codimension one
subvariety of $X$. A divisor $\mathfrak{D}$ is a locally finite formal linear combination
\begin{equation}\label{divisor} \mathfrak{D} = \sum_i a_i\,\mathfrak{D}_i \end{equation}
where the $a_i$ are integers, and the $\mathfrak{D}_i$ are prime divisors. A prime divisor $\mathfrak{D}$ can
be descrived by a collection $\{(U_\alpha,f_\alpha)\}$, where $\{U_\alpha\}$ is an open cover of $X$, and the
$\{f_\alpha\}$ are holomorphic functions on $U_\alpha$ such that $f_\alpha=0$ is the equation of
$\mathfrak{D}\cap U_\alpha$ in $U_\alpha$. As a consequence, the functions $g_{\alpha\beta}
=f_\alpha/f_\beta$ are holomorphic nowhere vanishing functions
$$ g_{\alpha\beta} \colon U_\alpha\cap U_\beta \to \mathbb C^\ast$$
that on triple intersections  $U_\alpha\cap U_\beta \cap U_\gamma $ satisfy the cocycle condition
$$ g_{\alpha\beta} g_{\beta\gamma} = g_{\alpha\gamma} $$
and therefore define a line bundle $\mathcal L(\mathfrak{D})$. If $\mathfrak{D}$ is a divisor as in
\eqref{divisor} then one sets
$$ \mathcal L(\mathfrak{D}) = \bigotimes_i \mathcal L(\mathfrak{D}_i)^{a_i}.$$
The inverse correspondence (from line bundles to divisors) is described as follows. If $s$ is a nonzero
meromorphic section of a line bundle $\mathcal L$, and $V$ is a codimension one subvariety of $X$ over which
$s$ has a zero or a pole, denoted by $\operatorname{ord}_{V}(s)$ the order of the zero, or minus the order of
the pole; then
$$ \mathfrak{D} = \sum_{V} \operatorname{ord}_{V}(s) \cdot V $$
is a divisor, whose associated line bundle $\mathcal L(\mathfrak{D})$ is isomorphic to $\mathcal L$.
\subsubsection{The exceptional divisor}
It is easy to work out the exceptional divisor in the Eguchi-Hanson case by performing the following
holomorphic coordinate transformation:
\begin{equation}\label{scorrecco}
    Z^1\,\to\, \left(1-\xi_1\right)\, \xi_2 \quad ; \quad Z^2 \, \to \,-\left(1+\xi_1\right) \, \xi_2
\end{equation}
Upon the substitution (\ref{scorrecco}) and the identification $\ell \, = \, m^2$ the K\"ahler potential
(\ref{raviolo}) becomes:
\begin{eqnarray}\label{sindrollo}
    \mathcal{K}_{EH} & = & \mathcal{K}_0 + m^2 \left(\mathcal{K}_\mathrm{E} +
    \log|W|^2\right)\nonumber\\
    \mathcal{K}_0 &=& \sqrt{m^4+4\left(1+|\xi_1|^2\right)^2\,|\xi_2|^4} \, - \, m^2 \,
    \log\left( m^2 \, + \, \sqrt{m^4+4\left(1+|\xi_1|^2\right)^2\,|\xi_2|^4}\right)\nonumber\\
    \mathcal{K}_\mathrm{E} & = & \log\left(1+|\xi_1|^2\right)\nonumber\\
    W & \equiv & \sqrt{2}\, \xi_2
\end{eqnarray}
Inspecting eq.\,(\ref{sindrollo}) we realize that $\mathcal{K}_\mathrm{E}$ is the standard K\"ahler potential
of a $\mathbb{P}^1$ written in the affine coordinate $\xi_1$. This suggests that the Eguchi-Hanson manifold
is covered by two open charts:
\begin{eqnarray}
  U_N &=& \left\{\xi_1^{N} ,\xi_2^{N}\right\} \nonumber\\
  U_S&=& \left\{\xi_1^{S} ,\xi_2^{S}\right\}\label{cromatto}
\end{eqnarray}
with transition function:
\begin{equation}\label{ciulifusco}
    \left\{\xi_1^{N} ,\xi_2^{N}\right\}\, = \, \left\{\frac{1}{\xi_1^{S}} ,\xi_2^{S}\,\xi_1^{S}\right\}
\end{equation}
Under the transformation (\ref{cromatto}) the function $\mathcal{K}_0$ is invariant, while
$\mathcal{K}_\mathrm{E}$ transforms as follows:
\begin{equation}\label{sumascet}
    \mathcal{K}_\mathrm{E}\left(\xi^N,\bar{\xi^N} \right)\, = \,
    \mathcal{K}_\mathrm{E}\left(\xi^S,\bar{\xi^S} \right)\, -\,
    \log|\xi_1^S|^2
\end{equation}
Therefore we can introduce a line bundle $\mathcal{L}\stackrel{\pi}{\rightarrow} \mathcal{M}_{EH}$ defined by
two local trivializations, one based on $U_N$, the other on $U_S$ with transition function:
\begin{equation}\label{sanculino}
    g_{NS} \quad : \quad W_N(\xi^N) \, =\, \xi_1^S W_S(\xi^S)
\end{equation}
a  fiber metric on such a bundle is obtained by defining the
following invariant norm for the bundle sections:
\begin{equation}\label{cristobolo}
    \parallel W\parallel^2 \equiv e^{\mathcal{K}_\mathrm{E}} \, |W|^2
\end{equation}
The corresponding first Chern class is:
\begin{equation}\label{c1}
    c_1(\mathcal{L}) \, = \, \omega^{(1,1)} \, \equiv \, \frac{i}{2\pi} \,
    \partial {\bar \partial}\log\parallel W\parallel^2 \quad \stackrel{W\to 0}{\longrightarrow}
     \quad \frac{i}{2\pi} \, \frac{d\xi_1 \wedge
    d{\bar \xi_1}}{\left(1+|\xi_1|^2\right)^2} \, \equiv \, \omega^{(1,1)}_\mathcal{D}
\end{equation}
The divisor $\mathfrak{D}$ related with this line bundle is obviously obtained  as the vanishing locus of the
global section $W=\xi_2 \, = \,0$. The cohomology class of $\omega^{(1,1)}$ is that of the Poincar\'e dual
$\omega^{(1,1)}_\mathcal{D}$ of the vanishing section $W$, namely of the  divisor $\mathfrak{D}$:
\begin{equation}\label{paratillus}
    \left[\omega^{(1,1)}\right] \, = \, \left[\omega^{(1,1)}_\mathfrak{D}\right]
\end{equation}
What we have discussed so far is just an explicit illustration of
the well known fact that the Eguchi-Hanson manifold is the total
space of the fiber bundle
$\mathcal O_{\mathbb P^1}(-2)$.
\par
Since the function $\mathcal{K}_0$ is invariant, it is clear that its contribution $\partial{\bar
\partial}\mathcal{K}_0$ to the K\"ahler 2-form is cohomologous to zero which implies:
\begin{equation}\label{comollogia}
    \left[\mathbb{K}_{EH} \right] \, = \, m^2 \left[ \omega^{(1,1)}_\mathfrak{D}\right]
\end{equation}
Finally it is instructive to compare the above complex description of the Eguchi-Hanson space with its
description in terms of quasi polar coordinates. To this effect it suffices to rewrite the coordinate
transformation (\ref{pariginodisera}) in terms of the $x_i$ coordinates. We have:
\begin{equation}\label{xipolli}
    \xi_1 \, = \, e^{i \phi } \cot \left(\frac{\theta }{2}\right)\, , \, \xi_2 \, = \,
    \frac{1}{2} \sqrt{1-\cos (\theta )} \sqrt[4]{r^4-m^4} e^{\frac{1}{2} i (\psi -\phi )}
\end{equation}
As we see the locus $\xi_2 \, = \, 0$ corresponds to $r=m$ and $\psi \, = \, \mbox{any value}$.
\subsection{Comparison with the two-center Gibbons-Hawking metric}
Finally we derive the map between the manifold with a two center Gibbons-Hawking (GH) metric
\cite{Gibbons:1979zt,Gibbons:1979xm} and the Eguchi-Hanson space. We begin with a conceptual discussion about
the parameters of GH-metrics (Appendix \ref{gibbone} is devoted to a concise description of GH-metrics and
also provides new formulae concerning their explicit description in terms of complex coordinates. There we
also address the problem of deriving the corresponding K\"ahler potential).
\par
The Gibbons-Hawking multi-center metrics have a number of parameters that can be counted in the following
way. Let $n$ be the number of centers. Each center has 3-coordinates, hence a priori we have $3n$ parameters.
Yet, using the Euclidean group of translations and rotations, which is a symmetry of the $3d$ laplacian, we
can always bring a center to a reference point, say the origin $\pmb{ x}=0$. So we are left with $3(n-1)$
parameters. Furthermore, once a center is fixed, another center lies somewhere on a two-sphere around the
first center and we can use the rotation group to bring it to a preferred direction. This kills two
additional parameters. In this way we have:
\begin{equation}
\# \text{ of} \text{ effective } \text{ parameters } \text{ in a}  \text{ GH } \text{metric} = 3n -5
\label{limonata}
\end{equation}
\par
From the point of view of the Kronheimer construction, the $n$-center metric corresponds to the resolution $Y
\rightarrow \frac{\mathbb{C}^2}{\mathbb{Z}_m}$ via a  hyperK\"ahler  quotient. In this case the gauge group
is $\mathrm{U(1)}^{n-1}$ and we have indeed $3(n-1)$ parameters. Two parameters corresponding to one complex
moment map level can be disposed of by a redefinition of the complex coordinates for the resolved manifold Y.
Hence also on the side of the  hyperK\"ahler  quotient we have:
\begin{equation}
\# \text{ of } \text{ effective} \text{ parameters } \text{ in a }  \text{  hyperK\"ahler  } \text{ quotient
} \text{ resolution }\text{  }\text{of } \frac{\mathbb{C}^2}{\mathbb{Z}_n} = 3n -5\label{cedrata}
\end{equation}
In the Eguchi-Hanson case $n=2$ and there is only one effective parameter on both sides of the
correspondence, namely the parameter $m^2$ that we have associated with real moment map level. The level of
the holomorphic moment map corresponds to the two parameters that can be disposed of by a coordinate
transformation and was set to zero.
\par
From the GH-side, the removal of the spurious parameters can be conventionally performed by aligning the two
centers on the z-axis at symmetrical positions with respect to the origin $z=0$. Hence referring to eqs.
(\ref{GWmetricca})and (\ref{laplaccio}) we set:
\begin{equation}\label{potoEH}
    \mathcal{V}_{EH} \, = \, \frac{1}{\sqrt{\left(\frac{m^2}{8}+z\right)^2+x^2+y^2}}
    +\frac{1}{\sqrt{\left(z-\frac{m^2}{8}\right)^2+x^2+y^2}}
\end{equation}
and we obtain the following connection one-form :
\begin{eqnarray}\label{sassoletto}
    \omega_{EH}& = &\left(m^2 \left(\frac{1}{\sqrt{\left(m^2-8 z\right)^2+64 \left(x^2+y^2\right)}}
    -\frac{1}{\sqrt{\left(m^2+8 z\right)^2+64
   \left(x^2+y^2\right)}}\right)\right.\nonumber\\
   &&\left.-\frac{8 z}{\sqrt{\left(m^2-8 z\right)^2+64 \left(x^2+y^2\right)}}
   -\frac{8 z}{\sqrt{\left(m^2+8 z\right)^2+64
   \left(x^2+y^2\right)}}+2\right) \times \frac{y \,\mathrm{d}x - x \,\mathrm{d}y}{x^2+y^2}\nonumber\\
\end{eqnarray}
which satisfies with $\mathcal{V}_{EH}$ the duality relation (\ref{cagnesco}). The one-form $\omega_{EH}$
agrees with eq.\,(\ref{prepotenzialno}) if we set:
\begin{eqnarray}
    \partial_z\mathcal{F}_{EH} & = & \int dz \, \mathcal{V}_{EH}\nonumber\\
    & = & \log \left(\sqrt{\left(z-\frac{m^2}{8}\right)^2+x^2+y^2}-\frac{m^2}{8}+z\right)+\log
   \left(\sqrt{\left(\frac{m^2}{8}+z\right)^2+x^2+y^2}+\frac{m^2}{8}+z\right)\nonumber\\
   \label{partiallo}
\end{eqnarray}
The metric:
\begin{equation}\label{crapato}
    ds^2_{two-center} \, = \,\frac{1}{\mathcal{V}_{EH}}\left(\mathrm{d}\tau
    +\omega_{EH}\right)^2+\mathcal{V}_{EH}\left(\mathrm{d}x^2 +\mathrm{d}y^2+\mathrm{d}z^2\right)
\end{equation}
is exactly mapped into the Eguchi-Hanson metric (\ref{carnepuzzosa}) by the following coordinate
transformation:
\begin{eqnarray}
&& x\to \frac{1}{8} \sin (\theta ) \sqrt{r^4-m^4} \cos (\psi )\quad ,\quad y\to \frac{1}{8} \sin (\theta )
\sqrt{r^4-m^4} \sin (\psi )\nonumber\\
&&z\to \frac{1}{8} r^2 \cos
   (\theta )\quad ,\quad \tau \to 2 \psi +2 \phi \label{puchero}
\end{eqnarray}
It is also interesting to work out the explicit form, in the present case of  the complex coordinates
$\mathfrak{h}$ and $\mathfrak{z}$ introduced in eqs.\,(\ref{gomorrus}) and (\ref{miciogatto}) within the
framework of the general discussion. After some algebra one finds:
\begin{equation}\label{binasato}
    \mathfrak{h}\, = \,\frac{64 \, e^{2 i (\psi +\phi )}}{(\cos (\theta )+1)^2 \left(r^4-m^4\right)}\quad ,
    \quad\mathfrak{z}\, = \,\frac{1}{8}\, i\, e^{-i \psi } \sin (\theta )
   \sqrt{r^4-m^4}
\end{equation}
As one realizes, both these coordinates are singular on the exceptional divisor $r=m$ and they are not
convenient to describe it. The relation with the good coordinates $\xi_{1,2}$ is actually antiholomorphic and
it would be difficult to be guessed a priori:
\begin{equation}\label{frakkarrosto}
    \bar{\xi }_1\, = \,  -\frac{i}{\mathfrak{z} \sqrt{\mathfrak{h}}}\quad ,\quad \bar{\xi }_2\, = \,
     -\frac{\sqrt{2}\,\, \mathfrak{z} \, \sqrt[4]{\mathfrak{h}}}{m}
\end{equation}
In terms of the GH-coordinates, by inspecting eq.\,(\ref{puchero}) we readily retrieve the image of exceptional
divisor inside the GH space. It is given by the locus:
\begin{equation}\label{ecceziunalo}
    \mathfrak{D}_E \, =\, \left\{x=y=0 \, , \, -\frac{m^2}{8} \le z \le \frac{m^2}{8} \, , \, 0 \le \tau \le 2\pi\right\}
\end{equation}
namely the  product of the segment joining the two centers on the z-axis with the circle spanned by the
$\tau$-angle. (Actually a detailed analysis of the metric shows that it degenerates at the ends of the cylinder, so that
the latter may be thought of as a sphere.) This observation  may be useful in order to find the exceptional divisors in the more complicated
multi-center cases.
\section{The  generalized Kronheimer construction for
$\frac{\mathbb{C}^3}{\Gamma}$ and the Tautological Bundles}
In the present section we aim at extracting a general scheme from the
detailed discussions presented  in the
previous sections. Our final goal is to establish all the algorithmic steps
that  give a precise meaning to each of
the lines appearing in the conceptual diagram of eq.\,(\ref{salumaio}).
\subsection{Construction of the space $\mathcal{N}_{\zeta }\equiv \mu ^{-1}(\zeta )$ }
Summarizing the points of our construction we have the following
situation. We have considered the moment map
\begin{equation}
\mu : \mathcal{S}_{\Gamma }\longrightarrow  \mathbb{F}_{\Gamma}{}^*
\end{equation}
where \(\mathbb{F}_{\Gamma }{}^*\) is the dual  of the Lie algebra
of the maximal compact subgroup \(\mathcal{F}_{\Gamma }\) of the
quiver group \(\mathcal{G}_{\Gamma }\). Next we have considered the
center of the above Lie algebra $\mathfrak{z}$[\(\mathbb{F}_{\Gamma
}\)]$\subset $\(\mathbb{F}_{\Gamma }\) and its dual
\(\mathfrak{z}\left[\mathbb{F}_{\Gamma } \right]{}^*\). The moment
map can be restricted to the subspace:
\begin{equation}
\mathcal{D}_{\Gamma } \subset  \mathcal{S}_{\Gamma } \quad ; \quad \mathcal{D}_{\Gamma}\equiv
 \left\{\left.p\in  \mathcal{S}_{\Gamma }\right| p\land p = 0\right\}
\end{equation}
which is just the orbit, with respect to the quiver group \(\mathcal{G}_{\Gamma }\), of a locus
$\mathcal{E}_{\Gamma }\subset \mathcal{S}_{\Gamma}$ of complex dimension three obtained in the following way.
\par
Consider the following subspace of \(\mathcal{S}_{\Gamma }^{[0,0]}\)\(\subset \mathcal{S}_{\Gamma }\)
\begin{equation}
\mathcal{S}_{\Gamma }^{[0,0]} =\left\{\left.p\in  \mathcal{S}_{\Gamma } \right| p\land p =0 \quad; \quad \mu
(p)=0\right\}
\end{equation}
whose elements are triples of $|\Gamma |\times |\Gamma |$ complex matrices (A,B,C) satisfying, by the above
definition, in addition to the invariance constraint (\ref{gammazione}-\ref{carnevalediPaulo})  also the
following two ones:
\begin{equation}
[A,B]=[B,C]=[C,A]=0 \quad ; \quad \text{Tr}\left[T_I\,\left( \left[A,A^{\dagger }\right]+\left[B,B^{\dagger
}\right]+\left[C,C^{\dagger }\right]\right)\right]\,=\,0 \quad ; \quad I\,=\, 1,\dots ,\, |\Gamma|-1
\end{equation}
Since the action of the compact group \(\mathcal{F}_{\Gamma }\) leaves both the first and the second
constraint invariant, it follows that it maps the locus { } \(\mathcal{S}_{\Gamma }^{[0,0]}\) into itself
\begin{equation}
\mathcal{F}_{\Gamma } \quad :\quad \mathcal{S}_{\Gamma }^{[0,0]}\, \to\, \mathcal{S}_{\Gamma}^{[0,0]}
\end{equation}
The locus \(\mathcal{E}_{\Gamma }\) is defined as the quotient:
\begin{equation}
\mathcal{E}_{\Gamma } \equiv  \frac{\mathcal{S}_{\Gamma }^{[0,0]}}{\mathcal{F}_{\Gamma}}
\end{equation}
which turns out to be of complex dimension three and to be isomorphic to the singular orbifold :
\begin{equation}
\frac{\mathcal{S}_{\Gamma }^{[0,0]}}{\mathcal{F}_{\Gamma }} \, \simeq\,  \frac{\mathbb{C}^3}{\Gamma}
\end{equation}
Choosing a representative in each equivalence class \(\frac{\mathcal{S}_{\Gamma
}^{[0,0]}}{\mathcal{F}_{\Gamma }}\) simply amounts to a choice of local coordinates on
\(\frac{\mathbb{C}^3}{\Gamma }\) which will be promoted to a system of local coordinates on the manifold
\(\mathcal{M}_{\zeta}\) of the final resolved singularity.
\par We have a canonical algorithm to construct a
canonical coordinate system for { }\(\mathcal{E}_{\Gamma }\) which originates from Kronheimer and from the
1994 paper by Anselmi, Bill\`o, Fr\`e, Girardello and Zaffaroni on ALE manifolds and conformal field theories
\cite{mango}. The construction is the following. We begin with the locus \(L_{\Gamma }\) $\subset
$\(\mathcal{S}_{\Gamma }\) defined as the set of triples (\(A_d\),\(B_d\),\(C_d\)) such that the invariance
constraint (\ref{carnevalediPaulo}) is satisfied with respect to $\Gamma $ and they are diagonal in the
natural basis of the regular representation. We have shown on the basis of several examples that :
\begin{equation}
\mathcal{D}_{\Gamma } = \text{Orbit}_{\mathcal{G}_{\Gamma}}\left(L_{\Gamma }\right) \label{quagliastro}
\end{equation}
We obtain an explicit parameterization of the locus \(\mathcal{E}_{\Gamma }\) by solving the algebraic
equation for the hermitian matrix $\mathcal{V}_0\,\in \, \exp\left[\mathbb{K}_{\Gamma }\right]$, such that
\begin{equation}
\forall p\, \in  \, L_{\Gamma }\quad : \quad \mu \left(\mathcal{V}_0.p\right) = 0
\end{equation}
The important thing is that the solution for the above equation is a constant matrix \(\mathcal{V}_0\),
indipendent from the point p $\in $ \(L_{\Gamma }\). Then we fix the coordinates of our manifold by
parameterizing
\begin{equation}
p \in \mathcal{E}_{\Gamma } \Rightarrow  p= \left(
\begin{array}{c}
 A_0 \\
\begin{array}{c}
 B_0 \\
 C_0 \\
\end{array}
 \\
\end{array}
\right)=\left(
\begin{array}{c}
 \mathcal{V}_0{}^{-1}A_d\mathcal{V}_0 \\
\begin{array}{c}
 \mathcal{V}_0{}^{-1}B_d\mathcal{V}_0 \\
 \mathcal{V}_0{}^{-1}C_d\mathcal{V}_0 \\
\end{array}
 \\
\end{array}
\right)\text{    }\text{where} \left(
\begin{array}{c}
 A_d \\
\begin{array}{c}
 B_d \\
 C_d \\
\end{array}
 \\
\end{array}
\right)\in L_{\Gamma }
\end{equation}
It follows that equation (\ref{quagliastro}) can be substituted by
\begin{equation}
\mathbb{V}_{|\Gamma |+2} \equiv \text{  }\mathcal{D}_{\Gamma } =
\text{Orbit}_{\mathcal{G}_{\Gamma}}\left(\mathcal{E}_{\Gamma }\right)
\end{equation}
We can also introduce a subspace \(\mathcal{D}_{\Gamma }{}^0\)$\subset $\(\mathbb{V}_{|\Gamma |+2}\) which is
the orbit of \(\mathcal{E}_{\Gamma}\) under the compact subgroup \(\mathcal{F}_{\Gamma }\):
\begin{equation}
\mathcal{D}_{\Gamma }{}^0\,=\, \text{Orbit}_{\mathcal{F}_{\Gamma}}\left(\mathcal{E}_{\Gamma }\right)
\end{equation}
This being the case we consider the restriction  of the moment map to \(\mathcal{D}_{\Gamma }\)
\begin{equation}
\mu  : \mathcal{D}_{\Gamma } \longrightarrow  \mathbb{F}_{\Gamma}{}^*
\end{equation}
and given an element
\begin{equation}
\zeta \in \mathfrak{z}\left[\mathbb{F}_{\Gamma }\right]{}^*
\end{equation}
we define:
\begin{equation}
\mathcal{N}_{\zeta }\equiv  \mu ^{-1}(\zeta ) \subset  \mathcal{D}_{\Gamma }\quad:\quad\mathcal{N}_{\zeta} \,
=\, \left\{\left.p\in \mathcal{D}_{\Gamma } \right| \mu (p) = \zeta \right\}
\end{equation}
\\
Obviously we have:
\begin{equation}
\mathcal{N}_0\equiv  \mu ^{-1}(0) =\text{  }\mathcal{D}_{\Gamma}{}^0
\end{equation}
\subsection{The space $\mathcal{N}_{\zeta }$  as a principal fiber bundle}
The space \(\mathcal{N}_{\zeta }\) has a natural structure of an \(\mathcal{F}_{\Gamma }\) principal line
bundle over the quotient \(\mathcal{M}_{\zeta}\):
\begin{equation}
\mathcal{N}_{\zeta }\,\overset{\pi }{\longrightarrow }\,\mathcal{M}_{\zeta }\, \equiv \,\mathcal{N}_{\zeta
}/\!\!/\mathcal{F} _{\Gamma } \label{gualtiero}
\end{equation}
On the tangent space to the total space of the \(\mathcal{F} _{\Gamma }\)--bundle
\(\text{T$\mathcal{N}$}_{\zeta }\) we have a metric induced, as the pullback, by the inclusion map:
\begin{equation}
\iota  : \mathcal{N}_{\zeta } \longrightarrow  \mathcal{S}_{\Gamma}
\end{equation}
of the flat metric $g$ on \(\mathcal{S}_{\Gamma }\)
\begin{equation}
g_{\mathcal{N}} = \iota ^*\left(g_{\mathcal{S}_\Gamma}\right)
\end{equation}
Since the metric \(g_{\mathcal{S}_\Gamma}\) is K\"ahlerian we have a K\"ahler potential
\(\mathcal{K}_{\mathcal{S}_\Gamma}\) from which it derives and we define the function
\begin{equation}
\mathcal{K}_{\mathcal{N}} \equiv \iota ^*\left(\mathcal{K}_{\mathcal{S}_\Gamma}\right)
\end{equation}
This function is not the K\"ahler potential of \(\mathcal{N}_{\zeta
}\) which is not even K\"ahlerian (it has odd dimensions) but it
will be related to the K\"ahler potential of the final quotient
\(\mathcal{M}_{\zeta }\) by means of an argument due to \cite{HKLR},
that we spell out a few lines below. Let us denote by \(p\in
\mathcal{M}_{\zeta }\) a point of the base manifold and by \(\pi
^{-1}(p)\) the \(\mathcal{F} _{\Gamma }\)-fiber over that point.
\subsubsection{The natural connection and the tautological bundles}\label{ciulifischio}
We can determine a natural connection on the principal bundle (\ref{gualtiero}) through the following steps.
As it is observed in eq.\,(2.7) of the paper by Degeratu and Walpuski \cite{degeratu}, which agrees with the
formulae of the present paper, the quiver group has always the following form:
\begin{equation}
\mathcal{G}_{\Gamma }=\prod _{i=1}^r \text{GL}\left(\mathbb{C}^{\dim \left[\pmb{D}_i\right]}\right)
\end{equation}
where \(\pmb{D}_i\) are the nontrivial irreducible representations of the finite group $\Gamma $, with the
esclusion of \(\pmb{D}_0,\) the identity representation. It also follows that the compact gauge subgroup
\(\mathcal{F}_{\Gamma }\) has the corresponding following structure
\begin{equation}
\mathcal{F}_{\Gamma }=\prod _{i=1}^r \mathrm{U}\left(\dim \left[D_i\right]\right)
\end{equation}
Consequently, the principal bundle (\ref{gualtiero}) induces holomorphic vector bundles of rank \(\dim
\left[\pmb{D}_i\right]\) on which the compact structural group acts non-trivially only with its component
\(\mathrm{U}\left(\dim \left[\pmb{D}_i\right]\right)\). A natural connection on these bundles is obtained as
it follows
\begin{equation}
\mathbb{A} = \frac{\rm i}{2}\left(\mathcal{H}^{-1}\partial  \mathcal{H} -\mathcal{H}\bar{\partial }
\mathcal{H}^{-1}\right) + g^{-1}d g \in  \underset{i=1}{\overset{r}{\oplus }} \mathbf{u}\left(\dim
\left[D_i\right]\right) \label{rodolfo}
\end{equation}
where $\mathcal{H}$ is a hermitian fiber--metric on the direct sum
of the tautological vector bundles defined below:
\begin{equation}
\mathcal{R} \equiv \underset{i=1}{\overset{r}{\bigoplus }} \mathcal{R}_i \quad ; \quad \mathcal{R}_i
\overset{\pi }{\longrightarrow } \mathcal{M}_{\zeta }\quad ; \quad \forall p \in \mathcal{M}_{\zeta } \quad :
\quad \pi ^{-1}(p) \simeq  \mathbb{C}^{\dim \left[D_i\right]}
\end{equation}
By definition the matrix $\mathcal{H}$ must be of dimension
\begin{equation}
\dim  [\mathcal{H}]= n\times n \quad\quad\text{where}\quad\quad n=\sum _{i=1}^r \dim
\left[D_i\right]=\sum_{i=1}^r n_i
\end{equation}
In order to find the hermitian matrix $\mathcal{H}$, we argue in the following way. First we observe that in
the regular representation $R$ each irreducible representation \(\pmb{D}_i\) is contained exactly \(\dim
\left[\pmb{D}_i\right]\) times, so that the form of the matrix $\mathcal{V}$ corresponding to the hermitian
parameterization of the coset \(\frac{\mathcal{G}_{\Gamma }}{ \mathcal{F}_{\Gamma }}\) has always the
following form:
\begin{equation}
\mathcal{V}\, = \,\left(
\begin{array}{|c|c|c|c|c|}
\hline
 \mathfrak{H}_0 & 0 & 0 & \dots  & 0 \\
\hline
 0 & \mathfrak{H}_1\otimes  \pmb{\pmb{1}_{n_1\times  n_1}}  & 0 & \dots &  \vdots \\
\hline
 0 & 0 & \mathfrak{H}_2\otimes  \pmb{\pmb{1}_{n_2\times  n_2}} & \dots &  \vdots \\
\hline
 \vdots & \dots & \dots & \dots\dots &  \vdots \\
\hline
 \vdots & \dots & \dots & \dots &  0 \\
\hline
 0 & \dots & \dots &   0 & \mathfrak{H}_{r}\otimes  \pmb{\pmb{1}_{n_{r}\times  n_{r}}} \\
\hline
\end{array}
\right) \label{cosettusGF}
\end{equation}
where $n_i$ denotes the dimension of the $i$-th nontrivial representation of the discrete group $\Gamma$ and
from this we extract the block diagonal matrix:
\begin{equation}
\mathcal{H}\text{ }\equiv \text{ }\left(
\begin{array}{|c|c|c|c|c|}
\hline
 \mathfrak{H}_1  & 0 & \dots & \dots & 0 \\
\hline
 0 & \mathfrak{H}_2 & \dots & \dots & \vdots \\
\hline
 \vdots & \dots\dots & \dots\dots & \dots & \vdots \\
\hline
 \vdots & \dots\dots & \dots & \mathfrak{H}_{r-1} & 0 \\
\hline
 0 & \dots\dots & \dots & 0 & \mathfrak{H}_{r} \\
\hline
\end{array}
\right)\label{tautobundmetro}
\end{equation}
The hermitian matrix $\mathcal{H}$ is the fiber metric on the direct
sum:
\begin{equation}\label{direttosummo}
    \mathcal{R}\,=\,\bigoplus_{i=1}^{r} \, \mathcal{R}_i
\end{equation}
of the $r$ tautological bundles that, by construction, are holomorphic vector bundles with rank equal to the
dimension of the $r$ irreducible representations of $\Gamma$:
\begin{equation}\label{tautibundiEach}
    \mathcal{R}_i \, \stackrel{\pi}{\longrightarrow}\, \mathcal{M}_\zeta \quad\quad ;
    \quad \quad\forall p \in \mathcal{M}_\zeta \quad :\quad
    \pi^{-1}(p) \approx \mathbb{C}^{n_i}
\end{equation}
The compatible connection\footnote{Following standard mathematical nomenclature, we call compatible
connection on a holomorphic vector bundle,  one whose $(0,1)$ part is the Cauchy Riemann operator of the
bundle} on the holomorphic vector bundle $\mathcal{R}$ is given by:
\begin{eqnarray}\label{comancio}
    \vartheta & = & \mathcal{H}^{-1} \,\partial\mathcal{H}\, = \,\bigoplus_{i=1}^{r}\, \theta_i \nonumber\\
    \theta_i & = & \mathfrak{H}_i^{-1} \, \partial \mathfrak{H}_i \, \in \, \mathbb{GL}(n_i,\mathbb{C})
\end{eqnarray}
where $\mathbb{GL}(n_i,\mathbb{C})$ is the Lie algebra of $\mathrm{GL}(n_i,\mathbb{C})$ which is structural
group of the $i$-th tautological vector bundle. The natural connection of the $\mathcal{F}_\Gamma$ principal
bundle, mentioned in eq.\,(\ref{rodolfo}) is just, according to a universal scheme, the imaginary part of the
holomorphic connection $\vartheta$.
\subsubsection{The tautological bundles from the irrep viewpoint and the
K\"ahler potential}\label{realecompdiscus} From the analysis of the
above section we have reached a very elegant conclusion. Once the
matrix $\mathcal{V}$ is calculated as function of the level
parameters $\zeta $ and of the base-manifold coordinates
(\(z_m,\bar{z}_m\)) ($m=1,2,3$), we also have the block diagonal
hermitian matrix $\mathcal{H}$ which encodes the hermitian fiber
metrics \(\mathfrak{H}_i\)\(\left(\zeta ,z,\bar{z}\right)\) on each
of the tautological holomorphic bundles \(\mathfrak{V}_i\) whose
ranks are equal, one by one, { }to the dimensions \(n_i\) of the
irreps of  $\Gamma $.
The first Chern classes of these bundles are represented by the
differential $(1,1)$ forms:
\begin{equation}
\omega _i{}^{(1,1)}= \frac{i}{2\pi} \,\bar{\partial }\partial \text{Log}\left[\text{Det}\left[\mathfrak{H}_{i
}\right]\right] \label{bambolone}
\end{equation}
Let us recall  another remarkable group theoretical fact. The number \textit{r} of nontrivial irreps of
$\Gamma $ is equal to the number \textit{ r} of nontrivial conjugacy classes and to the number \textit{r} of
generators of the center of the compact Lie algebra \(\mathbb{F}_{\Gamma }\), hence also to the number
\textit{ r} of level parameters $\zeta $ and to number \textit{ r} of holomorphic tautological bundles. Now
we are in a position to derive in full generality  the formula for the K\"ahler potential and, hence, for the
K\"ahler metric of the resolved manifold \(\mathcal{M}_{\zeta }\) that we anticipated in (\ref{celeberro}) .
In view of the above discussion, we rewrite the latter as it follows:
\begin{equation}
\mathcal{K}_{\mathcal{M}_{\zeta }}= \mathcal{K}_{\mathcal{S}_{\Gamma }} \mid _{\mathcal{N}_{\zeta }}+
\zeta^i\mathfrak{C}_{\text{ij}}\text{Log}\left[\text{Det}\left[\mathfrak{H}_{j }\right]\right]
\label{criceto1}
\end{equation}
where $\mathcal{K}_{\mathcal{S}_{\Gamma }}$ is the K\"ahler
potential of the flat space \(\mathcal{S}_{\Gamma}\) and \(\mid
_{\mathcal{N}_{\zeta }}\) denotes its restriction to the level
surface \(\mathcal{N}_{\zeta }\), while \(\mathfrak{C}_{\text{IJ}}\)
is an r$\times $r constant matrix whose structure we  will define
and determine below. Why the matrix defined there yields the
appropriate K\"ahler potential is what we will now explain starting
from an argument introduced in 1987 by Hitchin, Karlhede,
Lindstr\"om and Ro\v{c}ek.
\paragraph{The HKLR differential equation and its solution} Before explaining the origin
of the matrix \(\mathfrak{C}_{\text{IJ}}\), we would like to stress that, conceptually it encodes a pairing
between the level parameters ( = generators of the Lie algebra center) and the tautological bundles ( =
irreps). If we could understand the relation between conjugacy classes with their ages and cohomology
classes, then we would have a relation between irreps and conjugacy classes and we could close the three-sided relation diagram among the center \(\mathfrak{z}\left[\mathbb{F}_{\Gamma }\right.\)] and the other two
items, irreps and conjugacy classes.  As we are going to show, this side of the relation is based on the
concept of weighted blowup. On the other hand, understanding the matrix \(\mathfrak{C}_{\text{IJ}}\), is a
pure Lie algebra theory issue, streaming from the HKLR argument.
\par
Hence,  continuing such an argument, let us  consider the flat K\"ahler manifold \(\mathcal{S}_{\Gamma }\) and
its K\"ahler potential
\begin{equation}
\mathcal{K}= \sum _{i=1}^3 \text{Tr}\left[A_i A_i^{\dagger } \right]\text{ }\text{where we have
defined}\text{  }A_i=\{A,B,C\} \label{criceto2}
\end{equation}
The exponential of the K\"ahler potential is also, by definition, the hermitian metric on the Hodge line
bundle:
\begin{eqnarray}
&&\mathcal{L}_{\text{Hodge} }\overset{\pi }{\longrightarrow } \mathcal{S}_{\Gamma }\quad \text{ where }\quad
\forall p\in \mathcal{S}_{\Gamma }\quad : \quad \pi ^{-1}(p) \approx \mathbb{C}^* \nonumber\\
&&\quad\left\| W\right\| ^2\equiv e^{\mathcal{K}_{\mathcal{S}}}W\bar{W} \label{criceto3}
\end{eqnarray}
Indeed, the second line of the above equation $\left\| W\right\| ^2$  defines the invariant norm of any
section of $\mathcal{L}_{\text{Hodge}}$.
\par
Let us know consider the action of the quiver group on \(\mathcal{S}_{\Gamma }\) and its effect on the fiber
metric h=\(e^{\mathcal{K}}\). The maximal compact subgroup \(\mathcal{F}_{\Gamma }\) is an isometry group for
the K\"ahler metric defined by (\ref{criceto2}). Hence we focus on the orthogonal (with respect to the Killing form) complement of
\(\mathcal{F}_{\Gamma }\). Let
\begin{equation}
\pmb{\Phi} \in \mathbb{K}_{\Gamma } \label{criceto4}
\end{equation}
be an element of the orthogonal subspace to the maximal compact subalgebra
\begin{equation}
\mathbb{G}_{\Gamma }=\mathbb{F}_{\Gamma }\oplus \mathbb{K}_{\Gamma } \label{criceto5}
\end{equation}
consider the one parameter subgroup generated by this Lie algebra element
\begin{equation}
g(\lambda ) \equiv  e^{\lambda \pmb{\Phi} } \label{criceto6}
\end{equation}
The action of this group on the K\"ahler potential is easily calculated
\begin{equation}
\mathcal{K}_{\mathcal{S}}(\lambda ) =\sum _{i=1}^3 \text{Tr}\left[A_i e^{2\lambda \pmb{\Phi} }A_i^{\dagger }
e^{-2\lambda \pmb{\Phi} }\right]\label{criceto6bis}
\end{equation}
Performing the derivative with respect to $\lambda $ we obtain
\begin{equation}
\partial _{\lambda }\mathcal{K}_{\mathcal{S}}(\lambda )\mid_{\lambda =0}
 =\sum _{i=1}^3 \text{Tr}\left(\pmb{\Phi} \left[A_i, A_i^{\dagger }
\right]\right) \label{criceto7}
\end{equation}
Then we utilize the fact that each element $\pmb{\Phi}\in \mathbb{K}_\Gamma $ is just equal to  ${\rm i}
\times $ \(\pmb{\Phi} _c\) where \(\pmb{\Phi} _c\) denotes an appropriate element of the compact subalgebra.
Hence the above equation becomes
\begin{equation}
\partial _{\lambda }\mathcal{K}_{\mathcal{S}}(\lambda )\mid _{\lambda =0} ={\rm i} \times \sum _{i=1}^3
\text{Tr}\left(\pmb{\Phi} _c\left[A_i, A_i^{\dagger } \right]\right) = i \mathfrak{P}_{\pmb{\Phi} _c}
\label{criceto8}
\end{equation}
Let us decompose the moment map along the standard basis of compact generators. We obtain:
\begin{eqnarray}
\mathfrak{P}_{\Phi }&=&\sum _{I=1}^{|\Gamma| -1} \pmb{\Phi} ^I\text{Tr}\left(\mathfrak{K}_I^c\, \left[A_i,
A_i^{\dagger}
\right]\right)\nonumber\\
&=&{\rm i}\sum _{I=1}^{|\Gamma| -1} \pmb{\Phi} _c^I \mathfrak{P}_I(p) \, = \, \sum _{I=1}^{|\Gamma| -1}
\pmb{\Phi} ^I\mathfrak{P}_I(p)=\sum_{I=1}^{|\Gamma| -1}\pmb{\Phi} ^I\text{Tr}\left(\mathfrak{K}_I\left[A_i,
A_i^{\dagger } \right]\right) \label{criceto9}
\end{eqnarray}
where p $\in $\(\mathcal{D}_{\Gamma } \subset \mathcal{S}_{\Gamma
}\) denotes the arbitrary point in the ambient space described by
the triple of matrices \(A_i\), \(\mathfrak{K}_I\)=i
\(\mathfrak{K}_I^c\) are the $|\Gamma |$-1 noncompact generators of
the quiver group \(\mathcal{G}_{\Gamma }\) that, by construction,
are just as many as the compact generators \(\mathfrak{K}_I^c\) of
the maximal compact subgroup \(\mathcal{F}_{\Gamma }\). Formally
integrating the above differential equation it follows that the
fiber of the metric Hodge line bundle (\ref{criceto3})
\begin{equation}
h(p) \equiv \text{Exp}\left[\mathcal{K}_{\mathcal{S}}(p)\right] \label{criceto10}
\end{equation}
transforms in the following way under the action of the quiver group
\begin{equation}
\forall g\in \mathcal{G}_{\Gamma }\text{           }g : h(p) \longrightarrow  h^g(p) \equiv
 h\left(e^{\text{Log}[g]}p\right)=e^{c(g,p)}h(p)
 \label{criceto11}
\end{equation}
where
\begin{equation}
\text{Log}[g]\in \mathbb{G}_{\Gamma } \label{criceto12}
\end{equation}
is an element of the quiver group Lie algebra and as such can be decomposed along a complete basis of
generators
\begin{equation}
\text{Log}[g]= \sum _{I=1}^7 \pmb{\Phi} ^I\mathfrak{K}_I+\pmb{\Phi} _c{}^I\mathfrak{K}^c{}_I
\label{criceto13}
\end{equation}
and the anomaly \(c(g,p)\)introduced in eq.\,(\ref{criceto11}) has, in force of the differential equation
discussed above the following form:
\begin{equation}
c(g,p)=\sum _{I=1}^7  \left(\pmb{\Phi} ^I+i \pmb{\Phi} _c{}^I\right) \mathfrak{P}_I(p) \label{criceto14}
\end{equation}
where \(\mathfrak{P}_I(p)\) are the moment maps at point p.
\par
Next consider the diagram
\begin{equation}
\mathcal{S}_{\Gamma }\overset{\text{       }\iota }{\text{   }\longleftarrow }\text{    }\mathcal{N}_{\zeta
}\text{      }\overset{\pi }{\longrightarrow }\text{    }\mathcal{M}_{\zeta }\text{  }\equiv \text{
}\mathcal{N}_{\zeta }/\mathcal{F} _{\Gamma } \label{criceto15}
\end{equation}
where \(\mathcal{N}_{\zeta }\) is the level surface and \(\mathcal{M}_{\zeta }\) the final K\"ahler threefold
with its associated Hodge line bundle whose curvature is the K\"ahler form \(\mathbf{K} _{\mathcal{M}}\)
\begin{equation}
\mathbf{K} _{\mathcal{M}} \equiv \frac{i}{2\pi }\bar{\partial }\partial  \mathcal{K}_{\mathcal{M}} =
\frac{i}{2\pi }\bar{\partial }\left( \frac{1}{h_{\mathcal{M}}} \partial h_{\mathcal{M}}\right)
\label{criceto19}
\end{equation}
\(\mathcal{K}_{\mathcal{M}}\) being the K\"ahler potential of the resolved variety. Following HKLR, we
require that
\begin{equation}
\pi ^*\mathbf{K} _{\mathcal{M}} = \iota ^*\mathbf{K} _{\mathcal{S}_{\Gamma }} \label{criceto20}
\end{equation}
where \(\mathbf{K} _{\mathcal{S}_{\Gamma }}\) is the K\"ahler form
of the flat K\"ahler manifold \(\mathcal{S}_{\Gamma }\)
=\(\text{Hom}_{\Gamma }\)(Q\(\otimes\)R,R). At the level of fiber
metric on the associated Hodge line bundles, eq.\,(\ref{criceto20})
amounts to stating that
\begin{equation}
\forall p\in \mathcal{M}_{\zeta }\quad:\quad h_{\mathcal{M}}\text{  }(p) = h_{\mathcal{S}_{\Gamma }}^g (p)
=h_{\mathcal{S}_{\Gamma }} (g.p)=e^{c(g,p)}\text{  }h_{\mathcal{S}_{\Gamma }} (p)\text{    }
\end{equation}
where g is an element of the quiver group that brings the point p $\in $ \(\mathcal{N}_{\zeta }\) on the
level surface of level $\zeta $ to the reference level surface \(\mathcal{N}_0\) which corresponds to the
singular orbifold \(\frac{\mathbb{C}^3}{\Gamma }\). Applying this to eq.\,(\ref{criceto14}) we obtain:
\begin{equation}
c(g,p) =\zeta ^I\pmb{\Phi} _I(p)=\zeta ^I*\text{Tr}\left[\mathfrak{K}_I\text{Log}[g]\right]=\sum _{i=1}^r
\zeta ^I*\text{Tr}\left[\mathfrak{K}_I^{\text{central}}\text{Log}[g]\right]
\end{equation}
since the only non-vanishing levels are located in the Lie Algebra center. On the oher hand we have g =
$\mathcal{H}$ :
\begin{equation}
\text{Tr}\left[\mathfrak{K}_I^{\text{central}}\text{Log}[\mathcal{H}]\right]\equiv\sum _{J=1}^r
\mathfrak{C}_{\text{IJ}}\text{Log}\left[\text{Det}\left[\mathfrak{H}_J\right]\right]
\end{equation}
The above formula defines the constant matrix $\mathfrak{C}_{\text{IJ}}$ and  justifies the final formula
(\ref{criceto1}). In appendix \ref{exemplaria} we will calculate an example of matrix $\mathfrak{C}_{IJ}$
for a simple case of a nonabelian group $\Gamma$ which leads to tautological bundles of rank larger then
one. In the case  of cyclic $\Gamma$ the center of the Lie Algebra $\mathbb{F}_\Gamma$ coincides
with the algebra itself and the matrix $\mathfrak{C}_{IJ}$ is just diagonal and essentially trivial.
\paragraph{Expansion to first order} In the Eguchi-Hanson example, which is the only one where the explicit form of
the closed forms can be derived, we have explicitly verified that the term of order one
\(\overset{(1)}{\omega }_i{}^{(1,1)}\) in the small $\zeta $ parameter expansion is
\begin{equation}
\omega _i^{(1,1)} = 0 +\sum _{n=1}^{\infty } \zeta ^n\overset{(n)}{\omega }_i{}^{(1,1)}
\end{equation}
cohomologous to the full form \(\omega _i^{(1,1)}\). Hence it suffices to solve the moment map equations at
first order in $\zeta $ (which is always possible) and we obtain a calculation of the cohomology classes of
the resolved variety according to the above displayed scheme. At the same time we obtain a calculation of the
K\"ahler potential to the very same order.
\par
We assume that this is a general feature applying to all cases.
\paragraph{Dolbeault cohomology} The objects
we are dealing with are Dolbeault cohomology classes of the final resolved manifold $\mathcal{M}_{\zeta}$
which is K\"ahler as a result of its K\"ahler quotient construction.
\par
When we say that $\omega^{p,q}$ is a harmonic representative of a nontrivial cohomology class in
$H^{1,1}\left(\mathcal{M}_{\zeta}\right)$ we are stating that:
\begin{itemize}
  \item The form is $\partial$-closed and $\overline{\partial}$-closed
  \begin{equation*}
    \partial \omega^{p,q} \, = \,\overline{\partial}\omega^{p,q} \, = \, 0
  \end{equation*}
   \item There do  not exist forms $\phi^{p-1,q}$ and $\phi^{p,q-1}$ such that:
   \begin{equation*}
    \omega^{p,q}\, = \, \partial \phi^{p-1,q} \, = \,\overline{\partial}\phi^{p,q-1}
  \end{equation*}
\end{itemize}
The reason why the $\omega _i^{(1,1)}$ are nontrivial  representatives of $(1,1)$ cohomology classes is that
they are obtained as $\overline{\partial}$ of connection one-forms $\theta^{(1,0)}$ that are not globally
defined.  Indeed if we introduce the curvatures and the first Chern classes of the  tautological vector
bundles  we have the elegant formula anticipated in eq.\,(\ref{bambolone}):
\begin{eqnarray}
  \Theta_i &=& \overline{\partial} \theta_i \nonumber\\
  \omega^{(1,1)}_i &\equiv & c_1(\mathcal{R}_i) \, = \, \mathrm{Tr}(\Theta_i) \, = \,\overline{\partial}\,\partial
  \log \left[\mbox{Det} \left(\mathfrak{H}_i\right) \right]
\end{eqnarray}
Comparing now with the definition of Dolbeault cohomology we see that $\omega^{(1,1)}_i$ are nontrivial
cohomology classes because either
\begin{equation}\label{nobuono}
    \theta^{(1,0)} \, \equiv \, \partial \log \left[\mbox{Det} \left(\mathfrak{H}_i\right) \right] \quad \mbox{or}\quad
\theta^{(0,1)} \, \equiv \, \overline{\partial} \log \left[\mbox{Det} \left(\mathfrak{H}_i\right) \right]
\end{equation}
are non-globally defined $1$-forms on the base manifold. This is so because they transform nontrivially from
one local trivialization of the bundle to the next one. The transition functions on the connections are
determined by the transition functions on the metric $\mathcal{H}$, namely on the coset representative. Here
comes the delicate point.
\par
Where from in the Kronheimer--like construction do we know that there are different local trivializations,
otherwise that the tautological bundles are nontrivial? Computationally we solve the algebraic equations for
$\mathcal{H}$ in terms of the coordinates $z_i$ $(i=1,2,3)$ parameterizing the locus $L_\Gamma$, which is
equivalent to the singular locus $\frac{\mathbb{C}^3}{\Gamma}$ and we find $\mathcal{H}=\mathcal{H}(\zeta,z)$
where $\zeta$ are the level parameters. In order to conclude that the tautological bundle is nontrivial we
should divide the locus $L_\Gamma$ into patches and find the transition functions of the connections
$\theta_i$ from one patch to the other. Obviously the transition function must be an element of the the
quiver group $g\in \mathcal{G}_{\Gamma}$. At the first sight it is not clear  how to implement such a
program, since we do not know how we should partition the locus $L_\Gamma$. Clearly the actual solution of
the algebraic equations is complicated and, as we very well know, we are able to implement it only by means
of a power series in $\zeta$, yet it is obvious that this is not a case by case study. As everything else in
the Kronheimer--like construction, it must be based on first principles and it is precisely these first
principles that we are going to find out. It  is  at this level that the issue of ages is going to come into
play in an algorithmic way. We begin by inspecting the only case where the final analytic form of all the
construction items is available in closed form, namely the Eguchi-Hanson case.
\subsection{What we see  in the Eguchi-Hanson case}
Let us briefly summarize what we have verified in the EH case. The space \(\mathcal{N}_{\zeta }\) has a
natural structure of  principal U(1)-bundle over the quotient \(\mathcal{M}_{\zeta }\), as the maximal
compact subgroup of the quiver group \(\mathcal{F} _{\Gamma }\) $\subset $\(\mathcal{G} _{\Gamma }\) in this
case is just U(1).
\begin{equation}
\text{                                                     } \mathcal{N}_{\zeta }\text{      }\overset{\pi
}{\longrightarrow }\text{    }\mathcal{M}_{\zeta }\text{  }\equiv \text{  }\mathcal{N}_{\zeta
}/\!\!/\mathcal{F} _{\Gamma }\text{    }
\end{equation}
As $\mathcal{N}_{\zeta }$ is a closed submanifold of $  \mathcal{S}_{\Gamma}$ it has an induced metric $g$.
The vertical tangent bundle to $\mathcal{N}_{\zeta }$ is  locally  generated by the vector field
\begin{equation}
\text{                                                                }V_{\text{vert}} = \frac{\partial
}{\partial \phi }
\end{equation}
Pointwise we can consider the space \(\text{T$\mathcal{N}$}_{\text{hor}}\)  orthogonal to the vertical
tangent bundle
\begin{equation}
\text{                                              }\text{T$\mathcal{N}$}_{\text{hor}} = \left\{\left.X \in
\text{T$\mathcal{N}$}_{\zeta } \right| < X,\frac{\partial }{\partial \phi }> \equiv  g\left(X,\frac{\partial
}{\partial \phi }\right)=0\right\}
\end{equation}
This assignment of a complement to the vertical tangent spaces is smooth and U(1)-invariant, and therefore
defines a connection on the principal bundle $\mathcal{N}_{\zeta }$, whose connection form $\pmb{A}$
satisfies
\begin{equation}
\text{                                              }\forall X \in  \text{T$\mathcal{N}$}_{\text{hor}}
:\text{                  }\pmb{A}(X) = 0\text{
 }; \pmb{A}\left(\frac{\partial }{\partial \phi }\right) = 1
\end{equation}
In the chosen coordinates we find:
\begin{equation}
\text{                                                   }\pmb{A}\pmb{ } \pmb{=}\text{    }\text{d$\phi
$}-\frac{\zeta  \text{d$\theta $}_1 \rho _1^2 }{2 \left(1+\rho _1^2\right) \sqrt{\zeta ^2+64 \left(1+\rho
_1^2\right){}^2 \rho _2^4}} -\frac{\zeta  \text{d$\theta $}_2}{2 \sqrt{\zeta ^2+64 \left(1+\rho
_1^2\right){}^2 \rho _2^4}}
\end{equation}
where:
\begin{equation}\label{groppo}
  z_{1,2}\, = \, \exp\left[{\rm i} \, \theta_{1,2} \right] \, \rho_{1,2}
\end{equation}
are the standard complex coordinates labeling the points of the locus $L_\Gamma$, namely parametrizing the two
matrices $A,B$ that solve the invariance constraint of $\Gamma$, defining
$\mbox{Hom}_\Gamma(\mathcal{Q}\times R,R)$, and are also diagonal in the natural basis of the regular
representation. In the split basis they turn out to be antidiagonal:
\begin{equation}\label{ABAB}
A\, = \,   \left(
     \begin{array}{cc}
       0 & Z_1 \\
       Z_1 & 0 \\
     \end{array}
   \right) \quad ; \quad B\, = \,   \left(
     \begin{array}{cc}
       0 & Z_2 \\
       Z_2 & 0 \\
     \end{array}
   \right)
\end{equation}
By means of the usual correspondence between $\mathrm{U(1)}$ bundles and line  bundles we conclude that this
connection  $\pmb{A}$  is the imaginary part of the connection theta of the corresponding bundle and we write
the equation:
\begin{equation}\label{cavicchio}
  \theta \, = \, \mathfrak{H}^{-1}\partial  \mathfrak{H}
\end{equation}
where the explicit solution of the algebraic moment map equations yields:
\begin{equation}\label{groppello}
  \mathfrak{H} \, = \, \frac{\sqrt[4]{\frac{\zeta +\sqrt{\zeta ^2+16 \left|Z_1^2\right|^2
  +\left|Z_2^2\right|{}^2}}{|Z_1|^2+|Z_2|^2}}}{\sqrt{2}}
\end{equation}
\paragraph{Curvature of the line bundle}
In this way we find that the tautological bundle has the following curvature:
\begin{equation}
\text{ } \Theta  = \bar{\partial }\partial \text{Log}[\mathfrak{H}]
\end{equation}
$\Theta $ is the first Chern class of the tautological line bundle implicitly defined by the above
construction
\begin{eqnarray}
&&\mathcal{L}\text{   } \overset{\pi }{\longrightarrow } \mathcal{M}_{\zeta}\nonumber \\
&&c_1(\mathcal{L}) = \left [\frac{i}{2\pi} \,\Theta  \right ] \in
 H^{1,1}\left( \mathcal{M}_{\zeta }\right)
\end{eqnarray}
where \(H^{1,1}\left( \mathcal{M}_{\zeta }\right)\) is the (1,1) cohomology group of the manifold
\(\mathcal{M}_{\zeta }\). On the other hand the very space of Eguchi-Hanson \(\mathcal{M}_{\zeta }\) is a
line bundle over \(\mathbb{P}_1\):
\begin{equation}
\mathcal{M}_{\zeta }\text{  }\overset{\pi_0}{\longrightarrow }\text{  }\mathbb{P}_1
\end{equation}
There is a (1,1)-form $\omega $ over \(\mathbb{P}_1\) which is the the first Chern class of the bundle
\(\mathcal{M}_{\zeta }\).
\begin{equation}
c_1\left(\mathcal{M}_{\zeta }\right) = \omega  \in  H^{1,1}\left( \mathbb{P}_1\right)
\end{equation}
We find that, as usual the pullback \(\pi _0{}^*\)of the projection \(\pi _0\) works in particular as
follows:
\begin{equation}
\pi _0{}^*: \text{        }T_{(1,1)}{}^*\mathbb{P}_1 \longrightarrow T_{(1,1)}{}^*\mathcal{M}_{\xi }
\end{equation}
We find that the (1,1)-form { }$\Theta $ which is defined over the whole \(\mathcal{M}_{\zeta }\) is the
pullback image of the first Chern class of the line bundle \(\mathcal{M}_{\zeta }\).
\begin{equation}
\pi _0{}^*\left[ c_1\left(\mathcal{M}_{\zeta}\right) \right]\text{   }=\text{   }c_1(\mathcal{L})
\end{equation}
The line bundle \(\mathcal{M}_{\zeta }\text{  }\overset{\pi _0}{\longrightarrow }\text{  }\mathbb{P}_1\) is by definition the one associated
with the vanishing locus   of the section \(\xi _2\).
\paragraph{What we have learned from this explicit case?} The above detailed analysis reveals that, according to
general lore, the cohomology classes constructed as first Chern classes of the tautological holomorphic vector bundles
defined by the K\"ahler quotient via  hermitian matrices
$\mathfrak{H}_i$, are naturally associated with  the components of the exceptional divisor. This latter is defined
as the vanishing locus of a global holomorphic section $W(p)$ of a line bundle:
\begin{eqnarray}\label{calendula}
    &&\mathcal{L}_\mathfrak{D} \, \stackrel{\pi}{\longrightarrow} \,\mathcal{M}_\zeta \nonumber\\
    &&\mathfrak{D}\subset \mathcal{M}_\zeta \quad ; \quad \mathfrak{D} \, =\, \left\{
    p \in \mathcal{M}_\zeta \, \mid \, W(p) \, = \, 0 \quad \mbox{where} \quad W \in
    \Gamma\left(\mathcal{L}_\mathfrak{D}\right)
    \right\}
\end{eqnarray}
\begin{figure}[ht]
\begin{center}
\includegraphics[height=10cm]{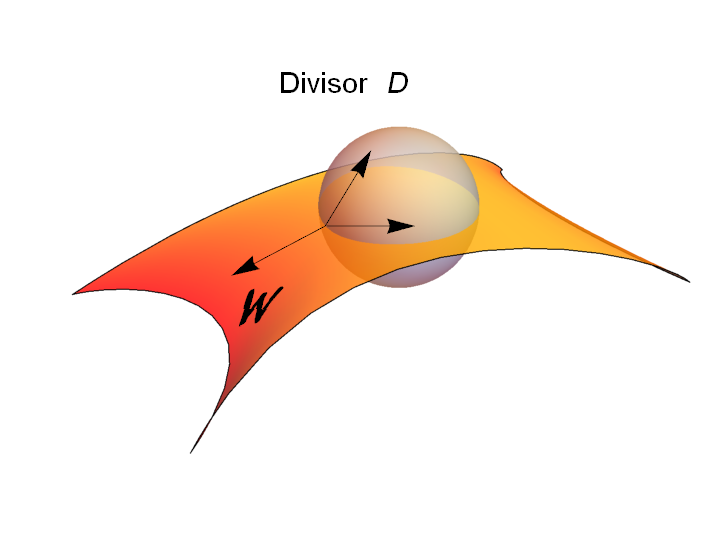}
\caption{\label{dividendo} In the Eguchi-Hanson case the exceptional divisor is a submanifold
$\mathfrak{D}\subset \mathcal{M}_\zeta$ of codimension one that is mapped into the singular point by the
resolving morphism $\mathcal{M}_\zeta \longrightarrow \frac{\mathbb{C}^2}{\Gamma}$. There is a projection
operation $\mathcal{M}_\zeta\stackrel{\pi}{\longrightarrow} \mathfrak{D}$ that makes $\mathcal{M}_\zeta$ the
total space of  a line bundle over the divisor. Accordingly we can choose a local coordinate system for
$\mathcal{M}_\zeta$  such that two coordinates span the divisor while the third, named $W$, transforms as if
it were a section of the mentioned line bundle.}
\end{center}
\end{figure}
The line bundle $\mathcal{L}_\mathfrak{D}$ is singled out by the divisor $\mathfrak{D}$ and for this reason it is
labeled by it. Its first Chern class $\omega_\mathfrak{D}^{(1,1)}$
is certainly a cohomology class and so it must be a linear combination of the
first Chern classes $\omega_i^{(1,1)}$ created by the K\"ahler quotient and associated with the hermitian matrices
$\mathfrak{H}_i(\zeta,p)$:
\begin{equation}\label{calzaturificio}
   \left[ \omega_\mathfrak{D}^{(1,1)}\right] \, = \, \mathrm{S}_{\mathfrak{D},i} \,
   \left[ \omega_i^{(1,1)}\right]
\end{equation}
The question is to know which is which and to determine the constant matrix $\mathrm{S}_{\mathfrak{D},i}$.
\par
Another point revealed by the analysis of the Eguchi-Hanson case is that, at least locally, the entire space
$\mathcal{M}_\zeta$ can be viewed as the total space of a line bundle over the divisor $\mathfrak{D}$:
\begin{eqnarray}\label{carnitinapotassica}
    && \mathcal{M}_\zeta \, \stackrel{\pi_d}{\longrightarrow} \,\mathfrak{D} \nonumber\\
    && \forall p \in \mathfrak{D}\quad ; \quad \pi_d^{-1}\left(p\right)\, \simeq \, \mathbb{C}^\star
\end{eqnarray}
Furthermore the matrix $\mathfrak{H}_i$ can be viewed as the invariant norm of a section of
the appropriate line bundle:
\begin{equation}\label{cosmitto}
    \mathfrak{H}_i(\zeta,z,\bar{z}) \, = \, H_i(\xi,\bar{\xi}, W,\bar{W}) \, |W|^2
\end{equation}
where $\xi$ denote the two coordinates spanning the divisor $\mathfrak{D}$ and $W$
(as in fig.\ref{dividendo}) spans the vertical fibers  out of the divisor. The projection $\pi_d$ corresponds
to setting $W\to 0$ and obtaining:
\begin{equation}\label{camalosto}
 \pi_d \, : \,   H(\xi,\bar{\xi}, W,\bar{W}) \longrightarrow h(\xi,\bar{\xi}) \equiv H(\xi,\bar{\xi}, 0,0)
\end{equation}
Just as in the case of Eguchi-Hanson, we expect that the two (1,1)-forms:
\begin{eqnarray}
  \Omega_i &=& \overline{\partial}\partial H_i(\xi,\bar{\xi}, W,\bar{W}) \nonumber\\
  \widehat{\Omega}_i &=& \overline{\partial}\partial h(\xi,\bar{\xi})\label{santeddu}
\end{eqnarray}
should be cohomologous:
\begin{equation}\label{commollo}
    \left[\Omega_i\right] \, = \, \left[\widehat{\Omega}_i\right]
\end{equation}
The form $\widehat{\Omega}_i$ is the first Chern class of the line bundle (\ref{carnitinapotassica}) while
$\Omega_i$ is the first Chern class of the line bundle (\ref{calendula}) that defines the divisor.
\paragraph{Divisors and conjugacy classes graded by age.}
Hence the question boils down to the following: \textit{What are the
components of the exceptional divisor of a crepant resolution of the
singularity $\mathbb C^3/\Gamma$,  and how many are they?} The
answer is provided by Theorem \ref{reidmarktheo} (Theorem 1.6 in
\cite{giapumckay}); they are the inverse images via the blowdown
morphism of the irreducible components of the fixed locus of the
action of $\Gamma$ on $\mathbb C^3$, and are in a one-to-one
correspondence with the junior conjugacy classes of $\Gamma$. The
irreducible components of the exceptional divisor may be compact
(corresponding to a component of the fixed locus which is just the
origin of $\mathbb C^3$) or noncompact (corresponding to fixed loci
of higher dimensions, i.e., curves).
\par
Let us consider the case of a cyclic group $\Gamma$, with only the origin as fixed locus, and choose a generator
$\gamma$ of $\Gamma$ of order $r$. As in eq.\,(\ref{gioffo}), we can write $\gamma=\frac1r(a_1,a_2,a_3)$.
As described in \cite{giapumckay}, Sections 2.3 and 2.4, the resolution of singularities is obtained by iterating the following construction, which uses toric geometry
(a general reference for toric geometry, which in particular explains how to perform a toric blowup by subdividing the fan of the toric variety one wants to blowup, is \cite{Fulton-toric}).
The fan of the toric variety $\mathbb C^3$ is the first octant of $\mathbb R^3$, with all its faces;
by adding the ray $\frac1r(a_1,a_2,a_3)$ we perform a blowup
$\mathbb{B}_{[a_1,a_2,a_3]} \to \mathbb C^3$ whose exceptional divisor  $F$ is the weighted projective space $\mathbb{WP}[a_1,a_2,a_3]$. The same procedure applied to $\mathbb C^3/\Gamma$
produces a partial desingularization $ W_\gamma \to \mathbb C^3/\Gamma$
which is the base of a cyclic covering $\mathbb{B}_{[a_1,a_2,a_3]} \to W_\gamma$, ramified along the exceptional divisor $E$ of $ W_\gamma \to \mathbb C^3/\Gamma$. The situation is summarised by the following diagram
\begin{equation}\label{blowup} \xymatrix@C+40pt{
F \ar@^{(->}[r] \ar[d] & \mathbb{B}_{[a_1,a_2,a_3]} \ar[r]^-{\mbox{\footnotesize weighted blowup}}\ar[d] &  \mathbb C^3 \ar[d] \\
E \ar@^{(->}[r] &W_\gamma \ar[r]^-{\mbox{\footnotesize  crepant resolution}} &  \mathbb C^3/\Gamma
}\,.
\end{equation}

The full desingularization is obtained by performing a multiple toric blowup, adding all rays corresponding to junior conjugacy classes.

\subsection{Steps of a weighted blowup} \label{steppisoffio}
\subsubsection{Weighted projective planes}
Let us define in a pedantic way the weighted blowup of the origin in $\mathbb{C}^3$. To this effect we begin
by recalling the definition of a weighted projective plane $\mathbb{WP}_{[a_1,a_2,a_3]}$, where
$[a_1,a_2,a_3]$ are the weights (good references for weighted projective spaces and line bundles on them are
\cite{Dolgy,Beltrametti-Robbiano,Rossi-Terracini}). We restrict our attention to the case where the weights are integers.
One defines an action of $\mathbb C^\ast$ on $\mathbb C^3-\{0\}$ letting
$$ (y_1,y_2,y_3) \to (y_1 \,\lambda^{a_1},y_2\,\lambda^{a_2},y_3\,\lambda^{a_2}),\qquad \lambda \in \mathbb C^\ast.$$
The weighted projective plane $\mathbb{WP}_{[a_1,a_2,a_3]}$ is the quotient of $\mathbb C^3-\{0\}$ under this action.
It is by construction a complex variety of dimension 2.
\par
To examine the properties of this space it is expedient to assume that the triple $(a_1,a_2,a_3)$ is reduced. One defines
$$ d_ i = \mbox{g.c.d.}\, (a_{i-1},a_{i+1}), \qquad  b_ i = \mbox{l.c.m.} (d_{i-1},d_{i+1}) $$
(where indices in the r.h.s.~are meant mod 3, i.e., $1-1=3$, etc.). The   triple $(a_1,a_2,a_3)$ is reduced
if $(b_1,b_2,b_3) = (1,1,1)$;
otherwise one defines $a'_i=a_i/b_i$. The numbers $a'_i$ are positive integers, the triple $(a'_1,a'_2,a'_3)$
is reduced, and  the weighted projective planes $\mathbb{WP}_{[a_1,a_2,a_3]}$ and $\mathbb{WP}_{[a'_1,a'_2,a'_3]}$
are isomorphic. Henceforth we shall  assume that the triple $(a_1,a_2,a_3)$ is reduced.
It turns out that $\mathbb{WP}_{[a_1,a_2,a_3]}$
is smooth if and only if $(a_1,a_2,a_3) = (1,1,1)$, in which case the weighted projective
plane is just $\mathbb P^2$.
\par
The same construction of line bundles on projective spaces (see
e.g.\cite{Hart}, Section II.5) produces on weighted projective
spaces rank one sheaves $\mathcal
O_{\mathbb{WP}_{[a_1,a_2,a_3]}}(i)$, with $i\in \mathbb Z$, that in
general are not locally free (i.e., they are not line bundles), but
only reflexive (i.e., they are isomorphic to their duals). It turns
out that  $\mathcal O_{\mathbb{WP}_{[a_1,a_2,a_3]}}(i)$ is locally
free if and only if $i$ is a multiple of $m = \mbox{l.c.m.}
(a_1,a_2,a_3)$ \cite{Beltrametti-Robbiano}.
\par
The weighted projective plane $\mathbb{WP}_{[a_1,a_2,a_3]}$ is covered by the open sets
$$ U _ i = \{ (y_1,y_2,y_3) \,\vert \, y_i\ne 0\}.$$
On this open cover the  line  bundle  $\mathcal
O_{\mathbb{WP}_{[a_1,a_2,a_3]}}(km)$ has  transition functions
\begin{equation}\label{transitionfunctions}
g_{ij}\colon U_i\cap U_j \to\mathbb C^\ast,\qquad
g_{ij}(y_1,y_2,y_3) =  y_j^{km/a_j}y_i^{-km/a_i}
\end{equation}
where $m = \mbox{l.c.m.}\ (a_1,a_2,a_3)$. In particular, (the
isomorphism class of) $\mathcal O_{\mathbb{WP}_{[a_1,a_2,a_3]}}(m)$
is the (very ample) generator of the Picard group of
$\mathbb{WP}_{[a_1,a_2,a_3]}$, the group of isomorphism classes of
line bundles on $\mathbb{WP}_{[a_1,a_2,a_3]}$, which is isomorphic
to $\mathbb Z$.
We conclude this brief introduction to weighted projective planes by
defining an orbifold K\"ahler metric for the spaces
$\mathbb{WP}_{[a_1,a_2,a_3]}$. Denoting again by $(y_1,y_2,y_3)$ a
set of homogeneous coordinates on $\mathbb{WP}_{[a_1,a_2,a_3]}$, and
$m = \mbox{l.c.m.}\,(a_1,a_2,a_3)$, one can check the 2-form
$$ \omega = \frac{i}{2\pi} \partial\bar\partial\log\sum_{i=1}^3 y_i^{m/a_i}\,\bar y_i^{m/a_i}$$
is invariant under rescaling of the homogeneous coordinates, and
therefore defines a 2-form on the smooth locus  of
$\mathbb{WP}_{[a_1,a_2,a_3]}$; this reduces to the usual
Fubini-Study form when the projective space is smooth.
\subsubsection{The weighted blowup and the tautological bundle}
The weighted blowup of $\mathbb{C}^3$, denoted
$\mathbb{B}_{[a_1,a_2,a_3]}$, is a subvariety
\begin{equation}\label{corleone}
    \mathbb{B}_{[a_1,a_2,a_3]} \, \subset \, \mathbb{C}^3 \times \mathbb{WP}_{[a_1,a_2,a_3]}
\end{equation}
defined by the equations
\begin{equation}\label{eqblowup} z_1y_2^{a_1a_3}=z_2y_1^{a_2a_3},\qquad z_2y_3^{a_1a_2}=z_3y_2^{a_1a_3},\qquad
z_1y_3^{a_1a_2}=z_3y_1^{a_2 a_3}
\end{equation}
where $\{z_1,z_2,z_3\}$ are standard coordinates in $\mathbb C^3$,
and $\{y_1,y_2,y_3\}$ are homogenous cooordinates in $
\mathbb{WP}_{[a_1,a_2,a_3]}$. Actually the three equations are not
independent (regarding them as a linear system in the unknowns $z$,
the associated matrix has rank at most 2) and therefore the locus
$\mathbb{B}_{[a_1,a_2,a_3]}$ is 3-dimensional.  The projections of $
\mathbb{C}^3 \times \mathbb{WP}_{[a_1,a_2,a_3]}$ onto its factors
define projections
 \begin{eqnarray}
 p \quad   &:& \quad \mathbb{B}_{[a_1,a_2,a_3]}  \to  \mathbb{C}^3 \nonumber\\
 \pi \quad   &:& \quad \mathbb{B}_{[a_1,a_2,a_3]} \to \mathbb{WP}_{[a_1,a_2,a_3]}
 \nonumber
  \label{portapannolini}
\end{eqnarray}
From eq.~\eqref{eqblowup} we see that the fibers of $\pi$ are
isomorphic to $\mathbb C$; indeed, by comparing with
eq.~\eqref{transitionfunctions}, we see that
$\mathbb{B}_{[a_1,a_2,a_3]}$ is the total space of the line bundle
$\mathcal O_{\mathbb{WP}_{[a_1,a_2,a_3]}}(-1)$ over the base
$\mathbb{WP}_{[a_1,a_2,a_3]}$, and $\pi$ is the bundle projection.
On the other hand, the morphism $p$ is   birational, as it is an
isomorphism away from the fiber $p^{-1}(0)$, while the fiber itself
--- the exceptional divisor $F$ of the blowup  ---  is isomorphic to
$\mathbb{WP}_{[a_1,a_2,a_3]}$.

\begin{figure}\label{figure1}
\begin{center}
\begin{tikzpicture}
  \path [fill=pink] (0,0) to (4,2.3) to  (4,6.9) to (0,0) ;
    \path [fill=yellow] (8,0) to (4,2.3) to  (4,6.9) to (8,0) ;
      \path [fill=green] (0,0) to (4,2.3)  to  (8,0) to (0,0);
\draw [fill] (0,0) circle (3pt);
    \draw [fill]  (8,0) circle (3pt);
      \draw [fill] (4,6.9) circle (3pt);
        \draw [fill] (4,2.3) circle (3pt);
\draw (0,0) -- (8,0); \draw (0,0) -- (4,6.9); \draw (8,0) --
(4,6.9); \draw (0,0) -- (4,2.3); \draw (8,0) -- (4,2.3); \draw
(4,6.9) -- (4,2.3); \node at (-0.3,0) {$e_1$}; \node at (8.3,0)
{$e_2$}; \node at (4,7.2) {$e_3$}; \node at (4.2,2.5) {$v$}; \node
at (4,1) {$\sigma_3$}; \node at (3,3) {$\sigma_2$}; \node at (5,3)
{$\sigma_1$};
\end{tikzpicture}
\caption{\label{fan} Representation of the fan of
$\mathbb{B}_{[a_1,a_2,a_3]}$. The vector $v$ \emph{is not} on the
plane singled out by $e_1$, $e_2$, $e_3$.}
\end{center}
\end{figure}
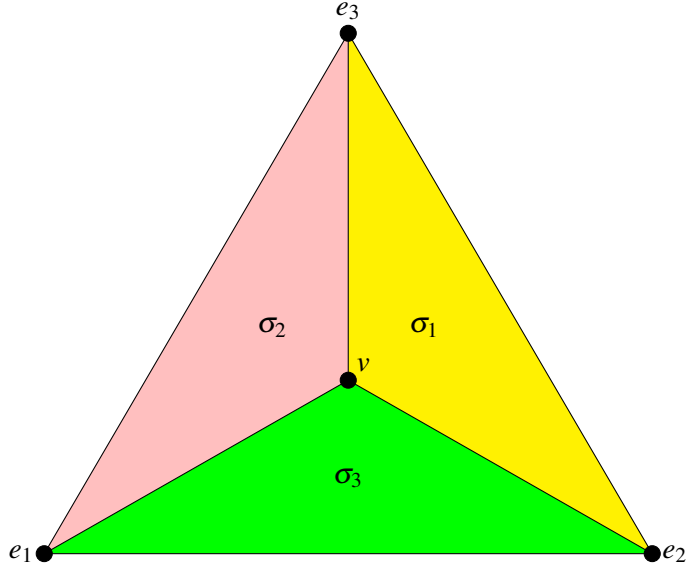

The blowup $\mathbb{B}_{[a_1,a_2,a_3]}$ is nicely described in terms
of toric geometry \cite{Fulton-toric}. Denoting by $\{e_i\}$ the
standard basis of $\mathbb R^3$, the variety
$\mathbb{B}_{[a_1,a_2,a_3]}$ is associated with the fan given by the
one-dimensional cones (rays) generated by
$$ e_1,\qquad  e_2, \qquad e_3, \qquad v = a_1e_1+a_2e_2+a_3e_3.$$
The fan has three 3-dimensional cones $\sigma_i$, corresponding to 3
open affine toric varieties $U_i$ which cover
$\mathbb{B}_{[a_1,a_2,a_3]}$  (see Figure \ref{fan}). It turns out
that $U_i$ is smooth if and only if $a_i=1$, so that
 $\mathbb{B}_{[a_1,a_2,a_3]}$ is smooth if and only if $a_1=a_2=a_3=1$ (in which case the exceptional divisor is a $\mathbb P^2$).
 Moreover, unless again $a_1=a_2=a_3=1$, $F$ is a Weil divisor, so that its associated rank one sheaf (it ideal sheaf,
 i.e., the sheaf of functions  $\mathbb{B}_{[a_1,a_2,a_3]}$ that vanish on $F$), is not locally free, but only reflexive. We shall
denote the dual of this sheaf as $\mathcal
O_{\mathbb{B}_{[a_1,a_2,a_3]}}(F)$. Although this in general is not
locally free, it is still true that its first Chern class if
Poincar\'e dual to $F$.

%
%

Let us assume that we are interested in desingularizing a quotient
$\mathbb C^3/\Gamma$, where $\Gamma$ is cyclic, and the
representation of $\Gamma$ on $\mathbb C^3$ has just one junior
class $\frac1r(a_1,a_2,a_3)$, which is compact, i.e., all $a_i$ are
strictly positive. The geometric constructions in this section show
that the orbifold K\"ahler form on $\mathbb{WP}_{[a_1,a_2,a_3]}$
induce by pullback a K\"ahler form on the smooth locus of the blowup
$\mathbb{B}_{[a_1,a_2,a_3]}$. With reference to diagram
\eqref{blowup}, the action of the group $\Gamma$ leaves the
exceptional divisor $F$ pointwise fixed, so that the form descends
to the desingularization $\mathcal M_\zeta$ of $\mathbb C^3/\Gamma$.
However, this K\"ahler form does not appear to coincide with the
K\"ahler form built on $\mathcal M_\zeta$ by means of the K\"ahler
reduction (cf.~Section \ref{exemplaria}). This comes to no surprise
as it is known that there are several different metrics on a
weighted projective space which all reduce to the standard
Fubini-Study metric in the smooth case (see \cite{Ross-Thomas}). We
shall analyze the relation between the naive metric introduced above
and the one obtain by K\"ahler reduction in a future work.
\subsubsection{Pairing between irreps and conjugacy class in the K\"ahler quotient resolution: open questions}
According to \cite{giapumckay}, in the crepant resolution:
\begin{equation}\label{sandalorotto}
    \mathcal{M}_\zeta \, \longrightarrow \, \frac{\mathbb{C}^3}{\Gamma}
\end{equation}
we obtain a component of the exceptional divisor
$\mathfrak{D}^{(E)}$ for each junior conjugacy class of the group
$\Gamma$, namely we have:
\begin{equation}\label{semifreddocaffe}
    \mathfrak{D}^{(E)} = \bigcup_{i=1}^{ \mbox{$\#$ of junior
    classes}} \,\mathfrak{D}_{[a_1,a_2,a_3]_i}
\end{equation}
When there is just one  junior class,  the procedure
described in previous subsections, which is graphically summarized
in  the diagram (\ref{blowup}), is exhaustive and we easily identify the
exceptional divisor with a single projective plane
$\mathbb{WP}_{[a_1,a_2,a_3]}$. Indeed, the divisor $F$ is the weighted projective plane
$\mathbb{WP}_{[a_1,a_2,a_3]}$  by construction, and the action of $\Gamma$ leaves it pointwise fixed, so
that $E$ is isomorphic to $F$.

Utilizing the correspondence between
line bundles and divisors,
we can conclude that the exceptional divisor
$\mathbb{WP}_{[a_1,a_2,a_3]}$ uniquely identifies a line-bundle,
\textit{i.e.}, the tautological line bundle
$\mathcal{T}_{[a_1,a_2,a_3]}$, whose first Chern class is
necessarily given by:
\begin{equation}\label{crescenzio}
    c_1\left(\mathcal{T}_{[a_1,a_2,a_3]}\right) \, = \, \frac{\rm i}{2\pi} \, \overline{\partial}\,\partial \, \log
    \left(H_{[a_1,a_2,a_3]}\right)
\end{equation}
where $H_{[a_1,a_2,a_3]}$ is a suitable hermitian fiber-metric. The
most  interesting issue is to relate such an invariant fiber metric
and the (1,1)-form $c_1\left(\mathcal{T}_{[a_1,a_2,a_3]}\right)$
with the real functions $\frak{H}_i$ in eq.(\ref{tautobundmetro})
and the corresponding (1,1)-forms  (\ref{bambolone}), that are
produced by the K\"ahler quotient construction.
\par
In the next section such a relation will be explicitly analyzed in
the context of a simple master example that indeed is characterized
by a unique \textit{junior conjugacy class}.
\par
The construction of the exceptional divisor and the structure of the
blowup in cases with several junior classes is more complicated and
it is still under investigation. The general pairing rules between
irreps and conjugacy classes will be discussed and elucidated in a
future publication by the present authors. In the next section we
briefly outline the general problem before presenting our explicit
results for the above mentioned one junior class master model.
\section{Analysis of the (1,1)-forms: irreps versus conjugacy classes that is cohomology versus homology }
In the present section we plan to analyze in full detail, within the
scope of a one junior class model, the relation between the above
extensively discussed $\omega^{(1,1)}_\alpha$ forms ($\alpha =
1,\dots, r = \#$ of nontrivial irreps), with the exceptional
divisors generated by the blowup of the singularity, together with
the other predictions of the fundamental theorem \ref{reidmarktheo}
which associates cohomology classes of $\mathcal{M}_\zeta$ with
conjugacy classes of $\Gamma$. The number of nontrivial conjugacy
classes and the number of nontrivial irreps are equal to each other
so that we use $r$ in both cases, yet what is the actual pairing is
not clear a priori and it is not intrinsic to group theory, as we
have stressed several times. In this section we want to explore this
pairing and to do that in an explicit way we need explicit
calculable examples. These are very few because of the bottleneck
constituted by the solution of the moment map equations, that are
algebraic of higher degree and only seldom admit explicit analytic
solutions. For this reason we introduce here the full-fledged
construction of one of those rare examples, where the moment map
equations are solved in terms of radicals. As anticipated above this
model has the additional nice feature that the number of junior
conjugacy classes is just one. It will be the master model for our
explicit analysis. The construction of other examples is confined to
appendices. In particular appendix \ref{exemplaria} contains two
Abelian examples where $\Gamma$ is a cyclic group, respectively
$\Gamma\, = \, \mathbb{Z}_3$ and $\Gamma\, = \,\mathbb{Z}_7$. In
both cases there analyzed the chosen group $\Gamma$ is a subgroup of
the maximal simple group $\mathrm{L_{168}} \, \simeq \,
\mathrm{PSL(2,7)}$ whose action on $\mathbb{C}^3$ was considered by
Markushevich in \cite{marcovaldo}. The master model that we discuss
here is also based on $\Gamma =\mathbb{Z}_{3}$ but the action of the
latter  on $\mathbb{C}^3$ is differently constructed. The radically
different results of appendix \ref{frugifera} from those of the next
section \ref{masterdegree} should advise the reader of the relevance
of the embedding:
\begin{equation}\label{relevancia}
   \iota \quad : \quad  \Gamma \, \hookrightarrow \, \mathrm{SU(3)}
\end{equation}
In appendix \ref{nonabbello} we also provide the construction of the simplest possible nonabelian model that
corresponds to the choice $Dih_3 \, \sim \, S_3$. Also for this toy model the moment maps are algebraic of
higher degree and an analytic solution is out of question. Nevertheless the nonabelian cases are our main
final target and we will return to them in future publications.
\par
It is also important to stress that aim of the Kronheimer-like
construction is not only the calculation of cohomology but also the
actual determination of the K\"ahler potential (yielding the
K\"ahler metric), which is encoded  in formula (\ref{criceto1}).
From this point of view one of the $\mbox{Det}\mathfrak{H}_i$ may
lead to a corresponding
$\omega_i^{(1,1)}=\ft{i}{2\pi}\bar{\partial}\partial\mbox{Det}\mathfrak{H}_i$
that is either exact or cohomologous to another one, yet its
contribution to the K\"ahler potential, which is very important in
physical applications, can not be neglected. It is only the
cohomology class of the K\"ahler 2-form that is affected by the
triviality of one or more of the $\omega_i^{(1,1)}$; the
contributions to the K\"ahler potential that give rise to exact form
deformations of the K\"ahler 2-form are equally important as others.
\par
Having anticipated these general considerations  we turn to our master model.
\subsection{The master model $\frac{\mathbb{C}^3}{\Gamma}$ with generator
$\{\xi,\xi,\xi\}$}\label{masterdegree} In this section we develop all the calculations for the K\"ahler
quotient resolution  of the quotient singularity \(\frac{\mathbb{C}^3}{\mathbb{Z}_3}\) in the case where the
generator $Y$ of \(\mathbb{Z}_3\) is of the following  form:
\begin{equation}
Y=\left(
\begin{array}{ccc}
\xi  & 0 & 0 \\
0 & \xi  & 0 \\
0 & 0 & \xi  \\
\end{array}
\right)
\end{equation}
 $\xi $ being a primite cubic root of unity \(\xi ^3\) =1.\\
The equation \(p\wedge p\)=0 which is a set of quadrics has solutions arranged in various branches. There is
a unique, principal branch of the solution that has maximal dimension \(\mathcal{D}_{\Gamma }^0\) and is
indeed isomorphic to the \(\mathcal{G}_{\Gamma }\) orbit of the singular locus \(L_{\Gamma }\) . This
principal branch is the algebraic variety \(\mathbb{V}_{|\Gamma |+2}\) mentioned in eq.\,(\ref{belgone}), of
which we perform the K\"ahler quotient with respect to the group \(\mathcal{F}_{\Gamma }\)
\begin{equation}
\mathcal{F}_{\Gamma } =\underset{\mu =1}{\overset{r+1}{\otimes }}\mathrm{U}\left(n_{\mu }\right)\cap
\text{SU}(|\Gamma |) = \mathrm{U(1)\otimes U(1)}
\end{equation}
in order to obtain the crepant resolution together with its K\"ahler metric. In the above formula \(n_{\mu
}\) = $\{$1,1,1$\}$ are the dimensions of the irreducible representations of $\Gamma $=\(\mathbb{Z}_3\) and
r+1=3 is the number of conjugacy classes of the group ($r$ is the number of nontrivial
representations). \\
To make a long story short, exactly as in the Kronheimer case we are able to retrieve the algebraic equation
of the singular locus from traces and determinants of the quiver matrices restricted to \(L_{\Gamma }.\)
Precisely for the \(\mathbb{Z}_3\) case under consideration we obtain
\begin{equation}
\mathcal{I}_1=\text{Det}\left[A_o\right] ; \mathcal{I}_2=\text{Det}\left[B_o\right] ;\text{
}\mathcal{I}_3=\text{Det}\left[C_o\right] ; \mathcal{I}_4 =
\frac{1}{3}\text{Tr}\left[A_oB_oC_o\right]\label{osteoporo1}
\end{equation}
and we find the relation
\begin{equation}
\mathcal{I}_1\mathcal{I}_2\mathcal{I}_3=\mathcal{I}_4{}^3
\end{equation}
which reproduces the \(\mathbb{C}^3\) analogue of eqs. (\ref{wgammus}-\ref{identifyAk}) applying to the
\(\mathbb{C}^2\) case of Kronheimer and Arnold.
\par
The main difference, as we have several time observed, is that now the same eqs. remain true, with no
deformation for the entire \(\mathcal{G}_{\Gamma }\) = \(\mathbb{C}^*\) \(\times\) \(\mathbb{C}^*\), orbit of
the locus \(L_{\Gamma }\), namely for the entire { }\(\mathbb{V}_{|\Gamma |+2}\) = \(\mathbb{V}_5\) variety
of which we construct the K\"ahler quotient with respect to the compact subgroup \(\mathrm{U(1)}\times
\mathrm{U(1)}\) $\subset $ \(\mathbb{C}^* \times \mathbb{C}^*\). This is in line with the many times
emphasized feature  that in the $\mathbb{C}^3$ case there is no deformation of the complex structure.
\subsubsection{The actual calculation of the K\"ahler quotient and of the K\"ahler potential}
The calculation of the final form of the K\"ahler potential is reduced to the solution of a set of two algebraic
equations. The solutions of such equations are accessible in this particular case  since they  reduce to a
single cubic for which we have  Cardano{'}s formula. For this reason the present case is the
three-dimensional analogue of the Eguchi-Hanson space where everything is explicitly calculable and all
theorems admit explicit testing and illustration.
\par
By calculating the ages we determine the number of \(\omega ^{(q,q)}\) harmonic forms (where $q=1,2$).
According to  theorem \ref{reidmarktheo} all these forms (and their dual cycles in homology) should be in
one-to-one correspondence with the \textit{r} nontrivial conjugacy classes of $\Gamma $. On the other hand
the K\"ahler quotient construction associates one level parameter $\zeta $ to each generator of the center
$\mathfrak{z}$(\(\mathcal{F}_{\Gamma }\)) of the group \(\mathcal{F}_{\Gamma }\), two $\zeta $.s in this
case, that are in one-to-one correspondence with the \textit{r} nontrivial irreducible representation of
$\Gamma $. The number is the same, but what is the pairing between \pmb{irreps} and \pmb{conjugacy classes}?
More precisely how do we see the homology cycles that are created when each of the \textit{r} level
parameters $\zeta $ departs from its original zero value? Using the explicit expression of the functions
\(\mathfrak{H}_{1,2}\) defined in eqs. (\ref{cosettusGF}-\ref{bambolone}) we arrive at  the calculation of
the \(\omega ^{(1,1)}{}_{i=1,2}\) forms that encode the first Chern classes of the two tautological bundles.
The expectation from the age argument is that these two 2-forms should be cohomologous corresponding to just
the unique predicted class of type (1,1) since \(h^{1,1}\)=1. On the other hand we should be able to
construct an \(\omega ^{(2,2)}\) form representing the unique class that is Poincar\`e  dual to the
exceptional divisor.
\par
In this case we can successfully answer both questions and this is very much illuminating.
\paragraph{Ages.} Indeed taking the explicit generator
\begin{equation}
Y=\left(
\begin{array}{ccc}
 (-1)^{2/3} & 0 & 0 \\
 0 & (-1)^{2/3} & 0 \\
 0 & 0 & (-1)^{2/3} \\
\end{array}
\right)
\end{equation}
 we easily calculate the $\{a_1,a_2,a_3\}$ vectors respectively associated to each of the three conjugacy
 classes and we obtain:
\begin{equation}
a-\text{vectors} =\{\{0,0,0\},\ft 13 \, \{1,1,1\},\ft 13 \,\{2,2,2\}\} \label{fistulario}
\end{equation}
from which we conclude that, apart from the class of the identity, there is just one junior and one senior
class.
\par
Hence we conclude that the Hodge numbers of the resolved variety should be \\
\(h^{(0,0) }=1 ;\,\, \)\(h^{(1,1) }=1\) ; \(h^{(2,2) }=1\). \\
If we follow the weighted blowup procedure described in section \ref{steppisoffio}  using the weights of the
unique junior class $\{1,1,1\}$, we see that the projection $\pi$ of eq.\,(\ref{portapannolini}) yields
\begin{equation}\label{pisastore}
\pi :\mathbb{B}_{(1,1,1) }\longrightarrow  \mathbb{W}\mathbb{P}_{(1,1,1)}\sim \mathbb{P}^2
\end{equation}
So the blowup replaces the singular point $0 \in \mathbb{C}^3$ with a $\mathbb{P}^2$, which is compact. As a result,
also the expectional divisor in the resolution $\mathcal M_\zeta$ is compact.
By Poincar\'e duality this entrains the existence of a harmonic (2,2)--form associated with the unique senior
class.
\subsubsection{The quiver matrix}
\begin{figure}
\centering \vskip -2.8 cm
\includegraphics[height=6cm]{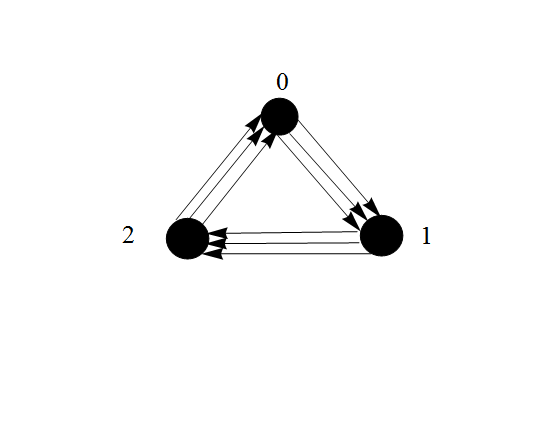}
\vskip -2.4cm \caption{ \label{z3maccaius} The quiver diagram of the cyclic group with generator $Y \, = \,
\mbox{diag}\{\xi,\xi.\xi\}$. }
\end{figure}

In this case, the quiver matrix defined by eq.\,(\ref{quiverro2}) is the following one :
\begin{equation}
A_{ij} =\left(
\begin{array}{ccc}
 0 & 3 & 0 \\
 0 & 0 & 3 \\
 3 & 0 & 0 \\
\end{array}
\right)
\end{equation}
and it has the graphical representation displayed in fig. \ref{z3maccaius}

\subsubsection{The space $\mathcal{S}_{\Gamma } \, = \, \mathrm{Hom}_\Gamma(\mathcal{Q}\otimes R, R)$
in the natural basis} Solving the invariance constraints (\ref{carnevalediPaulo}) in the natural basis of the
regular representation we find the triples of matrices $\{$A,B,C$\}$ spanning the locus \(\mathcal{S}_{\Gamma
}\). They are as follows:
\begin{eqnarray}
A&=&\left(
\begin{array}{ccc}
 \alpha _{1,1} & \alpha _{1,2} & \alpha _{1,3} \\
 (-1)^{2/3} \alpha _{1,3} & (-1)^{2/3} \alpha _{1,1} & (-1)^{2/3} \alpha _{1,2} \\
 -(-1)^{1/3} \alpha _{1,2} & -(-1)^{1/3} \alpha _{1,3} & -(-1)^{1/3} \alpha _{1,1} \\
\end{array}
\right)\nonumber\\
B &=&\left(
\begin{array}{ccc}
 \beta _{1,1} & \beta _{1,2} & \beta _{1,3} \\
 (-1)^{2/3} \beta _{1,3} & (-1)^{2/3} \beta _{1,1} & (-1)^{2/3} \beta _{1,2} \\
 -(-1)^{1/3} \beta _{1,2} & -(-1)^{1/3} \beta _{1,3} & -(-1)^{1/3} \beta _{1,1} \\
\end{array}
\right)\nonumber\\
C&=&\left(
\begin{array}{ccc}
 \gamma _{1,1} & \gamma _{1,2} & \gamma _{1,3} \\
 (-1)^{2/3} \gamma _{1,3} & (-1)^{2/3} \gamma _{1,1} & (-1)^{2/3} \gamma _{1,2} \\
 -(-1)^{1/3} \gamma _{1,2} & -(-1)^{1/3} \gamma _{1,3} & -(-1)^{1/3} \gamma _{1,1} \\
\end{array}
\right) \label{naturaliaz3}
\end{eqnarray}
\paragraph{The locus $L_\Gamma$.} The locus \(L_{\Gamma }\) $\subset $ \(\mathcal{S}_{\Gamma }\) is easily described by the equation:
\begin{eqnarray} A_0 &=&\left(
\begin{array}{ccc}
 \alpha _{1,1} & 0 & 0 \\
 0 & (-1)^{2/3} \alpha _{1,1} & 0 \\
 0 & 0 & -(-1)^{1/3} \alpha _{1,1} \\
\end{array}
\right)\nonumber\\
B_0&=&\left(
\begin{array}{ccc}
 \beta _{1,1} & 0 & 0 \\
 0 & (-1)^{2/3} \beta _{1,1} & 0 \\
 0 & 0 & -(-1)^{1/3} \beta _{1,1} \\
\end{array}
\right)\nonumber\\
C_0&=&\left(
\begin{array}{ccc}
 \gamma _{1,1} & 0 & 0 \\
 0 & (-1)^{2/3} \gamma _{1,1} & 0 \\
 0 & 0 & -(-1)^{1/3} \gamma _{1,1} \\
\end{array}
\right)
\end{eqnarray}
\subsubsection{The space $\mathcal{S}_{\Gamma }$ in the split basis}
Solving the invariance constraints in the split basis of the regular representation we find another
representation of the  triples of matrices $\{A,B,C\}$ that span  the space $\mathcal{S}_{\Gamma }$. They are
as follows:
\begin{eqnarray}
A&=&\left(
\begin{array}{ccc}
 0 & 0 & m_{1,3} \\
 m_{2,1} & 0 & 0 \\
 0 & m_{3,2} & 0 \\
\end{array}
\right)\nonumber\\
B&=&\left(
\begin{array}{ccc}
 0 & 0 & n_{1,3} \\
 n_{2,1} & 0 & 0 \\
 0 & n_{3,2} & 0 \\
\end{array}
\right)\nonumber\\
C&=&\left(
\begin{array}{ccc}
 0 & 0 & r_{1,3} \\
 r_{2,1} & 0 & 0 \\
 0 & r_{3,2} & 0 \\
\end{array}
\right)
\end{eqnarray}
\subsubsection{The equation $p \wedge p=0$ and the characterization of the variety
 $\mathbb{V}_5\,=\,\mathcal{D}_{\Gamma }$}
Here we are concerned with the solution of eq.\,(\ref{poffarbacchio}) and the characterization of the locus
$\mathcal{D}_{\Gamma}$.
\par
Differently from the more complicated cases of larger groups, in the present abelian case of small order, we
can explicitly solve the quadratic equations provided by the commutator constraints and we discover that
there is a principal branch of the solution, named \(\mathcal{D}_{\Gamma }^0\) that has indeed dimension
5=$|\Gamma |$+2. In addition however there are several other branches with smaller dimension. These branches
describe different components of the locus $\mathcal{D}_\Gamma$. Actually as already pointed out they are all
contained in the \(\mathcal{G}_{\Gamma }\) orbit of the subspace \(L_{\Gamma }\) defined above. The quadratic
equations defining \(\mathcal{D}_{\Gamma }\) have a set of 14 different solutions realized by a number \(n_{i
}\) of constraints on the 9 parameters. Hence there are 14 branches \(\mathcal{D}_{\Gamma }^i\)(i=0,1,...16)
of dimensions:
\begin{equation}
\dim  _{\mathbb{C}} \mathcal{D}_{\Gamma }^i = 9 -n_{i }
\end{equation}
The full dimension table of these branches is displayed below
$$\{5,4,4,4,4,3,3,3,3,3,3,3,2,2\}$$
As we see, there is a unique branch that has the maximal dimension 5 =$|$\(\left.\mathbb{Z}_3\right|+2\).
This is the principal branch \(\mathcal{D}_{\Gamma }^0\). It can be represented by the substitution:
\begin{equation}
n_{2,1}\to \frac{m_{2,1} n_{1,3}}{m_{1,3}},\quad n_{3,2}\to \frac{m_{3,2} n_{1,3}}{m_{1,3}},\quad r_{2,1}\to
\frac{m_{2,1} r_{1,3}}{m_{1,3}},\quad r_{3,2}\to \frac{m_{3,2} r_{1,3}}{m_{1,3}}
\end{equation}
In this way we have reached a complete resolution of the following problem. We have an explicit
parametrization of the variety \(V_{|\Gamma |+2}\). This variety is described by the following three matrices
depending on the 5 complex variables \(\omega _i\) (i=1,...,5):
\begin{eqnarray}
A&=&\left(
\begin{array}{ccc}
 0 & 0 & \omega _1 \\
 \omega _2 & 0 & 0 \\
 0 & \omega _3 & 0 \\
\end{array}
\right)\nonumber\\
B&=&\left(
\begin{array}{ccc}
 0 & 0 & \omega _4 \\
 \frac{\omega _2 \omega _4}{\omega _1} & 0 & 0 \\
 0 & \frac{\omega _3 \omega _4}{\omega _1} & 0 \\
\end{array}
\right)\nonumber\\
C&=&\left(
\begin{array}{ccc}
 0 & 0 & \omega _5 \\
 \frac{\omega _2 \omega _5}{\omega _1} & 0 & 0 \\
 0 & \frac{\omega _3 \omega _5}{\omega _1} & 0 \\
\end{array}
\right)
\end{eqnarray}
\subsubsection{The quiver group}
Our next point is the derivation of the group \(\mathcal{G}_{\Gamma }\) defined in eqs. (\ref{gstorto}) and
(\ref{carciofillo}), namely:
\begin{equation}
\mathcal{G}_{\Gamma } = \left\{ g\in  \text{SL}(|\Gamma |,\mathbb{C})\left| \forall \gamma \in \Gamma  :
\left[D_R(\gamma ),D_{\text{def}}(g)\right]\right.=0\right\}
\end{equation}
Let us proceed to this construction. In the diagonal basis of the regular representation this is a very easy
task, since the group is simply given by the diagonal 3$\times $3 matrices with determinant one. We introduce
such matrices
\begin{equation}
\mathfrak{g}\,\in \,\mathcal{G}_{\Gamma}\quad:\quad\mathfrak{g}=\left(
\begin{array}{ccc}
 a_1 & 0 & 0 \\
 0 & a_2 & 0 \\
 0 & 0 & a_3 \\
\end{array}
\right)
\end{equation}
\subsubsection{ $\mathbb{V}_5$ as the orbit under $\mathcal{G}_{\Gamma }$ of the locus $L_{\Gamma }$}
In this section we want to verify and implement eq.\,(\ref{belgone}), namely we aim at showing that
\(\mathbb{V}_{5 }\)=\(\mathcal{D}_{\Gamma }=\text{Orbit}_{\mathcal{G}_{\Gamma }}\left(L_{\Gamma }\right)\).
To this effect we rewrite the locus \(L_{\Gamma }\) in the diagonal split basis of the regular
representation. The change of basis is performed by the character table of the cyclic group \(\mathbb{Z}_3\).
The result is displayed below:
\begin{eqnarray}
A_0 &=&\left(
\begin{array}{ccc}
 0 & 0 & \alpha _{1,1} \\
 \alpha _{1,1} & 0 & 0 \\
 0 & \alpha _{1,1} & 0 \\
\end{array}
\right)\nonumber\\
B_0&=&\left(
\begin{array}{ccc}
 0 & 0 & \beta _{1,1} \\
 \beta _{1,1} & 0 & 0 \\
 0 & \beta _{1,1} & 0 \\
\end{array}
\right)\nonumber\\
C_0&=&\left(
\begin{array}{ccc}
 0 & 0 & \gamma _{1,1} \\
 \gamma _{1,1} & 0 & 0 \\
 0 & \gamma _{1,1} & 0 \\
\end{array}
\right)
\end{eqnarray}
Eventually the complex parameters
\begin{equation}
z_1\equiv \alpha _{1,1};  \quad z_2\equiv \beta _{1,1};  \quad  z_3\equiv \gamma _{1,1}
\end{equation}
will be utilized as complex coordinates of the resolved variety when the level parameters \(\zeta _{1,2}\)
are switched on. Starting from the above the orbit is given by:
\begin{equation}\label{colapasta}
    \mbox{Orbit}_{\mathcal{G}_\Gamma} \, \equiv \, \left\{\left\{\mathfrak{g}\, A_0 \,\mathfrak{g}^{-1},
    \mathfrak{g}\, B_0 \,\mathfrak{g}^{-1},\mathfrak{g}\, C_0 \,\mathfrak{g}^{-1} \right\}\quad \mid \quad
    \forall \,\mathfrak{g}\in\mathcal{G}_\Gamma \, , \, \forall \,\{A_0,B_0,C_0\} \in L_\Gamma\,\right\} \,
    \supset
    \, \mathcal{D}_\Gamma^0
\end{equation}
and the correspondence between the parameters of the principal branch \(\mathcal{D}_{\Gamma }^0\) and the
parameters spanning \(\mathcal{G}_{\Gamma }\) and \(L_{\Gamma }\) is provided below:
\begin{equation}
a_1\to \frac{\omega _2^{1/3}}{\omega _1^{1/3}},a_2\to \frac{\omega _3^{1/3}}{\omega _2^{1/3}},a_3\to
\frac{\omega _1^{1/3}}{\omega _3^{1/3}},z_1\to \omega _1^{1/3} \omega _2^{1/3} \omega _3^{1/3},z_2\to
\frac{\omega _2^{1/3} \omega _3^{1/3} \omega _4}{\omega _1^{2/3}},z_3\to \frac{\omega _2^{1/3} \omega
_3^{1/3} \omega _5}{\omega _1^{2/3}}
\end{equation}
Branches of smaller dimension  of the solution are all contained in the
\(\text{Orbit}_{\mathcal{G}_{\Gamma }}\left(L_{\Gamma }\right)\) and
correspond to the orbits of special points of \(L_{\Gamma }\) where
some of the \(z_i\) vanish or satisfy special relations among
themselves. Hence, indeed we have:
$$\mbox{Orbit}_{\mathcal{G}_\Gamma} \, = \,\mathcal{D}_\Gamma$$
\subsubsection{The compact gauge group $\mathcal{F}_{\Gamma } = \mathrm{U(1)}^2$}
We introduce a basis for the generators of the compact subgroup $\mathrm{U(1)}^2 =\mathcal{F}_{\Gamma}
\subset \mathcal{G}_{\Gamma }$ provided by the set of two generators displayed here below
\begin{equation}
T^1=\left(
\begin{array}{|c|c|c|}
\hline
 i & 0 & 0 \\
\hline
 0 & -i & 0 \\
\hline
 0 & 0 & 0 \\
\hline
\end{array}
\right)\quad  ; \quad T^2=\left(
\begin{array}{|c|c|c|}
\hline
 0 & 0 & 0 \\
\hline
 0 & i & 0 \\
\hline
 0 & 0 & -i \\
\hline
\end{array}
\right) \label{birraperoni}
\end{equation}
whose trace-normalization is the \(A_2\) Cartan matrix
\begin{equation}
\text{Tr}\left(T^iT^j\right) =\mathfrak{C}^{\text{ij}}\text{  }= \left(
\begin{array}{cc}
 2 & -1 \\
 -1 & 2 \\
\end{array}
\right)
\end{equation}
\subsubsection{Calculation of the K\"ahler potential and of the moment maps}
Naming \(\Delta _i\) the moduli of the coordinates \(z_i\) and \(\theta _i\) their phases according to
\(z_i\) = \(e^{\text{i$\theta $}_i}\)\(\Delta _i\) and considering a generic element \(\mathfrak{g}_R\) of
the quiver group that is real and hence is a representative of a coset class in \(\frac{\mathcal{G}_{\Gamma
}}{\mathcal{F}_{\Gamma }}\):
\begin{equation}
\mathfrak{g}_R \,= \,\left(
\begin{array}{|c|c|c|}
\hline
 e^{\lambda _1} & 0 & 0 \\
\hline
 0 & e^{-\lambda _1+\lambda _2} & 0 \\
\hline
 0 & 0 & e^{-\lambda _2} \\
\hline
\end{array}
\right)\quad ;\quad\lambda _{1,2}\in \mathbb{R}
\end{equation}
The triple of matrices $\{$A,B,C$\}$=\(\left\{\mathfrak{g}_R A_0 \mathfrak{g}_R{}^{-1}, \mathfrak{g}_R B_0
\mathfrak{g}_R{}^{-1},\mathfrak{g}_R C_0 \mathfrak{g}_R{}^{-1}\right\}\) have the following explicit
appearance:
\begin{eqnarray}
A&=&\left(
\begin{array}{ccc}
 0 & 0 & e^{i \theta _1-\lambda _1-\lambda _2} \Delta _1 \\
 e^{i \theta _1+2 \lambda _1-\lambda _2} \Delta _1 & 0 & 0 \\
 0 & e^{i \theta _1-\lambda _1+2 \lambda _2} \Delta _1 & 0 \\
\end{array}
\right)\nonumber\\
B&=&\left(
\begin{array}{ccc}
 0 & 0 & e^{i \theta _2-\lambda _1-\lambda _2} \Delta _2 \\
 e^{i \theta _2+2 \lambda _1-\lambda _2} \Delta _2 & 0 & 0 \\
 0 & e^{i \theta _2-\lambda _1+2 \lambda _2} \Delta _2 & 0 \\
\end{array}
\right)\nonumber\\
C&=&\left(
\begin{array}{ccc}
 0 & 0 & e^{i \theta _3-\lambda _1-\lambda _2} \Delta _3 \\
 e^{i \theta _3+2 \lambda _1-\lambda _2} \Delta _3 & 0 & 0 \\
 0 & e^{i \theta _3-\lambda _1+2 \lambda _2} \Delta _3 & 0 \\
\end{array}
\right)
\end{eqnarray}
Calculating the K\"ahler potential we find
\begin{equation}
\mathcal{K}_{\mathcal{S}}|_{\mathcal{D}}\, = \,\left(\text{Tr}\left[A\, \,
A^\dagger\right]+\text{Tr}\left[B\, \, B^\dagger\right]+\text{Tr}\left[C\, \, C^\dagger\right]\right)  =e^{-2
\left(\lambda _1+\lambda _2\right)} \left(1+e^{6 \lambda _1}+e^{6 \lambda _2}\right) \left(\Delta _1^2+\Delta
_2^2+\Delta _3^2\right)
\end{equation}
We have used the above notation since \(\text{Tr}\left[A,A^{\dagger }\right]+\text{Tr}\left[B,B^{\dagger
}\right]+\text{Tr}\left[C,C^{\dagger }\right]\) is the K\"ahler potential of the  ambient space
\(\mathcal{S}_{\Gamma }\) restricted to the orbit \(\mathcal{D}_{\Gamma }\). Indeed since
\(\mathcal{F}_{\Gamma }\) is an isometry of \(\mathcal{S}_{\Gamma }\), the dependence in
\(\mathcal{K}_{\mathcal{S}}|_{\mathcal{D}}\) is only on the real part of the quiver group, namely on the real
factors \(\lambda _{1,2}\). Just as it stands, \(\mathcal{K}_{\mathcal{S}}|_{\mathcal{D}}\) cannot work as
K\"ahler potential of a complex K\"ahler metric. Yet, when the real factors \(\lambda _{1,2}\) will be turned
into functions of the complex coordinates \(z_i\), then \(\mathcal{K}_{\mathcal{S}}|_{\mathcal{D}}\) will be
enabled to play the role of a contribution to the K\"ahler potential of the resolved manifold
$\mathcal{M}_\zeta$.
\par
Next we calculate the moment maps according to the formulas:
\begin{eqnarray}
\mathfrak{P}^1 &\equiv& - i\,\text{Tr}\left[ T^1 \left(\left[A,A^{\dagger }\right]+\left[B,B^{\dagger
}\right]+\left[C,C^{\dagger }\right]\right)\right] = e^{-2 \left(\lambda _1+\lambda _2\right)} \left(1-2 e^{6
\lambda _1}+e^{6 \lambda _2}\right)
\left(\Delta _1^2+\Delta _2^2+\Delta _3^2\right)\nonumber\\
\mathfrak{P}^2 & \equiv& - i\,\text{Tr}\left[ T^2 \left(\left[A,A^{\dagger }\right]+\left[B,B^{\dagger
}\right]+\left[C,C^{\dagger }\right]\right)\right] = e^{-2 \left(\lambda _1+\lambda _2\right)} \left(1+e^{6
\lambda _1}-2 e^{6 \lambda _2}\right) \left(\Delta _1^2+\Delta _2^2+\Delta _3^2\right)
\end{eqnarray}
\subsubsection{Solution of the moment map equations}
In order to solve the moment map equations it is convenient to introduce the new variables
\begin{equation}
\Upsilon _{1,2} = \exp\left[2 \,\lambda _{1,2}\right]
\end{equation}
and to redefine the moment maps with indices lowered by means of the inverse of the Cartan matrix mentioned
above
\begin{equation}
\mathfrak{P}_i = \left(\mathfrak{C}^{-1}\right){}_{\text{ij}} \mathfrak{P}^j
\end{equation}
In this way imposing the level condition
\begin{equation}
\mathfrak{P}_i  = - \zeta _i
\end{equation}
where \(\zeta _{1,2}\) $>$ 0 are the two level parameters, we obtain the final pair of algebraic equations
for the factors \(\Upsilon _{1,2}\)
\begin{equation}
\left\{\frac{\Sigma  \left(-1+\Upsilon _1^3\right)}{\Upsilon _1 \Upsilon _2},\frac{\Sigma \left(-1+\Upsilon
_2^3\right)}{\Upsilon _1 \Upsilon _2}\right\} = \left\{\zeta _{1,}\zeta _2\right\}
\end{equation}
where we have introduced the shorthand notation:
\begin{equation}
\Sigma  = \sum_{i=1}^3 \,|z_i|^2
\end{equation}
The above algebraic system composed of two cubic equations is simple enough in order to find all of its nine
roots by means of Cardano's formula. The very pleasant property of these solutions is that one and only one
of the nine branches satisfies the correct boundary conditions, namely provides real  \(\Upsilon _i\)($\zeta
$,$\Sigma $) that are positive for all values of $\Sigma $ and $\zeta $ and reduce to 1 when $\zeta
$$\rightarrow $0.
\par
The complete solution of the algebraic equations can be written in the following way. For the first factor we
have:
\begin{equation}
\Upsilon _1 =\frac{1}{6^{1/3}}\left(\frac{N}{\Sigma ^3\Pi  ^{\frac{1}{3}}}\right)^{\frac{1}{3}}
\end{equation}
where
\begin{eqnarray}
N &=& 2\times 2^{1/3} \zeta _1^3 \zeta _2^2+6 \Sigma ^3\Pi  ^{\frac{1}{3}}+2 \zeta _1^2 \left(3\times 2^{1/3}
\Sigma ^3+\zeta _2\Pi  ^{\frac{1}{3}}\right)+\zeta _1\left(6\times 2^{1/3} \Sigma ^3 \zeta _2+2^{2/3}\Pi
^{\frac{2}{3}}\right)\nonumber\\
\Pi & =& 27 \,\Sigma ^6+9 \,\Sigma ^3 \,\zeta _1^2 \zeta _2+9\, \Sigma ^3 \,\zeta _1 \zeta _2^2+2 \,\zeta
_1^3 \zeta
_2^3+3 \sqrt{3}\, \Sigma ^3\,\mathfrak{R} \nonumber\\
\mathfrak{R} &=&\sqrt{27 \,\Sigma ^6+6 \,\Sigma ^3 \zeta _1 \zeta _2^2-\zeta _1^4 \zeta _2^2-4 \,\Sigma ^3
\zeta _2^3+\zeta _1^3 \left(-4\, \Sigma ^3+2 \zeta _2^3\right)+\zeta _1^2 \left(6\,\Sigma ^3 \zeta _2-\zeta
_2^4\right)}\label{cardanone1}
\end{eqnarray}
For the second factor we have
\begin{equation}
\Upsilon _2= \frac{-\frac{M^{8/3}}{\Sigma ^5}+\frac{18 M^{5/3}}{\Sigma ^2}-72\, M^{2/3} \Sigma +36
\left(\frac{M}{\Sigma ^3}\right)^{2/3} \zeta _1^3-36 \left(\frac{M}{\Sigma ^3}\right)^{2/3} \zeta _1^2 \zeta
_2+6 \left(\frac{M}{\Sigma ^3}\right)^{5/3} \zeta _1^2 \zeta _2}{36\times 6^{2/3}\, \Sigma ^2 \zeta _1}
\end{equation}
where
\begin{equation}
M \,= \,\frac{6 \,\Sigma ^3 \Pi ^{1/3}+2^{2/3} \Pi ^{2/3} \zeta _1+6\times 2^{1/3} \Sigma ^3 \,\zeta
_1^2+6\times 2^{1/3} \Sigma ^3 \,\zeta _1 \zeta _2+2 \Pi ^{1/3} \zeta _1^2 \zeta _2+2\times 2^{1/3} \zeta
_1^3 \zeta _2^2}{\Omega ^{1/3}}
\end{equation}
\subsection{Discussion of cohomology in the master model}
Since the two scale factors \(\Upsilon _{1,2}\) are functions only of $\Sigma $, the two (1,1)-forms,
relative to the two tautological bundles, respectively associated with the first and second nontrivial
irreps of the cyclic group, defined in eq.\,(\ref{bambolone}) take the following general appearance:
\begin{eqnarray}
\omega _{1,2}^{(1,1)} &=&\frac{i}{2\pi }\left(\frac{d}{\text{d$\Sigma $}} \text{Log}\left[\Upsilon
_{1,2}(\Sigma )\right] d\bar{z}^i\wedge \text{dz}^i +\frac{d^2}{\text{d$\Sigma $}^2} \text{Log}\left[\Upsilon
_{1,2}(\Sigma )\right] z^j\bar{z}^i \text{dz}^i\wedge d\bar{z}^j\right)\nonumber\\
 &=& \frac{i}{2\pi }\left( f_{1,2} \Theta
\, + \, g_{1,2} \Psi\right)
\end{eqnarray}
where we have introduced the short hand notation
\begin{equation}\label{sciortando}
 \Theta \, = \,  \sum_{i=1}^3 d\bar{z}^i\wedge \text{dz}^i \quad ;\quad
 \Psi \, = \, \quad \sum_{i,j=1}^3 z^j\, \bar{z}^i {dz}^i\wedge d\bar{z}^j
\end{equation}
Indeed in the present case the fiber metrics $\mathfrak{H}_{1,2}$
are one-dimensional and given by $\mathfrak{H}_{1,2} \, = \,
\sqrt{\Upsilon _{1,2}}$. The most relevant point is that the two
functions $f_{1,2}$ and $g_{1,2}$ being the derivatives (first and
second) of $\Upsilon _{1,2}$ depend only on the variable $\Sigma$.
\par
It follows that a triple wedge product of the two--forms $\omega _{a}^{(1,1)}$ (a=1,2) has always the
following structure:
\begin{equation}\label{strutturone}
    \omega _{a}^{(1,1)}\wedge \omega _{b}^{(1,1)}\wedge \omega _{b}^{(1,1)} \, = \left(\frac{i}{2 \,\pi}\right)^3\,
    \left(f_a\,f_b\,f_c + 2\,
    \Sigma \, \left(g_a \, f_b \, f_c +g_b\,f_c\,f_a +g_c\,f_a\,f_b \right) \right)\times \mbox{Vol}
\end{equation}
where
\begin{equation}\label{voluminoso}
    \mbox{Vol} \, = \, dz_1\wedge dz_2\wedge dz_3\wedge d\bar{z}_1\wedge d\bar{z}_2\wedge d\bar{z}_3
\end{equation}
This structure enables us to calculate intersection integrals of the considered forms very easily. It
suffices to change variables as we explain below. The equations
\begin{equation}\label{varolegione}
    \Sigma \, = \, \sum_{i=1}^3 |z_i|^2 \, = \, \rho^2
\end{equation}
define 5-spheres of radius $\rho$. Introducing the standard Euler angle parametrization of a $5$-sphere, the
volume form (\ref{voluminoso}) reduces to:
\begin{equation}\label{esile}
 \mbox{Vol} \,= \,   8 i \varrho ^5 \cos ^4\left(\theta _1\right) \cos
   ^3\left(\theta _2\right) \cos ^2\left(\theta _3\right)
   \cos \left(\theta _4\right) \, \prod_{i=1}^5 d\theta_i
\end{equation}
The integration on the Euler angles can be easily performed and we obtain:
\begin{equation}\label{integratato}
   \prod_{i}^4 \, \int_{\ft{\pi}{2}}^{\ft{\pi}{2}} \,d\theta_i \, \int_{0}^{2\pi} \, d\theta_5 \, \left(8 i \varrho ^5
   \cos ^4\left(\theta _1\right) \cos^3\left(\theta _2\right) \cos ^2\left(\theta _3\right)
   \cos \left(\theta _4\right) \,\right) \, = \,8 i \pi ^3 \varrho ^5
\end{equation}
Hence defining the intersection integrals:
\begin{equation}\label{interseziunka}
    \mathcal{I}_{abc}\, = \, \int_{\mathcal{M}} \, \omega _{a}^{(1,1)}\wedge \omega _{b}^{(1,1)}\wedge \omega _{b}^{(1,1)}
\end{equation}
we arrive at
\begin{eqnarray}
\mathcal{I}_{\text{abc}}&=&\left(\frac{i}{2\pi }\right)^3\times 8 i \pi ^3 \times \int_0^{\infty } \left(6
\varrho ^5f_af_b f_c+2\varrho ^7 \left(f_b f_c g_a+f_a f_c g_b+f_a f_b g_c\right)\right) \, d\rho\nonumber\\
&=& \int_0^{\infty } \left(6 \varrho ^5f_af_b f_c+2\varrho ^7 \left(f_b f_c g_a+f_a f_c g_b+f_a f_b
g_c\right)\right) \, d\rho
\end{eqnarray}
We have performed the numerical integration of these functions and we have found the following results
\begin{equation}
\begin{array}{lcl}
\left(\zeta _1>0,\zeta _2=0\right)&:&\mathcal{I}_{111} =\frac{1}{8}\\
\left(\zeta _1=0,\zeta _2\geq 0\right)&:&\mathcal{I}_{111} = 0\\
\left(\zeta _1>0,\zeta _2>0\right)&:&\mathcal{I}_{111} = 1\\
\end{array}
\end{equation}
From this we reach the following conclusion. Since the corresponding integral is nonzero it follows that:
\begin{equation}
\omega _S^{(2,2)} \equiv  \omega _1^{(1,1)}\wedge \omega _1^{(1,1)}
\end{equation}
is closed but not exact and by Poincar\'e duality it is the Poincar\'e dual of some cycle S $\in $
\(H_2\)($\mathcal{M}$) such that:
\begin{equation}
\int _S \iota ^*\omega _1^{(1,1)} = \int _{\mathcal{M}}\omega _1^{(1,1)}\wedge \omega _S^{(2,2)}
\end{equation}
where
\begin{equation}
\iota  : S \longrightarrow \mathcal{M}
\end{equation}
is the inclusion map. Since \(H_c^2(\mathcal{M})\)=
\(H^2\)($\mathcal{M}$) and both have dimension 1 it follows that dim
$H_2$($\mathcal{M}$) = 1, so that every nontrivial cycle S is
proportional (as homology class) via some coefficient $\alpha $ to a
single cycle $\mathcal{C}$, namely we have S= $\alpha $
$\mathcal{C}$. Then we can interpret eq.\,(29) as follows
\begin{equation}
\text{                                              }\int _{\alpha  \mathcal{C}} \iota ^*\omega _1^{(1,1)} =
\alpha \int _{\mathcal{M}}\omega _1^{(1,1)}\wedge \omega _{\mathcal{C}}^{(2,2)}
\end{equation}
If we choose as fundamental cycle, that one for which
\begin{equation}
\int _{\mathcal{C}} \iota ^*\omega _1^{(1,1)}= 1
\end{equation}
we conclude that
\begin{equation}
\alpha  =\left\{
\begin{array}{c|c}
 1 & \text{case} \left\{\zeta _1>0,\zeta _2>0\right\} \\
\hline
 \frac{1}{8} & \text{case} \left\{\zeta _1>0,\zeta _2=0\right\} \\
\end{array}
\right.
\end{equation}
Next we have calculated the intersection integral $\mathcal{I}_{211}$ and we have found:
\begin{equation}
\begin{array}{lcl}
\left(\zeta _1>0,\zeta _2=0\right)&:&\mathcal{I}_{211} =0\\
\left(\zeta _1=0,\zeta _2\geq 0\right)&:&\mathcal{I}_{211} = 0\\
\left(\zeta _1>0,\zeta _2>0\right)&:&\mathcal{I}_{211} = 1\\
\end{array}
\end{equation}
\paragraph{Conclusions on cohomology.}
We have two cases.
\begin{description}
\item[
\(\text{case} \left\{\zeta _1>0,\zeta _2>0\right\}.\)] The the first Chern classes of the two tautological
bundles are cohomologous:
\begin{equation}
\left[\omega _1^{(1,1)} \right] = \left[\omega _2^{(1,1)}\right]=\left[\omega ^{(1,1)}\right]
\end{equation}
\item[\(\text{case} \left\{\zeta _1>0,\zeta _2=0\right\}.\)]
The the first Chern class of the first tautological bundle is nontrivial and generates \,
$H_c^{(1,1)}(\mathcal{M})$ = $H^{1,1}(\mathcal{M})$.
\begin{equation}
\left[\omega _1^{(1,1)} \right] =\text{  }\text{nontrivial}
\end{equation}
The the first Chern class of the second tautological bundle is trivial , namely
\begin{equation}
\omega _2^{(1,1)}\text{  }=\text{  }\text{exact form}
\end{equation}
\end{description}
Obviously since there is symmetry in the exchange of the first and second scale factors, exchanging \(\zeta
_1\)$\leftrightarrow $\(\zeta _2\), the above conclusion is reversed in the case \(\left\{\zeta _1=0,\zeta
_2>0\right\}\).
\par
In passing we have also proved that the unique (2,2)-class is just the square of the unique (1,1)-class
\par
\begin{equation}
\left[\omega ^{(2,2)} \right] =\left[\omega ^{(1,1)}\right]\wedge \left[\omega ^{(1,1)}\right]
\end{equation}
\subsubsection{The exceptional divisor}
Finally let us discuss how we retrieve the exceptional divisor $\mathbb{P}^2$ predicted by the weighted
blowup argument. As we anticipated in eqs. (\ref{cosmitto}-\ref{camalosto}),  replacing the three coordinates
$z_i$ with
\begin{equation}\label{paffuto}
  z_1 \, = \, W \quad ; \quad z_2 \, = \, W \, \xi_1 \quad ; \quad \quad z_3 \, = \, W \, \xi_2
\end{equation}
which is the appropriate change  for one of the three standard open charts of $\mathbb{P}^2$, we obtain
\begin{equation}\label{ripulisti}
\mathfrak{H}_1 (\Sigma)\, = \, \frac{1}{|W|^2 \, H_1(\xi,\bar{\xi},W,\bar{W})}
\end{equation}
where the function $H_1(\xi,\bar{\xi},W,\bar{W})$ has the property that:
\begin{equation}\label{canizzo}
 \lim_{W\to 0} \, \log[H_1(\xi,\bar{\xi},W,\bar{W})]\, = \, - \log[1+|\xi_1|^2 + |\xi_2|^2 ] + \log[\mbox{const}]
\end{equation}
From the above result we conclude that the exceptional divisor
$\mathfrak{D}^{(E)}$ is indeed the locus $W=0$ and that on this
locus the first Chern class of the first tautological bundle reduces
to the K\"ahler 2-form of the Fubini-Study K\"ahler metric on
$\mathbb{P}^2$. Indeed we can write:
\begin{equation}\label{fubinistudia}
  c_1\left(\mathcal{L}_1\right)|_{\mathfrak{D}^{(E)}} \, = \, -\,\frac{i}{2\pi} \, \bar{\partial} \, \partial \, \log[1+|\xi_1|^2 + |\xi_2|^2 ]
\end{equation}
From this point of view this master example is the perfect three-dimensional analogue of the Eguchi-Hanson
space, the $\mathbb{P}^1$ being substituted by a $\mathbb{P}^2$.
\subsection{The model $\frac{\mathbb{C}^3}{\mathbb{Z}_4}$}
Another interesting model is the case $\frac{\mathbb{C}^3}{\mathbb{Z}_4}$, where the group $\mathbb{Z}_4$ is generated by
\begin{equation}\label{clinica}
Y \, = \, \left( \begin{array}{ccc}
     i & 0& 0 \\
     0 & i & 0 \\
     0 & 0 & -1
   \end{array}\right)
\end{equation}
\subsubsection{Construction of the blowups according to conjugacy classes}
Calculating the ages we find two junior conjugacy classes,
namely \(\frac{1}{4}\) $\{$1, 1, 2$\}$ and \(\frac{1}{4}\) $\{$2,2,0$\}$, and one senior class
\(\frac{1}{4}\)$\{$3,3,2$\}$. Hence we expect two cohomology classes  of type (1,1) and one cohomology class of type  (2,2).
The fact that there is a senior cohomology class means that one of the two exceptional divisors is compact, and the other is not.
One of the consequences of all this is that out of the three tautological line bundles
that we construct solving the moment map equation one must be dependent on the other two. \\
According to the theory of weighted blowup we introduce the following two sets of coordinates
corresponding to the two junior classes\\
\begin{equation}
\Psi _{[1,1,2]}=W\left(1,\Psi _1,\sqrt{\Psi _3}\right) ; \Phi _{[2,2,0]}=W^2\left(1,\Phi _1\right) \times \Phi_2
\end{equation}
In the first case $\Psi_1$ and $\Psi_3$ are the inhomogenous coordinates on $\mathbb{WP}^{1,1,2}$, while in the second case
$\Phi _1 $ is the inhomogeneous coordinate on $\mathbb{P}^1$, while $\Phi_2$ spans $\mathbb{C}$.
\par
The three moment maps $\mathfrak{P}_i$ can be calculated exactly with the same procedure
as in the previous master case and as in the cyclic examples of appendices \ref{exemplaria}.
It is also convenient to rearrange the moment maps as follows
\begin{equation}
\Pi _1=\mathfrak{P}_1-\mathfrak{P}_3\text{   };
\text{  }\Pi _2=\mathfrak{P}_1+\mathfrak{P}_3-\mathfrak{P}_2\text{    };
\text{   }\Pi _3 = -\mathfrak{P}_2
\end{equation}
This is done by the following nonsingular matrix
\begin{equation}
S =\left(
\begin{array}{c|c|c}
 1 & 0 & -1 \\
\hline
 1 & -1 & 1 \\
 \hline
 0 & 1 & 0 \\
\end{array}
\right)\text{         };\text{               }\Pi =S.\mathfrak{P}
\end{equation}
Furthermore it is convenient to introduce reduced variables \(\mathfrak{H} _i\)=\(\sqrt{X_i}\)
and new level parameters \(\kappa _i\) = \(S_{\text{ij}}\)\(\zeta
_j\)
\begin{equation}
\left\{\kappa _1 ,\kappa _2 ,\kappa _3 \right\}=\left\{\zeta _1-\zeta _3,\zeta _1-\zeta _2+\zeta
_3,\zeta _2\right\}
\end{equation}
In terms of these variables the moment map equation hence takes the following form:
\begin{doublespace}
\noindent\(\pmb{\text{}}\)
\end{doublespace}
\begin{equation}
\left(
\begin{array}{c}
 -\frac{\left(X_1^2-X_3^2\right) \left(X_1 X_3 \left(\Delta _1^2+\Delta _2^2\right)+\left(1+X_2^2\right)
 \Delta _3^2\right)}{X_1 X_2 X_3} \\
 \frac{\left(X_2+X_2^3-X_1 X_3 \left(X_1^2+X_3^2\right)\right) \left(\Delta _1^2+\Delta _2^2\right)}{X_1 X_2 X_3} \\
 -\frac{\left(-1+X_2^2\right) \left(X_2 \left(\Delta _1^2+\Delta _2^2\right)+\left(X_1^2+X_3^2\right)
 \Delta _3^2\right)}{X_1 X_2 X_3} \\
\end{array}
\right)=\left(
\begin{array}{c}
 \kappa _1 \\
 \kappa _2 \\
 \kappa _3 \\
\end{array}
\right)
\end{equation}
where $\Delta_i \, = \, \mid z_i\mid$ are the moduli of the three complex coordinates.
\subsubsection{Proof that one tautological line bundle depends from the other two}
By this we mean that the isomorphism classes of these bundles are
not linearly independent in the Picard group. We shall indeed show
that the differential forms representing the first Chern classes of
the three bundles satisfy a linear relation. The strategy we adopt
to check this fact is the following. We observe that the equations
are written in such a way that they depend only on two variables Z=
$|$\(z_1\)\(|^2\)+$|$\(z_2\)\(|^2\) and U=$|$\(z_3\)\(|^2\). Hence
instead of solving the equations for \(X_{1,2,3}\), in terms of the
levels \(\kappa _{1,2,3 }\)and of U and Z, we rather do the reverse
and we solve them for U and Z in terms of the levels \(\kappa
_{1,2,3 }\) and of \(X_{1,2,3. }\). Just because there are three
independent equations for two variables, by substituting back we
obtain a condition that has to be satisfied by \(
\mathfrak{H}_{1,2,3}=\sqrt{X_{1,2,3}} \) and \(\kappa _{1,2,3 }\),
among themselves which is the following one:
\begin{equation}
\mathfrak{H} _3\,= \,\mathfrak{H} _1\sqrt[4]{\frac{\kappa _1-\kappa _2+\kappa _3+\left(-\kappa
_1+\kappa _2+\kappa _3\right) \mathfrak{H} _2^4}{-\kappa _1-\kappa _2+\kappa _3+\left(\kappa _1+\kappa _2+\kappa _3\right)
\mathfrak{H} _2^4}}
\label{salamandra}
\end{equation}
which implies
\begin{equation}
\text{Log}\left[\mathfrak{H} _3 \right]=\text{Log}\left[\mathfrak{H} _1\right]+\frac{1}{4} \text{Log}\left[-a+\mathfrak{H}
_2^4\right]-\frac{1}{4} \text{Log}\left[-b+\mathfrak{H} _2^4\right]
\end{equation}
where $a$, $b$ are two constants
\begin{equation}
a=-\frac{\kappa _1-\kappa _2+\kappa _3}{-\kappa _1+\kappa _2+\kappa _3}\quad ; \quad
b=-\frac{-\kappa _1-\kappa _2+\kappa _3}{\kappa _1+\kappa _2+\kappa _3}
\end{equation}
Expanding in power series of a small parameter we have:
\begin{eqnarray}
&&\kappa _1= \epsilon  k_1+\mathcal{O}[\epsilon ^2] \quad , \quad \kappa _2= \epsilon  k_2+\mathcal{O}[\epsilon^2 ] \quad
, \quad \kappa _3= \epsilon k_3+\mathcal{O}[\epsilon ^2] \nonumber\\
&&\mathfrak{H} _1= 1-\epsilon  \omega _1+\mathcal{O}[\epsilon ]^2\nonumber\\
&&\mathfrak{H} _2= 1-\epsilon  \omega
_2+O[\epsilon ]^2\nonumber\\
&&\mathfrak{H} _3= 1-\epsilon  \omega _3+\mathcal{O}[\epsilon^2 ]
\end{eqnarray}
we calculate to that order the first Chern classes of the line bundles and we find:
\begin{eqnarray}
\omega _1^{(1,1)} &= &\frac{i}{2\pi} \,\partial \bar{\partial }\text{Log}\left[1 - \epsilon \, \omega_1
+\mathcal{O}[\epsilon^2 ]\right]
\nonumber \\
\omega _2^{(1,1)} &= &\frac{i}{2\pi} \,\partial \bar{\partial }\text{Log}\left[1 - \epsilon \, \omega_2+\mathcal{O}[\epsilon^2 ]\right] \\
\omega _3^{(1,1)} &= &\frac{i}{2\pi} \,\partial \bar{\partial }\text{Log}\left[1 - \epsilon \,
\omega_3+\mathcal{O}[\epsilon^2 ]\right]
\end{eqnarray}
where
\begin{eqnarray}
\omega _1&=&\frac{i}{2\pi} \,\frac{ 2 k_1 \left(\left|z_1|^2+\left|z_2|^2\right.\right.\right)+2 k_3
\left(\left|z_1|^2+\left|z_2|^2\right.\right.\right)+k_2 \left(\left|z_1|^2+\left|z_2|^2+2
\right|z_3|^2\right.\right)} {\left(16 \left(\left|z_1|^2+\left|z_2|^2\right.\right.\right)
\left(\left|z_1|^2+\left|z_2|^2+2 \right|z_3|^2\right.\right)\right)}
\nonumber\\
\omega _2&=&\frac{i}{2\pi} \,\frac{\epsilon  k_3}{4 \left(\left|z_1|^2+\left|z_2|^2+2
\right|z_3|^2\right.\right)}
\nonumber\\
\omega _3& =& \frac{i}{2\pi} \,\frac{ \left(-2 k_1 \left(\left|z_1|^2+\left|z_2|^2\right.\right.\right)+2 k_3
\left(\left|z_1|^2+\left|z_2|^2\right.\right.\right)+k_2 \left(\left|z_1|^2+\left|z_2|^2+2
\right|z_3|^2\right.\right) \right)}{16
\left(\left|z_1|^2+\left|z_2|^2\right.\right.\right) \left(\left|z_1|^2+\left|z_2|^2+2 \right|z_3|^2\right.\right)}\nonumber\\
\label{peshicod}
\end{eqnarray}
This solution perfectly agrees with the prediction on the relation between Chern classes at this order that
follows from the
relation (\ref{salamandra}) between $\mathfrak{H}$ factors:
\begin{equation}
\text{Log}\left[\mathfrak{H} _3 \right]\,=\, \frac{1}{4} \text{Log}\left[1+\frac{2 k_1}{-k_1+k_2+k_3}\right]
+\left(\omega _1-\frac{k_1
\omega _2}{k_3}\right) \epsilon +O[\epsilon^2 ]
\label{sandramilo}
\end{equation}
which implies
\begin{equation}\label{caraffa}
\omega _3=\text{  }\left(\omega _1-\frac{k_1 \omega _2}{k_3}\right)
\end{equation}
\subsubsection{Special solution of the moment  map equations}
If we choose the following special values of the level parameters
\begin{equation}
\kappa _3= -\kappa , \kappa _2= 0, \kappa _1= \kappa
\label{spechoice}
\end{equation}
we obtain a solution of the moment map equations by setting:
\begin{equation}
\mathfrak{H}_3=\mathfrak{H}_2,\mathfrak{H}_1=1 \label{fistacchio}
\end{equation}
where the \(\mathfrak{H}_2\) satisfies the following quartic equation:
\begin{equation}
U+Z \mathfrak{H}_2+\kappa  \mathfrak{H}_2^2-Z \mathfrak{H}_2^3-U \mathfrak{H}_2^4 = 0
\end{equation}
Indeed with the choices (\ref{spechoice}) and (\ref{fistacchio}) the moment map equations reduce to the
above quartic algebraic constraint.
\par
Thanks to Cardano's formula we have four roots only one of which has the correct property that it reduces
to 1 when the level parameter $\kappa
$ goes to zero. Such a solution has the following explicit appearance:
\begin{equation}
\mathfrak{H}_2=\frac{1}{2\sqrt{3}}\sqrt{\frac{\sqrt{6}\sqrt{A}+\sqrt{3}\sqrt{B}-3Z}{U}}
\end{equation}
\begin{eqnarray}
A&=&8 \kappa  U+2^{2/3} \sqrt[3]{\Lambda _2+\sqrt{\Lambda _2^2-4 \Lambda _1^3}} U+\frac{2 \sqrt[3]{2}
\Lambda _1 U}{\sqrt[3]{\Lambda _2+\sqrt{\Lambda
_2^2-4 \Lambda _1^3}}}\nonumber\\
&&-\frac{3 \sqrt{3} \Lambda _3}{\sqrt{8 \kappa  U-2\ 2^{2/3} \sqrt[3]{\Lambda _2+\sqrt{\Lambda _2^2-4
\Lambda _1^3}} U-\frac{4
\sqrt[3]{2} \Lambda _1 U}{\sqrt[3]{\Lambda _2+\sqrt{\Lambda _2^2-4 \Lambda _1^3}}}+3 Z^2}}+3 Z^2\nonumber\\
\end{eqnarray}
\begin{equation}
B=8 \kappa  U-2\ 2^{2/3} \sqrt[3]{\Lambda _2+\sqrt{\Lambda _2^2-4 \Lambda _1^3}} U-\frac{4 \sqrt[3]{2}
\Lambda _1 U}{\sqrt[3]{\Lambda _2+\sqrt{\Lambda
_2^2-4 \Lambda _1^3}}}+3 Z^2
\end{equation}
where \(\Lambda _{1,2,3}\) is a short hand for three polynomials in U and Z that are specified below:
\begin{equation}
\Lambda _1= -12 U^2+3 Z^2+\kappa ^2,\Lambda _2= 72 U^2 \kappa +9 Z^2 \kappa +2 \kappa ^3,\Lambda _3
=Z\left(-8 U^2+Z^2+4 U \kappa \right)
\end{equation}
\\
On the other hand Z and U are short hand notations for:
\begin{equation}
Z = \left|z_1|^2+\left|z_2|^2\right.\right.\text{  }; \text{  }U = \left|z_3|^2 \right.
\end{equation}
\\
In this way the calculation of the first Chern classes from the irrep side is explicit in this case:
\begin{equation}
\omega _1^{(1,1)} = 0
\end{equation}
\begin{equation}
\omega _3^{(1,1)} = \omega _2^{(1,1)} =\frac{i}{2\pi}\bar{\partial }\partial
\text{Log}\left[\mathfrak{H}_2\right]
\end{equation}
\section{Conclusions}
As we emphasized in the introduction the present paper focuses on
the resolution of $\mathbb{C}^3/\Gamma$ singularities, $\Gamma
\subset \mathrm{SU(3)}$ being a finite group, in the perspective of
applications to the duality between superconformal Chern-Simons
gauge theories in three space--time dimensions and M2-brane
solutions of $D=11$ supergravity. In many regards this perspective
is the M-theory analogue of what was done about 18 years ago in
\cite{Bertolini:2002pr,Bertolini:2001ma} where the smooth analogue
of fractional three-branes was considered, replacing the
6-dimensional singular transverse dimensions $\mathbb{C}  \times
\mathbb{C}^2/\mathbb{Z}_k$ to a fractional three-brane with their
smooth resolution $\mathbb{C}\times ALE_k$. Here the analogous
scenario is:
\begin{equation}\label{rapitallus}
    \mathbb{C} \times \frac{\mathbb{C}^3}{\Gamma} \, \longrightarrow \, \mathbb{C} \times \mathcal{M}_\zeta
\end{equation}
The main point of this paper is that the generalization to the case
of three-folds of the Kronheimer construction of ALE manifolds
encodes precisely, in a well-defined geometrical language, all the
necessary information  and the needed steps that lead to a dual
superconformal Chern-Simons gauge theory on the M2-brane
world-volume. A comment is in order to appreciate the difference
between  our results and those obtained in the ABJM set up
\cite{Aharony:2008ug}. The orbifold of the seven sphere considered
in \cite{Aharony:2008ug} corresponds to an identical phase action of
the cyclic group $\mathbb{Z}_k$ on all the four homogeneous
coordinates of $\mathbb{P}^3$ base of the Hopf-fibration of
$\mathbb{S}^7$. The resulting   quotient
$\mathbb{C}^4/\mathbb{Z}_k$ pertaining to the ABJM case is not of
the type considered in this paper, since that representation of $\Gamma=\mathbb{Z}_k$ is
not in $\mathrm{SU(4)}$ (unless $k=4$), but only in $\mathrm{U(4)}$.
The resolution of such a
singularity  does not have the properties predicted by the
Ito--Reid theorem, in particular, it is not crepant. The quantization of the
levels of the Chern-Simons terms in terms of $k$, which is the
characterizing features of the ABJM-case is to be rediscussed from
scratch in the case of the theories addressed in the present paper.
At the present stage we do not know what the analogue feature might
be in our case.
\par
 A part from the above mentioned unsolved problem, we already listed the impressive translation
vocabulary between geometry and field theory in the introduction and
we do not repeat it once again. In this final section we rather
summarize the achieved results, the open problems and the
perspectives for further investigations.
\par\begin{description}
      \item[A)] One important issue concerns the difference between hyperK\"ahler  and K\"ahler quotients, which is also the
difference between $\mathcal{N}=4/\mathcal{N}=3$ supersymmetry and
$\mathcal{N}=2$ supersymmetry in three space--time dimensions. As we
extensively discussed in the main text, there is an   analogue
of the holomorphic part of moment map equation that is provided by
the equation $\pmb{p}\wedge\pmb{p}=0$, defining the subspace
$\mathcal{D}_\Gamma \subset \mathrm{Hom}_\Gamma(Q\times R, R)$ (see
eqs. (\ref{ganimusco}-\ref{poffarbacchio})) and leads to a precise
suggestion for the superpotential $\mathcal{W}$ in the corresponding
superconformal field theory. A striking feature of this
generalization that cannot escape notice is the following. In the
hyperK\"ahler case the holomorphic moment map equation is of the form
$[p_1,p_2]=0$ (eventually deformed by the $\zeta^+$ parameters). In
the K\"ahler case it is $\epsilon^{xyz}p_x\cdot p_y \, = \, 0$ with
$x,y,z=1,2,3$. It looks like we have a \textit{reversed
correspondence} with division algebras, the hyperK\"ahler case
corresponding to $\mathbb{C}$ (one imaginary unit, hence one matrix
condition), the K\"ahler case corresponding instead to quaternions
(three imaginary units, hence three matrix conditions). This
naturally leads to a wild conjecture. What about a further
generalization of the Kronheimer construction corresponding to
octonions? We might consider finite groups $\Gamma\subset
\mathrm{SU(4)}$ that are also subgroups of $\mathrm{G_{2(-14)}}$,
namely preserve the octonion structure constants $a^{ijk}$, and we
might consider 7-tuples of matrices $p_i$ ($i=1,\dots,7)$ each
belonging to $\mathrm{Hom(R,R)}$, where R denotes the regular
representation of $\Gamma$. Because of our hypothesis there is a
natural $7$-dimensional representation $\mathcal{Q}$ of $\Gamma$
(corresponding to the embedding $\Gamma\hookrightarrow
\mathrm{G_{2(-14)}}$) and we can define the space
$\widehat{\mathcal{S}}_\Gamma \, = \,
\mathrm{Hom}_\Gamma(\mathcal{Q}\otimes R,R)$. Next one can deem of
the analogue  subspace $\widehat{D}_\Gamma \subset
\widehat{S}_\Gamma$ singled out by the $7$ conditions:
\begin{equation}\label{cartuccia}
    a^{ijk}\, p_j \cdot p_k \, = \, 0
\end{equation}
 We might perform some quotient of this space  with respect to
a suitable compact quiver group $\mathcal{F}_\Gamma$. What the
result might be of such a construction is totally unexplored both
from the mathematical and from the physical point of view. Yet it is
particularly inspiring that the mentioned conditions for the finite
group $\Gamma$ are  satisfied by $\mathrm{L_{168}}$ and by all of
its subgroups. Careful consideration of this possibility is
certainly an interesting line  for further research.
\item[B)]As we repeatedly stressed, the relation between the first Chern classes
of the tautological bundles associated, by means of the generalized
Kronheimer construction, with the nontrivial irreps of the discrete
group $\Gamma$ and the components of the exceptional divisor
produced by the blowup is of central relevance on both sides of the
correspondence Geometry/Superconformal Gauge-Theory. In the present
paper this relation has been explored and fully established in one
of the few cases where the algebraic moment map equations admit an
explicit analytic solution, \textit{i.e} for the master case treated
in section \ref{masterdegree}, where there is just one junior class
and the blowup produces just one component of the exceptional
divisor which is a $\mathbb{P}^2$. The extension of this in depth
analysis to cases where the junior classes are several and moreover
to cases where $\Gamma$ is nonabelian is one principal direction for
further investigation and we plan to address such a question in the
nearest future. This issue was fully solved in the abelian case in
the paper \cite{Craw-JA} using different techniques and relying on
results presented in \cite{Craw-Ishii}.
\item[C)] In relation with the above issue a very important point concerns the possibility of analyzing
algebraic moment map equations by means of a series development in the neighborhood of zero for the level
parameters $\zeta_I$. Since intersection integrals in cohomology  do not depend on the value of $\zeta_I$,
except for a jump when $\zeta_I=0$ for some value of $I$,   we can advocate the use of infinitesimal
$\zeta_I$. A rigorous investigation of this possibility is of utmost relevance since the solution of moment
map equations at first order in the $\zeta_I$ is always accessible. This might prove the winning weapon to
discriminate among all the available cohomology patterns. We  plan to address this question in the nearest
future.
\item[D)] In the long run the main target is to promote the superconformal Chern-Simons gauge theory on the
brane from the theory of one M2-brane to the theory of several M2-branes which means to add color indices to
the scalar fields. These latter  are the coordinates of $\mathcal{S}_\Gamma \equiv \mbox{Hom}_\Gamma\left(
\mathcal{Q}\otimes R,R\right)$ and in the case where $\Gamma$ is a cyclic $\mathbb{Z}_k$,  the gauge group
$\mathcal{F}_\Gamma$ produced by the Kronheimer construction is just a product of $\mathrm{U(1)}$.s. Adding
color indices typically amounts to promote the $\mathrm{U(1)}$'s to $\mathrm{U(N)}$, where $\mathrm{N}$ is
the number of colors. It is a completely open question to establish the enlargement by colors of those nonabelian gauge groups $\mathcal{F}_\Gamma =\prod_{\mu=1}^r \mathrm{U(n_\mu)}$ that are produced by a nonabelian discrete group $\Gamma$.
\item[E)] Last but not least in the list of open problems is  a detailed analysis of the relation  between the algebraic blowup procedure
of singular orbifolds $\mathbb{C}^3/\Gamma$, as that utilized by Markushevich for the case
$\Gamma=\mathrm{L_{168}}$ \cite{marcovaldo}, and the K\"ahler quotient algorithm discussed in the present
paper. We plan to come back to this point in the nearest future.
\end{description}

\section*{Acknowledgements}
We acknowledge important clarifying discussions with our long time collaborators and friends Pietro Antonio Grassi,
Dimitri Markushevich, Aleksander Sorin and Mario Trigiante. U.B.'s research is partially supported by
PRIN 2015 ``Geometria delle variet\`a algebriche'' and INdAM-GNSAGA. This work was completed while U.B. was visiting the
Instituto de Matem\'atica e Estat\'istica of the University of S\~ao Paulo, Brazil, supported by the FAPESP grant 2017/22091-9. 
He likes to thank FAPESP for providing support and his hosts, in particular P.~Piccione, for their hospitality.
\newpage
\appendix
\section{Age grading for various groups}
Since our main interest is to chart the possible M2-brane solutions and their dual Chern-Simons gauge
theories that can be obtained by various choices of the discrete group $\Gamma,$ and since several
supergravity arguments point to the fundamental relevance of the group $\mathrm{L_{168}}$ (see
\cite{miol168,futuro}), in this appendix we just provide the calculation of ages for the maximal subgroups of
$\mathrm{L_{168}}$ and their further subgroups starting from its embedding in $\mathrm{SU(3)}$ considered by
Markushevich \cite{marcovaldo}.
\subsection{Ages for $\Gamma\subset \mathrm{L_{168}}$}
Starting from the construction of the irreducible three-dimensional complex
representation discussed in \cite{miol168,futuro},   we have computed the ages of the various conjugacy classes
for the holomorphic action of the group $\mathrm{L_{168}}$  on  $\mathbb{C}^3$.
\par
In order to be able to compare with Markusevich{'}s paper \cite{marcovaldo}, it is important to note that the
form given by Markusevich of the generators which he calls $\tau $ , $\chi $ and $\omega $, respectively of
order 7, 3 and 2, does not correspond to the standard generators in the presentation of the group \(L_{168}\)
utilized
 by one of us in the recent paper \cite{miol168}.  Yet there is no
problem since we have a translation vocabulary.
Setting:\\
\begin{equation}
R = \omega .\chi \quad; \quad S \, = \, \chi .\tau \quad; \quad T \, =\,\chi ^2. \omega\label{agnaturio}
\end{equation}
these new generators satisfy the standard relations of the presentation displayed below:
\begin{equation}\label{abstroL168}
\mathrm{L_{168}} \, = \, \left(R,S,T \,\parallel \, R^2 \, = \, S^3 \, = \, T^7 \, = \, RST \, = \,
\left(TSR\right)^4 \, = \, \mathbf{e}\right)
\end{equation}
From now on we utilize the abstract notation in terms of $\rho=R$,$\sigma=S$,$\tau=T$.
\par
We begin by constructing explicitly the group \(L_{168}\) in Markushevich's basis substituting the analytic form
of the generators which follows from the identification (\ref{agnaturio}). We find
\\
\(\epsilon \to  \left(
\begin{array}{ccc}
 1 & 0 & 0 \\
 0 & 1 & 0 \\
 0 & 0 & 1 \\
\end{array}
\right)\\
\\
\rho \to  \left(
\begin{array}{ccc}
 -\frac{2 \text{Cos}\left[\frac{\pi }{14}\right]}{\sqrt{7}} & -\frac{2 \text{Cos}\left[\frac{3 \pi
 }{14}\right]}{\sqrt{7}} & \frac{2 \text{Sin}\left[\frac{\pi
}{7}\right]}{\sqrt{7}} \\
 -\frac{2 \text{Cos}\left[\frac{3 \pi }{14}\right]}{\sqrt{7}} & \frac{2 \text{Sin}\left[\frac{\pi
 }{7}\right]}{\sqrt{7}} & -\frac{2 \text{Cos}\left[\frac{\pi
}{14}\right]}{\sqrt{7}} \\
 \frac{2 \text{Sin}\left[\frac{\pi }{7}\right]}{\sqrt{7}} & -\frac{2 \text{Cos}\left[\frac{\pi
}{14}\right]}{\sqrt{7}} & -\frac{2 \text{Cos}\left[\frac{3
\pi }{14}\right]}{\sqrt{7}} \\
\end{array}
\right)\\
\\
\sigma \to  \left(
\begin{array}{ccc}
 0 & 0 & -(-1)^{1/7} \\
 (-1)^{2/7} & 0 & 0 \\
 0 & (-1)^{4/7} & 0 \\
\end{array}
\right)\\
\\
\tau \to  \left(
\begin{array}{ccc}
 \frac{i+(-1)^{13/14}}{\sqrt{7}} & -\frac{(-1)^{1/14} \left(-1+(-1)^{2/7}\right)}{\sqrt{7}} &
 \frac{(-1)^{9/14} \left(1+(-1)^{1/7}\right)}{\sqrt{7}}
\\
 \frac{(-1)^{11/14} \left(-1+(-1)^{2/7}\right)}{\sqrt{7}} & \frac{i+(-1)^{5/14}}{\sqrt{7}} &
 \frac{(-1)^{3/14} \left(1+(-1)^{3/7}\right)}{\sqrt{7}}
\\
 -\frac{(-1)^{11/14} \left(1+(-1)^{1/7}\right)}{\sqrt{7}} & -\frac{(-1)^{9/14}
 \left(1+(-1)^{3/7}\right)}{\sqrt{7}} & -\frac{-i+(-1)^{3/14}}{\sqrt{7}}
\\
\end{array}
\right)\)\\
\\
\begin{equation}
\label{birillus1}
\end{equation}
 We remind the reader that $\rho $,$\sigma $,$\tau $ are the abstract names
for the generators of \(L_{168}\) whose 168 elements are written as words in these letters (modulo
relations). Substituting these letters with explicit matrices that satisfy the defining relation of the group
one obtains an explicit representation of the latter. In the present case the substitution \ref{birillus1}
produces the irreducible 3-dimensional representation \(\text{DA}_3\).
\subsubsection{ The case of the full
group $\Gamma=\mathrm{L_{168}}$}
\par
Utilizing this explicit representation it is straightforward to calculate the age of each conjugacy class and
we obtain the result displayed in the following table.
\begin{equation}\label{vecchierello168}
\mbox{
\begin{tabular}{||c|c|c|c|c|c|c||}
\hline \hline
Conjugacy class of $L_{168}$
&$\mathcal{C}_1$&$\mathcal{C}_2$&$\mathcal{C}_3$&$\mathcal{C}_4$&$\mathcal{C}_5$&$\mathcal{C}_6$\\
  \hline
  \hline
  representative of the class  & $\mathbf{e}$ & $R$ & $S$ &$TSR$ & $T$ & $SR$ \\
  \hline
  order of the elements in the class & 1 & 2 & 3 & 4 & 7 & 7 \\
  \hline
  age &0 & 1 & 1 & 1 & 1 & 2 \\
  \hline
  number of elements in the class & 1 & 21 & 56 & 42 & 24 & 24  \\
  \hline
  \hline
\end{tabular}}
\end{equation}
\subsubsection{ The case of the maximal subgroup $\Gamma = \mathrm{G_{21}} \subset \mathrm{L_{168}}$}
\par
In order to obtain the ages for the conjugacy classes of the maximal subgroup $\mathrm{G_{21}}$, we just need
to obtain the explicit three-dimensional form of its generators $\mathcal{X}$ and $\mathcal{Y}$ satisfying
the defining relations:
\begin{equation}
\mathcal{X}^3 \, = \, \mathcal{Y}^7 \, = \,\mathbf{ 1} \quad ; \quad \mathcal{X}\mathcal{Y}  = \mathcal{Y}^2
\mathcal{X} \label{faloluna}
\end{equation}
This latter is determined by the above explicit form of the $\mathrm{L_{168}}$ generators, by recalling the
embedding relations:
\begin{equation}\label{guftollo}
    \mathcal{Y} \, = \, \rho\,\sigma\,\tau^3\, \sigma\, \rho \quad ; \quad  \mathcal{X} \, = \, \sigma \, \rho\, \sigma\,
    \rho\, \tau^2
\end{equation}
In this way we obtain the following explicit result:
\begin{eqnarray}
 \mathcal{Y} \, \to \, \mathrm{Y} &=& \left(
\begin{array}{ccc}
 -(-1)^{3/7} & 0 & 0 \\
 0 & (-1)^{6/7} & 0 \\
 0 & 0 & -(-1)^{5/7} \\
\end{array}
\right)\nonumber \\
 \mathcal{Y} \, \to \,  \mathrm{X} &=& \left(
\begin{array}{ccc}
 0 & 1 & 0 \\
 0 & 0 & 1 \\
 1 & 0 & 0 \\
\end{array}
\right)\label{pesciYX}
\end{eqnarray}
Hence, for the action on $\mathbb{C}^3$ of the maximal subgroup $\mathrm{G_{21}}\subset \mathrm{L_{168}}$  we
obtain the following ages of its conjugacy classes:
\begin{equation}\label{vecchioG21}
\begin{array}{||c|c|c|c|c|c||}
\hline \hline \mbox{Conjugacy} \mbox{ Class of $\mathrm{G_{21}}$} & C_1 & C_2 & C_3 & C_4 & C_5 \\ \hline
\mbox{representative of the class} & e & \mathcal{Y} & \mathcal{X}^2 \mathcal{Y}\mathcal{X}\mathcal{Y}^2 &
\mathcal{Y}\mathcal{X}^2 & \mathcal{X}
\\
\hline \mbox{order of the elements in the class} & 1 & 7 & 7 & 3& 3 \\
\hline \mbox{age} & 0 & 2 & 1 & 1 & 1 \\
\hline \mbox{number of elements in the class} & 1 & 3 & 3 & 7 & 7 \\
\hline
\end{array}
\end{equation}
\subsubsection{The case of the two maximal octahedral subgroups}
For the other two maximal subgroups $\mathrm{O_{24A}}$ and $\mathrm{O_{24B}}$ we find instead an identical
result. This  is retrieved from the two embedding conditions of the generators $S$ and $T$, satisfying
the defining relations:
\begin{equation}
S^2 \, = \, T^3\,  = \,(ST)^4 \, = \,  \mathbf{1}
\end{equation}
\paragraph{Subgroup $\mathrm{O_{24A}}$}
\begin{equation} \label{O24Aembed}
  T \, = \, \rho\,\sigma \, \rho \, \tau^2 \, \sigma\, \rho \, \tau \quad ; \quad S =
  \tau^2 \, \sigma \, \rho \, \tau \, \sigma^2
\end{equation}
\paragraph{Subgroup $\mathrm{O_{24B}}$}
\begin{equation} \label{O24Bembed}
  T \, = \, \rho\, \tau\, \sigma\, \rho\, \tau^2\, \sigma\, \rho\, \tau \quad ; \quad S =
  \sigma\, \rho\, \tau\, \sigma\, \rho\, \tau
\end{equation}
In this way we get:
\begin{equation}\label{o24Avecchione}
  \begin{array}{|c|c|c|c|c|c|} \hline
 \mbox{Conjugacy Class of the $\mathrm{O_{24A}}$} & C_1 & C_2 & C_3 & C_4 & C_5 \\
\hline
 \mbox{representative of the class} & e & T & STST  & S &  ST  \\
\hline
 \mbox{order of the elements in the class} & 1 & 3 & 2 & 2 & 4 \\
\hline  \mbox{age} & 0 & 1 & 1 & 1 & 1 \\
\hline \mbox{number of elements in the class} & 1 & 8 & 3 & 6 & 6 \\
\hline
\end{array}
\end{equation}
and
\begin{equation}\label{o24Bvecchione}
  \begin{array}{|c|c|c|c|c|c|} \hline
 \mbox{Conjugacy Class of the $\mathrm{O_{24B}}$} & C_1 & C_2 & C_3 & C_4 & C_5 \\
\hline
 \mbox{representative of the class} & e & T & STST  & S &  ST  \\
\hline
 \mbox{order of the elements in the class} & 1 & 3 & 2 & 2 & 4 \\
\hline  \mbox{age} & 0 & 1 & 1 & 1 & 1 \\
\hline \mbox{number of elements in the class} & 1 & 8 & 3 & 6 & 6 \\
\hline
\end{array}
\end{equation}
\subsubsection{The case of the cyclic subgroups $\mathbb{Z}_3$ and $\mathbb{Z}_7$}
Last we consider the age grading for the quotient singularities $\mathbb{C}^3/\mathbb{Z}_3$ and
$\mathbb{C}^3/\mathbb{Z}_7$. As generators of the two cyclic groups we respectively choose the matrices
$\mathrm{X}$ and $\mathrm{Y}$ displayed in eq.\,(\ref{pesciYX}). In other words we utilize either one of the
two generators of the maximal subgroup $\mathrm{G_{21}}\subset \mathrm{L_{168}}$.
\paragraph{The $\Gamma =\mathbb{Z}_3$ case.} The first step consists of diagonalizing the action of the
generator $\mathrm{X}$. Introducing the unitary matrix:
\begin{equation}\label{qumatta}
    \mathfrak{q}\, = \, \left(
\begin{array}{ccc}
 \frac{1}{\sqrt{3}} & \frac{1}{\sqrt{3}} & \frac{1}{\sqrt{3}} \\
 \frac{-1+i \sqrt{3}}{2 \sqrt{3}} & \frac{-1-i \sqrt{3}}{2 \sqrt{3}} & \frac{1}{\sqrt{3}} \\
 \frac{-1-i \sqrt{3}}{2 \sqrt{3}} & \frac{-1+i \sqrt{3}}{2 \sqrt{3}} & \frac{1}{\sqrt{3}} \\
\end{array}
\right)
\end{equation}
we obtain:
\begin{equation}\label{xtilda}
    \widetilde{\mathrm{X}} \, \equiv \,\mathfrak{q}^\dagger \, \mathrm{X} \, \mathfrak{q} \, = \, \left(
\begin{array}{ccc}
 e^{\frac{2 i \pi }{3}} & 0 & 0 \\
 0 & e^{-\frac{2 i \pi }{3}} & 0 \\
 0 & 0 & 1 \\
\end{array}
\right)
\end{equation}
This shows that the quotient singularity $\mathbb{C}^3/\mathbb{Z}_3$ is actually of the form
$\mathbb{C}^2/\mathbb{Z}_3\times \mathbb{C}$ since it suffices to change basis of $\mathbb{C}^3$ by
introducing the new complex coordinates:
\begin{equation}\label{ciangiotto}
    \tilde{z}_a \, = \, \mathfrak{q}_{a}^{\phantom{a}b} \, z_b
\end{equation}
It follows that in the resolution of the singularity we will obtain:
\begin{equation}\label{ghinotucco}
    ALE_{\mathbb{Z}_3} \times \mathbb{C} \, \rightarrow \, \frac{\mathbb{C}^3}{\mathbb{Z}_3}
\end{equation}
Yet, as we discuss more extensively below, the starting setup $\mathbb{C}^3/\Gamma$  produces a special type
of ALE-manifold where all the holomorphic moment map levels are frozen to zero and only the K\"ahler quotient
parameters are switched on.
\par
Eq.\,(\ref{xtilda}) corresponds also to the decomposition of the three-dimensional representation of $Z_3$
into irreducible representations of $\mathbb{Z}_3$. From the diagonalized form (\ref{xtilda}) of the
generator we immediately obtain the ages of the conjugacy classes:
\begin{equation}\label{z3vecchione}
  \begin{array}{|c|c|c|c|} \hline
 \mbox{Conjugacy Class of $\mathbb{Z}_3$} & C_1 & C_2 & C_3  \\
\hline
 \mbox{representative of the class} & e & \mathrm{X} & \mathrm{X}^2   \\
\hline
 \mbox{order of the elements in the class} & 1 & 3 & 3  \\
\hline  \mbox{age} & 0 & 1 & 1  \\
\hline \mbox{number of elements in the class} & 1 & 1 & 1  \\
\hline
\end{array}
\end{equation}
\paragraph{The $\Gamma =\mathbb{Z}_7$ case.} In the $\mathbb{Z}_7$ case, the generator $Y$ is already diagonal and,
as we see,  none of the three complex coordinates is invariant under the action of the group. Hence differently
from the previous case we  obtain:
\begin{equation}\label{ghinotuccone}
    \mathcal{M}_{\mathbb{Z}_7} \, \rightarrow \, \frac{\mathbb{C}^3}{\mathbb{Z}_7}
\end{equation}
where the resolved smooth manifold is not the direct product of $\mathbb{C}$ with an ALE-manifold:
\begin{equation}\label{cominato}
 \mathcal{M}_{\mathbb{Z}_7} \, \neq \,  ALE_{\mathbb{Z}_7} \times \mathbb{C}
\end{equation}
 From the explicit diagonal  form (\ref{pesciYX}) of the
generator we immediately obtain the ages of the conjugacy classes:
\begin{equation}\label{z7vecchione}
  \begin{array}{|c|c|c|c|c|c|c|c|} \hline
 \mbox{Conjugacy Class of $\mathbb{Z}_7$} & C_1 & C_2 & C_3 & C_4 & C_5 & C_6 & C_7  \\
\hline
 \mbox{representative of the class} & e & \mathrm{Y} & \mathrm{Y}^2 & \mathrm{Y}^3 & \mathrm{Y}^4 & \mathrm{Y}^5 & \mathrm{Y}^6\\
\hline
 \mbox{order of the elements in the class} & 1 & 7 & 7 & 7 & 7 & 7 & 7 \\
\hline  \mbox{age} & 0 & 2 & 2 & 1 & 2 & 1 & 1  \\
\hline \mbox{number of elements in the class} & 1 & 1 & 1 & 1 & 1 & 1 & 1 \\
\hline
\end{array}
\end{equation}
\section{ The McKay quiver of various groups}
Since it is of central relevance to the resolution of the singularity by means of a K\"ahler quotient based
on a generalized Kronheimer construction, it is convenient to calculate also the Mckay quiver matrices
associated to the subgroups considered in the previous appendix.
\subsection{The McKay quiver of $\mathrm{L_{168}}$} 
\begin{figure}
\centering
\includegraphics[height=8cm]{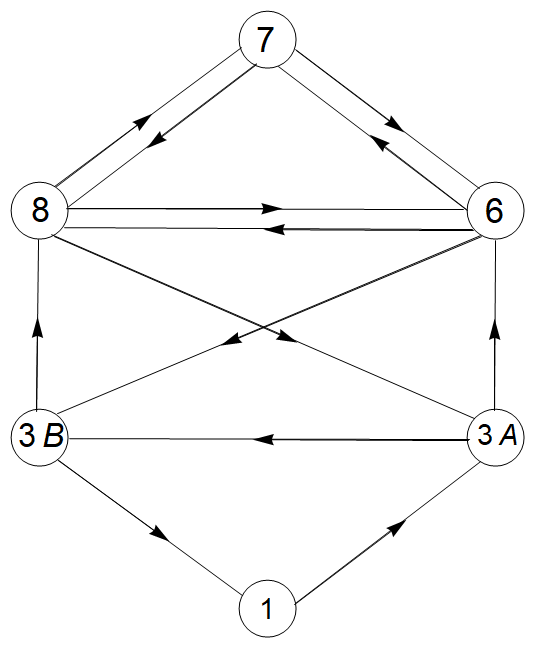}
\caption{ \label{quivettoPL27} The quiver diagram of the finite
group $\mathrm{L_{168}}\subset \mathrm{SU(3)}$ }
\end{figure}

We calculate the McKay matrix defined by
\begin{equation}\label{quiverro}
    \mathcal{Q}\otimes \mathrm{D}_i \, = \, \bigoplus_{j=1}^6 \, \mathcal{A}_{ij}\,\mathrm{D}_j
\end{equation}
where $\mathcal{Q}$ is the three-dimensional complex representation defining the action of $\mathrm{L_{168}}$
on $\mathbb{C}^3$ while $\mathrm{D}_i$ denote the 6 irreducible representation ordered in the standard way we
have so far adopted, namely:
\begin{equation}\label{orinnno}
    \mathrm{D}_i \, = \, \left\{\mathrm{D_1,D_6,D_7,D_8,D_3,D_{\bar 3}} \right\}
\end{equation}
We find the following matrix:
\begin{equation}\label{Amatricia}
    \mathcal{A}\, = \, \left(
\begin{array}{cccccc}
 0 & 0 & 0 & 0 & 1 & 0 \\
 0 & 0 & 1 & 1 & 0 & 1 \\
 0 & 1 & 1 & 1 & 0 & 0 \\
 0 & 1 & 1 & 1 & 1 & 0 \\
 0 & 1 & 0 & 0 & 0 & 1 \\
 1 & 0 & 0 & 1 & 0 & 0 \\
\end{array}
\right)
\end{equation}
The matrix $\mathcal{A}$ admits the  graphical representation
displayed in fig.\ref{quivettoPL27}, named the \textit{McKay quiver}
of the quotient $\mathbb{C}^3/\mathrm{L_{168}}$ \footnote{The
authors are grateful to their friend Massimo Bianchi who noticed
that the McKay quiver presented in the ArXiv version of the present
paper was probably mistaken since the number of lines entering and
outgoing from some of the nodes was not equal. This caused a
revision of  our calculation of the quiver matrix (\ref{Amatricia})
where there was indeed a trivial, yet hidden, mistake. The mistake
was spotted by Bianchi because of his experience with quiver gauge
theories discussed in the several papers
\cite{Bianchi:2000de,Bianchi:2009bg,Bianchi:2007wy}. What is not
widely appreciated in this literature and in general in the physical
community is that the quivers utilized in such a gauge--theory
capacity and the McKay quivers, group theoretically defined as we do
in the present paper, are just one and the same thing. This point
will be addressed in full generality in a forthcoming paper
\cite{conmasbia}.}
\subsection{The McKay quiver of $\mathrm{G_{21}}$} 
\begin{figure}
\centering
\includegraphics[height=8cm]{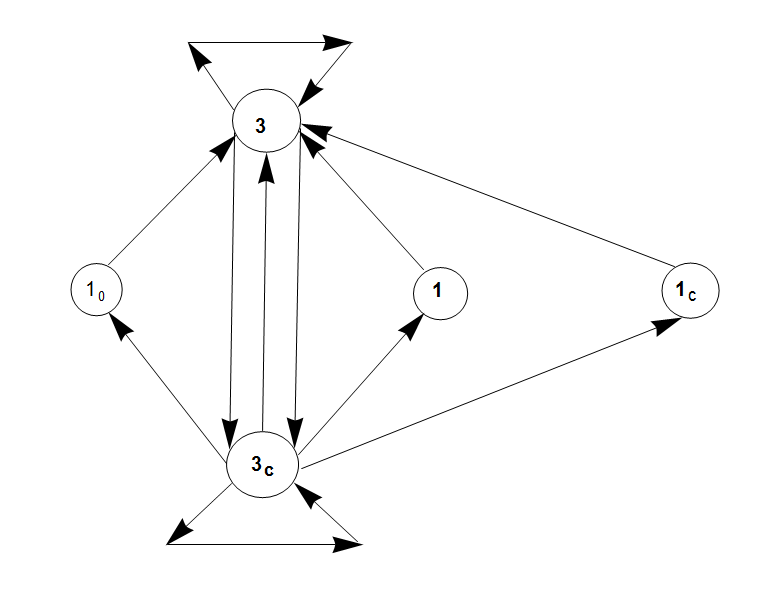}
\caption{ \label{quivetto21} The quiver diagram of the finite group $\mathrm{G_{21}}\subset \mathrm{L_{168}}$
}
\end{figure}

We calculate the McKay matrix defined by
\begin{equation}\label{quiverro21}
    \mathcal{Q}\otimes \mathrm{D}_i \, = \, \bigoplus_{j=1}^5 \, \mathcal{A}_{ij}\,\mathrm{D}_j
\end{equation}
where $\mathcal{Q}$ is the three-dimensional complex representation defining the action of $\mathrm{G_{21}}$
on $\mathbb{C}^3$ while $\mathrm{D}_i$ denote the 5 irreducible representations ordered in the standard way
we have so far adopted, namely:
\begin{equation}\label{orinnno21}
    \mathrm{D}_i \, = \, \left\{\mathrm{D_{0},D_1,D_{\bar 1},D_3,D_{\bar 3}} \right\}
\end{equation}
We find the following matrix:
\begin{equation}\label{Amatricia21}
    \mathcal{A}\, = \, \left(
\begin{array}{ccccc}
 0 & 0 & 0 & 1 & 0 \\
 0 & 0 & 0 & 1 & 0 \\
 0 & 0 & 0 & 1 & 0 \\
 0 & 0 & 0 & 1 & 2 \\
 1 & 1 & 1 & 1 & 1 \\
\end{array}
\right)
\end{equation}
The matrix $\mathcal{A}$ admits the  graphical representation presented in fig. \ref{quivetto21}, named the
\textit{McKay quiver} of the quotient $\mathbb{C}^3/\mathrm{G_{21}}$.

\subsection{The McKay quiver for $\mathbb{Z}_3$}
Next we calculate the McKay matrix for the case where $\mathcal{Q}$ is the three-dimensional representation
of the group $Z_3$ generated by $\mathrm{X}$ given as in eq.\,(\ref{pesciYX})
\begin{figure}
\centering
\includegraphics[height=8cm]{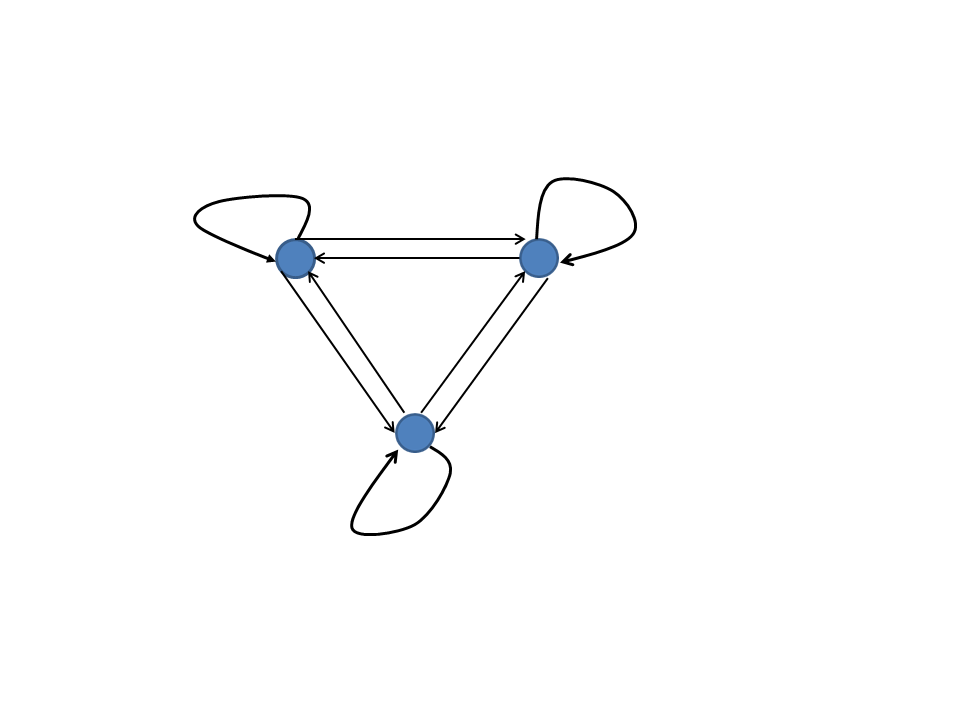}
\caption{ \label{quivettoz3} The quiver diagram of the finite group $\mathbb{Z}_3\subset \mathrm{L_{168}}$.
The three vertices correspond to the three irreducible representations, $\mathbf{1}$, $\psi$ and $\psi^2,$
where $\psi$ is a primitive cubic root of unity. It is not necessary to mark the names of the representation
since the quiver diagram is completely symmetric. In each vertex converge three lines and three lines depart
from each vertex. It is interesting to compare this quiver diagram with that of the master model presented in
fig.\ref{z3maccaius}.}
\end{figure}
and $D_{i}$ are the three irreducible one-dimensional representations of $\mathbb{Z}_3$. The result is
displayed below.
\begin{eqnarray}
  \mathcal{Q}\otimes \mathrm{D}_i &=& \bigoplus_{j=1}^3 \mathcal{A}_{ij} \, \mathrm{D}_j \\
  \mathcal{A}_{ij} &=& \left(
\begin{array}{ccc}
1 &  1 & 1 \\
 1  & 1 & 1 \\
 1 & 1  & 1 \\
\end{array}
\right) \label{komsomolt3}
\end{eqnarray}
The matrix $A_{ij}$ in eq.\,(\ref{komsomolt3}) admits the graphical representation  shown in
eq.\,(\ref{komsomolt3}).
\begin{figure}
\centering
\includegraphics[height=8cm]{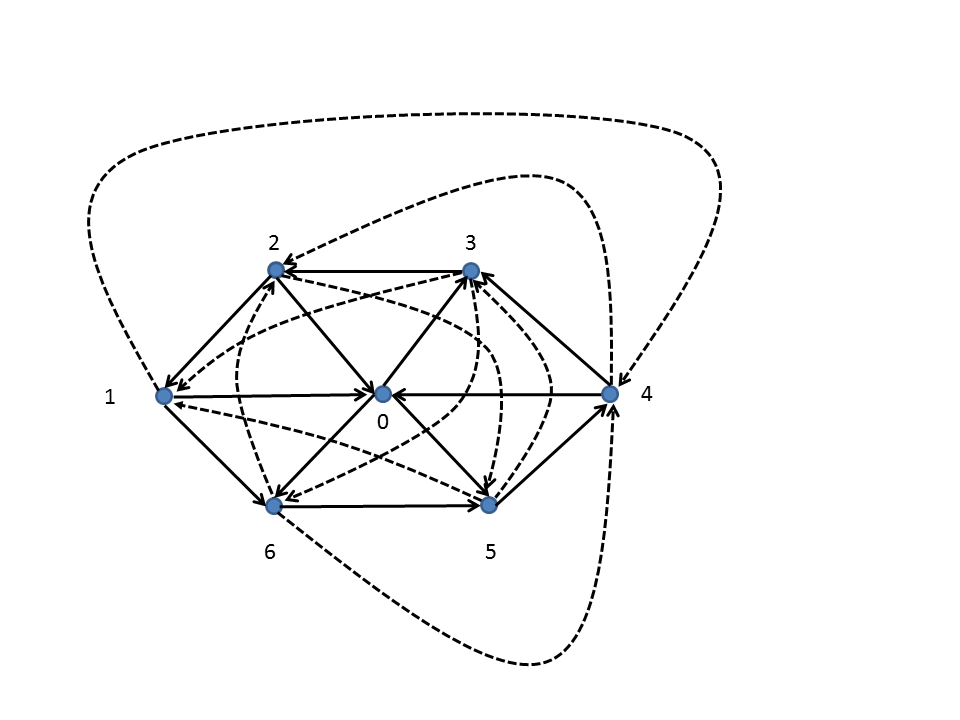}
\caption{ \label{quivettoz7} The quiver diagram of the finite group $\mathbb{Z}_7\subset \mathrm{L_{168}}$.
The seven vertices correspond to the seven irreducible representations, $\mathbf{1}$, marked $0$, and
$\psi,\dots,\psi^6$, marked $1,2,\dots,6$, where $\psi$ is a primitive seventh root of the unity.  In each
vertex converge three lines and three lines depart from each vertex. }
\end{figure}
\subsection{The McKay quiver for $\mathbb{Z}_7$}
Then we calculate the McKay matrix for the case where $\mathcal{Q}$ is the three-dimensional representation
of the group $\mathbb{Z}_7$ generated by $\mathrm{Y}$ as given  in eq.\,(\ref{pesciYX}) and $D_{i}$ are the
seven irreducible one-dimensional representations of $\mathbb{Z}_7$. The result is displayed below.
\begin{eqnarray}
  \mathcal{Q}\otimes \mathrm{D}_i &=& \bigoplus_{j=1}^7 \mathcal{A}_{ij} \, \mathrm{D}_j \\
  \mathcal{A}_{ij} &=& \left(
\begin{array}{ccccccc}
 0 & 0 & 0 & 1 & 0 & 1 & 1 \\
 1 & 0 & 0 & 0 & 1 & 0 & 1 \\
 1 & 1 & 0 & 0 & 0 & 1 & 0 \\
 0 & 1 & 1 & 0 & 0 & 0 & 1 \\
 1 & 0 & 1 & 1 & 0 & 0 & 0 \\
 0 & 1 & 0 & 1 & 1 & 0 & 0 \\
 0 & 0 & 1 & 0 & 1 & 1 & 0 \\
\end{array}
\right) \label{komsomolt7}
\end{eqnarray}
The McKay matrix in eq.\,(\ref{komsomolt7}) admits the graphical representation of fig.\ref{quivettoz7}.
\newpage
\section{Gibbons-Hawking metrics and the resolution of $\mathbb{C}^2/\Gamma$ singularities}
\label{gibbone} As an illustration of  the above outlined generalized Kronheimer construction
which resolves the quotient singularities $\mathbb{C}^3/\Gamma$, we intend to discuss the abelian cases
$\Gamma \, = \, \mathbb{Z}_3$ and $\mathbb{Z}_7$, looking at the case where such finite groups are subgroups
of $\mathrm{L_{168}} \subset \mathrm{SU(3)}$.  In this way, by steps of increasing complexity, we
approach the discussion of the nonabelian cases like $\mathrm{G_{21}}$. When $\Gamma=\mathbb{Z}_3\subset\mathrm{L_{168}} \subset \mathrm{SU(3)} $ we
already pointed out that the singularity is actually of the type mentioned in eq.\,(\ref{ghinotucco}).  This is
quite useful for our purposes since the $ALE_{\mathbb{Z}_k}$ manifolds admit another well known
representation with which we can compare the Kronheimer construction in order to get orientation in our main
task of understanding the cohomology of the resolved K\"ahler manifold. The representation we are alluding to
is that of the Gibbons-Hawking multicenter metrics that are known to be hyperK\"ahlerian and indeed
equivalent to $ALE_{\mathbb{Z}_k}$. The comparison between these two forms of the same metrics is very useful
in order to get  cues  about the mechanisms by means of which the moment map parameters blowup the
singularities in the purely K\"ahlerian case. Hence let us start with the general form of the GH-metrics.
\par
Let the $x,y,z$ be the real coordinates of $\mathbb{R}^3$ to which we adjoin an angle $\tau$ spanning a
circle $\mathbb{S}^1$. A general GH-metric has the following form:
\begin{equation}\label{GWmetricca}
  ds^2_{\mathrm{GH}} \, = \,  \frac{(\text{d$\tau $}+\omega )^2}{\mathcal{V}}+\mathcal{V}
  \left(\text{dx}^2+\text{dy}^2+\text{dz}^2\right)
\end{equation}
where $\mathcal{V} \, = \, \mathcal{V}(x,y,z)$ is a harmonic function on $\mathbb{R}^3$:
\begin{equation}\label{laplaccio}
    \frac{\partial^2\,\mathcal{V}}{\partial x^2}\,+\,\frac{\partial^2\,\mathcal{V}}{\partial y^2}\,
    +\,\frac{\partial^2\,\mathcal{V}}{\partial z^2} \, = \, 0
\end{equation}
and
\begin{equation}\label{omeghetto}
    \omega \, = \, \omega_x \,dx + \omega_y \,dy + \omega_z \,dz
\end{equation}
is a one-form whose external derivative is requested to be Hodge dual, in the flat metric
$ds^2_{\mathbb{R}^3}\, = \, \text{dx}^2+\text{dy}^2+\text{dz}^2$ of $\mathbb{R}^3$, to the gradient of
$\mathcal{V}$:
\begin{equation}\label{cagnesco}
    \star_{\tiny{\mathbb{R}^3}} \,\text{d}\omega \, = \, \text{d}\mathcal{V}
\end{equation}
Without loss of generality we can choose an axial gauge for the connection $\omega$ by setting:
\begin{equation}\label{axialus}
    \omega_z \, = \, 0
\end{equation}
The four-dimensional  Riemannian space $\mathcal{M}_{\mathrm{GH}}$, whose metric is provided by
eq.\,(\ref{GWmetricca}), is  a $\mathrm{U(1)}$-bundle over $\mathbb{R}^3$. Actually we can easily prove that
$\mathcal{M}_{\mathrm{GH}}$ is K\"ahlerian by means of the following argument. Consider the following
two-form:
\begin{equation}\label{kallerformGW}
    \mathbb{K}_{\mathrm{GH}}\, = \, 2\left( \left(\mathrm{d}\tau +\omega\right)\wedge \mathrm{d}z \, - \, \mathcal{V} \,
    \mathrm{d}x\wedge \mathrm{d}y\right)
\end{equation}
which is closed in force of eqs.(\ref{laplaccio}) and (\ref{omeghetto}):
\begin{equation}\label{chiusuralampo}
    \mathrm{d}\mathbb{K}_{\mathrm{GH}}\, = \,0
\end{equation}
From eq.\,(\ref{GWmetricca}) we easily workout the components of the metric in the $x,y,z,\tau$ coordinate
basis:
\begin{equation}\label{coniglione1}
   g_{ij} \, = \,  \left(
\begin{array}{cccc}
 \mathcal{V}+\frac{\omega _x^2}{\mathcal{V}} & \frac{\omega _x \omega _y}{\mathcal{V}} & 0 & \frac{\omega _x}{\mathcal{V}} \\
 \frac{\omega _x \omega _y}{\mathcal{V}} & \mathcal{V}+\frac{\omega _y^2}{\mathcal{V}} & 0 & \frac{\omega _y}{\mathcal{V}} \\
 0 & 0 & \mathcal{V} & 0 \\
 \frac{\omega _x}{\mathcal{V}} & \frac{\omega _y}{\mathcal{V}} & 0 & \frac{1}{\mathcal{V}} \\
\end{array}
\right)
\end{equation}
and of its inverse:
\begin{equation}\label{coniglione2}
   g^{ij} \, = \,  \left(
\begin{array}{cccc}
 \frac{1}{\mathcal{V}} & 0 & 0 & -\frac{\omega _x}{\mathcal{V}} \\
 0 & \frac{1}{\mathcal{V}} & 0 & -\frac{\omega _y}{\mathcal{V}} \\
 0 & 0 & \frac{1}{\mathcal{V}} & 0 \\
 -\frac{\omega _x}{\mathcal{V}} & -\frac{\omega _y}{\mathcal{V}} & 0 & \frac{\mathcal{V}^2+\omega _x^2+\omega _y^2}{\mathcal{V}} \\
\end{array}
\right)
\end{equation}
Similarly, from eq.\,(\ref{kallerformGW}) we work out the components of the form $\mathbb{K}_{\mathrm{GH}}$:
\begin{equation}\label{Kij}
    K_{ij} \, =\, \left(
\begin{array}{cccc}
 0 & -\mathcal{V} & \omega _x & 0 \\
 \mathcal{V} & 0 & \omega _y & 0 \\
 -\omega _x & -\omega _y & 0 & -1 \\
 0 & 0 & 1 & 0 \\
\end{array}
\right)
\end{equation}
Raising the second index of the antisymmetric tensor $K_{ij}$ with the inverse metric $g^{j\ell}$ we obtain a
mixed tensor
\begin{equation}\label{complessostrutto}
  J_{i}^{\phantom{i}\ell} \, \equiv \, K_{ij} \, g^{j\ell} \, = \,   \left(
\begin{array}{cccc}
 0 & -1 & \frac{\omega _x}{\mathcal{V}} & \omega _y \\
 1 & 0 & \frac{\omega _y}{\mathcal{V}} & -\omega _x \\
 0 & 0 & 0 & -\mathcal{V} \\
 0 & 0 & \frac{1}{\mathcal{V}} & 0 \\
\end{array}
\right)
\end{equation}
which satisfies the property:
\begin{equation}\label{cinesino}
    J_{i}^{\phantom{i}\ell} \, J_{\ell}^{\phantom{\ell}\mathrm{m}}\, = \, - \, \delta_{i}^{\mathrm{m}}
\end{equation}
Hence $J$ is a \textit{almost complex structure} which is proved to
be a \textit{complex structure} by verifying that its Nijenhuis
tensor vanishes:
\begin{equation}\label{cardenson}
    N_{ij}^\ell \, \equiv \, \partial_{[i} J_{j]}^{\phantom{j]}\ell} -
    J_{i}^{\phantom{i}\mathrm{m}}\,J_{j}^{\phantom{j}\mathrm{n}}\, \partial_{[\mathrm{m}}
    J_{\mathrm{n}]}^{\phantom{m]}\ell}\, = \, 0
\end{equation}
It follows that $\mathcal{M}_{\mathrm{GH}}$ is a complex manifold, the metric (\ref{kallerformGW}) being
hermitian with respect to J since the matrix  $K_{ij} \, \equiv \, J_{i}^{\phantom{i}\ell} \,g_{\ell j}$ is
by construction antisymmetric and, as such, it defines a K\"ahler $2$-form. Thus we have a K\"ahler form
which is closed and this, by definition, implies that the complex manifold $\mathcal{M}_{\mathrm{GH}}$ is a
K\"ahler manifold.
\par
\subsection{Integration of the complex structure and the issue of the K\"ahler potential}
The first task  to put the K\"ahler metric of a $2n$-dimensional real manifold into a standard
complex form derived from a K\"ahler potential is that of deriving a suitable set of complex coordinates
$Z_\mu$ that are eigenstates of the complex structure. This means to find a complete set of $n$ complex
solutions of the following differential equation:
\begin{equation}\label{complessastrutequa}
    J_{i}^{\phantom{i}\ell} \, \partial_\ell Z \, = \, i \, \partial_i Z
\end{equation}
In the case of the complex structure in equation (\ref{complessostrutto}) a basis of the eigenspace
pertaining to the eigenvalue ${\rm i}$ is easily provided by the following two vectors
\begin{eqnarray}
    \mathbf{v}_1 & = & \left\{-i \omega _y,i \omega _x,i \mathcal{V},1\right\}\nonumber\\
    \mathbf{v}_2& = & \{i,1,0,0\} \nonumber\\
    J \,\mathbf{v}_{1,2} & = & i \mathbf{v}_{1,2} \label{autovettore}
\end{eqnarray}
The second eigenvector $\mathbf{v}_2$ inserted into equation (\ref{complessastrutequa}) immediately singles
out one of the two complex coordinates:
\begin{equation}\label{gomorrus}
    \mathfrak{z} \, \equiv \, y+ i\, x
\end{equation}
 In order to integrate eq.\,(\ref{complessastrutequa}) utilizing the first eigenvector $\mathbf{v}_1$, a very useful
tool is provided by a recent observation made by Ortin et al in \cite{ortinus} who pointed out that a
convenient way of automatically realizing conditions (\ref{laplaccio}) and (\ref{cagnesco}) is obtained by
setting:
\begin{equation}\label{prepotenzialno}
    \omega_x \, = \, \frac{\partial^2 \mathcal{F}}{\partial y \partial z}\quad ; \quad
    \omega_y \, = \, - \, \frac{\partial^2 \mathcal{F}}{\partial x \partial z}
    \quad ; \quad \mathcal{V}\, = \, \frac{\partial^2 \mathcal{F}}{\partial z^2}
\end{equation}
where $\mathcal{F}(x,y,z)$ is a harmonic prepotential:
\begin{equation}\label{laplaccio2}
    \frac{\partial^2\,\mathcal{F}}{\partial x^2}\,+\,\frac{\partial^2\,\mathcal{F}}{\partial y^2}\,
    +\,\frac{\partial^2\,\mathcal{F}}{\partial z^2} \, = \, 0
\end{equation}
Using the prepotential $\mathcal{F}$ the differential equation to be satisfied by the searched for complex
coordinate $\mathfrak{w}$ is the following one:
\begin{equation}\label{bidonerotto}
    \left\{i\,\frac{\partial^2 \mathcal{F}}{\partial z \partial z}, i\,\frac{\partial^2 \mathcal{F}}{\partial y \partial z},
    i \,\frac{\partial^2 \mathcal{F}}{\partial z^2}, 1  \right\}\, = \,
    \left\{ \partial_x \mathfrak{w},\partial_y \mathfrak{w},\partial_z \mathfrak{w},\partial_\tau
    \mathfrak{w}\right\}
\end{equation}
In view of eq.\,(\ref{prepotenzialno}) we can set:
\begin{equation}\label{rovigo}
    \mathcal{F}(x,y,z)\, = \, \int \, \mathrm{d}z \, \int \, \mathrm{d}z \, \mathcal{V}(x,y,z)
\end{equation}
and the differential equation (\ref{bidonerotto})
is immediately integrated by setting:
\begin{equation}\label{deltadelpo}
    \mathfrak{w} \, = \, \tau + i\, \partial_z \mathcal{F}
    \, = \, \tau + i\, \int \, \mathcal{V}\, \mathrm{d}z
\end{equation}
Obviously, whenever a complex coordinate has been found, any  invertible holomorphic function of the same is an equally
good complex coordinate. Hence in addition to $\mathfrak{z}$, defined in eq.\,(\ref{gomorrus}), we choose the
second complex coordinate as follows:
\begin{equation}\label{miciogatto}
    \mathfrak{h} \, = \, \exp\left[i\mathfrak{w}\right] \, = \, e^{i\tau} \rho \quad ; \quad \rho \,= \,
    \exp\left[ - \, \int \, \mathcal{V} \, \mathrm{d}z\right]
\end{equation}
Using the above implicit definition of the complex coordinates one can transform the K\"ahler 2-form
(\ref{kallerformGW}) to the complex coordinates obtaining:
\begin{equation}\label{simonebolione}
    \mathbb{K}_{\mathrm{GH}} \, = \, K_{\mathfrak{h}\bar{\mathfrak{h}}} \, d\mathfrak{h} \wedge
    d\bar{\mathfrak{h}}+K_{\mathfrak{h}\bar{\mathfrak{z}}} \, d\mathfrak{h} \wedge
    d\bar{\mathfrak{z}}+K_{\mathfrak{z}\bar{\mathfrak{h}}} \, d\mathfrak{z} \wedge
    d\bar{\mathfrak{h}}+K_{\mathfrak{z}\bar{\mathfrak{z}}} \, d\mathfrak{z} \wedge
    d\bar{\mathfrak{z}}
\end{equation}
where
\begin{eqnarray}
  K_{\mathfrak{h}\bar{\mathfrak{h}}}  &=& i \, \frac{1}{\mathfrak{h} \, \bar{\mathfrak{h}} \, \mathcal{V}}
  \, =\, \partial_{\mathfrak{h}}\bar{\partial}_{\bar{\mathfrak{h}}} \mathcal{K}\nonumber \\
 K_{\mathfrak{h}\bar{\mathfrak{z}}} &=& \frac{i \omega_x +\omega_y}{\mathfrak{h} \,  \mathcal{V}}
 \, =\, \partial_{\mathfrak{h}}\bar{\partial}_{\bar{\mathfrak{z}}} \mathcal{K}\nonumber \\
 K_{\mathfrak{z}\bar{\mathfrak{h}}} &=& \frac{i \omega_x -\omega_y}{\bar{\mathfrak{h}} \,  \mathcal{V}}
 \, =\, \partial_{\mathfrak{z}}\bar{\partial}_{\bar{\mathfrak{h}}} \mathcal{K}\nonumber\\
 K_{\mathfrak{z}\bar{\mathfrak{z}}} &=& i \frac{ \omega_x^2 +\omega_y^2 + \mathcal{V}^2}{  \mathcal{V}}
 \, =\, \partial_{\mathfrak{z}}\bar{\partial}_{\bar{\mathfrak{z}}} \mathcal{K}\label{secularia}
\end{eqnarray}
The problem of deriving the K\"ahler potential
$\mathcal{K}(\mathfrak{h},\mathfrak{z},\bar{\mathfrak{h}},\bar{\mathfrak{z}})$  corresponding to the
GH-metric is reduced to the nontrivial task  of inverting the coordinate transformation encoded in eqs.
(\ref{deltadelpo}) and (\ref{gomorrus}) and then solving the system of coupled differential equations encoded
in eqs. (\ref{secularia}). Typically this is far from being  a nontrivial task, but in some simple cases it can be
done. The primary illuminating example is provided by the Eguchi-Hanson metric corresponding to
$ALE_{\mathbb{Z}_2}$.
\newpage
\section{Abelian Examples of generalized Kronheimer constructions}
\label{exemplaria}
\subsection{Construction \`a la Kronheimer of the crepant resolution $\mathcal{M}_{\mathbb{Z}_3}\,
\rightarrow \,\frac{\mathbb{C}^3}{\mathbb{Z}_3}$ with $\mathbb{Z}_3 \subset
\mathrm{L_{168}}$}\label{frugifera}
Our next exercise is the resolution of the singularity $\frac{\mathbb{C}^3}{\mathbb{Z}_3}$ where the group $\mathbb{Z}_3$ is embedded in $G_{21}\subset
L_{168}$. We will see that here the algebraic moment map equations are of higher order, actually reduce to a
single equation of the sixth order which cannot be explicitly solved and no analytically close
expression of either the K\"ahler potential or the first Chern classes of the tautological bundles can be
written down. Yet from another view point we know that the  resolved manifold is the product
$\mathbb{C}\times ALE_{\mathbb{Z}_3}$, the second factor being equivalent to a Gibbons-Hawking space for
which an explicit expression of the metric exists although written in different coordinates.
\par
Following the general strategy outlined in the main text sections, the first step consists in deriving the
invariant space $\mathcal{S}_\Gamma\, = \, \text{Hom}_{\Gamma }(R, \mathcal{Q}\otimes R)$ made of
triples $\{A,B,C\}$ of $3\times 3$ matrices that satisfy eq.\,(\ref{gammazione}), namely:
\begin{equation}\label{gerticco}
   \mathrm{X}\, \left(\begin{array}{c}
               A \\
               B \\
               C
             \end{array}
   \right) \, = \, \left(
\begin{array}{c}
R(\mathrm{X}^{-1})\, A \, R(\mathrm{X}) \\
R(\mathrm{X}^{-1})\, B \, R(\mathrm{X}) \\
R(\mathrm{X}^{-1})\, C \, R(\mathrm{X}) \\
\end{array}
\right)
\end{equation}
where $\mathrm{X}$ is the $\mathbb{Z}_3$-generator displayed in eq (\ref{pesciYX}) and $R(\mathrm{X})$ is its
representation in the natural basis of the regular representation. Quite exceptionally this latter coincides
with $\mathrm{X}$ as a matrix:
\begin{equation}\label{cagnotto}
    R(\mathrm{X})\, = \, \mathrm{X}
\end{equation}
The constraint (\ref{gerticco}) reduces the number of parameters from 27 to 9. Explicitly we have:
\begin{eqnarray}
  \left(A,B,C\right) &\in & \text{Hom}_{\Gamma }(R, \mathcal{Q}\otimes R) \nonumber\\
  \null &\Downarrow &\null \nonumber\\
  A &=& \left(
\begin{array}{ccc}
 \alpha _{1,1} & \alpha _{1,2} & \alpha _{1,3} \\
 \alpha _{2,1} & \alpha _{2,2} & \alpha _{2,3} \\
 \alpha _{3,1} & \alpha _{3,2} & \alpha _{3,3} \\
\end{array}
\right) \nonumber\\
  B &=& \left(
\begin{array}{ccc}
 \alpha _{2,2} & \alpha _{2,3} & \alpha _{2,1} \\
 \alpha _{3,2} & \alpha _{3,3} & \alpha _{3,1} \\
 \alpha _{1,2} & \alpha _{1,3} & \alpha _{1,1} \\
\end{array}
\right) \nonumber\\
  C &=& \left(
\begin{array}{ccc}
 \alpha _{3,3} & \alpha _{3,1} & \alpha _{3,2} \\
 \alpha _{1,3} & \alpha _{1,1} & \alpha _{1,2} \\
 \alpha _{2,3} & \alpha _{2,1} & \alpha _{2,2} \\
\end{array}
\right)\label{cratinus}
\end{eqnarray}
If we perform the diagonalization of the regular representation by means of the matrix (\ref{qumatta}), the
result of the same constraint (\ref{gerticco}) can be expressed in the following way:
\begin{eqnarray}
  \left(A,B,C\right) &\in & \text{Hom}_{\Gamma }(R, \mathcal{Q}\otimes R) \nonumber\\
  \null &\Downarrow &\null \nonumber\\
  A &=& \left(
\begin{array}{ccc}
 m_{1,1} & m_{1,2} & m_{1,3} \\
 m_{2,1} & m_{2,2} & m_{2,3} \\
 m_{3,1} & m_{3,2} & m_{3,3} \\
\end{array}
\right) \nonumber\\
  B &=& \left(
\begin{array}{ccc}
 m_{1,1} & -\sqrt[3]{-1} m_{1,2} & (-1)^{2/3} m_{1,3} \\
 (-1)^{2/3} m_{2,1} & m_{2,2} & -\sqrt[3]{-1} m_{2,3} \\
 -\sqrt[3]{-1} m_{3,1} & (-1)^{2/3} m_{3,2} & m_{3,3} \\
\end{array}
\right) \nonumber\\
  C &=& \left(
\begin{array}{ccc}
 m_{1,1} & (-1)^{2/3} m_{1,2} & -\sqrt[3]{-1} m_{1,3} \\
 -\sqrt[3]{-1} m_{2,1} & m_{2,2} & (-1)^{2/3} m_{2,3} \\
 (-1)^{2/3} m_{3,1} & -\sqrt[3]{-1} m_{3,2} & m_{3,3} \\
\end{array}
\right)\label{stupidus}
\end{eqnarray}
in terms of a new set of nine parameters $m_{i,j}$.
\subsubsection{Characterization of the locus
$\mathcal{D}_{\mathbb{Z}_3}$ and of the variety $\mathbb{V}_{3+2}$} Next we define the reduction of the
invariant space $\text{Hom}_{\Gamma }(R, \mathcal{Q}\otimes R)$ to the locus cut out by the holomorphic
constraint (\ref{poffarbacchio}) which was named $\mathcal{D}_\Gamma$:
\begin{equation}
\text{
}\mathcal{D}_{{\mathbb{Z}_3} } \equiv  \left\{\text{  }p=\left(
\begin{array}{c}
 A \\
 B \\
 C \\
\end{array}
\right)\in  \text{Hom}_{\Gamma }(R, \mathcal{Q}\otimes R)\text{  }/\text{    }[A,B]=[B,C]=[C,A] =0 \right \}
\label{pidocchio}
\end{equation}
Differently from the more complicated cases of maximal subgroups of \(L_{168}\), in the present abelian case
we can explicitly solve the quadratic equations provided by the commutator constraints and we discover that
there is a principal branch of the solution, named \(\mathcal{D}_{\Gamma }^0\) that has indeed dimension
$5=|\Gamma |+2$. However, in addition to that there are 25 more  branches of the solution with smaller
dimension. This clarifies a point stated by us in our previous general discussion. There is always a unique
principal branch of the solution with the maximal number $|\Gamma |+2$ of free parameters and  we are  able
to show that such principal branch \(\mathcal{D}_{\Gamma }^0\) is indeed the orbit with respect to the group
\(\mathcal{G}_{\Gamma }\) of the subspace \(L_{\Gamma}\) defined in eq.\,(\ref{thespacel}). Hence the variety
\(\mathbb{V}_{|\Gamma |+2}\) of which we are supposed to take the K\"ahler quotient is not defined by
eq.\,(\ref{pidocchio}), rather by the principal branch of that variety.
\par
In the present case, the principal branch of the solution to eq.\,(\ref{pidocchio}) can be expressed in the
following way:
\begin{equation}\label{principalbranca}
   m_{1,1}= \omega _1, \, m_{1,2}= \omega _2, \, m_{1,3}= \omega _3, \, m_{2,1}= \omega _4, \, m_{2,2}=
   \omega _1, \, m_{2,3}= \omega _5, \, m_{3,1}=
   \frac{\omega _2 \omega _4}{\omega _3}, \, m_{3,2}= \frac{\omega _2 \omega _4}{\omega _5}, \, m_{3,3}= \omega _1
\end{equation}
where $\omega_{1,\dots,5}$ are five free complex parameters. Substituting eq.\,(\ref{principalbranca}) in
eq.\,(\ref{stupidus}) we obtain the explicit parameterization of the locus $\mathcal{D}^0_{\mathbb{Z}_3}$ in
terms of the 5 parameters $\omega_i$.
\subsubsection{Derivation of the quiver group $\mathcal{G}_\Gamma$} Our
next point is the derivation of the group \(\mathcal{G}_{\Gamma }\) defined in eq.\,(\ref{caliente}), namely:
\begin{equation}
\text{                                           }\mathcal{G}_{\mathbb{Z}_3 } = \left\{ g\in
\text{SL}(3,\mathbb{C})\quad \left| \quad\forall \gamma \in \mathbb{Z}_3  : \;\;\left[D_R(\gamma
),D_{\text{def}}(g)\right]\right.=0\right\}
\end{equation}
Let us proceed to this construction. In the diagonal basis of the regular representation this is a very easy
task, since the group is simply given by the diagonal 3$\times $3 matrices with determinant one. We introduce
these matrices:
\begin{equation}
g=\left(
\begin{array}{|c|c|c|}
\hline
 a_1 & 0 & 0 \\
\hline
 0 & a_2 & 0 \\
\hline
 0 & 0 & a_3 \\
\hline
\end{array}
\right) \label{3adiago}
\end{equation}
subject to the constraint \(\prod _{i=1}^3 a_i\)=1.
\par
Next we want to show that\\
\begin{equation}
\text{}\mathcal{D}_{\mathbb{Z}_3 }^0\text{ }= \text{Orbit}_{\mathcal{G}_{\mathbb{Z}_3
}}\left(L_{\mathbb{Z}_3}\right) \label{VorbitaL}
\end{equation}
\par
To this effect we rewrite the locus
\begin{eqnarray}
L_{\mathbb{Z}_3} & \equiv & \left\{ A_0,B_0,C_0\right\}\nonumber\\
A_0 &=& \left(
\begin{array}{ccc}
 \alpha _{1,1} & 0 & 0 \\
 0 & \alpha _{2,2} & 0 \\
 0 & 0 & \alpha _{3,3} \\
\end{array}
\right) \; ; \; B_0=\left(
\begin{array}{ccc}
 \alpha _{2,2} & 0 & 0 \\
 0 & \alpha _{3,3} & 0 \\
 0 & 0 & \alpha _{1,1} \\
\end{array}
\right)\; ; \;  C_0=\left(
\begin{array}{ccc}
 \alpha _{3,3} & 0 & 0 \\
 0 & \alpha _{1,1} & 0 \\
 0 & 0 & \alpha _{2,2} \\
\end{array}
\right)
\end{eqnarray}
 in the diagonal basis of the regular representation. The change of basis is performed by the
normalized character table of the cyclic group \(\mathbb{Z}_3\)
\begin{equation}
\pmb{\chi} \, = \, \frac{1}{\sqrt{3}} \,\left( \begin{array}{ccc}
 1 & 1 & 1 \\
 1 & e^{\frac{2 i \pi }{3}} & e^{-\frac{2 i \pi }{3}} \\
 1 & e^{-\frac{2 i \pi }{3}} & e^{\frac{2 i \pi }{3}} \\
\end{array}
\right)
\end{equation}
by setting $\tilde{A}_0 \, = \,\pmb{\chi}^\dagger \, A_0 \, \pmb{\chi}$ and similarly for $\tilde{B}_0$ and
$\tilde{C}_0$. Next let us setup the orbit of interest to us:
\begin{eqnarray}\label{rutilio}
    \mbox{Orbit}_{\mathcal{G}_{\mathbb{Z}_3}}\left(L_{\mathbb{Z}_3}\right)\, \equiv\, \left\{ g^{-1}\pmb{\chi}^\dagger \,
     A_0 \, \pmb{\chi} \, g,\, g^{-1}\pmb{\chi}^\dagger \,
     B_0 \, \pmb{\chi} \, g,\, g^{-1}\pmb{\chi}^\dagger \,
     C_0 \, \pmb{\chi} \, g\right\}
\end{eqnarray}
where $g$ is given in eq.\,(\ref{3adiago}). The space (\ref{rutilio}) is clearly 5-dimensional, the five
parameters being, $\alpha_{1,1},\alpha_{2,2},\alpha_{3,3}$ and $a_1,a_2,a_3$ with constraint $\prod_{i=1}^3
a_i \, =\,1$. We can easily verify that the identification (\ref{VorbitaL}) holds true upon use of the
following identification:
\begin{eqnarray}
a_1& = & \frac{\omega _4^{1/3}}{\omega _3^{1/3}}\nonumber\\
a_2& = & \frac{\omega _2^{1/3}}{\omega _5^{1/3}}\nonumber\\
a_3& = & \frac{\omega _3^{1/3} \omega _5^{1/3}}{\omega _2^{1/3} \omega _4^{1/3}}\nonumber\\
\alpha _{1,1}& = & \omega _1+\frac{\omega _2^{1/3} \omega _3^{1/3} \omega _4^{2/3}}{\omega
_5^{1/3}}+\frac{\omega _2^{2/3} \omega _4^{1/3} \omega _5^{1/3}}{\omega
_3^{1/3}}\nonumber\\
\alpha _{2,2}& = & \frac{1}{6} \left(6 \omega _1+\frac{3 i \left(i+\sqrt{3}\right) \omega _2^{1/3} \omega
_3^{1/3} \omega _4^{2/3}}{\omega _5^{1/3}}-\frac{6
(-1)^{1/3} \omega _2^{2/3} \omega _4^{1/3} \omega _5^{1/3}}{\omega _3^{1/3}}\right)\nonumber\\
\alpha _{3,3}& = & \frac{-2 \sqrt{3} \omega _2^{1/3} \omega _3^{2/3} \omega _4^{2/3}+\left(-3
i+\sqrt{3}\right) \omega _1 \omega _3^{1/3} \omega _5^{1/3}+\left(3 i+\sqrt{3}\right) \omega _2^{2/3} \omega
_4^{1/3} \omega _5^{2/3}}{\left(-3 i+\sqrt{3}\right) \omega _3^{1/3} \omega _5^{1/3}}
\end{eqnarray}
between the orbit parameters and those that parameterize, according to (\ref{principalbranca}), the
principal branch $\mathcal{D}^0_{\mathbb{Z}_3}$ of the locus (\ref{poffarbacchio}).
\par
In conclusion the variety $\mathbb{V}_5$ of which we are supposed to take the K\"ahler quotient in order to
obtain the resolution $\mathcal{M}_\zeta \rightarrow \frac{\mathbb{C}^3}{\mathbb{Z}_3}$ is defined in the
following way:
\begin{eqnarray}
  \mathbb{V}_5 &\equiv &  \mbox{Orbit}_{\mathcal{G}_{\mathbb{Z}_3}} \, = \, \left\{ A,B,C\right\} \\
  A &=& \left(
\begin{array}{ccc}
 \omega _1 & \omega _2 & \omega _3 \\
 \omega _4 & \omega _1 & \omega _5 \\
 \frac{\omega _2 \omega _4}{\omega _3} & \frac{\omega _2 \omega _4}{\omega _5} & \omega _1 \\
\end{array}
\right)\nonumber\\
  B &=& \left(
\begin{array}{ccc}
 \omega _1 & -\sqrt[3]{-1} \omega _2 & (-1)^{2/3} \omega _3 \\
 (-1)^{2/3} \omega _4 & \omega _1 & -\sqrt[3]{-1} \omega _5 \\
 -\frac{\sqrt[3]{-1} \omega _2 \omega _4}{\omega _3} & \frac{(-1)^{2/3} \omega _2 \omega _4}{\omega _5} & \omega _1 \\
\end{array}
\right)\nonumber\\
  C &=& \left(
\begin{array}{ccc}
 \omega _1 & (-1)^{2/3} \omega _2 & -\sqrt[3]{-1} \omega _3 \\
 -\sqrt[3]{-1} \omega _4 & \omega _1 & (-1)^{2/3} \omega _5 \\
 \frac{(-1)^{2/3} \omega _2 \omega _4}{\omega _3} & -\frac{\sqrt[3]{-1} \omega _2 \omega _4}{\omega _5} & \omega _1 \\
\end{array}
\right) \label{garrumovsky}
\end{eqnarray}
\subsubsection{The algebraic equation of the orbifold locus} Let us now consider the action of the
$\mathbb{Z}_3$ group on $\mathbb{C}^3$ as defined  by the generator $\mathrm{X}$ in eq.\,(\ref{pesciYX}). If
$\{z_1,z_2,z_3\}$ are the coordinates of a point in $\mathbb{C}^3$, we
 see that there are four invariant polynomials:
 \begin{eqnarray}
   J_1 &=& \frac{1}{216} \left(2 \sqrt{3} z_1-\left(\sqrt{3}+3 i\right)
   z_2-\left(\sqrt{3}-3 i\right) z_3\right){}^3 \, \equiv \, w_1^3\nonumber\\
   J_2 &=& \frac{1}{216} \left(2 \sqrt{3} z_1-\left(\sqrt{3}-3 i\right)
   z_2-\left(\sqrt{3}+3 i\right) z_3\right){}^3 \, \equiv \, w_2^3\nonumber\\
   J_3 &=& \frac{\left(z_1+z_2+z_3\right){}^3}{3 \sqrt{3}}\, \equiv \, w_3^3 \nonumber\\
   J_4 &=& \frac{z_1^3-3 z_2 z_3 z_1+z_2^3+z_3^3}{3 \sqrt{3}}\, \equiv \, w_1\,w_2\,w_3\label{invariantuli}
 \end{eqnarray}
which satisfy the following equation:
\begin{equation}\label{equatiaz3}
    J_1\,J_2\,J_3 \, - \, J_4^3 \, = \, 0
\end{equation}
This equation can be regarded as the cubic equation which cuts out in $\mathbb{C}^4$ the locus corresponding to
the singular orbifold $\mathbb{C}^3/{\mathbb{Z}_3}$. As we know this latter is equivalent to
$\mathbb{C}\times \mathbb{C}^3/{\mathbb{Z}_3}$. How do we retrieve this fact in the present language? It is
a simple matter. Consider the new coordinates of $\mathbb{C}^3$ that diagonalize the action of $\mathrm{X}$
and are implicitly defined already in eq.\,(\ref{invariantuli}):
\begin{eqnarray}
  \mathbf{w} &\equiv& \pmb{\chi}^\dagger \,\mathbf{z} \nonumber\\
  w_1 &=& \ft 1 6 \,\left(2 \sqrt{3} z_1-\left(\sqrt{3}+3 i\right)
   z_2-\left(\sqrt{3}-3 i\right) z_3\right) \nonumber\\
  w_2 &=& \ft{1}{6} \left(2 \sqrt{3} z_1-\left(\sqrt{3}-3 i\right)
   z_2-\left(\sqrt{3}+3 i\right) z_3\right) \nonumber\\
  w_3 &=& \ft {1}{\sqrt{3}}\left(z_1+z_2+z_3\right)\label{wudoppi}
\end{eqnarray}
We have the correspondence:
\begin{equation}\label{catinus}
    \mathrm{X} \, \mathbf{z} \, \Leftrightarrow \, \widetilde{\mathrm{X}} \,\mathbf{w} \quad \mbox{where}
    \quad \widetilde{\mathrm{X}} \, = \,  \left(
                                                  \begin{array}{ccc}
                                                    e^{\ft{2\pi}{3} \, i} & 0 & 0 \\
                                                    0 & e^{- \,\ft{2\pi}{3} \, i} & 0 \\
                                                    0 & 0 & 1 \\
                                                  \end{array}
                                                \right)
\end{equation}
which implies that in terms of the $w$-variables we have the following $\mathbb{Z}_3$ invariants:
\begin{equation}\label{ninnananna}
    \mathfrak{j}_1 = w_1^3 \quad ; \quad \mathfrak{j}_2 = w_2^3\quad ; \quad \mathfrak{j}_3 = w_1 \, w_2 ;
    \quad ; \quad \mathfrak{j}_4 = w_3
\end{equation}
and we can write:
\begin{equation}\label{morfinat}
    J_1 \, = \, \mathfrak{j}_1 \quad ; \quad J_2 \, = \, \mathfrak{j}_2\quad ; \quad J_3 \, = \,
    \mathfrak{j}_4^3 \quad ; \quad  J_4 \, = \,
    \mathfrak{j}_3 \, \mathfrak{j}_4
\end{equation}
Regarding $J_i$ and $\mathfrak{j}_i$ as the coordinates of two copies of $\mathbb{C}^4$ we can regard
eq.\,(\ref{morfinat}) as a morphism:
\begin{equation}\label{craccolino}
    \mu \, : \, \mathbb{C}^{3} \times \mathbb{C}\, \rightarrow \, \mathbb{C}^{4}
\end{equation}
Under such a morphism the algebraic equation (\ref{equatiaz3}) is mapped into:
\begin{equation}\label{kastriula}
    \mathfrak{j}_4^3 \, \left(\mathfrak{j}_1 \, \mathfrak{j}_2 \, - \, \mathfrak{j}_3^3\right)\, = \,0
\end{equation}
and in the expression in bracket we recognize the equation of the $\frac{\mathbb{C}^2}{\mathbb{Z}_3}$
orbifold as described in eq.\,(\ref{transformillina}) while discussing the standard Kronheimer construction
(it suffices to identify $x=\mathfrak{j}_1, y=\mathfrak{j}_2, z =\mathfrak{j}_3$).
\subsubsection{Map of the variety $\mathbb{V}_5$ into the algebraic locus corresponding to the orbifold}
Having established the above relations we verify that, in complete analogy with the standard Kronheimer construction,
we can reproduce the defining equation (\ref{equatiaz3}) in
terms of invariants of the three matrices (\ref{garrumovsky}) spanning the $\mathbb{V}_5$ variety. It
suffices to identify:
\begin{eqnarray}
  J_1 & = & \mbox{Det}[ A ]\, = \, \omega _1^3-3 \omega _2 \omega _4 \omega _1+\frac{\omega _2^2
   \omega _4 \omega _5}{\omega _3}+\frac{\omega _2 \omega _3
   \omega _4^2}{\omega _5}\nonumber\\
  J_2 &=& \mbox{Det}[ B ]\, = \, \omega _1^3-3 \omega _2 \omega _4 \omega _1+\frac{\omega _2^2
   \omega _4 \omega _5}{\omega _3}+\frac{\omega _2 \omega _3
   \omega _4^2}{\omega _5}\nonumber\\
   J_3 &=& \mbox{Det}[ C ]\, = \, \omega _1^3-3 \omega _2 \omega _4 \omega _1+\frac{\omega _2^2
   \omega _4 \omega _5}{\omega _3}+\frac{\omega _2 \omega _3
   \omega _4^2}{\omega _5}\nonumber\\
  J_4 &=& \mbox{Tr}[A\,B\,C] \, = \left(\omega _1^3-3 \omega _2 \omega _4 \omega _1+\frac{\omega _2^2
   \omega _4 \omega _5}{\omega _3}+\frac{\omega _2 \omega _3
   \omega _4^2}{\omega _5}\,\right)^3 \label{pagnuflo}
\end{eqnarray}
Eq.\,(\ref{pagnuflo}) describes an explicit inclusion map of the variety $\mathbb{V}_5$ into the algebraic
locus $\mathbb{C}^3/\mathbb{Z}_3$:
\begin{equation}\label{imbecillus}
    \mathbb{V}_5\, \rightarrow \, \frac{\mathbb{C}^3}{\mathbb{Z}_3}
\end{equation}
\subsubsection{The K\"ahler quotient} The next step consists in performing the K\"ahler quotient of the
K\"ahler manifold $\mathbb{V}_5$ with respect to the compact subgroup of the quiver group
$\mathcal{G}_{\mathbb{Z}_3}$, which, as we several times emphasized, is the \textit{gauge group} of the
corresponding three-dimensional Chern-Simons gauge theory:
\begin{equation}\label{gaugogz3}
    \mathcal{F}_{\mathbb{Z}_3} \, \equiv \, \mathcal{G}_{\mathbb{Z}_3}\bigcap \mathrm{SU(3)}
\end{equation}
A  generic element $g \in \mathcal{F}_{\mathbb{Z}_3} $ is  of the form (\ref{3adiago}) with:
\begin{equation}\label{ciulifischio2}
    a_i \, = \, \exp\left[ i \, \theta_i\right] \quad ; \quad \sum_{i=1}^3 \, \theta_i \, = \, 2\,\pi \, n
    \quad \left(n\in \mathbb{Z}\right)
\end{equation}
The K\"ahler structure of $\mathbb{V}_5$ is provided by the pullback on the $\mathbb{V}_5$ surface of the
K\"ahler potential of the entire flat K\"ahler manifold $\mbox{Hom}_{\mathbb{Z}_3} \left(R,\mathcal{Q}\otimes
R\right)$, namely we have:
\begin{eqnarray}\label{kallerV}
    \mathcal{K}_\mathbb{V} & \equiv &\mbox{Tr} \;\left(\left[ A^\dagger\, , \, A\right] + \left[ B^\dagger\, , \, B\right] +
    \left[ C^\dagger\, , \, C\right]\right)\nonumber\\
    & = &3 \left(3 \omega _1 \bar{\omega }_1+\omega _3 \bar{\omega
   }_3+\omega _4 \bar{\omega }_4+\omega _2 \bar{\omega }_2
   \left(\frac{\omega _4 \bar{\omega }_4}{\omega _3 \bar{\omega
   }_3}+\frac{\omega _4 \bar{\omega }_4}{\omega _5 \bar{\omega
   }_5}+1\right)+\omega _5 \bar{\omega }_5\right)
\end{eqnarray}
$\mathcal{K}_\mathbb{V} $  is obviously invariant under the unitary transformations of the \textit{gauge
group} :
\begin{equation}\label{constituzia}
\forall \, g \, \in  \mathcal{F}_{\mathbb{Z}_3}\, : \quad   \left\{A,\,B,\,C\right\} \, \rightarrow \,
\left\{g^\dagger A g,\, g^\dagger B g,\, g^\dagger C g\right\}
\end{equation}
which, for that reason, is an \textit{isometry group} of the corresponding K\"ahler metric on
$\mbox{Hom}_{\mathbb{Z}_3} \left(R,\mathcal{Q}\otimes R\right)$ and  of its restriction to $\mathbb{V}_5$.
The last point follows from the fact that, by construction,
 $\mathcal{F}_{\mathbb{Z}_3}$ maps
$\mathbb{V}_5\subset\mbox{Hom}_{\mathbb{Z}_3} \left(R,\mathcal{Q}\otimes R\right) $ into itself.
\par
A basis of two linearly independent generators of the Lie algebra $\mathbb{F}_{\mathbb{Z}_3}$ is provided by
the following two matrices:
\begin{equation}\label{effiunodue}
    \mathfrak{f}_1 \, = \, \left(
\begin{array}{ccc}
 i & 0 & 0 \\
 0 & -i & 0 \\
 0 & 0 & 0 \\
\end{array}
\right) \quad ; \quad  \mathfrak{f}_2 \, = \, \left(
\begin{array}{ccc}
 0 & 0 & 0 \\
 0 & i & 0 \\
 0 & 0 & -i \\
\end{array}
\right)
\end{equation}
and the moment maps corresponding to the two isometries generated by them are defined as follows:
\begin{equation}\label{cineseturco}
\mathfrak{P}_A \, = \, -\, i \, \mbox{Tr} \left(\mathfrak{f}_A \left(\left[ A^\dagger\, , \, A\right] +
\left[ B^\dagger\, , \, B\right] + \left[ C^\dagger\, , \, C\right]\right) \right),\quad \left( A=1,2\right)
\end{equation}
Explicitly we find:
\begin{eqnarray}
  \mathfrak{P}_1(\omega,\bar{\omega}) &=& 3 \omega _3 \bar{\omega }_3-6 \omega _4 \bar{\omega }_4+3 \omega
   _2 \bar{\omega }_2 \left(-\frac{\omega _4 \bar{\omega
   }_4}{\omega _3 \bar{\omega }_3}+\frac{\omega _4 \bar{\omega
   }_4}{\omega _5 \bar{\omega }_5}+2\right)-3 \omega _5
   \bar{\omega }_5 \nonumber\\
  \mathfrak{P}_2(\omega,\bar{\omega}) &=& 3 \left(\omega _3 \bar{\omega }_3+\omega _4 \bar{\omega
   }_4+\omega _2 \bar{\omega }_2 \left(-\frac{\omega _4
   \bar{\omega }_4}{\omega _3 \bar{\omega }_3}-\frac{2 \omega _4
   \bar{\omega }_4}{\omega _5 \bar{\omega }_5}-1\right)+2 \omega
   _5 \bar{\omega }_5\right) \label{micromiceto}
\end{eqnarray}
\paragraph{Choosing a gauge.} As we emphasized several times,  the quiver group $\mathcal{G}_{\mathbb{Z}_3}$ leaves the
surface $[A,B]=[B,C]=[C,A]=0$ invariant, namely it maps $V_5$ into itself, yet it does not leave the
moment maps $\mathfrak{P}_A$ invariant since it is not an isometry. The latter are invariant only under the
compact gauge subgroup. This property is very important since it is the key instrument to obtain the lifting
from a zero level to a prescribed one of the level surfaces, thus providing the algorithm to perform the
K\"ahler quotient explicitly.
\par
To this effect we consider the action of an element:
\begin{equation}\label{racconigi}
    g \, = \, \left(
                \begin{array}{ccc}
                  \mu_1 & 0 & 0 \\
                  0 & \frac{\mu_2}{\mu_1} & 0 \\
                  0 & 0 & \frac{1}{\mu_2} \\
                \end{array}
              \right) \, \in \, \mathcal{G}_{\mathbb{Z}_3} \quad ; \quad \mu_{1,2} \, \in \, \mathbb{C}
\end{equation}
on the coordinates $\omega_i$. Such an action is easily worked out to be the following one:
\begin{equation}\label{saccofagioli}
    \left\{\omega _1\to \omega _1,\omega _2\to \frac{\mu _2 \omega
   _2}{\mu _1^2},\omega _3\to \frac{\omega _3}{\mu _1 \mu _2},\omega
   _4\to \frac{\mu _1^2 \omega _4}{\mu _2},\omega _5\to \frac{\mu
   _1 \omega _5}{\mu _2^2}\right\}
\end{equation}
Relying on this we can introduce three complex coordinates $u_i$ ($i=1,2,3$) parameterizing the locus
$L_{\mathbb{Z}_3}$ and we identify the remaining two complex coordinates as parameters of the quiver group.
With some ingenuity we have  singled out the following transformation:
\begin{equation}\label{secchiobucato}
   \omega _1\to u_1\, , \quad\omega _2\to \frac{\mu _2 u_2^2}{\mu _1^2
   u_3}\, , \quad\omega _3\to \frac{u_3^2}{\mu _1 \mu _2 u_2}\, , \quad\omega _4\to
   \frac{\mu _1^2 u_3^2}{\mu _2 u_2}\, , \quad\omega _5\to \frac{\mu _1
   u_3^2}{\mu _2^2 u_2}
\end{equation}
If we separate the modulus from the phase of the complex  parameters $\mu_{1,2}$ by writing:
\begin{equation}\label{separatis}
    \mu_{1,2} \, = \, \Omega_{1,2} \, e^{i\theta_{1,2}} \, \quad \Omega_{1,2} \in \mathbb{R}_+
\end{equation}
the substitution (\ref{secchiobucato}) can be rewritten as follows:
\begin{equation}\label{thereisahole}
    \omega _1\to u_1\, , \quad\omega _2\to \frac{e^{2 \theta _2-2
   \theta _1} u_2^2 \Omega _2}{u_3 \Omega _1^2}\, , \quad\omega _3\to
   \frac{e^{-\theta _1-2 \theta _2} u_3^2}{u_2 \Omega _1 \Omega
   _2}\, , \quad\omega _4\to \frac{e^{2 \theta _1-2 \theta _2} u_3^2 \Omega
   _1^2}{u_2 \Omega _2}\, , \quad\omega _5\to \frac{e^{\theta _1-4 \theta
   _2} u_3^2 \Omega _1}{u_2 \Omega _2^2}
\end{equation}
which can be implemented into eqs.\,(\ref{micromiceto}) that define the moment maps. One easily verifies that
the phases $\theta_i$ disappear in the resulting expressions as a result of the invariance of the moment maps
with respect to the gauge-group. Neither do the moment maps depend from the phases of the complex coordinates
$u_i$. Setting
\begin{equation}\label{deltosi}
     u_{1,2,3} \, = \, \sqrt{\Delta_{1,2,3}} \, e^{i\phi_{1,2,3}} \, \quad \Delta_{1,2,3} \in \mathbb{R}_+
\end{equation}
and using the following convenient new basis for the gauge Lie algebra and for the moment maps:
\begin{equation}\label{novalabasa}
    \hat{\mathfrak{P}}_1 \, = \,\ft 13 \left(2\,\mathfrak{P}_1 + \mathfrak{P}_2 \right) \quad ;
    \quad \hat{\mathfrak{P}}_2 \, = \,\ft 13 \left(\mathfrak{P}_1 + 2\,\mathfrak{P}_2 \right)
\end{equation}
we obtain the following quite simple result:
\begin{equation}\label{chiavoso}
\left\{\hat{\mathfrak{P}}_1\, , \, \hat{\mathfrak{P}}_2\right\} \, = \,    \left\{-\frac{3 \left(\Omega
_1^6-1\right) \left(\Delta _2^3
   \Omega _2^4+\Delta _3^3 \Omega _1^2\right)}{\Delta _2 \Delta
   _3 \Omega _1^4 \Omega _2^2},\frac{3 \left(\Omega _1^4+\Omega
   _2^2\right) \left(\Delta _3^3-\Delta _2^3 \Omega
   _2^6\right)}{\Delta _2 \Delta _3 \Omega _1^2 \Omega
   _2^4}\right\}
\end{equation}
In order to study the moment map equations:
\begin{equation}\label{mommaqua}
   \left\{\hat{\mathfrak{P}}_1\, , \, \hat{\mathfrak{P}}_2\right\} \, = \, \left\{\zeta_1 \, , \,
   \zeta_2\right\}
\end{equation}
where $\zeta_{1,2}$ are the level parameters it is convenient to make the following change of
parameterization:
\begin{equation}\label{chucco}
    \Omega _1\to \sqrt[3]{\Upsilon _1} \sqrt[6]{\Upsilon_2}\quad ,\quad \Omega _2\to \sqrt[6]{\Upsilon _1}
    \sqrt[3]{\Upsilon_2}
\end{equation}
The remaining task is that of solving eq.\,(\ref{mommaqua}) as an algebraic equation for the parameters
$\Upsilon_{1,2}$ to be determined in terms of the coordinates $u_i$ and of the level parameters
$\zeta_{1,2}$.
\paragraph{Calibration of the solutions at $\zeta_{1,2}\, = \, 0$} Upon the substitution
(\ref{chucco}), eq.\,(\ref{mommaqua}) can be easily solved for $\zeta_{1,2}=0$ and we find the following six
solutions
\begin{equation}\label{seisoluzie}
    \begin{array}{|c|l|l|}
    \hline
1) & \Upsilon _1\to -1 & \Upsilon _2\to 1 \\
\hline 2) & \Upsilon _1\to -1 & \Upsilon _2\to -\frac{\Delta _3^3}{\Delta
   _2^3}
   \\
   \hline
3)& \Upsilon _1\to \frac{\Delta _2}{\Delta _3} &
   \Upsilon _2\to
   \frac{\Delta _3^2}{\Delta _2^2}\\
   \hline
4)& \Upsilon _1\to -\frac{\sqrt[3]{-1} \Delta _2}{\Delta
   _3} & \Upsilon _2\to -\frac{\sqrt[3]{-1} \Delta _3^2}{\Delta
   _2^2}\\
   \hline
5)& \Upsilon _1\to \frac{(-1)^{2/3} \Delta _2}{\Delta
   _3} & \Upsilon _2\to \frac{(-1)^{2/3} \Delta _3^2}{\Delta
   _2^2} \\
   \hline
6)& \Upsilon _1\to \frac{\Delta _2^3}{\Delta _3^3} & \Upsilon
   _2\to -\frac{\Delta _3^3}{\Delta _2^3} \\
   \hline
\end{array}
\end{equation}
Inspecting the above result we immediately see that   only  the solution 3) is acceptable since it is the only one
for which both $\Upsilon_1$ and $\Upsilon_2$ are real and positive. In terms of the original parameters
$\Omega_{1,2}$  solution 3) means:
\begin{equation}\label{cicirotta}
    \Omega_1 \, \to \, 1 \quad ; \quad \Omega_2 \, \to \, \frac{|u_3|}{|u_2|}
\end{equation}
and reinstalling the phases we can argue that it corresponds to the following complex transformation of the
quiver group:
\begin{equation}\label{canutolo}
    \mu_1  \, \to \, 1 \quad ; \quad \mu _2\to \frac{u_3}{u_2}
\end{equation}
Inserting eq.\,(\ref{canutolo}) in eq.\,(\ref{secchiobucato}) and replacing the $\omega_i$ accordingly in
eq.\,(\ref{garrumovsky}) we obtain a set of three matrices that we name $\hat{A}_0,\hat{B}_0,\hat{C}_0$.
Transforming these latter back to the natural basis of the regular representation by setting ${A}_0 \, \equiv
\, \pmb{\chi}\,\hat{A}_0\,\pmb{\chi}^\dagger $ and similarly for $B_0$ and $C_0$ we get: { \small
\begin{eqnarray}
{A}_0  &=& \left(
\begin{array}{ccc}
 u_1+u_2+u_3 & 0 & 0 \\
 0 & \frac{1}{2} \left(2 u_1+\left(-1-i \sqrt{3}\right) u_2+i
   \left(\sqrt{3}+i\right) u_3\right) & 0 \\
 0 & 0 & \frac{1}{2} \left(2 u_1+i \left(\sqrt{3}+i\right)
   u_2+\left(-1-i \sqrt{3}\right) u_3\right) \\
\end{array}
\right)  \nonumber\\
  {B}_0  &=& \left(
\begin{array}{ccc}
 u_1-\sqrt[3]{-1} u_2+(-1)^{2/3} u_3 & 0 & 0 \\
 0 & \frac{1}{2} \left(2 u_1+i \left(\sqrt{3}+i\right)
   u_2+\left(-1-i \sqrt{3}\right) u_3\right) & 0 \\
 0 & 0 & u_1+u_2+u_3 \\
\end{array}
\right) \nonumber\\
  {C}_0  &=& \left(
\begin{array}{ccc}
 u_1+(-1)^{2/3} u_2-\sqrt[3]{-1} u_3 & 0 & 0 \\
 0 & u_1+u_2+u_3 & 0 \\
 0 & 0 & \frac{1}{2} \left(2 u_1+\left(-1-i \sqrt{3}\right) u_2+i
   \left(\sqrt{3}+i\right) u_3\right) \\
\end{array}
\right)  \label{broccolus}
\end{eqnarray}}
Being diagonal and belonging to $\mbox{Hom}_{\mathbb{Z}_3}\left(R,\mathcal{Q}\otimes R\right)$, the triple
of matrices $\left\{A_0,B_0,C_0\right\}$  lies  in the locus $L_{\mathbb{Z}_3}$ which we have already
established to be isomorphic with the singular orbifold $\mathbb{C}^3/\mathbb{Z}_3$. The coordinates
$u_{1,2,3}$ just provide a linear parameterization of $\mathbb{C}^3/\mathbb{Z}_3$. This shows that at
zero-level of the momentum map we are back to the singular orbifold. The resolution of singularities occur at
the non-vanishing levels.
\subsubsection{Solutions of the moment map equations at $\zeta \neq 0$}
When $\zeta_{1,2} \neq 0$ we have been able to find the solutions of the moment map equation (\ref{mommaqua})
in an implicit way in terms of the roots of a sixth order equation whose coefficients  depend on the level
parameters $\zeta_{1,2}$ and on the moduli-square $\Delta_{1,2,3}$ of the complex coordinates $u_{1,2,3}$.
The resolving sixth-order equation has the form:
\begin{equation}
\sum _{n=0}^6 c_n \alpha ^n=0 \label{tarvisio}
\end{equation}
where the coefficients have the  form:
\begin{eqnarray}
c_1 & = &9 \Delta _3^6\nonumber\\
\nonumber\\
c_2 & = &3 \Delta _3^3 \left(-3 \Delta _2^3+6 \Delta _3^3+\Delta _2 \Delta _3 \left(2 \zeta _1-\zeta _2\right)\right)\nonumber\\
\nonumber\\
c_3 & = &\Delta _3^2 \left(-18 \Delta _2^3 \Delta _3+9 \Delta _3^4+\Delta _2^2 \zeta _1 \left(\zeta _1-\zeta
_2\right)+6 \Delta _2
\Delta _3^2 \left(2 \zeta _1-\zeta _2\right)\right)\nonumber\\
\nonumber\\
c_4 & = &\Delta _2 \Delta _3 \left(-18 \Delta _2^2 \Delta _3^2+3 \Delta _3^3 \left(2 \zeta _1-\zeta
_2\right)+\Delta _2^3 \left(-6
\zeta _1+3 \zeta _2\right)+\Delta _2 \Delta _3 \left(2 \zeta _1^2-2 \zeta _1 \zeta _2+\zeta _2^2\right)\right)\nonumber\\
\nonumber\\
c_5 & = &\Delta _2^2 \left(9 \Delta _2^4-18 \Delta _2 \Delta _3^3+\Delta _3^2 \zeta _1 \left(\zeta _1-\zeta
_2\right)+6 \Delta _2^2
\Delta _3 \left(-2 \zeta _1+\zeta _2\right)\right)\nonumber\\
\nonumber\\
c_6 & = &3 \Delta _2^3 \left(6 \Delta _2^3-3 \Delta _3^3+\Delta _2 \Delta _3 \left(-2 \zeta _1+\zeta _2\right)\right)\nonumber\\
\nonumber\\
c_7 & = &9 \Delta _2^6
\end{eqnarray}
Let us name $\alpha_i$ the 6 roots of equation (\ref{tarvisio}). For each $\alpha = \alpha_i$, the solution
of the momentum-map equations is given by:
\par
\begin{doublespace}
\noindent\({ \Upsilon_1 =}\alpha\)
\end{doublespace}
\begin{doublespace}
\noindent\({ \Upsilon_2=}\\
{\left(-81 (1+\alpha ) \Delta _2^{11}+81 \left(1+2 \alpha +\alpha ^2+\alpha ^3+\alpha ^4\right) \Delta _2^8
\Delta _3^3-54 \alpha ^3 (1+\alpha
)^2 \Delta _3^{10} \zeta _2-9 \alpha ^2 (1+\alpha )^2 (2+\alpha ) \Delta _2 \Delta _3^8 \left(2 \zeta _1-\zeta _2\right) \zeta _2+\right.}\\
{27 \Delta _2^9 \Delta _3 \left(\left(1+3 \alpha +2 \alpha ^2\right) \zeta _1-\left(-2+\alpha ^2\right) \zeta
_2\right)-27 \Delta _2^6 \Delta _3^4 \left(3 \alpha ^2 \zeta _1+3 \alpha ^4 \zeta _1+4 \zeta _2+\alpha
\left(3 \zeta _1+\zeta _2\right)+\alpha ^3 \left(3 \zeta _1+\zeta
_2\right)\right)-}\\
{9 \Delta _2^7 \Delta _3^2 \left(3 \alpha ^2 \zeta _1^2+\alpha ^3 \zeta _1 \left(\zeta _1-\zeta
_2\right)+\left(4 \zeta _1-\zeta _2\right)
\zeta _2+\alpha  \left(2 \zeta _1^2+5 \zeta _1 \zeta _2-4 \zeta _2^2\right)\right)+}\\
{9 \Delta _2^4 \Delta _3^5 \left(\alpha  \left(8 \zeta _1-5 \zeta _2\right) \zeta _2+\left(4 \zeta _1-\zeta
_2\right) \zeta _2+\alpha ^3
\left(3 \zeta _1^2+6 \zeta _1 \zeta _2-4 \zeta _2^2\right)+\alpha ^2 \left(\zeta _1^2+7 \zeta _1 \zeta _2-3 \zeta _2^2\right)
+\right.}\\
{\left.\alpha ^4 \left(2 \zeta _1^2+3 \zeta _1 \zeta _2-2 \zeta _2^2\right)\right)+}\\
{3 \alpha  \Delta _2^2 \Delta _3^6 \left(27 \alpha ^4 \Delta _3^3+2 \zeta _1 \zeta _2 \left(-\zeta _1+\zeta
_2\right)+\alpha  \zeta _2 \left(-8
\zeta _1^2+8 \zeta _1 \zeta _2-3 \zeta _2^2\right)-2 \alpha ^2 \zeta _2 \left(5 \zeta _1^2-5 \zeta _1 \zeta _2+\zeta _2^2\right)+
\right.}\\
{\left.\alpha ^3 \left(27 \Delta _3^3-\zeta _2 \left(-2 \zeta _1+\zeta _2\right){}^2\right)\right)-}\\
{3 \Delta _2^5 \Delta _3^3 \left(54 \alpha ^4 \Delta _3^3+27 \alpha ^5 \Delta _3^3+\zeta _1 \zeta _2 \left(-2
\zeta _1+\zeta _2\right)+\alpha ^3 \left(27 \Delta _3^3-\zeta _1^3+\zeta _1 \zeta _2^2\right)+\alpha ^2
\left(27 \Delta _3^3-\zeta _1^3-5 \zeta _1^2 \zeta _2+5 \zeta _1 \zeta
_2^2-2 \zeta _2^3\right)+\right.}\\
{\left.\alpha  \left(27 \Delta _3^3+\zeta _2 \left(-8 \zeta _1^2+9 \zeta _1 \zeta _2-2 \zeta _2^2\right)\right)\right)+}\\
{\Delta _2^3 \Delta _3^4 \left(54 \Delta _3^3 \zeta _2+81 \alpha ^4 \Delta _3^3 \left(\zeta _1+\zeta
_2\right)+27 \alpha ^5 \Delta _3^3 \left(\zeta
_1+\zeta _2\right)+\alpha ^2 \zeta _2 \left(54 \Delta _3^3-4 \zeta _1^3+6 \zeta _1^2 \zeta _2-4 \zeta _1 \zeta _2^2+
\zeta _2^3\right)+\right.}\\
{\left.\left.\left.\alpha  \zeta _2 \left(54 \Delta _3^3-\zeta _1 \left(2 \zeta _1^2-3 \zeta _1 \zeta
_2+\zeta _2^2\right)\right)+\alpha
^3 \left(27 \Delta _3^3 \left(2 \zeta _1+3 \zeta _2\right)-\zeta _1 \zeta _2 \left(2 \zeta _1^2-3 \zeta _1 \zeta _2+
\zeta _2^2\right)\right)\right)\right)\right/}\\
{\left(9 \Delta _2^6 \Delta _3 \zeta _2 \left(6 \Delta _2^3-6 \Delta _3^3+\Delta _2 \Delta _3 \left(-2 \zeta
_1+\zeta _2\right)\right)\right);}\)
\end{doublespace}
Instructed by the case of zeroth level we try to inspect the solution which at 0-th level reduces to
\begin{equation}
\left\{\Omega _1\to 1\, ,\,\Omega _2\to \sqrt{\frac{\Delta _3}{\Delta _2}}\right\}\text{
 }\Longleftrightarrow \left\{\Upsilon _1\to \frac{\Delta _2}{\Delta _3}\, ,\,\Upsilon _2\to \frac{\Delta _3^2}{\Delta _2^2}\right\}
 \label{krasaviza}
\end{equation}
To this effect  we consider a power series expansion of the solution for small moment maps, around the 0th
level solution:
\begin{equation}\label{espunzio}
    \Upsilon _1\to \frac{\Delta _2}{\Delta _3}+Y_1 \epsilon
   , \, \Upsilon _2\to \frac{\Delta _3^2}{\Delta _2^2}+Y_2 \epsilon
   , \, \zeta _1\to k_1 \epsilon , \, \zeta _2\to k_2 \epsilon
\end{equation}
where $\epsilon$ is an infinitesimal parameter. Inserting eq.\,(\ref{espunzio}) into the moment map equation we
obtain the approximate solution:
\begin{equation}\label{caramellinabis}
  \Omega _1\to 1-\frac{k_1 \epsilon }{18 \left(\Delta
   _2+\Delta _3\right)}\,,\quad\Omega _2\to \frac{18  \Delta _3  \left(\Delta
   _2+\Delta _3\right)-\Delta _3 k_2 \epsilon }{18 \sqrt{\Delta
   _2 \Delta _3} \left(\Delta _2+\Delta _3\right)}
\end{equation}
which in terms of the complex quiver group transformations can also be interpreted as follows:
\begin{equation}\label{corciofino}
    \mu _1\to 1-\epsilon \, \frac{k_1  }{18 \left(\Delta _2+\Delta
   _3\right)}\,,\quad\mu _2\to \frac{u_3}{u_2}\left(1 -\epsilon \,\frac{k_2
   }{18 \left(\Delta _2+\Delta _3\right)}\right)
\end{equation}
In some way, to be better clarified, the above deformation should describe the inflation of the two homology
cycles predicted by the general theorem.
\subsubsection{The formal solution for the K\"ahler potential} In
any case, assuming  the scalar factors $\Omega_{1,2} \, = \, \Omega_{1,2}\left(\pmb{u},\bar
{\pmb{u}},\pmb{\zeta}\right)$ known in terms of the coordinates and of the moment map parameters, we can
calculate the final K\"ahler potential. Substituting (\ref{secchiobucato}) into eq.\,(\ref{kallerV}) we obtain
the restriction to the level surface $\mathcal{N}$ of the original K\"ahler potential:
\begin{equation}\label{yyy}
    \mathcal{K}|_{\mathcal{N}}\, = \, \frac{3 u_3^2 \Omega _1^4 \bar{u}_3^2}{u_2 \Omega _2^2 \bar{u}_2}
    +\frac{3 u_3^2 \Omega _1^2 \bar{u}_3^2}{u_2 \Omega _2^4 \bar{u}_2}+\frac{3 u_2^2 \Omega _2^2
   \Omega _1^2 \bar{u}_2^2}{u_3 \bar{u}_3}+\frac{3 u_3^2 \bar{u}_3^2}{u_2 \Omega _2^2 \Omega _1^2 \bar{u}_2}
   +\frac{3 u_2^2 \Omega _2^4 \bar{u}_2^2}{u_3
   \Omega _1^2 \bar{u}_3}+\frac{3 u_2^2 \Omega _2^2 \bar{u}_2^2}{u_3 \Omega _1^4 \bar{u}_3}+9 u_1 \bar{u}_1
\end{equation}
Then the final K\"ahler potential of the resolved smooth manifold is:
\begin{equation}\label{finazzo}
    \mathcal{K}_\mathcal{M} \,= \, \mathcal{K}|_{\mathcal{N}}\, + \, \zeta^1\,\log\Omega_1 +\zeta^2 \,
    \log\Omega_2
\end{equation}
Note that when $\zeta_{1,2}\,=\,0$ we have (see eq.\,(\ref{krasaviza})):
\begin{equation}\label{caruscolo}
    \Omega _1 \, = 1 \quad ,\quad \Omega _2 \, = \,  \sqrt{\frac{u_3 \bar{u}_3}{u_2
   \bar{u}_2}}
\end{equation}
which inserted into eq.\,(\ref{yyy}) yields:
\begin{equation}\label{uuu}
\lim_{\pmb{\zeta}\to\pmb{0}}\, \mathcal{K}|_{\mathcal{M}} \, = \,   \lim_{\pmb{\zeta}\to\pmb{0}}\,
\mathcal{K}|_{\mathcal{N}} \, = \, 9 u_1 \bar{u}_1+9 u_2 \bar{u}_2+9 u_3 \bar{u}_3
\end{equation}
namely we obtain the K\"ahler potential of the flat $\mathbb{C}^3/\mathbb{Z}_3$ orbifold of which $u_{1,2,3}$
are the $\mathbb{Z}_3$ invariant coordinates.
\subsubsection{Comparison with GH metrics} Although for the case under study, the explicit form of the
K\"ahler potential is not available  in terms of radicals, since it involves the roots of a sextic equation,
yet the metric can be easily written in terms of real coordinates by utilizing the Gibbons-Hawking form of
ALE metrics. Let us recall eq.\,(\ref{cedrata}) which predicts the number of parameters in the  hyperK\"ahler
quotient resolution of a $\mathbb{C}^2/\mathbb{Z}_p$ singularity. For $p=3$ this number is $4$, yet in our
case the number of parameters is less  since we take only the K\"ahler quotient and we keep fixed the
analogue of the holomorphic moment map equations, namely the constraint $[A,B]=[B,C]=[C,A]$. Hence although,
as we have observed, the resolution of the singularity $\mathbb{C}^3/\mathbb{Z}_3$ reduces to a direct
product $\mathbb{C}\times\left(Y\,\rightarrow\,\mathbb{C}^2/\mathbb{Z}_3\right)$ yielding
$\mathbb{C}\times\ ALE_{\mathbb{Z}_3}$ (indeed, once diagonalized the action of $\mathbb{Z}_3$ is effective
only on two complex coordinates), the $ALE_{\mathbb{Z}_3}$ that we obtain in this process is a particular one
depending only on 2 of the 4 available parameters.  From the holomorphic point of view the missing parameters
are clearly localized. They are those of a holomorphic moment map level which is not switched on. The other
can always be suppressed by a coordinated change.  This is the reason why in this type of resolutions, the
algebraic equation is not touched (it is preserved identical to the case of the orbifold, by utilizing only
the locus $\mathcal{D}^0_{\Gamma}$ which is the orbit of the locus $L^{\Gamma}$ under the action of the
quiver group $\mathcal{G}_\Gamma$). In the case of Eguchi-Hanson, as we have seen, there is no loss of
generality, but for all the other cases we have a reduced number of moduli with respect to the general ALE.
This implies that also the GH-metric which is equivalent to these specialized ALE manifolds should  be in
some sense a special one and should depend only on (p-1) parameters. We conjecture that the GH multicenter
spaces equivalent to the special ALE of the above discussion are those where the centers are all aligned on
the same line, say along the z-axis. Let us follow this idea.
\subsubsection{The harmonic potential  and the connection one-form}
As for the harmonic potential $\mathcal{V}_{\mathbb{Z}_3}$ we adopt the following one:
\begin{equation}\label{camones}
    \mathcal{V}_{\mathbb{Z}_3} \, = \, \frac{1}{\sqrt{x^2+y^2+\left(z-\delta
   _2\right){}^2}}+\frac{1}{\sqrt{x^2+y^2+\left(z-\delta
   _3\right){}^2}}+\frac{1}{\sqrt{x^2+y^2+\left(z-\delta
   _1\right){}^2}}
\end{equation}
implying that the three centers of the metric are at $\mathbf{c}_i\,=\, \left\{0,0,\delta_i\right\}$. The
corresponding connection one-form is as follows:
\begin{equation}\label{crucion}
   \omega_{\mathbb{Z}_3} \, = \, \frac{\left(z-\delta _1\right) (x \mathrm{d}y-y
   \mathrm{d}x)}{\left(x^2+y^2\right) \sqrt{x^2+y^2+\left(z-\delta
   _1\right){}^2}}+\frac{\left(z-\delta _2\right) (x \mathrm{d}y-y
   \mathrm{d}x)}{\left(x^2+y^2\right) \sqrt{x^2+y^2+\left(z-\delta
   _2\right){}^2}}+\frac{\left(z-\delta _3\right) (x \mathrm{d}y-y
   \mathrm{d}x)}{\left(x^2+y^2\right) \sqrt{x^2+y^2+\left(z-\delta
   _3\right){}^2}}
\end{equation}
This information suffices to write the explicit form of the metric as given in eq.\,(\ref{GWmetricca}) and of
the the K\"ahler 2-form as given in eq.\,(\ref{kallerformGW}). Inspired by the example of the Eguchi-Hanson
case, we can now conjecture the location of the homology two-cycles within the GH-manifold. We embed two
$\mathbb{S}^2$ parameterized by the angles $\theta,\phi$ by setting:
\begin{eqnarray}\label{ciccorro}
D_E^1 & = & \left\{ x=y=0 \quad , \quad z=\ft 12 \,\left(\delta_1+\delta_2\right) + \ft 12
\,\left(\delta_1-\delta_2\right) \, \cos\theta \quad , \quad
\tau \, = \, \phi\right\}\nonumber\\
D_E^2 & = & \left\{ x=y=0 \quad , \quad z=\ft 12 \,\left(\delta_2+\delta_3\right) + \ft 12
\,\left(\delta_2-\delta_3\right) \, \cos\theta \quad , \quad
\tau \, = \, \phi\right\}\nonumber\\
\end{eqnarray}
\subsection{Construction \`a la Kronheimer of the crepant resolution $\mathcal{M}_{\mathbb{Z}_7}\,
\rightarrow \,\frac{\mathbb{C}^3}{\mathbb{Z}_7}$} 
Next we address  the case of the singularity $\frac{\mathbb{C}^3}{\mathbb{Z}_7}$, provided by the embedding
$\mathbb{Z}_7 \hookrightarrow \mathrm{SU(3)}$ encoded in the form of the generator $\mathrm{Y}$ of
eq.\,(\ref{pesciYX}). It is still an abelian case as the previous one, yet the resolution
$\mathcal{M}_{\mathbb{Z}_7} \rightarrow \frac{\mathbb{C}^3}{\mathbb{Z}_3}$ no longer factorizes:
\begin{equation}\label{nofattorio}
    \mathcal{M}_{\mathbb{Z}_7} \, \neq \, \mathbb{C}\times ALE_{\mathbb{Z}_7}
\end{equation}
although $ALE_{\mathbb{Z}_7}$ does independently exist. The reason is that in the quotient
$\frac{\mathbb{C}^3}{\mathbb{Z}_7}$ there is no linear subspace of $\mathbb{C}^3$ which is left invariant.
Hence this is a truly new case which displays intrinsically three-dimensional features. The most dramatic and
relevant of them is the prediction from theorem \ref{reidmarktheo} and from eq.\,(\ref{z7vecchione}) of the
existence of 3-harmonic (2,2)-forms, along side with three harmonic (1,1)-forms:
\begin{equation}\label{cannibalus}
    h^{1,1}\left(\mathcal{M}_{\mathbb{Z}_7}\right) \, = \, 3 \quad ;
    \quad  h^{2,2}\left(\mathcal{M}_{\mathbb{Z}_7}\right)\, = \, 3
\end{equation}
Notwithstanding these novelties, the construction of the resolution \`a la Kronheimer, which is the one
directly mirrored in the structure of the $D=3$ Chern-Simons gauge theory supposedly dual to one M2-brane
that probes the corresponding singularity, goes along the same lines as before.
\par
Following the general strategy, the first step in the construction consists of deriving the invariant space
$\mathcal{S}_\Gamma\, = \, \text{Hom}_{\Gamma }(R, \mathcal{Q}\otimes R)$ made of those triples $\{A,B,C\}$
of $7\times 7$ matrices that satisfy eq.\,(\ref{gammazione}), namely:
\begin{equation}\label{gerticco7}
   \mathrm{Y}\, \left(\begin{array}{c}
               A \\
               B \\
               C
             \end{array}
   \right) \, = \, \left(
\begin{array}{c}
R(\mathrm{Y}^{-1})\, A \, R(\mathrm{Y}) \\
R(\mathrm{Y}^{-1})\, B \, R(\mathrm{Y}) \\
R(\mathrm{Y}^{-1})\, C \, R(\mathrm{Y}) \\
\end{array}
\right)
\end{equation}
where $\mathrm{Y}$ is the $\mathbb{Z}_7$-generator displayed in eq\,(\ref{pesciYX}) and $R(\mathrm{Y})$ is its
representation in the natural basis of the regular representation:
\begin{equation}\label{cagnottoY}
    R(\mathrm{Y})\, = \, \left(
\begin{array}{ccccccc}
 0 & 0 & 0 & 0 & 0 & 0 & 1 \\
 1 & 0 & 0 & 0 & 0 & 0 & 0 \\
 0 & 1 & 0 & 0 & 0 & 0 & 0 \\
 0 & 0 & 1 & 0 & 0 & 0 & 0 \\
 0 & 0 & 0 & 1 & 0 & 0 & 0 \\
 0 & 0 & 0 & 0 & 1 & 0 & 0 \\
 0 & 0 & 0 & 0 & 0 & 1 & 0 \\
\end{array}
\right)
\end{equation}
The constraint (\ref{gerticco7}) reduces the number of parameters from 147 to 21. If we perform the
diagonalization of the regular representation $R_{diag}(\mathrm{Y})\equiv \pmb{\chi}^\dagger \,
R(\mathrm{Y})\,\pmb{\chi}$ by means of the  normalize character table:
\begin{equation}\label{lucardone}
    \pmb{\chi} \, = \, \frac{1}{\sqrt{7}}\left(
\begin{array}{ccccccc}
 1 & 1 & 1 & 1 & 1 & 1 & 1 \\
 1 & e^{\frac{2 i \pi }{7}} & e^{\frac{4 i \pi }{7}} & e^{\frac{6
   i \pi }{7}} & e^{-\frac{6 i \pi }{7}} & e^{-\frac{4 i \pi }{7}}
   & e^{-\frac{2 i \pi }{7}} \\
 1 & e^{\frac{4 i \pi }{7}} & e^{-\frac{6 i \pi }{7}} &
   e^{-\frac{2 i \pi }{7}} & e^{\frac{2 i \pi }{7}} & e^{\frac{6 i
   \pi }{7}} & e^{-\frac{4 i \pi }{7}} \\
 1 & e^{\frac{6 i \pi }{7}} & e^{-\frac{2 i \pi }{7}} & e^{\frac{4
   i \pi }{7}} & e^{-\frac{4 i \pi }{7}} & e^{\frac{2 i \pi }{7}}
   & e^{-\frac{6 i \pi }{7}} \\
 1 & e^{-\frac{6 i \pi }{7}} & e^{\frac{2 i \pi }{7}} &
   e^{-\frac{4 i \pi }{7}} & e^{\frac{4 i \pi }{7}} & e^{-\frac{2
   i \pi }{7}} & e^{\frac{6 i \pi }{7}} \\
 1 & e^{-\frac{4 i \pi }{7}} & e^{\frac{6 i \pi }{7}} & e^{\frac{2
   i \pi }{7}} & e^{-\frac{2 i \pi }{7}} & e^{-\frac{6 i \pi }{7}}
   & e^{\frac{4 i \pi }{7}} \\
 1 & e^{-\frac{2 i \pi }{7}} & e^{-\frac{4 i \pi }{7}} &
   e^{-\frac{6 i \pi }{7}} & e^{\frac{6 i \pi }{7}} & e^{\frac{4 i
   \pi }{7}} & e^{\frac{2 i \pi }{7}} \\
\end{array}
\right)
\end{equation}
the result of the  constraint (\ref{gerticco7}) with $R_{diag}(\mathrm{Y})$ in place of $R(\mathrm{Y})$ can
be expressed in the following way:
\begin{eqnarray}
  \left(A,B,C\right) &\in & \text{Hom}_{\Gamma }(R, \mathcal{Q}\otimes R) \nonumber\\
  \null &\Downarrow &\null \nonumber\\
  A &=& \left(
\begin{array}{ccccccc}
 0 & 0 & m_{1,3} & 0 & 0 & 0 & 0 \\
 0 & 0 & 0 & m_{2,4} & 0 & 0 & 0 \\
 0 & 0 & 0 & 0 & m_{3,5} & 0 & 0 \\
 0 & 0 & 0 & 0 & 0 & m_{4,6} & 0 \\
 0 & 0 & 0 & 0 & 0 & 0 & m_{5,7} \\
 m_{6,1} & 0 & 0 & 0 & 0 & 0 & 0 \\
 0 & m_{7,2} & 0 & 0 & 0 & 0 & 0 \\
\end{array}
\right) \label{stupidus7A}
\end{eqnarray}
\begin{eqnarray}
  B &=& \left(
\begin{array}{ccccccc}
 0 & 0 & 0 & 0 & n_{1,5} & 0 & 0 \\
 0 & 0 & 0 & 0 & 0 & n_{2,6} & 0 \\
 0 & 0 & 0 & 0 & 0 & 0 & n_{3,7} \\
 n_{4,1} & 0 & 0 & 0 & 0 & 0 & 0 \\
 0 & n_{5,2} & 0 & 0 & 0 & 0 & 0 \\
 0 & 0 & n_{6,3} & 0 & 0 & 0 & 0 \\
 0 & 0 & 0 & n_{7,4} & 0 & 0 & 0 \\
\end{array}
\right)\label{stupidus7B}
\end{eqnarray}
\begin{eqnarray}
  C &=& \left(
\begin{array}{ccccccc}
 0 & r_{1,2} & 0 & 0 & 0 & 0 & 0 \\
 0 & 0 & r_{2,3} & 0 & 0 & 0 & 0 \\
 0 & 0 & 0 & r_{3,4} & 0 & 0 & 0 \\
 0 & 0 & 0 & 0 & r_{4,5} & 0 & 0 \\
 0 & 0 & 0 & 0 & 0 & r_{5,6} & 0 \\
 0 & 0 & 0 & 0 & 0 & 0 & r_{6,7} \\
 r_{7,1} & 0 & 0 & 0 & 0 & 0 & 0 \\
\end{array}
\right)\label{stupidus7}
\end{eqnarray}
in terms of a new set of 21 parameters evident from the above formulae.
\subsubsection{Derivation of the
locus $\mathcal{D}_{\mathbb{Z}_7}$ and of the variety $\mathbb{V}_{7+2}$} Next we define the reduction of the
invariant space $\text{Hom}_{\Gamma }(R, \mathcal{Q}\otimes R)$ to the locus cut out by the holomorphic
constraint (\ref{poffarbacchio}) which was named $\mathcal{D}_\Gamma$:
\begin{equation}
\text{ }\mathcal{D}_{{\mathbb{Z}_7} } \equiv  \left\{\text{  }p=\left(
\begin{array}{c}
 A \\
 B \\
 C \\
\end{array}
\right)\in  \text{Hom}_{\mathbb{Z}_7 }(R, \mathcal{Q}\otimes R)\text{  }/\text{    }[A,B]=[B,C]=[C,A] =0
\right \} \label{pidocchio7}
\end{equation}
Differently from the more complicated cases of maximal subgroups of \(L_{168}\) and analogously to the
previous $\mathbb{Z}_3$ case, here we can explicitly solve the quadratic equations provided by the commutator
constraints and we discover that there is a unique principal branch of the solution, named
\(\mathcal{D}_{\Gamma }^0\), that has the maximal dimension $9=|\Gamma |+2$. However, in addition to that
there are 785 more  branches of the solution with smaller dimension. Also here, \(\mathcal{D}_{\Gamma }^0\)
is the orbit with respect to the group \(\mathcal{G}_{\Gamma }\) of the subspace \(L_{\Gamma}\) defined in
eq.\,(\ref{thespacel}). Hence the variety \(\mathbb{V}_{|\Gamma |+2}\) of which we are supposed to take the
K\"ahler quotient is not defined by eq.\,(\ref{pidocchio7}), rather by the principal branch of that variety.
\subsubsection{Derivation of the quiver group $\mathcal{G}_{\mathbb{Z}_7}$} In view of what we stated above
our next point is precisely the derivation of the group \(\mathcal{G}_{\Gamma }\) defined in eq.\,(\ref{caliente}), namely:
\begin{equation}
\text{                                           }\mathcal{G}_{\mathbb{Z}_7 } = \left\{ g\in
\text{SL}(7,\mathbb{C})\quad \left| \quad\forall \gamma \in \mathbb{Z}_3  : \;\;\left[D_R(\gamma
),D_{\text{def}}(g)\right]\right.=0\right\}
\end{equation}
Let us proceed to this construction. In the diagonal basis of the regular representation this is a very easy
task, since the group is simply given by the diagonal 7$\times $7 matrices with determinant one. A convenient
parametrization of such a group is the following one:
\begin{equation}\label{racconigi7}
    g \, = \, \left(
\begin{array}{ccccccc}
 \mu _1 & 0 & 0 & 0 & 0 & 0 & 0 \\
 0 & \frac{\mu _2}{\mu _1} & 0 & 0 & 0 & 0 & 0 \\
 0 & 0 & \frac{\mu _3}{\mu _2} & 0 & 0 & 0 & 0 \\
 0 & 0 & 0 & \frac{\mu _4}{\mu _3} & 0 & 0 & 0 \\
 0 & 0 & 0 & 0 & \frac{\mu _5}{\mu _4} & 0 & 0 \\
 0 & 0 & 0 & 0 & 0 & \frac{\mu _6}{\mu _5} & 0 \\
 0 & 0 & 0 & 0 & 0 & 0 & \frac{1}{\mu _6} \\
\end{array}
\right) \, \in \, \mathcal{G}_{\mathbb{Z}_3} \quad ; \quad \mu_{1,\dots,6} \, \in \, \mathbb{C}
\end{equation}
\par
The explicit form (\ref{racconigi7}) allows to construct the
$\text{Orbit}_{\mathcal{G}_{\mathbb{Z}_7}}\left(L_{\mathbb{Z}_7}\right)$ of the locus
\begin{eqnarray}
L_{\mathbb{Z}_7} & \equiv & \left\{ A_0,B_0,C_0\right\}
\end{eqnarray}
made by those triples of matrices belonging to $\mbox{Hom}_{\mathbb{Z}_7}\left(\mathcal{Q}\otimes R,R\right)$
that are diagonal in the natural basis of the regular representation. Considering
eqs.\,(\ref{stupidus7A}-\ref{stupidus7}) such a locus, which has complex dimension $3$, is obtained, by
setting:
\begin{eqnarray}
m_{1,3}\,=\,m_{2,4} \,=\,m_{3,5}\, = \, m_{4,6} \, = \, m_{5,7} \, = \,m_{6,1}\, = \,m_{7,2} &=& u_1 \nonumber\\
n_{1,5}\,=\,n_{2,6} \,=\,n_{3,7}\, = \, n_{4,1} \, = \, n_{5,2} \, = \, n_{6,3}\, = \,n_{7,4}&=& u_2 \nonumber\\
r_{1,2}\,=\,r_{2,3} \,=\,r_{3,4}\, = \, r_{4,5} \, = \, r_{5,6} \, = \, r_{6,7}\, = \,r_{7,1}&=& u_3
\label{fantozzi}
\end{eqnarray}
where $u_{1,2,3}$ are three complex parameter that will play the role of coordinates of the resolved
manifold. The six complex parameters of the quiver group plus these three make the total of nine parameters
of $\mathbb{V}_9 \equiv \text{Orbit}_{\mathcal{G}_{\mathbb{Z}_7}}\left(L_{\mathbb{Z}_7}\right)$, which, as we
have explicitly verified is the same as the principal branch $\mathcal{D}^0_{\mathbb{Z}_7}$ of the quadratic
locus $[A,B]=[B,C]=[C,A]=0$.
\par
In conclusion the variety $\mathbb{V}_9$ of which we are supposed to perform the K\"ahler quotient is spanned
by the following triple of matrices, depending on the 9 complex parameters $u_{1,2,3},\mu_{1,\dots,6}$:
\begin{eqnarray}
  \mathbb{V}_9 &=& \left\{A,B,C\right\}\nonumber \\
  A &=& \left(
\begin{array}{ccccccc}
 0 & 0 & \frac{u_1 \mu _3}{\mu _1 \mu _2} & 0 & 0 & 0 & 0 \\
 0 & 0 & 0 & \frac{u_1 \mu _1 \mu _4}{\mu _2 \mu _3} & 0 & 0 & 0 \\
 0 & 0 & 0 & 0 & \frac{u_1 \mu _2 \mu _5}{\mu _3 \mu _4} & 0 & 0 \\
 0 & 0 & 0 & 0 & 0 & \frac{u_1 \mu _3 \mu _6}{\mu _4 \mu _5} & 0 \\
 0 & 0 & 0 & 0 & 0 & 0 & \frac{u_1 \mu _4}{\mu _5 \mu _6} \\
 \frac{u_1 \mu _1 \mu _5}{\mu _6} & 0 & 0 & 0 & 0 & 0 & 0 \\
 0 & \frac{u_1 \mu _2 \mu _6}{\mu _1} & 0 & 0 & 0 & 0 & 0 \\
\end{array}
\right) \label{pantagrueloA}
\end{eqnarray}
\begin{eqnarray}
  B &=& \left(
\begin{array}{ccccccc}
 0 & 0 & 0 & 0 & \frac{u_2 \mu _5}{\mu _1 \mu _4} & 0 & 0 \\
 0 & 0 & 0 & 0 & 0 & \frac{u_2 \mu _1 \mu _6}{\mu _2 \mu _5} & 0 \\
 0 & 0 & 0 & 0 & 0 & 0 & \frac{u_2 \mu _2}{\mu _3 \mu _6} \\
 \frac{u_2 \mu _1 \mu _3}{\mu _4} & 0 & 0 & 0 & 0 & 0 & 0 \\
 0 & \frac{u_2 \mu _2 \mu _4}{\mu _1 \mu _5} & 0 & 0 & 0 & 0 & 0 \\
 0 & 0 & \frac{u_2 \mu _3 \mu _5}{\mu _2 \mu _6} & 0 & 0 & 0 & 0 \\
 0 & 0 & 0 & \frac{u_2 \mu _4 \mu _6}{\mu _3} & 0 & 0 & 0 \\
\end{array}
\right) \label{pantagrueloB}
\end{eqnarray}
\begin{eqnarray}
  C &=& \left(
\begin{array}{ccccccc}
 0 & \frac{u_3 \mu _2}{\mu _1^2} & 0 & 0 & 0 & 0 & 0 \\
 0 & 0 & \frac{u_3 \mu _1 \mu _3}{\mu _2^2} & 0 & 0 & 0 & 0 \\
 0 & 0 & 0 & \frac{u_3 \mu _2 \mu _4}{\mu _3^2} & 0 & 0 & 0 \\
 0 & 0 & 0 & 0 & \frac{u_3 \mu _3 \mu _5}{\mu _4^2} & 0 & 0 \\
 0 & 0 & 0 & 0 & 0 & \frac{u_3 \mu _4 \mu _6}{\mu _5^2} & 0 \\
 0 & 0 & 0 & 0 & 0 & 0 & \frac{u_3 \mu _5}{\mu _6^2} \\
 u_3 \mu _1 \mu _6 & 0 & 0 & 0 & 0 & 0 & 0 \\
\end{array}
\right)\label{pantagrueloC}
\end{eqnarray}
\subsubsection{The algebraic equation of the orbifold locus} Let us now consider the action of the
$\mathbb{Z}_3$ group on $\mathbb{C}^3$ as defined  by the generator $\mathrm{X}$ in eq.\,(\ref{pesciYX}). If
$\{z_1,z_2,z_3\}$ are the coordinates of a point in $\mathbb{C}^3$, we
 see that there are four invariant polynomials:
 \begin{eqnarray}
   J_1 &=&  z_1^7\nonumber\\
   J_2 &=& z_2^7\nonumber\\
   J_3 &=&  z_3^7 \nonumber\\
   J_4 &=&  z_1\,z_2\,z_3\label{invariantuli7}
 \end{eqnarray}
which satisfy the following equation:
\begin{equation}\label{equatiaz7}
    J_1\,J_2\,J_3 \, - \, J_4^7 \, = \, 0
\end{equation}
This equation can be regarded as the cubic equation which cuts out in $\mathbb{C}^4$ the locus corresponding to
the singular orbifold $\mathbb{C}^3/{\mathbb{Z}_7}$.
\subsubsection{Map of the variety $\mathbb{V}_9$ into the algebraic locus corresponding to the orbifold}
Having established the above relations we verify that, in a completely analogous way to what happens in the
case of the standard Kronheimer construction, we can reproduce the defining equation (\ref{equatiaz7}) in
terms of invariants of the three matrices (\ref{pantagrueloA}-\ref{pantagrueloC}) spanning the $\mathbb{V}_9$
variety. It suffices to identify:
\begin{eqnarray}
  J_1 & = & \mbox{Det}[ A ]\, = \, u_1^7\nonumber\\
  J_2 &=& \mbox{Det}[B ]\, = \,u_2^7\nonumber\\
   J_3 &=& \mbox{Det}[ C ]\, = \, u_3^7\nonumber\\
  J_4 &=& \mbox{Det}[ A\,B\,C ]\, = \, u_1 \, u_2 \, u_3 \label{pagnuflo7}
\end{eqnarray}
Eq.\,(\ref{pagnuflo7}) describes an explicit  map of the variety $\mathbb{V}_9$ into the algebraic locus
$\mathbb{C}^3/\mathbb{Z}_7$:
\begin{equation}\label{imbecillus7}
    \mathbb{V}_9\, \rightarrow \, \frac{\mathbb{C}^3}{\mathbb{Z}_7}
\end{equation}
\subsubsection{The K\"ahler quotient} The next step consists of performing the K\"ahler quotient of the
K\"ahler manifold $\mathbb{V}_9$ with respect to the compact subgroup of the quiver group
$\mathcal{G}_{\mathbb{Z}_7}$, which, as we several times emphasized, is the \textit{gauge group} of the
corresponding three-dimensional Chern-Simons gauge theory:
\begin{equation}\label{gaugogz7}
    \mathcal{F}_{\mathbb{Z}_7} \, \equiv \, \mathcal{G}_{\mathbb{Z}_7}\bigcap \mathrm{SU(7)}
\end{equation}
A  generic element $g \in \mathcal{F}_{\mathbb{Z}_7} $, is  of the form (\ref{racconigi7}) with:
\begin{equation}\label{ciulifischio7}
    \mu_i \, = \, \exp\left[ i \, \theta_i\right]
\end{equation}
The K\"ahler structure of $\mathbb{V}_9$ is provided by the pullback on the $\mathbb{V}_9$ surface of the
K\"ahler potential of the entire flat K\"ahler manifold $\mbox{Hom}_{\mathbb{Z}_3} \left(R,\mathcal{Q}\otimes
R\right)$, namely we have:
\begin{eqnarray}\label{kallerV7}
    \mathcal{K}_{\mathbb{V}_9} & \equiv &\mbox{Tr} \;\left(\left[ A^\dagger\, , \, A\right] + \left[ B^\dagger\, , \, B\right] +
    \left[ C^\dagger\, , \, C\right]\right)\nonumber\\
    & = &\frac{\mu _2 \mu _4 u_3 \bar{\mu }_4 \bar{\mu }_2 \bar{u}_3}{\mu
   _3^2 \bar{\mu }_3^2}+\frac{\mu _2 \mu _5 u_1 \bar{\mu }_5
   \bar{\mu }_2 \bar{u}_1}{\mu _3 \mu _4 \bar{\mu }_3 \bar{\mu
   }_4}+\frac{\mu _2 \mu _6 u_1 \bar{\mu }_6 \bar{\mu }_2
   \bar{u}_1}{\mu _1 \bar{\mu }_1}+\frac{\mu _2 \mu _4 u_2 \bar{\mu
   }_4 \bar{\mu }_2 \bar{u}_2}{\mu _1 \mu _5 \bar{\mu }_1 \bar{\mu
   }_5}\nonumber\\
   &&+\frac{\mu _2 u_2 \bar{\mu }_2 \bar{u}_2}{\mu _3 \mu _6
   \bar{\mu }_3 \bar{\mu }_6}+\frac{\mu _2 u_3 \bar{\mu }_2
   \bar{u}_3}{\mu _1^2 \bar{\mu }_1^2}+\frac{\mu _5 u_2 \bar{\mu
   }_5 \bar{u}_2}{\mu _1 \mu _4 \bar{\mu }_1 \bar{\mu
   }_4}\nonumber\\
   &&+\frac{\mu _3 \mu _5 u_3 \bar{\mu }_3 \bar{\mu }_5
   \bar{u}_3}{\mu _4^2 \bar{\mu }_4^2}+\mu _1 \mu _6 u_3 \bar{\mu
   }_1 \bar{\mu }_6 \bar{u}_3+\frac{\mu _4 \mu _6 u_2 \bar{\mu }_4
   \bar{\mu }_6 \bar{u}_2}{\mu _3 \bar{\mu }_3}+\frac{\mu _3 \mu _6
   u_1 \bar{\mu }_3 \bar{\mu }_6 \bar{u}_1}{\mu _4 \mu _5 \bar{\mu
   }_4 \bar{\mu }_5}\nonumber\\
   &&+\frac{\mu _4 \mu _6 u_3 \bar{\mu }_4 \bar{\mu
   }_6 \bar{u}_3}{\mu _5^2 \bar{\mu }_5^2}+\frac{\mu _1 \mu _3 u_2
   \bar{\mu }_1 \bar{\mu }_3 \bar{u}_2}{\mu _4 \bar{\mu
   }_4}\nonumber\\
   &&+\frac{\mu _1 \mu _5 u_1 \bar{\mu }_1 \bar{\mu }_5
   \bar{u}_1}{\mu _6 \bar{\mu }_6}+\frac{\mu _4 u_1 \bar{\mu }_4
   \bar{u}_1}{\mu _5 \mu _6 \bar{\mu }_5 \bar{\mu }_6}+\frac{\mu _5
   u_3 \bar{\mu }_5 \bar{u}_3}{\mu _6^2 \bar{\mu }_6^2}+\frac{\mu
   _3 u_1 \bar{\mu }_3 \bar{u}_1}{\mu _1 \mu _2 \bar{\mu }_1
   \bar{\mu }_2}\nonumber\\
   &&+\frac{\mu _1 \mu _4 u_1 \bar{\mu }_1 \bar{\mu }_4
   \bar{u}_1}{\mu _2 \mu _3 \bar{\mu }_3 \bar{\mu }_2}+\frac{\mu _1
   \mu _6 u_2 \bar{\mu }_1 \bar{\mu }_6 \bar{u}_2}{\mu _2 \mu _5
   \bar{\mu }_5 \bar{\mu }_2}+\frac{\mu _3 \mu _5 u_2 \bar{\mu }_3
   \bar{\mu }_5 \bar{u}_2}{\mu _2 \mu _6 \bar{\mu }_6 \bar{\mu
   }_2}+\frac{\mu _1 \mu _3 u_3 \bar{\mu }_1 \bar{\mu }_3
   \bar{u}_3}{\mu _2^2 \bar{\mu }_2^2}
\end{eqnarray}
$\mathcal{K}_{\mathbb{V}_9} $  is obviously invariant under the unitary transformations of the \textit{gauge
group} :
\begin{equation}\label{constituzia7}
\forall \, g \, \in  \mathcal{F}_{\mathbb{Z}_7}\, : \quad   \left\{A,\,B,\,C\right\} \, \rightarrow \,
\left\{g^\dagger A g,\, g^\dagger B g,\, g^\dagger C g\right\}
\end{equation}
which, for that reason, is an \textit{isometry group} of the corresponding K\"ahler metric on
$\mbox{Hom}_{\mathbb{Z}_7} \left(R,\mathcal{Q}\otimes R\right)$ and  of its restriction to $\mathbb{V}_9$.
The last point follows from the fact that, by construction,
 $\mathcal{F}_{\mathbb{Z}_7}$ maps
$\mathbb{V}_9\subset\mbox{Hom}_{\mathbb{Z}_3} \left(R,\mathcal{Q}\otimes R\right) $ into itself.
\par
A basis of six linearly independent generators of the Lie algebra $\mathbb{F}_{\mathbb{Z}_3}$ is provided by
the following six matrices:
\begin{eqnarray}
    \mathfrak{f}_1 & = & \mbox{diag}\left(i,-i,0,0,0,0,0\right)\nonumber\\
    \mathfrak{f}_2 & = & \mbox{diag}\left(0,i,-i,0,0,0,0\right)\nonumber\\
    \mathfrak{f}_3 & = & \mbox{diag}\left(0,0,i,-i,0,0,0\right)\nonumber\\
    \mathfrak{f}_4 & = & \mbox{diag}\left(0,0,0,i,-i,0,0\right)\nonumber\\
    \mathfrak{f}_5 & = & \mbox{diag}\left(0,0,0,0,i,-i,0\right)\nonumber\\
    \mathfrak{f}_6 & = & \mbox{diag}\left(0,0,0,0,0,i,-i\right)\label{effiunosei}
\end{eqnarray}
and the moment maps corresponding to the  isometries generated by them are defined as follows:
\begin{equation}\label{cineseturcobis}
\mathfrak{P}_A \, = \, -\, i \, \mbox{Tr} \left(\mathfrak{f}_A \left(\left[ A^\dagger\, , \, A\right] +
\left[ B^\dagger\, , \, B\right] + \left[ C^\dagger\, , \, C\right]\right) \right),\quad \left( A=1,\dots,
6\right)
\end{equation}
We do not write the explicit form of $\mathfrak{P}_A$ since it is rather involved. First we change basis for
the generators of the Lie algebra $\mathbb{F}_{\mathbb{Z}_7}$ observing that they are not orthogonal with
respect to the Killing form defined by the trace. Indeed we have:
\begin{equation}\label{carnevale}
    \kappa_{AB} \, \equiv \, \mbox{Tr} \left(\mathfrak{f}_A^\dagger \, \mathfrak{f}_B^{\phantom{\dagger}}
    \right)\,= \, \left(
\begin{array}{cccccc}
 2 & -1 & 0 & 0 & 0 & 0 \\
 -1 & 2 & -1 & 0 & 0 & 0 \\
 0 & -1 & 2 & -1 & 0 & 0 \\
 0 & 0 & -1 & 2 & -1 & 0 \\
 0 & 0 & 0 & -1 & 2 & -1 \\
 0 & 0 & 0 & 0 & -1 & 2 \\
\end{array}
\right)
\end{equation}
which is nothing else but the Cartan matrix of the $A_6$ Lie algebra. Consequently we define a new basis of
moment maps, dual to the above one:
\begin{equation}\label{inversione}
    \mathfrak{P}^A\, = \, \left(\kappa^{-1}\right)^{AB} \, \mathfrak{P}_B
\end{equation}
and we express them in terms of the following variables:
\begin{equation}\label{insalatagreca}
    \Delta_{1,2,3} \, \equiv \, |u_{1,2,3}|^2 \quad ; \quad \mu_i \, = \, {\bar \mu}_i \, = \, \Omega_i \,
    \in \, \mathbb{R}_+ \quad (i=1,\dots \, 6)
\end{equation}
The choice (\ref{insalatagreca}) streams from the following three facts that we find it convenient to recall
once again:
\begin{description}
  \item[A)] When $\mu_i \, = \,1$ ($i=1,\dots,6)$, namely when the quiver group element with which we have rotated the locus
  $L_{\mathbb{Z}_7}$ is the identity, the moment maps $\mathfrak{P}^A$ are all zero.
  \item[B)] The moment maps $\mathfrak{P}^A$ are invariant under the action of the gauge group
  $\mathcal{F}_{\mathbb{Z}_7}\subset\mathcal{G}_{\mathbb{Z}_7}$.
  \item[C)] It follows from A) and B) that the only transformations of the quiver group that lift the levels
  of the moment maps from zero are those in the coset
  $\frac{\mathcal{G}_{\mathbb{Z}_7}}{\mathcal{F}_{\mathbb{Z}_7}}$.
\end{description}
In this way we obtain the following result:
\begin{eqnarray}
\mathfrak{P}^1 & = &\frac{\Omega _1^2 \left(\frac{\Delta _2 \left(\Omega _5^2-\Omega
   _1^4 \Omega _3^2\right)}{\Omega _4^2}+\Delta _1
   \left(\frac{\Omega _3^2}{\Omega _2^2}-\frac{\Omega _1^4 \Omega
   _5^2}{\Omega _6^2}\right)\right)+\Delta _3 \left(\Omega
   _2^2-\Omega _1^6 \Omega _6^2\right)}{\Omega _1^4}\label{PZ7A1}\\
 \mathfrak{P}^2 & = &\frac{\frac{\Delta _3 \left(\Omega _3^2-\Omega _2^4 \Omega
   _6^2\right) \Omega _1^4}{\Omega _2^4}+\Delta _1
   \left(-\frac{\Omega _5^2 \Omega _1^4}{\Omega _6^2}-\Omega _2^2
   \Omega _6^2+\frac{\Omega _4^2 \Omega _1^4+\Omega _3^4}{\Omega
   _2^2 \Omega _3^2}\right)+\Delta _2 \left(\left(\frac{\Omega
   _6^2}{\Omega _2^2 \Omega _5^2}-\frac{\Omega _3^2}{\Omega
   _4^2}\right) \Omega _1^4+\frac{\Omega _5^4-\Omega _2^2 \Omega
   _4^4}{\Omega _4^2 \Omega _5^2}\right)}{\Omega _1^2}\label{PZ7A2}\\
 \mathfrak{P}^3 & = &\Delta _3 \left(\frac{\Omega _2^2 \Omega _4^2}{\Omega _3^4}-\Omega
   _1^2 \Omega _6^2\right)+\Delta _1 \left(\left(\frac{\Omega
   _4^2}{\Omega _2^2 \Omega _3^2}-\frac{\Omega _5^2}{\Omega
   _6^2}\right) \Omega _1^2+\frac{\Omega _2^2 \Omega _5^2}{\Omega
   _3^2 \Omega _4^2}-\frac{\Omega _2^2 \Omega _6^2}{\Omega
   _1^2}\right)\nonumber\\
   &&+\Delta _2 \left(\left(\frac{\Omega _6^2}{\Omega
   _2^2 \Omega _5^2}-\frac{\Omega _3^2}{\Omega _4^2}\right) \Omega
   _1^2+\frac{\Omega _2^4-\Omega _3^4 \Omega _5^2}{\Omega _2^2
   \Omega _3^2 \Omega _6^2}+\frac{\Omega _5^4-\Omega _2^2 \Omega
   _4^4}{\Omega _4^2 \Omega _5^2 \Omega _1^2}\right)\label{PZ7A3}\\
 \mathfrak{P}^4 & = &\Delta _3 \left(\frac{\Omega _3^2 \Omega _5^2}{\Omega _4^4}-\Omega
   _1^2 \Omega _6^2\right)+\Delta _1 \left(\left(\frac{\Omega
   _5^2}{\Omega _3^2 \Omega _4^2}-\frac{\Omega _6^2}{\Omega
   _1^2}\right) \Omega _2^2+\frac{\Omega _3^2 \Omega _6^2}{\Omega
   _4^2 \Omega _5^2}-\frac{\Omega _1^2 \Omega _5^2}{\Omega
   _6^2}\right)\nonumber\\
   &&+\Delta _2 \left(\frac{\Omega _1^2 \Omega
   _6^2}{\Omega _2^2 \Omega _5^2}+\frac{\Omega _5^4-\Omega _2^2
   \Omega _4^4}{\Omega _1^2 \Omega _4^2 \Omega _5^2}+\frac{\Omega
   _2^4-\Omega _4^2 \Omega _6^4 \Omega _2^2-\Omega _3^4 \Omega
   _5^2}{\Omega _2^2 \Omega _3^2 \Omega _6^2}\right)\label{PZ7A4}\\
 \mathfrak{P}^5 & = &\frac{\frac{\Delta _3 \left(\Omega _4^2-\Omega _1^2 \Omega
   _5^4\right) \Omega _6^4}{\Omega _5^4}+\Delta _1
   \left(-\frac{\Omega _2^2 \Omega _6^4}{\Omega _1^2}+\frac{\Omega
   _3^2 \Omega _6^4}{\Omega _4^2 \Omega _5^2}-\Omega _1^2 \Omega
   _5^2+\frac{\Omega _4^2}{\Omega _5^2}\right)+\Delta _2
   \left(-\frac{\Omega _4^2 \Omega _6^4}{\Omega _3^2}+\frac{\Omega
   _1^2 \Omega _6^4-\Omega _3^2 \Omega _5^4}{\Omega _2^2 \Omega
   _5^2}+\frac{\Omega _2^2}{\Omega _3^2}\right)}{\Omega _6^2}\label{PZ7A5}\\
 \mathfrak{P}^6 & = &\frac{\Omega _6^2 \left(\Delta _1 \left(\frac{\Omega _4^2}{\Omega
   _5^2}-\frac{\Omega _2^2 \Omega _6^4}{\Omega
   _1^2}\right)+\frac{\Delta _2 \left(\Omega _2^2-\Omega _4^2
   \Omega _6^4\right)}{\Omega _3^2}\right)+\Delta _3 \left(\Omega
   _5^2-\Omega _1^2 \Omega _6^6\right)}{\Omega _6^4}\label{PZ7A6}
\end{eqnarray}
and we are  led to the following system of six higher order algebraic equations:
\begin{equation}\label{momentmapeque}
    \mathfrak{P}^A(\Delta,\Omega)\, = \, \zeta^A
\end{equation}
that has to be solved for $\Omega_A$ in terms of $\Delta_{1,2,3}$ and of the level parameters $\zeta^A$. At
vanishing level $\pmb{\zeta}=\pmb{0}$ we know  that the solution is $\pmb{\Omega} \,=\,\pmb{1}$, hence the best
we can do, since we deal with higher order equations that admit no solution by radicals, is to attempt a power
series solution in terms of the levels $\pmb{\zeta}$. Formally, however, the problem of the K\"ahler quotient
is solved.  Assuming  the scalar factors $\Omega_{1,\dots ,6} \, = \, \Omega_{1,\dots,
6}\left(\pmb{|u|^2},\pmb{\zeta}\right)$ to be known in terms of the coordinates and of the moment map
parameters, we can calculate the final K\"ahler potential. Substituting (\ref{insalatagreca}) into
eq.\,(\ref{kallerV7}) we obtain the restriction to the level surface $\mathcal{N}$ of the original K\"ahler
potential:
\begin{eqnarray}\label{yyy7}
    \mathcal{K}|_{\mathcal{N}}& = & \frac{\Delta _3 \Omega _4^2 \Omega _2^2}{\Omega _3^4}+\frac{\Delta
   _1 \Omega _5^2 \Omega _2^2}{\Omega _3^2 \Omega
   _4^2}+\frac{\Delta _1 \Omega _6^2 \Omega _2^2}{\Omega
   _1^2}+\frac{\Delta _2 \Omega _4^2 \Omega _2^2}{\Omega _1^2
   \Omega _5^2}+\frac{\Delta _2 \Omega _2^2}{\Omega _3^2 \Omega
   _6^2}+\frac{\Delta _3 \Omega _2^2}{\Omega _1^4}+\frac{\Delta _2
   \Omega _5^2}{\Omega _1^2 \Omega _4^2}+\frac{\Delta _3 \Omega
   _3^2 \Omega _5^2}{\Omega _4^4}\nonumber\\
   &&+\Delta _3 \Omega _1^2 \Omega
   _6^2+\frac{\Delta _2 \Omega _4^2 \Omega _6^2}{\Omega
   _3^2}+\frac{\Delta _1 \Omega _3^2 \Omega _6^2}{\Omega _4^2
   \Omega _5^2}+\frac{\Delta _3 \Omega _4^2 \Omega _6^2}{\Omega
   _5^4}+\frac{\Delta _2 \Omega _1^2 \Omega _3^2}{\Omega
   _4^2}+\frac{\Delta _1 \Omega _1^2 \Omega _5^2}{\Omega
   _6^2}+\frac{\Delta _1 \Omega _4^2}{\Omega _5^2 \Omega
   _6^2}\nonumber\\
   &&+\frac{\Delta _3 \Omega _5^2}{\Omega _6^4}+\frac{\Delta _1
   \Omega _3^2}{\Omega _1^2 \Omega _2^2}+\frac{\Delta _1 \Omega
   _1^2 \Omega _4^2}{\Omega _3^2 \Omega _2^2}+\frac{\Delta _2
   \Omega _1^2 \Omega _6^2}{\Omega _5^2 \Omega _2^2}+\frac{\Delta
   _2 \Omega _3^2 \Omega _5^2}{\Omega _6^2 \Omega
   _2^2}+\frac{\Delta _3 \Omega _1^2 \Omega _3^2}{\Omega _2^4}
\end{eqnarray}
 Then the final K\"ahler potential of the resolved smooth manifold is:
\begin{equation}\label{finazzo7}
    \mathcal{K}_\mathcal{M} \,= \, \mathcal{K}|_{\mathcal{N}}\, + \,\sum^{6}_{A=1} \zeta^A\,\log\Omega_A
\end{equation}
When $\zeta^{A}\,=\,0$ we have, as we already said, $\Omega _A \, = \,1$ ($A=1,\dots ,6$ which, inserted into
eq.\,(\ref{yyy7}), yields:
\begin{equation}\label{uuu7}
\lim_{\pmb{\zeta}\to\pmb{0}}\, \mathcal{K}|_{\mathcal{M}} \, = \,   \lim_{\pmb{\zeta}\to\pmb{0}}\,
\mathcal{K}|_{\mathcal{N}} \, = \, 7 u_1 \bar{u}_1+7 u_2 \bar{u}_2+7 u_3 \bar{u}_3
\end{equation}
namely we obtain the K\"ahler potential of the flat $\mathbb{C}^3/\mathbb{Z}_7$ orbifold of which $u_{1,2,3}$
are the $\mathbb{Z}_7$ invariant coordinates.
\section{A non-abelian example: crepant resolution \`a la Kronheimer of  $\mathcal{M}_{\mathrm{Dih_3}}$}
\label{nonabbello} In the ADE classification of SU(2) subgroups we do not find the dihedral \(\text{groups}
\text{Dih}_m\) whose order is $|$\(\text{Dih}_m\)$|$ $= 2m.$ We rather find their binary extensions
\(\text{Dih}_m{}^b\) whose order is $|$\(\text{Dih}_m{}^b\)$|$ = 4m.
\par
The \(\text{Dih}_m{}^b\) groups correspond to the Lie algebras \(D_{m+2 }\) in the McKay correspondence. The
smallest nonabelian case corresponds to \(\text{Dih}_3{}^b\) with 12 elements, since \(\text{Dih}_2{}^b\)
with eight elements is abelian. The matrices occurring in the generalized Kronheimer construction are already
rather big and the degree of the resolving algebraic equations is expected to be rather high.
\par
On the other hand we can easily embed the groups \(\text{Dih}_m\) into SU(3). Hence we address immediately
the case of the dihedral groups in \(\frac{\mathbb{C}^3} {\text{Dih}_m}\) which allows to consider the  case
$m=3$ with 6 elements. As it is well known \(\text{Dih}_3\) is isomorphic to \(S_3\), the symmetric group on
three elements and this is the smallest nonabelian group. It remains to understand what is the generalized
Weyl group corresponding to the quiver matrix generated by this case which, as we show below is the following
one:
\begin{equation}
 \text{CQ}\, = \, \left(
\begin{array}{ccc}
 0 & 1 & 1 \\
 1 & 0 & 1 \\
 1 & 1 & 2 \\
\end{array}
\right) \label{dih3quivmat}
\end{equation}
The corresponding extended diagram is the following one:
\begin{equation}
\text{CE}\,=\,\left(
\begin{array}{ccc}
 3 & -1 & -1 \\
 -1 & 3 & -1 \\
 -1 & -1 & 1 \\
\end{array}
\right)\label{extdih3quivmat}
\end{equation}
\subsection{Definition of $\mathrm{Dih}_m \subset\mathrm{SU(3)}$}
The presentation of the dihedral group \(\text{Dih}_m\) is the following one
\begin{equation}
A^m = 1 \quad ;\quad  B^2 =1 \quad ;\quad (AB)^2 = \mathbf{1}
\end{equation}
we introduce the following representation of the generators as matrices acting on \(\mathbb{C}^3\)
\begin{equation}
A \,= \,\left(
\begin{array}{ccc}
 e^{\frac{2i*\pi }{m}} & 0 & 0 \\
 0 & e^{-\frac{2i*\pi }{m}} & 0 \\
 0 & 0 & 1 \\
\end{array}
\right)\label{ArepC3}
\end{equation}
\begin{equation}
B \, = \, \left(
\begin{array}{ccc}
 0 & i & 0 \\
 -i & 0 & 0 \\
 0 & 0 & -1 \\
\end{array}
\right)\label{BrepC3}
\end{equation}
\subsection{Abstract structure of the Dihedral Groups $\mathrm{Dih}_m$
with $m  =  \mathrm{odd}$ }
\subsubsection{Conjugacy classes}
The total number of conjugacy classes is
\begin{equation}
\ell  = \frac{m+3}{2}
\end{equation}
that are enumerated as follows\footnote{In table
\ref{clasconiugdihm} we denote by $E$ the identity element of the
group.}:
\begin{eqnarray}
&&\mathcal{C}_1 = \langle E\rangle \nonumber\\
&& \mathcal{C}_2 = \left\langle B,AB,A^2B,\text{...}A^{m-1}B\right\rangle ,\nonumber\\
&&\mathcal{C}_3 =<A,A^{m-1}>,\mathcal{C}_4 =<A^2,A^{m-2}>,\nonumber\\
&& \mathcal{C}_5 =<A^3,A^{m-3}>,\nonumber\\
&&\text{...},\nonumber\\
&&\mathcal{C}_{\frac{m+3}{2}} =<A^{\frac{m-1}{2}},A^{\frac{m+1}{2}}>
\label{clasconiugdihm}
\end{eqnarray}
Hence we have one class of population 1, one class of population $m$ and \(\frac{m-1}{2}\) classes of
population 2
\begin{equation}
\left|\text{Dih}_m\right|=2m = 1+ m + 2\times \frac{m-1}{2}\label{verif}
\end{equation}
\subsubsection{Irreps}
Accordingly we expect \(\ell  = \frac{m+3}{2}\) irreducible representations. They are as follows:
\begin{enumerate}
\item
The one dimensional identity representation \(\mathbb{D}_0\)
\item
The alternating one-dimensional representation \(\mathbb{D}_1\), obtained by setting { }A $\rightarrow $ 1, {
}B $\rightarrow $ -1.
\item
The \(\frac{m-1}{2}\) two-dimensional representations obtained in the following way:
\begin{equation}
\mathbb{D}_{k+1}[A] = \left(
\begin{array}{cc}
 e^{\frac{2i*\pi }{m}k} & 0 \\
 0 & e^{-\frac{2i*\pi }{m}k} \\
\end{array}
\right)\quad ; \quad \mathbb{D}_{k+1}[B]=\left(
\begin{array}{cc}
 0 & 1 \\
 1 & 0 \\
\end{array}
\right)\quad ;\quad \left(k,1,2,\text{...},\frac{m-1}{2}\right);
\end{equation}
\end{enumerate}
\subsubsection{Characters}
In this way we obtain the following character table:
\begin{equation}
\begin{array}{c|ccccccc}
 0 & \mathcal{C}_1 & \mathcal{C}_2 & \mathcal{C}_3 & \mathcal{C}_4 & \mathcal{C}_5 & \text{...} & \mathcal{C}_6 \\
\hline
 \times  & E & B & A & A^2 & A^3 & \text{...} & A^{\frac{m-1}{2}} \\
 \mathbb{D}_1 & 1 & 1 & 1 & 1 & 1 & \text{...} & 1 \\
 \mathbb{D}_2 & 1 & -1 & 1 & 1 & 1 & \text{...} & 1 \\
 \mathbb{D}_3 & 2 & 0 & 2\,\cos\left[\frac{2 \pi }{m}\right] & 2\,\cos\left[\frac{4 k \pi }{m}\right] &
 2\,\cos\left[\frac{6 k \pi }{m}\right]
& \text{...} & 2\,\cos\left[\frac{(-1+m) \pi }{m}\right] \\
 \mathbb{D}_4 & 2 & 0 & 2\,\cos\left[\frac{4 \pi }{m}\right] & 2\,\cos\left[\frac{8 \pi }{m}\right] &
 2\,\cos\left[\frac{12 \pi }{m}\right]
& \text{...} & 2\,\cos\left[\frac{2 (-1+m) \pi }{m}\right] \\
 \text{...} & \text{...} & \text{...} & \text{...} & \text{...} & \text{...} & \text{...} & \text{...} \\
 \mathbb{D}_{\frac{m-1}{2}} & 2 & 0 & 2\,\cos\left[\frac{(-1+m) \pi }{m}\right] &
 2\,\cos\left[\frac{2 (-1+m) \pi }{m}\right] & 2\,\cos\left[\frac{3
(-1+m) \pi }{m}\right] & \text{...} & 2\,\cos\left[\frac{(-1+m)^2 \pi }{2 m}\right] \\
\end{array}
\end{equation}
\subsection{The quiver matrix of $\mathrm{Dih_3}$ acting on $\mathbb{C}^3$}

Using the above character table (case $m=3$) we easily derive the following decomposition into irreps:
\begin{equation}
 \mathcal{Q}\otimes  \mathbb{D}_i = \underset{j=1}{\overset{\ell = 3
 }{\oplus }} \text{QC}_{\text{ij}} \mathbb{D}_j
\end{equation}
where $\mathcal{Q}$ is the three-dimensional representation of the dihedral group defined by eqs.\,(\ref{ArepC3},\ref{BrepC3}) and \(\mathbb{D}_i\) are the irreducible representations listed above. The
matrix $\text{QC}_{\text{ij}}$ is that anticipated in equation (\ref{dih3quivmat})
\subsection{Ages}
In a similar easy way we derive the age grading of the three conjugacy classes and the associated triple of
integer numbers $a_i$ that define the weights in the weighted blowup procedure.  We find:
\begin{enumerate}
\item age = 0 \hskip 2cm ; \hskip2cm  $1\{0,0,0\}$
\item
age = 1 \hskip 2cm ; \hskip 2cm  $\frac{1}{2}\{1,1,0\}$
\item
age = 1 \hskip 2cm ; \hskip 2cm  $\frac{1}{3}\{0,2,1\}$
\end{enumerate}
Hence apart from the age = 0 class of the identity we find two junior classes and no senior one. In force of
the fundamental theorem 41 we conclude that the Hodge numbers of the resolved variety $\mathcal{M}$ are as
follows $h^{0,0}=1$, $h^{1,1} \, = \, 2$, $h^{2,2} \, = \, 0$. Indeed no (1,1)-class has compact support.
\subsection{The regular representation construction and decomposition}
The next item that we need is the regular representation of the  group \(\text{Dih}_{3 }\) and its
block diagonalization into irreducible subspaces corresponding to the irreducible representations. As it is
well known the regular representation of any finite group contains as many copies of each irrep
\(\mathbb{D}_i\) as it is its dimension. Hence ordering the three irreducible representations of
\(\text{Dih}_{3 }\) according to  the above presented scheme, namely
\(\mathbb{D}_1\),\(\mathbb{D}_2,\mathbb{D}_{3 }\) with
\begin{equation}
\dim  \mathbb{D}_1 = 1 \quad ; \quad \dim \mathbb{D}_2 = 1 \quad ; \quad \dim  \mathbb{D}_3 = 2
\end{equation}
and naming R the regular representation we have:
\begin{equation}
R = \mathbb{D}_1\oplus \mathbb{D}_2\oplus \mathbb{D}_3\oplus \hat{\mathbb{D}}_3
\end{equation}
With some effort one derives the matrix $\mathbf{m}$ that performs the change of basis from the natural basis
of the regular representation whose axes are the group elements to the block diagonal basis where each block
correspond to one irreducible representation. The explicit form of $\mathbf{m}$ is displayed below:
\begin{equation}
\mathbf{m}=\left(
\begin{array}{cccccc}
 1 & 1 & 0 & -\frac{1}{2} i \left(-i+\sqrt{3}\right) & \frac{1}{2} i \left(i+\sqrt{3}\right) & 0 \\
 1 & -1 & \frac{1}{2} i \left(i+\sqrt{3}\right) & 0 & 0 & -\frac{1}{2} i \left(-i+\sqrt{3}\right) \\
 1 & -1 & 1 & 0 & 0 & 1 \\
 1 & -1 & -\frac{1}{2} i \left(-i+\sqrt{3}\right) & 0 & 0 & \frac{1}{2} i \left(i+\sqrt{3}\right) \\
 1 & 1 & 0 & 1 & 1 & 0 \\
 1 & 1 & 0 & \frac{1}{2} i \left(i+\sqrt{3}\right) & -\frac{1}{2} i \left(-i+\sqrt{3}\right) & 0 \\
\end{array}
\right)
\end{equation}
Using $\mathbf{m}$ we obtain the explicit form of the 6 group elements of the dihedral group in the block
diagonal form of the 6-dimensional regular representation. They are displayed below: \vskip 0.2cm
\noindent\(\text{R[}1\text{] = }\left(
\begin{array}{cccccc}
 1 & 0 & 0 & 0 & 0 & 0 \\
 0 & 1 & 0 & 0 & 0 & 0 \\
 0 & 0 & 1 & 0 & 0 & 0 \\
 0 & 0 & 0 & 1 & 0 & 0 \\
 0 & 0 & 0 & 0 & 1 & 0 \\
 0 & 0 & 0 & 0 & 0 & 1 \\
\end{array}
\right)\) \vskip 0.2cm \noindent\(\text{R[}2\text{] = }\left(
\begin{array}{cccccc}
 1 & 0 & 0 & 0 & 0 & 0 \\
 0 & -1 & 0 & 0 & 0 & 0 \\
 0 & 0 & 0 & \frac{1}{2} i \left(i+\sqrt{3}\right) & 0 & 0 \\
 0 & 0 & -\frac{1}{2} i \left(-i+\sqrt{3}\right) & 0 & 0 & 0 \\
 0 & 0 & 0 & 0 & 0 & \frac{1}{2} i \left(i+\sqrt{3}\right) \\
 0 & 0 & 0 & 0 & -\frac{1}{2} i \left(-i+\sqrt{3}\right) & 0 \\
\end{array}
\right)\) \vskip 0.2cm \noindent\(\text{R[}3\text{] = }\left(
\begin{array}{cccccc}
 1 & 0 & 0 & 0 & 0 & 0 \\
 0 & -1 & 0 & 0 & 0 & 0 \\
 0 & 0 & 0 & -\frac{1}{2} i \left(-i+\sqrt{3}\right) & 0 & 0 \\
 0 & 0 & \frac{1}{2} i \left(i+\sqrt{3}\right) & 0 & 0 & 0 \\
 0 & 0 & 0 & 0 & 0 & -\frac{1}{2} i \left(-i+\sqrt{3}\right) \\
 0 & 0 & 0 & 0 & \frac{1}{2} i \left(i+\sqrt{3}\right) & 0 \\
\end{array}
\right)\) \vskip 0.2cm \noindent\(\text{R[}4\text{] = }\left(
\begin{array}{cccccc}
 1 & 0 & 0 & 0 & 0 & 0 \\
 0 & -1 & 0 & 0 & 0 & 0 \\
 0 & 0 & 0 & 1 & 0 & 0 \\
 0 & 0 & 1 & 0 & 0 & 0 \\
 0 & 0 & 0 & 0 & 0 & 1 \\
 0 & 0 & 0 & 0 & 1 & 0 \\
\end{array}
\right)\) \vskip 0.2cm \noindent\(\text{R[}5\text{] = }\left(
\begin{array}{cccccc}
 1 & 0 & 0 & 0 & 0 & 0 \\
 0 & 1 & 0 & 0 & 0 & 0 \\
 0 & 0 & \frac{1}{2} i \left(i+\sqrt{3}\right) & 0 & 0 & 0 \\
 0 & 0 & 0 & -\frac{1}{2} i \left(-i+\sqrt{3}\right) & 0 & 0 \\
 0 & 0 & 0 & 0 & \frac{1}{2} i \left(i+\sqrt{3}\right) & 0 \\
 0 & 0 & 0 & 0 & 0 & -\frac{1}{2} i \left(-i+\sqrt{3}\right) \\
\end{array}
\right)\) \vskip 0.2cm \noindent\(\text{R[}6\text{] = }\left(
\begin{array}{cccccc}
 1 & 0 & 0 & 0 & 0 & 0 \\
 0 & 1 & 0 & 0 & 0 & 0 \\
 0 & 0 & -\frac{1}{2} i \left(-i+\sqrt{3}\right) & 0 & 0 & 0 \\
 0 & 0 & 0 & \frac{1}{2} i \left(i+\sqrt{3}\right) & 0 & 0 \\
 0 & 0 & 0 & 0 & -\frac{1}{2} i \left(-i+\sqrt{3}\right) & 0 \\
 0 & 0 & 0 & 0 & 0 & \frac{1}{2} i \left(i+\sqrt{3}\right) \\
\end{array}
\right)\)
\begin{equation}
\null\label{regdiagdih}
\end{equation}
\subsection{The invariant space $\mathcal{S}_{\Gamma} =\text{Hom}_{\Gamma}(R, \mathcal{Q}\otimes R)$}
Imposing the invariance constraint (\ref{gammazione}) and (\ref{carnevalediRio}) and using the explicit form
of the regular representation derived above we obtain the triples of matrices spanning \(\mathcal{S}_{\Gamma
}\). They depend on $3 \times
 6 =18$ parameters and their explicit form  written in the split basis where the regular representation is
block diagonal is displayed below:
\begin{eqnarray}
A &=&\left(
\begin{array}{cccccc}
 0 & 0 & m_{1,3} & 0 & m_{1,5} & 0 \\
 0 & 0 & m_{2,3} & 0 & m_{2,5} & 0 \\
 0 & 0 & 0 & m_{3,4} & 0 & m_{3,6} \\
 m_{4,1} & m_{4,2} & 0 & 0 & 0 & 0 \\
 0 & 0 & 0 & m_{5,4} & 0 & m_{5,6} \\
 m_{6,1} & m_{6,2} & 0 & 0 & 0 & 0 \\
\end{array}
\right) \nonumber\\
B &=& {\scriptsize
\left(
\begin{array}{cccccc}
 0 & 0 & 0 & -\frac{1}{2} \left(-i-\sqrt{3}\right) m_{1,3} & 0 & -\frac{1}{2} \left(-i-\sqrt{3}\right) m_{1,5} \\
 0 & 0 & 0 & -\frac{1}{2} \left(i+\sqrt{3}\right) m_{2,3} & 0 & -\frac{1}{2} \left(i+\sqrt{3}\right) m_{2,5} \\
 -\frac{1}{2} \left(-i-\sqrt{3}\right) m_{4,1} & -\frac{1}{2} \left(i+\sqrt{3}\right) m_{4,2} & 0 & 0 & 0 & 0 \\
 0 & 0 & -\frac{1}{2} \left(-i-\sqrt{3}\right) m_{3,4} & 0 & -\frac{1}{2} \left(-i-\sqrt{3}\right) m_{3,6} & 0 \\
 -\frac{1}{2} \left(-i-\sqrt{3}\right) m_{6,1} & -\frac{1}{2} \left(i+\sqrt{3}\right) m_{6,2} & 0 & 0 & 0 & 0 \\
 0 & 0 & -\frac{1}{2} \left(-i-\sqrt{3}\right) m_{5,4} & 0 & -\frac{1}{2} \left(-i-\sqrt{3}\right) m_{5,6} & 0 \\
\end{array}
\right)} \nonumber\\
 C &=&\left(
\begin{array}{cccccc}
 0 & r_{1,2} & 0 & 0 & 0 & 0 \\
 r_{2,1} & 0 & 0 & 0 & 0 & 0 \\
 0 & 0 & r_{3,3} & 0 & r_{3,5} & 0 \\
 0 & 0 & 0 & -r_{3,3} & 0 & -r_{3,5} \\
 0 & 0 & r_{5,3} & 0 & r_{5,5} & 0 \\
 0 & 0 & 0 & -r_{5,3} & 0 & -r_{5,5} \\
\end{array}
\right)    \label{ABCdiedro}
\end{eqnarray}
\subsubsection{The locus $L_{\Gamma} \subset \mathcal{S}_\Gamma$}
The  triples of matrices $\left\{A_0, B_0, C_0\right\}$ spanning the locus \(L_{\Gamma
}\subset\mathcal{S}_{\Gamma }\) defined in eq.\,(\ref{thespacel}) depend on three complex parameters
\(z_1\),\(z_2\),\(z_3\) that can be regarded as a set of global coordinates for the orbifold locus
\(\frac{\mathbb{C}^3}{\text{Dih}_3}\). They are displayed below written in the split basis of the regular
representation. They are the image after the change of basis of those matrices that belong to
$\mathcal{S}_\Gamma$ and are diagonal in the natural basis:
\begin{eqnarray}
A_0&=&\left(
\begin{array}{cccccc}
0 & 0 & \frac{z_2}{\sqrt{15}} & 0 & \frac{z_1}{\sqrt{15}} & 0 \\
0 & 0 & -\frac{z_2}{\sqrt{15}} & 0 & \frac{z_1}{\sqrt{15}} & 0 \\
0 & 0 & 0 & 0 & 0 & \frac{2 z_2}{\sqrt{15}} \\
\frac{2 z_1}{\sqrt{15}} & \frac{2 z_1}{\sqrt{15}} & 0 & 0 & 0 & 0 \\
0 & 0 & 0 & \frac{2 z_1}{\sqrt{15}} & 0 & 0 \\
\frac{2 z_2}{\sqrt{15}} & -\frac{2 z_2}{\sqrt{15}} & 0 & 0 & 0 & 0 \\
\end{array}
\right) \nonumber\\
B_0 &=&\left(
\begin{array}{cccccc}
0 & 0 & 0 & -\frac{e^{i \frac{7}{6}\pi } z_2}{2 \sqrt{15}} & 0 & -\frac{e^{i \frac{7}{6}\pi } z_1}{2 \sqrt{15}} \\
0 & 0 & 0 & -\frac{e^{i \frac{7}{6}\pi } z_2}{2 \sqrt{15}} & 0 & \frac{e^{i \frac{7}{6}\pi } z_1}{2 \sqrt{15}} \\
-\frac{e^{i \frac{7}{6}\pi } z_1}{\sqrt{15}} & \frac{e^{i \frac{7}{6}\pi } z_1}{\sqrt{15}} & 0 & 0 & 0 & 0 \\
0 & 0 & 0 & 0 & -\frac{e^{i \frac{7}{6}\pi } z_2}{\sqrt{15}} & 0 \\
-\frac{e^{i \frac{7}{6}\pi } z_2}{\sqrt{15}} & -\frac{e^{i \frac{7}{6}\pi } z_2}{\sqrt{15}} & 0 & 0 & 0 & 0 \\
0 & 0 & -\frac{e^{i \frac{7}{6}\pi } z_1}{\sqrt{15}} & 0 & 0 & 0 \\
\end{array}
\right) \nonumber\\
C_0 &=&\left(
\begin{array}{cccccc}
0 & \frac{z_3}{\sqrt{6}} & 0 & 0 & 0 & 0 \\
\frac{z_3}{\sqrt{6}} & 0 & 0 & 0 & 0 & 0 \\
0 & 0 & -\frac{z_3}{\sqrt{6}} & 0 & 0 & 0 \\
0 & 0 & 0 & \frac{z_3}{\sqrt{6}} & 0 & 0 \\
0 & 0 & 0 & 0 & \frac{z_3}{\sqrt{6}} & 0 \\
0 & 0 & 0 & 0 & 0 & -\frac{z_3}{\sqrt{6}} \\
\end{array}\right)
\label{A0B0C0diedro}
\end{eqnarray}
\subsection{The complex quiver group $\mathcal{G}_{\Gamma}$}
Our next point is the derivation of the group \(\mathcal{G}_{\Gamma }\) discussed in section \ref{quiverino}
of the main text and defined by the following condition
\begin{equation}
\mathcal{G}_{\Gamma } = \left\{ g\in  \text{SL}(|\Gamma |,\mathbb{C})\quad \left|\quad \forall \gamma \in
\Gamma \,\, :\, \, \left[D_R(\gamma ),D_{\text{def}}(g)\right]\right.=0\right\}
\end{equation}
Abstractly the quiver group turns out to have has the following structure:
\begin{equation}
\mathcal{G}_{\Gamma } = \mathbb{C}^*\times \mathbb{C}^*\times \text{SL}(2,\mathbb{C})
\end{equation}
Indeed a generic group element of the quiver group depends on 5 complex parameters \(\chi _i\) (i=1,..,5) and
it has the following form:
\begin{equation}
\mathfrak{g}\text{  }\in  \mathcal{G}_{\Gamma }\text{    }:\text{   }\mathfrak{g}\text{  }= \left(
\begin{array}{cccccc}
 \frac{1}{\chi _1 \left(\chi _3 \chi _4-\chi _2 \chi _5\right){}^2} & 0 & 0 & 0 & 0 & 0 \\
 0 & \chi _1 & 0 & 0 & 0 & 0 \\
 0 & 0 & \chi _2 & 0 & \chi _3 & 0 \\
 0 & 0 & 0 & \chi _2 & 0 & \chi _3 \\
 0 & 0 & \chi _4 & 0 & \chi _5 & 0 \\
 0 & 0 & 0 & \chi _4 & 0 & \chi _5 \\
\end{array}
\right)
\end{equation}
\subsubsection{Construction of the $\mathbb{V}_{|\Gamma |+2}$ manifold as an
orbit of the quiver group} According to the discussion of the main text and with eq.\,(\ref{gneccoD}) we
construct the variety \(\mathbb{V}_{|\Gamma |+2\text{  }}\)= { }\(\mathbb{V}_{8}\) as the orbit
\(\mathcal{D}_{\Gamma }\text{  }= \text{Orbit}_{\mathcal{G}_{\Gamma}}\left(L_{\Gamma}\right)\) of the locus
\(L_{\Gamma }\) with respect to the quiver group. Therefore the coordinates on \(\mathbb{V}_{8 }\)  are
$\{$\(z_1\),\(z_2\),\(z_3\),\(\chi _1\),...\(\chi _5\)$\}$.
\subsubsection{Reduction to the compact gauge group}
The next task is to derive the maximal compact subgroup of the quiver group
\begin{equation}
 \mathcal{F}_{\Gamma }\subset \mathcal{G}_{\Gamma }
\end{equation}
which, as emphasized several times, is the gauge group of the Chern-Simons theory on the brane volume and the
group with respect to which we perform the K\"ahler quotient. In the present case we have:
\begin{equation}
\mathcal{F}_{\Gamma }\text{  }= \mathrm{U(1)}\times  \mathrm{U(1)}\times \text{SU}(2)\subset
\mathbb{C}^*\times \mathbb{C}^*\times \text{SL}(2,\mathbb{C})\text{  }= \mathcal{G}_{\Gamma }
\end{equation}
Let us name the corresponding Lie algebras according to the following obvious nomenclature and perform the
following orthogonal split
\begin{eqnarray}
 \mathbb{G}_{\Gamma }&=&\mathbb{F}_{\Gamma }\oplus \mathbb{K}_{\Gamma }\nonumber\\
\left[\mathbb{F}_{\Gamma } ,\mathbb{F}_{\Gamma } \right] &\subset &\mathbb{F}_{\Gamma }\quad; \quad
\left[\mathbb{F}_{\Gamma } ,\mathbb{K}_{\Gamma } \right] \subset \mathbb{K}_{\Gamma }\quad; \quad
\left[\mathbb{K}_{\Gamma } ,\mathbb{K}_{\Gamma } \right] \subset \mathbb{F}_{\Gamma }\label{ortogunalno}
\end{eqnarray}
The main final item is encoded in the coset manifold:
\begin{equation}
\mathcal{V}_{\Gamma } \equiv \frac{\mathcal{G}_{\Gamma }}{\mathcal{F}_{\Gamma }} \quad ; \quad\dim
_{\mathbb{R}}\mathcal{V}_{\Gamma } = \dim _{\mathbb{C}} \mathcal{G}_{\Gamma } = \dim _{\mathbb{R}}
\mathcal{F}_{\Gamma } = 5 \quad(\text{in our case})
\end{equation}
for whose solving element $ \exp[ \mho  ]$ we will find a system of algebraic equations encoding the complete
resolution of the problem of the K\"ahler quotient and the definition of all the tautological bundles.
\par
Schematically we will have the following equation. Let $\mathfrak{P}$ represent the moment map
\begin{equation}
\mathfrak{P} : \mathcal{S}_{\Gamma } \longrightarrow  \mathbb{F}_{\Gamma}
\end{equation}
and let us denote by $\mathit{g}$.p the action of the quiver group on the space \(\mathcal{S}_{\Gamma }\):
\begin{eqnarray}
\mathcal{G}_{\Gamma } & : &\mathcal{S}_{\Gamma }\longrightarrow
 \mathcal{S}_{\Gamma }\nonumber\\
\forall \mathit{g} \in \mathcal{G}_{\Gamma }\text{  },\text{  }\Omega_\mathit{g} & : & p \longrightarrow
\text{ }\mathit{g}.p\in \mathcal{S}_{\Gamma }
\end{eqnarray}
The fundamental property of the compact subgroup is the following one:
\begin{equation}
\mathfrak{P}(\mathit{g}.p) = \mathfrak{P}(p)\quad\quad \text{iff}\quad\quad\mathit{g} \in \mathcal{F}_{\Gamma
}\subset \mathcal{G}_{\Gamma}
\end{equation}
In view of this, provided we have chosen some parametrization of the coset \(\mathcal{V}_{\Gamma }\), given
by coordinates \(\mho _i\) (i=1,..,\(\dim _{\mathbb{R}}\mathcal{V}_{\Gamma }\)) we  find an equation of the
following form:
\begin{equation}
\mathfrak{P}\left(\text{Exp}[\mho ].p_0\right) = \zeta \text{  }
\end{equation}
where
\begin{equation}
\mathfrak{P}\left(p_0\right)= 0\text{     }\Leftrightarrow \text{
 }p_o \in  L_{\Gamma }
\end{equation}
\subsubsection{Generators of the Lie Algebra $\mathbb{F}_{\Gamma }= \uu(1)\oplus\uu(1)\oplus \su(2)$ }
Here we restart our analysis at the level of Lie algebra rather than at the level of Lie Groups and we derive
the form of the generators of \(\mathcal{F}_{\Gamma }\). This is very important for the explicit discussion
of the K\"ahler quotient. Then we construct the complementary subspace \(\mathbb{K}_{\Gamma }\).
\par
A generic element of the compact Lie algebra \(\mathbb{F}_{\Gamma }\) can be written as the following
6$\times $6 matrix depending on five real parameters \(\nu _1\),...,\(\nu _5\):
\begin{equation}\label{fgotichino}
    \mathfrak{f} \in  \mathbb{F}_{\Gamma \text{  }}\text{  }:\text{  }\mathfrak{f} =\text{  }\left(
\begin{array}{cccccc}
 i \left(-\nu _1-2 \nu _2-2 \nu _3\right) & 0 & 0 & 0 & 0 & 0 \\
 0 & i \nu _1 & 0 & 0 & 0 & 0 \\
 0 & 0 & i \nu _2 & 0 & \nu _4+i \nu _5 & 0 \\
 0 & 0 & 0 & i \nu _2 & 0 & \nu _4+i \nu _5 \\
 0 & 0 & -\nu _4+i \nu _5 & 0 & i \nu _3 & 0 \\
 0 & 0 & 0 & -\nu _4+i \nu _5 & 0 & i \nu _3 \\
\end{array}
\right)
\end{equation}
Hence the generators are provided by the matrices multiplying each of the parameters \(\nu _1\),...,\(\nu
_5\).
\subsubsection{Generators in the complementary subspace $\mathbb{K}_{\Gamma }$ }
The explicit form of a matrix belonging to the complementary subspace \(\mathbb{K}_{\Gamma }\) is given here
below and depends on the five real parameters \(\psi _i\) (i=1,..,5)
\begin{equation}
\mathfrak{k} \in  \mathbb{K}_{\Gamma}\text{  }:\text{  }\mathfrak{k}=\left(
\begin{array}{cccccc}
 -\psi _1-2 \psi _2-2 \psi _3 & 0 & 0 & 0 & 0 & 0 \\
 0 & \psi _1 & 0 & 0 & 0 & 0 \\
 0 & 0 & \psi _2 & 0 & \psi _4+i \psi _5 & 0 \\
 0 & 0 & 0 & \psi _2 & 0 & \psi _4+i \psi _5 \\
 0 & 0 & \psi _4-i \psi _5 & 0 & \psi _3 & 0 \\
 0 & 0 & 0 & \psi _4-i \psi _5 & 0 & \psi _3 \\
\end{array}
\right)
\end{equation}
A basis of the subspace \(\mathbb{K}_{\Gamma }\) is given by the matrices that multiply each \(\psi _i\).
\par
The above explicit matrices exemplify the discussion in section \ref{realecompdiscus} of the main text.
\subsubsection{Exponentiation of the orthogonal subspace $\mathbb{K}_{\Gamma }$}
In this subsection we derive the hermitian matrix \(\mathcal{V} =\mathcal{V}^{\dagger }\) =
Exp[\(\mathbb{K}_{\Gamma }\)] that can be used as a coset representative for the coset
\(\frac{\mathcal{G}_{\Gamma }}{\mathcal{F}_{\Gamma }}\).
\par
One parametrization of the coset manifold \(\frac{\mathcal{G}_{\Gamma }}{\mathcal{F}_{\Gamma }}\) is provided
$\exp[\mathbb{K}_{\Gamma }] = \mathcal{V}$ which depends on five real parameters, namely the three scale
factors \(\Upsilon _1,\Upsilon _2,\Upsilon _3\) , the hyperbolic angle $\varrho $ and the elliptic angle
$\phi $. Indeed taking the product of the exponentiation of the various generators we get that $\mathcal{V}
\in \text{Exp}\left[\mathbb{K}_{\Gamma }\right]$ has the following form:
\begin{equation}
 \mathcal{V} =\left(
\begin{array}{cccccc}
 \frac{\text{Cosh}[\varrho ]^4}{\Upsilon _1 \Upsilon _2^2 \Upsilon _3^2} & 0 & 0 & 0 & 0 & 0 \\
 0 & \Upsilon _1 & 0 & 0 & 0 & 0 \\
 0 & 0 & \Upsilon _2 & 0 & e^{i \phi } \sqrt{\Upsilon _2 \Upsilon _3} \text{Tanh}[\varrho ] & 0 \\
 0 & 0 & 0 & \Upsilon _2 & 0 & e^{i \phi } \sqrt{\Upsilon _2 \Upsilon _3} \text{Tanh}[\varrho ] \\
 0 & 0 & e^{-i \phi } \sqrt{\Upsilon _2 \Upsilon _3} \text{Tanh}[\varrho ] & 0 & \Upsilon _3 & 0 \\
 0 & 0 & 0 & e^{-i \phi } \sqrt{\Upsilon _2 \Upsilon _3} \text{Tanh}[\varrho ] & 0 & \Upsilon _3 \\
\end{array}
\right)
\end{equation}
Subsequently suitably renaming the parameters we can rewrite the above hermitian matrix $\mathcal{V}$ in a
different more friendly way which is the following one:
\begin{eqnarray}
&&\mathcal{V}\,=\,  \nonumber\\
&&\left(
\begin{array}{cccccc}
 \frac{1}{\left(-1+X^2+Y^2\right)^2 \Upsilon _1 \Upsilon _2^2 \Upsilon _3^2} & 0 & 0 & 0 & 0 & 0 \\
 0 & \Upsilon _1 & 0 & 0 & 0 & 0 \\
 0 & 0 & \Upsilon _2 & 0 & (X+i Y) \sqrt{\Upsilon _2 \Upsilon _3} & 0 \\
 0 & 0 & 0 & \Upsilon _2 & 0 & (X+i Y) \sqrt{\Upsilon _2 \Upsilon _3} \\
 0 & 0 & (X-i Y) \sqrt{\Upsilon _2 \Upsilon _3} & 0 & \Upsilon _3 & 0 \\
 0 & 0 & 0 & (X-i Y) \sqrt{\Upsilon _2 \Upsilon _3} & 0 & \Upsilon _3 \\
\end{array}
\right)\nonumber\\
\label{cosettondih3}
\end{eqnarray}
The 5 real parameters \(\Upsilon _1\),\(\Upsilon _2 ,\Upsilon _3\) $>$0 and $X,Y$ play a fundamental role in
the elaboration of the tautological bundles.
\subsubsection{Calculation of the center of the compact Lie algebra}
 We are interested
in the calculation of the center   of the compact Lie algebra
\(\mathfrak{z}\left[\mathbb{F}_{\Gamma }\right]\) since it is only
the moment maps in this center that can be assigned non vanishing
values. The abstract form of the searched for center is the
following:
\begin{equation}
\mathfrak{z}\left[\mathbb{F}_{\Gamma }\right]= \uu(1)\oplus \uu(1)
\end{equation}
To find the explicit immersion of
\(\mathfrak{z}\left[\mathbb{F}_{\Gamma }\right]\) in
\(\mathbb{F}_{\Gamma }\) we impose the condition that an element of
the center  should commute with all the generators of
\(\mathbb{F}_{\Gamma }\). In this way we can reorganize the listing
of the 5 generators into 2 belonging to the center for which we can
lift the level of the moment map from 0 to a finite value $\zeta $
and 3 whose moment map must remain at level 0. Henceforth we
introduce a new basis for the complete Lie Algebra
\(\mathbb{F}_{\Gamma }\) which is reorganized as follows:
\paragraph{The two central generators}
\begin{equation}
\mathfrak{z}_1=\left(
\begin{array}{cccccc}
 -i \sqrt{\frac{5}{6}} & 0 & 0 & 0 & 0 & 0 \\
 0 & \frac{i}{\sqrt{30}} & 0 & 0 & 0 & 0 \\
 0 & 0 & \frac{i}{\sqrt{30}} & 0 & 0 & 0 \\
 0 & 0 & 0 & \frac{i}{\sqrt{30}} & 0 & 0 \\
 0 & 0 & 0 & 0 & \frac{i}{\sqrt{30}} & 0 \\
 0 & 0 & 0 & 0 & 0 & \frac{i}{\sqrt{30}} \\
\end{array}
\right)\quad;\quad\mathfrak{z}_2=\left(
\begin{array}{cccccc}
 0 & 0 & 0 & 0 & 0 & 0 \\
 0 & -\frac{2 i}{\sqrt{5}} & 0 & 0 & 0 & 0 \\
 0 & 0 & \frac{i}{2 \sqrt{5}} & 0 & 0 & 0 \\
 0 & 0 & 0 & \frac{i}{2 \sqrt{5}} & 0 & 0 \\
 0 & 0 & 0 & 0 & \frac{i}{2 \sqrt{5}} & 0 \\
 0 & 0 & 0 & 0 & 0 & \frac{i}{2 \sqrt{5}} \\
\end{array}
\right)
\end{equation}

\paragraph{The three generators of $\su(2)$}

\begin{equation}
J_1=\left(
\begin{array}{cccccc}
 0 & 0 & 0 & 0 & 0 & 0 \\
 0 & 0 & 0 & 0 & 0 & 0 \\
 0 & 0 & \frac{i}{2} & 0 & 0 & 0 \\
 0 & 0 & 0 & \frac{i}{2} & 0 & 0 \\
 0 & 0 & 0 & 0 & -\frac{i}{2} & 0 \\
 0 & 0 & 0 & 0 & 0 & -\frac{i}{2} \\
\end{array}
\right) \, ;\, J_2=\left(
\begin{array}{cccccc}
 0 & 0 & 0 & 0 & 0 & 0 \\
 0 & 0 & 0 & 0 & 0 & 0 \\
 0 & 0 & 0 & 0 & \frac{1}{2} & 0 \\
 0 & 0 & 0 & 0 & 0 & \frac{1}{2} \\
 0 & 0 & -\frac{1}{2} & 0 & 0 & 0 \\
 0 & 0 & 0 & -\frac{1}{2} & 0 & 0 \\
\end{array}
\right)\, ;\, J_3=\left(
\begin{array}{cccccc}
 0 & 0 & 0 & 0 & 0 & 0 \\
 0 & 0 & 0 & 0 & 0 & 0 \\
 0 & 0 & 0 & 0 & \frac{i}{2} & 0 \\
 0 & 0 & 0 & 0 & 0 & \frac{i}{2} \\
 0 & 0 & \frac{i}{2} & 0 & 0 & 0 \\
 0 & 0 & 0 & \frac{i}{2} & 0 & 0 \\
\end{array}
\right)
\end{equation}
that satisfy the standard commutation relations:
\begin{equation}\label{jeioni}
    \left[J_i \, ,\, J_j\right] \, = \, \epsilon_{ijk} \, J_k
\end{equation}
According to the general discussion of section \ref{realecompdiscus} as a basis of $\mathbb{K}_\Gamma$ we
can just take the same 5 generators listed above, each multiplied by an $i$ factor.
\subsection{Fixing the zero level point and the moment maps} In the previously analyzed abelian cases the point in the
\(\text{Orbit}_{\mathcal{G}_{\Gamma }}\)(\(L_{\Gamma }\)) where the moment map vanishes and which therefore
corresponds to the orbifold, was just the locus \(L_{\Gamma }\) itself, defined as that one where, in the
natural basis of the Regular Representation the three matrices A,B,C are diagonal. There is nothing magic in
that point apart from the fact of being a convenient starting point to calculate the entire orbit. What has
an intrinsic geometrical significance in the $\text{Orbit}_{\mathcal{G}_{\Gamma }} (L_{\Gamma } )$ = $
\mathbb{V}_{|\Gamma |+2}$ is the point \(p_0\) where the moment map vanishes for all components. This is the
orbifold limit \(\frac{\mathbb{C}^3}{\Gamma }\)$\Leftrightarrow $\(p_{\Gamma }^0\) $\in $
\(\text{Orbit}_{\mathcal{G}_{\Gamma }}\)(\(L_{\Gamma }\))
\begin{equation}
\mathfrak{P}\left(p_{\Gamma }^0\right) = 0
\end{equation}
The present section is devoted to the determination of \(p_{\Gamma }^0\)=$\{$\(A_0\),\(B_0\),\(C_0\)$\}$.
\subsubsection{Construction of the moment maps} In order to construct the moment maps we need the matrices
$\{$A, B, C$\}$ in the orbit and their hermitian conjugate \(A^{\dagger }\), \(B^{\dagger }\), \(C^{\dagger}
\). Then the K\"ahler potential of the ambient space is defined as follows:
\begin{equation}
\mathcal{K}\text{   }= \text{Tr}\left( A^{\dagger } A\right)+\text{  }\text{Tr}\left( B^{\dagger }
B\right)+\text{Tr}\left( C^{\dagger } C\right)\label{cellerinoDih}
\end{equation}
while the moment maps are obtained from
\begin{equation}
\mathfrak{P}_I = -i \left(\text{Tr}\left(T_I\left[A,A^{\dagger
}\right]\right)+\text{Tr}\left(T_I\left[B,B^{\dagger }\right]\right)+\text{Tr}\left(T_I\left[C,C^{\dagger
}\right]\right)\right)\text{  };\text{ }(I=1,\text{...},5)\label{momentidih}
\end{equation}
where \(T_I\) are the generators of the compact subgroup.
\par
Calculating the moment maps at p $\in $ \(L_{\Gamma }\) we find that they are not zero, yet it suffices to
apply a transformation of exp[\(\mathbb{K}_{\Gamma }\)] as in equation (\ref{cosettondih3}) with the
following parameters:
\begin{equation}
\Upsilon _1=\left(\sqrt[6]{\frac{ 1}{4}}\right),\quad \Upsilon _2=\text{  }\sqrt[6]{2},\quad \Upsilon _3-
\frac{1}{\left(\sqrt[6]{ \frac{ 1}{4}}\right)* \sqrt[6]{2}},\quad \varrho = 0,\quad \phi = 0
\end{equation}
and the moment map vanishrs. In this way we have found the point \(p_{\Gamma }^0\) $\in $
\(\text{Orbit}_{\mathcal{G}_{\Gamma }}\)(\(L_{\Gamma }\)) mentioned above which exactly corresponds to the
orbifold limit \(\frac{\mathbb{C}^3}{\Gamma }\). The explicit form of the triple \(p_{\Gamma
}^0\)=$\{$\(A^0\),\(B^0\),\(C^0\)$\}$ is displayed below (where $\xi=\ft 76 \pi$):
\begin{eqnarray}
A^0&=&\left(
\begin{array}{cccccc}
 0 & 0 & \sqrt{\frac{2}{15}} z_2 & 0 & \sqrt{\frac{2}{15}} z_1 & 0 \\
 0 & 0 & -\sqrt{\frac{2}{15}} z_2 & 0 & \sqrt{\frac{2}{15}} z_1 & 0 \\
 0 & 0 & 0 & 0 & 0 & \frac{2 z_2}{\sqrt{15}} \\
 \sqrt{\frac{2}{15}} z_1 & \sqrt{\frac{2}{15}} z_1 & 0 & 0 & 0 & 0 \\
 0 & 0 & 0 & \frac{2 z_1}{\sqrt{15}} & 0 & 0 \\
 \sqrt{\frac{2}{15}} z_2 & -\sqrt{\frac{2}{15}} z_2 & 0 & 0 & 0 & 0 \\
\end{array}
\right)\nonumber\\ B^0&=&\left(
\begin{array}{cccccc}
 0 & 0 & 0 & -\frac{e^{i \xi } z_2}{\sqrt{30}} & 0 & -\frac{e^{i \xi } z_1}{\sqrt{30}} \\
 0 & 0 & 0 & -\frac{e^{i \xi } z_2}{\sqrt{30}} & 0 & \frac{e^{i \xi } z_1}{\sqrt{30}} \\
 -\frac{e^{i \xi } z_1}{\sqrt{30}} & \frac{e^{i \xi } z_1}{\sqrt{30}} & 0 & 0 & 0 & 0 \\
 0 & 0 & 0 & 0 & -\frac{e^{i \xi } z_2}{\sqrt{15}} & 0 \\
 -\frac{e^{i \xi } z_2}{\sqrt{30}} & -\frac{e^{i \xi } z_2}{\sqrt{30}} & 0 & 0 & 0 & 0 \\
 0 & 0 & -\frac{e^{i \xi } z_1}{\sqrt{15}} & 0 & 0 & 0 \\
\end{array}
\right)\nonumber\\
C^0&=&\left(
\begin{array}{cccccc}
 0 & \frac{z_3}{\sqrt{6}} & 0 & 0 & 0 & 0 \\
 \frac{z_3}{\sqrt{6}} & 0 & 0 & 0 & 0 & 0 \\
 0 & 0 & -\frac{z_3}{\sqrt{6}} & 0 & 0 & 0 \\
 0 & 0 & 0 & \frac{z_3}{\sqrt{6}} & 0 & 0 \\
 0 & 0 & 0 & 0 & \frac{z_3}{\sqrt{6}} & 0 \\
 0 & 0 & 0 & 0 & 0 & -\frac{z_3}{\sqrt{6}} \\
\end{array}
\right)\label{p0matrozze}
\end{eqnarray}
The above matrices depend on new triple of complex parameters \(z_i\) that will eventually be interpreted as
the complex coordinates of the resolved variety just as they are the coordinates of the locus {
}\(\frac{\mathbb{C}^3}{\Gamma }\) when the moment map is not lifted from its zero value. Note that we have
named $A^0,B^0,C^0$ the matrices at the point $p_0$ in order to distinguish them from the matrices
$A_0,B_0,C_0$ that correspond to the locus $L_\Gamma$.
\subsubsection{Construction of the moment maps starting from the right zero-point}
Next we reconsider the orbit starting from the triple of matrices that have zero-moment map:
\begin{equation}
p_0 =\left(
\begin{array}{c}
 A_0 \\
\begin{array}{c}
 B_0 \\
 C_0 \\
\end{array}
 \\
\end{array}
\right)\quad \quad ;\quad \quad p =\left(
\begin{array}{c}
 A \\
\begin{array}{c}
 B \\
 C \\
\end{array}
 \\
\end{array}
\right) = \left(
\begin{array}{c}
  \mathcal{V}^{-1} A_0\mathcal{V} \\
\begin{array}{c}
  \mathcal{V}^{-1}B_0\mathcal{V} \\
 \mathcal{V}^{-1}C_0\mathcal{V} \\
\end{array}
 \\
\end{array}
\right)
\end{equation}
and we also utilize a polar parametrization of the coordinates
\begin{equation}
z_i=\Delta _i \,  \text{Exp}\left[i \,  \theta _i\right]
\end{equation}
\par
Inserting the newly calculated matrices in the formula (\ref{cellerinoDih}) for the K\"ahler potential  we
obtain a complicated K\"ahler potential that for p = \(p_0\) reduces to:
\begin{equation}
\mathcal{K}_0 =\Delta _1^2+\Delta _2^2+\Delta _3^2 = \sum _{i=1}^3 \left|z_i|^2\right.
\end{equation}
This shows that at zero level the space is indeed every where flat except for the singular fixed point.
\par
Next we construct the explicit form of the moment maps for a generic point of the orbit and we obtain five
algebraic functions $\mathfrak{P}_I(\Upsilon,X,Y)$ of the five parameters $\Upsilon_{1,2,3}$,$X,Y$ that
depend on the three moduli $\Delta_{1,2,3}$ and the three phases $\theta_{1,2,3}$ of the complex coordinates.
Equating the five moment maps to $\{\zeta_1,\zeta_2,0,0,0\}$ we have an algebraic system that determines the
parameters $\Upsilon,X,Y$ in terms of the levels $\zeta_{1,2}$ and the coordinates $z_i$.
\paragraph{Higher degree algebraic system.} The explicit expressions of the moment maps are rather formidable and it is
difficult to display them on paper since they are very large. Furthermore the degree of the equations is
certainly higher than the fourth and there is no hope to solve them by radicals. Yet we know that there is
just one and only one solution that has the correct reality property, namely $\Upsilon_{1,2,3}$,$X,Y$ are all
real and $\Upsilon_{1,2,3}$ are also positive. A convenient way to verify such property of the system is
provided by considering small deformations, namely level parameters $\zeta_{1,2}$ infinitesimally close to
zero.
\paragraph{First order solution of the algebraic equations}
We are supposed to solve the moment map equations. We recall that the last three moment maps have to be zero,
while the first two have to be lifted to an arbitrary level $\zeta $. We consider the solution by power
series in the neighborhood of the identity element, namely we set:
\begin{equation}
\Upsilon _1\to 1+\epsilon \omega _1,\quad \Upsilon _2\to 1+\epsilon  \omega _2,\quad \Upsilon _3\to
1+\epsilon \omega _3,\quad X\to x \epsilon ,\quad Y\to y \epsilon
\end{equation}
and
\begin{equation}
\zeta _1\to  \epsilon  c \quad , \quad \zeta _2\to  \epsilon  d
\end{equation}
where $\epsilon $ is an infinitesimal parameter. At first order in $\epsilon $ the moment maps are:
\begin{eqnarray}
\mathfrak{P}_1 &=& \epsilon \left(-\frac{4}{3} i \left(2 \left(x \text{Cos}\left[\theta _1-\theta _2\right]-y
\text{Sin}\left[\theta _1-\theta _2\right]\right) \Delta _1 \Delta _2\right.\right.\nonumber\\
&& \left.\left.+\Delta _1^2 \left(\omega _1+\omega _2+\omega _3\right)+\left(\Delta _2^2+2 \Delta _3^2\right)
\left(\omega _1+\omega _2+\omega
_3\right)\right)\right)+\mathcal{O}\left(\epsilon ^2\right)\nonumber\\
\mathfrak{P}_2 &=& \epsilon  \left(-\frac{2}{3} i \left(8 \left(x \text{Cos}\left[\theta _1-\theta
_2\right]-y \text{Sin}\left[\theta _1-\theta _2\right]\right)
\Delta _1 \Delta _2+\right.\right.\nonumber\\
&&\left.\left.8 \Delta _3^2 \left(\omega _1+\omega _2+\omega _3\right)+\Delta _1^2 \left(4 \omega _1+13
\left(\omega _2+\omega _3\right)\right)+\Delta _2^2 \left(4 \omega _1+13 \left(\omega _2+\omega
_3\right)\right)\right)\right)+\mathcal{O}\left(\epsilon ^2\right)\nonumber\\
\mathfrak{P}_3 &=& \epsilon  \left(-i \left(\Delta _1^2+\Delta _2^2\right) \left(\omega _2-\omega
_3\right)\right)+\mathcal{O}\left(\epsilon ^2\right)\nonumber\\
\mathfrak{P}_4 &=& \epsilon  \left(-\frac{4}{3} i \left(y \Delta _1^2+y \left(\Delta _2^2+2 \Delta
_3^2\right) +\Delta _1 \Delta _2 \left(y \text{Cos}\left[\theta _1-\theta _2\right]-x \text{Sin}\left[\theta
_1-\theta
_2\right] \right.\right.\right.\nonumber\\
&&\left.\left.\left. -\text{Sin}\left[\theta _1-\theta _2\right] \omega
_1-\text{Sin}\left[\theta_1-\theta_2\right] \omega_2-\text{Sin}\left[\theta _1-\theta _2\right]
\omega_3\right)\right)\right)
+\mathcal{O}\left(\epsilon ^2\right)\nonumber\\
\mathfrak{P}_5 &=& \epsilon  \left(-\frac{4}{3} i \left(x \Delta _1^2+x \left(\Delta _2^2+2 \Delta
_3^2\right) -\Delta _1 \Delta _2 \left(x \text{Cos}\left[\theta _1-\theta _2\right]+y \text{Sin}\left[\theta
_1-\theta
_2\right]-\text{Cos}\left[\theta _1-\theta _2\right] \omega _1\right.\right.\right.\nonumber\\
&&\left.\left.\left.-\text{Cos}\left[\theta _1-\theta _2\right] \omega _2-\text{Cos}\left[\theta _1-\theta
_2\right] \omega
_3\right)\right)\right)+\mathcal{O}\left(\epsilon ^2\right)\nonumber\\
\end{eqnarray}
Equating the 5-vector of these moment maps to the 5-vector $\{$c,d,0,0,0$\}$ we obtain the solution:
\begin{eqnarray}
\omega _1&\to& \frac{1}{12} i \left(\frac{4 c-2 d}{\Delta _1^2+\Delta _2^2}+\frac{9 c \left(\Delta
_1^4+\left(\Delta _2^2+2 \Delta _3^2\right){}^2+\Delta _1^2 \left(\Delta _2^2+4 \Delta
_3^2\right)\right)}{\Delta _1^6-2 \text{Cos}\left[3 \left(\theta _1-\theta _2\right)\right] \Delta _1^3
\Delta _2^3+6 \Delta _1^4 \Delta _3^2+6 \Delta _1^2 \Delta _3^2 \left(\Delta _2^2+2 \Delta _3^2\right)
+\left(\Delta _2^2+2 \Delta _3^2\right){}^3}\right)\nonumber\nonumber\\
\omega _2&\to& -\frac{i (2 c-d)}{12 \left(\Delta _1^2+\Delta _2^2\right)}\nonumber\\
\omega _3&\to& -\frac{i (2 c-d)}{12 \left(\Delta _1^2+\Delta _2^2\right)}\nonumber\\
x&\to& -\frac{3 i c \Delta _1 \Delta _2 \left(\text{Cos}\left[\theta _1-\theta _2\right] \Delta
_1^2+\text{Cos}\left[2 \left(\theta _1-\theta _2\right)\right] \Delta _1 \Delta _2+\text{Cos}\left[\theta
_1-\theta _2\right] \left(\Delta _2^2+2 \Delta _3^2\right)\right)}{4 \left(\Delta _1^6-2 \text{Cos}\left[3
\left(\theta _1-\theta _2\right)\right] \Delta _1^3 \Delta _2^3+6 \Delta _1^4 \Delta _3^2+6 \Delta _1^2
\Delta _3^2 \left(\Delta _2^2+2 \Delta _3^2\right)+\left(\Delta
_2^2+2 \Delta _3^2\right){}^3\right)}\nonumber\\
y&\to& \left(3 i c \text{Sin}\left[\theta _1-\theta _2\right] \Delta _1 \Delta _2\right) \times \left(4
\left(\Delta _1^4+2 \text{Cos}\left[\theta _1-\theta _2\right] \Delta _1^3 \Delta _2+2 \text{Cos}\left[\theta
_1-\theta _2\right] \Delta _1 \Delta _2 \left(\Delta _2^2+2 \Delta
_3^2\right)\right.\right.\nonumber\\
&&\left.\left.+\left(\Delta _2^2+2 \Delta _3^2\right){}^2+\Delta _1^2 \left(\left(1+2 \text{Cos}\left[2
\left(\theta _1-\theta _2\right)\right]\right) \Delta _2^2+4 \Delta _3^2\right)\right)\right)^{-1}
\end{eqnarray}
This clearly shows that the 5-fields \(\Upsilon _1,\Upsilon _2,\Upsilon _3,X,Y\) are all activated and
equally necessary in the solution of the moment map equation as soon as we move out of level zero.
\subsection{The tautological bundles and their Chern classes}
Assuming that we have solved the algebraic equations for the fields \(\Upsilon _1,\Upsilon _2,\Upsilon
_3,X,Y\) (at first order in the level parameters we have done it, and for many considerations this might turn
out to be sufficient) we can now utilize the present example in order to illustrate the discussion of the
main text in section \ref{ciulifischio}. With reference to the matrix $\mathcal{H}$ in
eq.\,(\ref{tautobundmetro}) we have determined it. Indeed in our case of \(\text{Dih}_3\) there are two non
trivial irreducible representations and correspondingly two tautological bundles, respectively of rank 1 and
of rank 2. The matrix $\mathcal{H}$ has the following appearance:
\begin{equation}
\mathcal{H}= \left(
\begin{array}{ccc}
 \Upsilon _1 & 0 & 0 \\
 0 & \Upsilon _2 & (X-i Y) \sqrt{\Upsilon _2 \Upsilon _3} \\
 0 & (X+i Y) \sqrt{\Upsilon _2 \Upsilon _3} & \Upsilon _3 \\
\end{array}
\right)
\end{equation}
which corresponds to
\begin{equation}
\mathfrak{H}_1=\Upsilon _1\quad \quad ;\quad \quad \mathfrak{H}_{2 } \,=\,
\left(
\begin{array}{cc}
 \Upsilon _2 & (X-i Y) \sqrt{\Upsilon _2 \Upsilon _3} \\
 (X+i Y) \sqrt{\Upsilon _2 \Upsilon _3} & \Upsilon _3 \\
\end{array}
\right)
\end{equation}
and we get
\begin{equation}
\text{Det}\left[\mathfrak{H}_{2 }\right] =\Upsilon _2 \Upsilon
_3\left(1 -X^2-Y^2\right)
\end{equation}
This allows us to calculate the first Chern classes explicitly according
to eq.\,(\ref{bambolone}) and the
K\"ahler potential according to eq.\,(\ref{criceto1})
\par
The provided illustration of the present  nonabelian case, which is
the smallest possible one, was finalized to show that everything in
the Kronheimer-like construction is fully algorithmic and uniquely
determined. The bottleneck however is localized in the system of
algebraic equations for the entries of the matrix $\mathcal{H}$
that, also in the simplest cases, are typically quite formidable and
of higher degree. Yet it appears and it is worth further
investigation that the relevant topological information is fully
encapsulated in the first order approximation of small level
parameters $\zeta$. We shall go back to this issue in future
publications.

\end{document}